\documentclass[reprint,showpacs]{revtex4-1}

\usepackage{amssymb,amsmath} 
\usepackage{graphicx}

\usepackage{appendix}
\usepackage[small,bf]{caption}

\usepackage{color}
\usepackage[usenames,dvipsnames,svgnames,table]{xcolor}
\usepackage[T1]{fontenc}
\begin{document}

\author{C. Patrick Royall}
\email{paddy.royall@bristol.ac.uk}
\affiliation{HH Wills Physics Laboratory, Tyndall Avenue, Bristol, BS8 1TL, UK}
\affiliation{School of Chemistry, University of Bristol, Cantock Close, Bristol, BS8 1TS, UK}
\affiliation{Centre for Nanoscience and Quantum Information, Tyndall Avenue, Bristol, BS8 1FD, UK}
\author{Stephen R. Williams}
\email{stephen.williams@anu.edu.au}
\affiliation{Research School of Chemistry, The Australian National University, Canberra, ACT 0200, Australia}

\title{The role of local structure in dynamical arrest}

\begin{abstract}
Amorphous solids, or glasses, are distinguished from crystalline solids by their lack of long-range structural order. At the level of two-body structural correlations, glassformers show no qualitative change upon vitrifying from a supercooled liquid. Nonetheless the dynamical properties of a glass are so much slower that it appears to take on the properties of a solid. While many theories of the glass transition focus on dynamical quantities, a solid's resistance to flow is often viewed as a consequence of its structure. Here we address the viewpoint that this remains the case for a glass. Recent developments using higher-order measures show a clear emergence of structure upon dynamical arrest in a variety of glass formers and offer the tantalising hope of a structural mechanism for arrest. However a rigorous fundamental identification of such a causal link between structure and arrest remains elusive. We undertake a critical survey of this work in experiments, computer simulation and theory and discuss what might strengthen the link between structure and dynamical arrest. We move on to highlight the relationship between crystallisation and glass-forming ability made possible by this deeper understanding of the structure of the liquid state, and emphasize the potential to design materials with optimal glassforming and crystallisation ability, for applications such as phase-change memory. We then consider aspects of the phenomenology of glassy systems where structural measures have yet to make a large impact, such as polyamorphism (the existence of multiple liquid states), aging (the time-evolution of non-equilibrium materials below their glass transition) and the response of glassy materials to external fields such as shear. 
\end{abstract}

%\begin{keyword}
%Geometric Frustration, Locally Favoured Structures, Model glassforming systems
%\end{keyword}

%\end{frontmatter}

\maketitle

\tableofcontents

\section{Introduction and motivation}

Glass has been an everyday material for 4000 years. A trip to any respectable glassblower provides a glimpse of how much information has been gleaned : trial and error, though not efficient, has over time provided a clear set of guidelines as to how to manipulate glass into all manner of shapes, colours, etc. However beyond (``common'') silica glass,  materials in glassy, or amorphous, states feature an ever increasing role in daily life. Ceramic materials are typically glasses of oxygen combined with one or more other elements. Polymeric glasses are better known as plastics ~\cite{berthier2011}. Colloidal suspensions provide an important class of glassforming materials, whose properties as model systems make them particularly appealing to experimentalists ~\cite{ivlev,cipeletti2005,hunter2012,gasser2009}.

Key emergent amorphous materials include metallic and chalcogenide glassformers. These have novel properties which make them technologically important. Metallic glass, for example, does not possess grain boundaries unlike normal polycrystalline metal. Since the grain size strongly affects the mechanical properties of a polycrystalline material, compared to conventional materials, metallic glass (which may be viewed as possessing an exceedingly small grain size) has superior mechanical properties, such as increased hardness and tensile strength ~\cite{greer2009,cheng2011,faupel2003,wondraczek2011}. A second example is chalcogenide glassformers. Here the optical and electronic contrast between amorphous and crystalline forms is exploited. Optical contrast underpins re-writeable optical media, while electronic contrast is the key ingredient in phase change memory, a potential replacement technology for current hard disks which is on the brink of commercialisation. Such rewriteable memory requires fast switching between amorphous and crystalline forms, which necessitates control of the delicate balance between crystallisation and glass formation~\cite{wuttig2007,lencer2008}.

Metallic and chalcogenide glassformers emphasize the notion of a glass as a ``failed crystal'', in that the material is cooled from the liquid phase without freezing, and so rests on the balance between vitrification and crystallisation. There are thus two sides to the challenge of understanding glasses, particularly with a view to developing new materials. On the one hand, we seek to understand how solidification can be achieved without freezing. Secondly, and in order to achieve conditions appropriate for the first criterion, we need to understand how to suppress freezing. Concerning the first, a number of excellent reviews ~\cite{berthier2011,cavagna2009,dyre2006,stillinger2013,debenedetti} and shorter perspectives ~\cite{ediger2012,biroli2013,debenedetti2001} of the glass transition from a theoretical viewpoint have appeared recently. More specific reviews focus on certain theoretical aspects such as the energy landscape ~\cite{goldstein1969,sciortino2005,heuer2008}, dynamic heterogeniety  ~\cite{ediger2000,berthier}, Mode-Coupling Theory  ~\cite{goetze,charbonneau2005}, random first-order theory ~\cite{lubchenko2007}, replica theory ~\cite{parisi2010}, dynamic facilitation ~\cite{chandler2010} and soft glassy rheology ~\cite{sollich2006} and the insights gained from phenomena such as phase behaviour, interaction potentials and glassy crystals ~\cite{angell2008}.

Regarding crystallisation, the situation is much less clear. Crystallisation is of course much studied ~\cite{debenedetti,sear2012,kelton,auer2004}, however the interplay of crystallisation and dynamical arrest is only beginning to be tackled. Here we shall discuss the current understanding of crystallisation in key glassformers. However we emphasize that the phase diagrams for the two most popular model Lennard-Jones glassformers, the Wahnstr\"{o}m ~\cite{wahnstrom1991} and Kob-Andersen ~\cite{kob1995a} models have not yet been determined. It is only recently that crystallisation in the the Wahnstr\"{o}m model has even been observed ~\cite{pedersen2010}, while for the Kob-Andersen model we have only predictions of conditions under which it is expected to crystallise ~\cite{toxvaerd2009}. This underlines how much further we need to go in order to understand the interplay between vitrification and crystallisation.

Here we focus on the role of structure assumed by the constituent particles in materials undergoing a glass transition. We shall begin by discussing the general phenomenology of dynamical arrest before briefly outlining those theories which relate to structure. We then proceed to describe in some detail means by which structural measures can be extracted in computer simulation and experiment before reviewing what contribution these have made and how our understanding of the glass transition has been changed by such measurements. In particular we enquire as to the progress towards a structural mechanism for dynamical arrest. Having discussed the role of structure in the fundamental aspects of dynamical arrest, we shall move on to consider some examples of more applied materials. At this point we note that, while the role of structure in glassforming liquids in more fundamental studies is controversial ~\cite{charbonneau2013pre}, studies using high-order structural measures of more applied systems such as metallic ~\cite{inoue2011} and chalcogenide ~\cite{wuttig2007} glassformers are quite popular.

We close this brief motivation by emphasising again the link between vitrification and crystallisation. Glasses are non-equilibrium states. Unlike supercooled liquids their properties are not stationary as a function of time and given enough time, most would crystallise. ``Enough time'' can here easily extend to durations which are so long that they are hopelessly inaccessible. It is an open question as to any link between liquid structure and a material's tendency to crystallise, but one which is just beginning to be addressed.

\textit{Aims of this review ---} The large range of materials that undergo dynamical arrest, coupled with the desire for some ``universal mechanism'' presents a considerable challenge. We have chosen to focus on our own field --- colloids and simulations of model systems. However, the similarity in behaviour with metallic glassformers is so clear that some work on metallic glass has also been reviewed. We also consider chalcogenide glassforming systems, because their applications rest on the competition between crystallisation and vitrification. Where possible, we have referenced relevant review papers, but in any case humbly ask for patience on the part of readers from these and other fields, including our own, regarding those papers we have missed, or where our opinion seems at odds with theirs. We nevertheless hope to convey the similar behaviour between these very different systems, and in particular that it might be possible to improve the transfer of ideas between these fields.

\begin{figure}
\includegraphics[width=80mm]{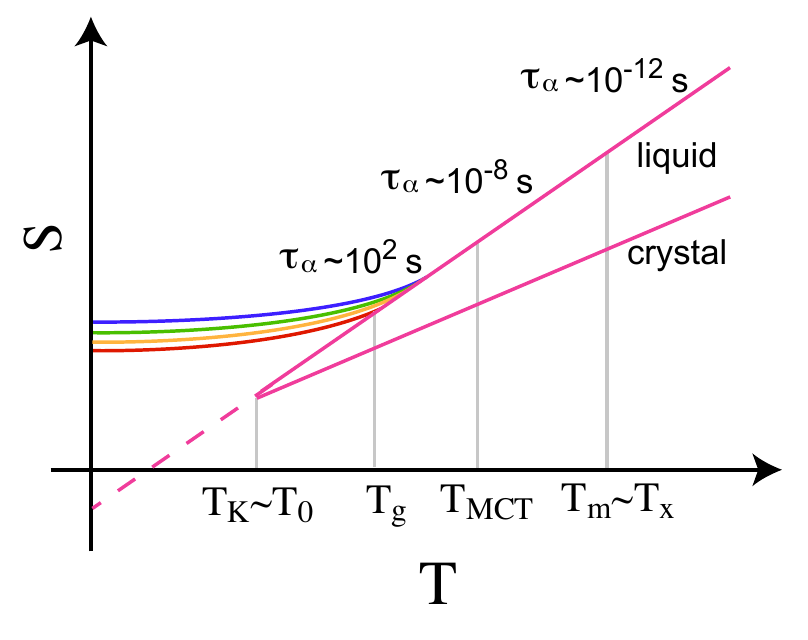}
\caption{\label{figCavagna} Entropy $S$ as a function of temperature. Typically, entropy of liquids falls faster than that of crystals as as function of temperature. This suggests that, at some low temperature --- the Kauzmann Temperature $T_{K}$ --- the liquid entropy would fall below that of the crystal. $T_K$ is similar to $T_0$, the temperature at which the structural relaxation time $\tau_\alpha$ is predicted to diverge by the Vogel-Fulcher-Tamman law Eq. ~\ref{eqVFT}. $T_{g}$ is the operational glass transition temperature where the structural relaxation time reaches 100 s. $T_{\mathrm{MCT}}$ is the mode-coupling transition (see section ~\ref{sectionMCT}), $T_m$ is the melting point and $T_{x}$ denotes a crossover temperature below which relaxation occurs through local fluctuations.}
\end{figure}

\textit{Is there a glass transition? ---} The nature of the glass transition, if any, is not understood. Crystallisation, by comparison is a first-order (discontinuous) phase transition between two \emph{equilibrium} thermodynamic phases - liquid and crystalline solid \cite{berthier2011,cavagna2009}. On timescales accessible to experiment and computer simulation, the glass transition takes the form of a \emph{continuous} increase in viscosity, or structural relaxation time. In normal liquids, the relaxation time occurs on the picosecond timescale. When it rises to 100 s, liquids are termed glasses. This is a purely operational definition (it is easier to time 100 s than a divergent timescale!) which can obscure the fact that at the temperature corresponding to a relaxation time of 100 s, $T_{g}$, precisely nothing happens in a thermodynamic sense. Put another way, $T_{g}$ could as well correspond to a relaxation time of 10 s or 1000 s. Thus whether we have a supercooled liquid or a glass depends, not necessarily on the material or state point, but \emph{upon how long we wait}.

That the structural relaxation time grows to unmanageable timescales has significant consequences. The relaxation time is an indicative measure of how long it takes a liquid to reach equilibrium. At least 10-100 relaxation times are required before one can say that some liquid is ``in equilibrium''. As the relaxation time reaches 100 s, a material will appear solid. It will no longer flow on short timescales. It is this gradual \emph{emergence} of solidity that characterises the glass transition which is so hard to understand, but it obscures a further fundamental question. That such glassy liquids cannot be equilibrated below $T_{g}$ means that it is very hard to answer questions about whether there is a true thermodynamic transition at some lower temperature $T_{0}<T_{g}$.

There are two main conjectures. Firstly, it is suggested that there is a thermodynamic transition at or around the Kauzmann temperature $T_K \sim T_0$ where extrapolation suggests that the liquid entropy would fall below that of the crystal, leading to a singularity $T_{K}<T_{g}$ (Fig. \ref{figCavagna}). Alternatively, there may be no transition until absolute zero (where in the classical approach we shall pursue, all motion ceases) ~\cite{berthier2011,cavagna2009,chandler2010}. For the present, despite some recent developments, these remain conjectures. We also note the mode-coupling transition at $T_\mathrm{MCT}$ (see section ~\ref{sectionMCT}) where a transition to a glass whose relaxation time diverges occurs \emph{in mean field}. Echoes of this mean field transition can be observed in experiment, and it also marks a crossover between single-particle and correlated motion.

In this discussion, we have identified \emph{three} relevant temperatures : $T_g$ the experimental transition, $T_0$ a possible thermodynamic transition and $T_\mathrm{MCT}$ the mode-coupling temperature. We shall encounter more temperatures as we continue, but an important point is that in considering dynamical arrest, one should think of glass transition\emph{s}, or that at least we must specify what we mean by ``glass transition''. Before proceeding, we briefly review the phenomenology of dynamical arrest. Glasses emerge from the liquid state and a number of theories of the glass transition may be viewed as extensions of liquid state theory. Most of our discussion will focus on glasses, but we shall also consider their soft matter relatives, gels.

\section{Phenomenology}
\label{sectionPhenomenology}

The glass transition can be approached from the low-temperature (solid) or high-temperature (liquid) side. Whilst an approach from the solid side is often carried out \cite{heuer2008,williams2010}, given that glasses are non-equilibrium states, it can be attractive to approach the transition from the supercooled liquid side. We therefore begin our description of the phenomenology of dynamical arrest starting from the liquid state. Our discussion reflects the fact that the glass transition  is often treated as a predominantly dynamic phenomenon (unlike the other mode of solidification, crystallisation). Thus we shall focus on dynamical quantities, leaving the role of structure to be introduced in section ~\ref{sectionTheories}.
We shall concentrate on simple liquids, by which we mean liquids comprised of spheres with no preferential orientation. The measures we discuss may also be applied to more realistic models of practical materials where directionality in bonding is important, which we consider in section ~\ref{sectionApplications}.

\subsection{Critique of experimental and simulation techniques}
\label{sectionCritique}

We shall be dealing with three sources of data : experiments on atomic and molecular systems, experiments on colloidal systems, and computer simulations. For the purposes of elucidating structure and its role in dynamical arrest, each has its own strengths and weakness, which we now outline.

\textit{Experiments on atomic and molecular systems --- } Of course this technique has the longest history in studies of the glass transition, and has provided many of the benchmarks by which the phenomenology is measured. Not least, the relaxation time of 100 s being identified as the experimental glass transition. In this way, molecular experiments define the nature of the problem, \emph{i.e.} the 14 decades in relaxation time that lie between a glass and its ``normal'' liquid. From a point of view of local structure, molecular systems are exceedingly challenging to work with. Only very recently has it become possible to resolve individual atoms ~\cite{ashtekar2011,huang2013} and then only on the top surface whose dynamics are generally not representative of the bulk material. This can be circumvented by imaging instead a 2d material ~\cite{huang2013}. Indirect measurements  of structure are possible and discussed in section ~\ref{sectionStructure}, as are measurements of dynamic lengthscales, which are important to distinguish between competing theoretical approaches ~\cite{ediger2000,donth1982,cicerone1995,berthier2005,tatsumi2012}. However, the challenges associated with determining the coordinates upon which many structural measures are predicated limit the use of atomic and molecular systems to tackle the role of structure in the glass transition.

\textit{Computer simulation --- } Molecular simulation resolves the lack of knowledge of the atomic coordinates. By their very nature, the coordinate of every constituent particle in the system is known at all times ~\cite{allen,frenkel}. However, simulation is limited by computer power, and in the case of the glass transititon, this limitation is severe. Only the first four or five of the 14 decades of dynamic slowing are accessible to simulation. This leaves the more interesting behaviour at deeper supercooling to the imagination, although (depending on the theory or approach one believes), some recent developments show promise to access states more deeply supercooled states than accessible to brute force simulation ~\cite{cammarota2012,speck2012,singh2013,speck2014}. Simulations use a variety of dynamic schemes. Molecular dynamics which integrates Newton's equations of motion is appropriate to atomic and molecular systems, Brownian dynamics follow the Langevin equation and is appropriate for colloids while dynamic Monte Carlo approaches Brownian dynamics in the limit of small step length. In supercooled liquids, the frequency of collisions is such that momentum is lost on timescales far shorter than the structural relaxation time. Under these conditions, the three methods give very similar dynamical behaviour ~\cite{berthier2007,sanz2010,lopez2012}.

The above comments hold for classical simulations of particles with simple, spherically symmetric interactions, such as the models mentioned below in section ~\ref{sectionCommonModelSystems}. For example many studies of oxide or chalcogenide glassforming systems are carried out with more sophisticated \emph{ab initio} techniques which (under a range of approximations) solve the electronic wavefunctions of the system using density functional theory ~\cite{car1985,kuhne2012}. As elegant and accurate as these methods are, the increased computation cost carries a high penalty. At best, around 100 ps for system sizes of order 100 atoms can be calculated, which severely limits the ability to tackle the time- and length-scales inherent in the glass transition ~\cite{cheng2011}.

\textit{Experiments on colloids ---} Colloidal dispersions follow the same rules of equilibrium statistical mechanics as do atoms and molecules, and thus these particles form crystals, liquids and glasses in much the same way ~\cite{pusey}. Colloids have three properties which make them useful to the experimentalist ~\cite{ivlev,cipeletti2005,hunter2012,gasser2009}. Firstly, their relatively large size means they are readily visualised in an optical microscope. Indeed with confocal microscopy, the structure of colloidal (supercooled) liquids can be accessed at the single-particle level, yielding data at the level otherwise available only to simulation ~\cite{vanblaaderen1995,crocker1995,royall2003}. Secondly, their large size leads to sluggish dynamics, so time-resolved data can be taken ~\cite{weeks2001,kegel2001} (ironically this very strength means that it is hard to model the kind of timescales associated with the molecular glass transition with colloids, see section ~\ref{sectionGTSM}). Thirdly, the interactions between the colloidal particles can be tuned, and only classical interactions need be considered. This means that they form sought-after model systems against which theories based on, for example hard spheres can be readily compared.

Colloidal dynamics are diffusive, and their motion is approximately described by Langevin  dynamics. The approximation here stems from hydrodynamic interactions between the colloids which are mediated by the solvent ~\cite{dhont}. However, at the packing fractions appropriate for the glass transition, the effects of such hydrodynamic interactions appear to be weak enough to be neglected, apart from a simple time scaling correction. One hydrodynamic effect which does come into play at high packing fraction is lubrication which would further slow down the dynamics. However, lubrication becomes important when particles are around $10^{-3}$ diameters apart. This corresponds to a volume fraction within $\sim 3\times10^{-3}$ of random close packing. This is dynamically inaccessible regardless of lubrication, thus this effect does not play a major role in vitrification. Given that systems obeying Langevin dynamics give similar behaviour to molecular dynamics under supercooled conditions ~\cite{berthier2007,lopez2012}, colloids form a good model system for atoms and molecules.

Although the interactions in colloidal model systems can be tuned, it is important to emphasise that precise control over their interactions is \emph{very limited}. More work has been carried out on colloidal ``hard'' spheres than any other system. Ironically, hard spheres do not exist in nature --- there is always \emph{some} softness ~\cite{royall2013myth}. How much softness and how important this is depends on the state point. However, the softness changes the value one should take for the particle diameter from the true value, and thus the \emph{effective} colloid volume fraction $\phi$ changes. A glance at Fig. ~\ref{figAngell} indicates that, for the glass transition, accurate knowledge of $\phi$ is absolutely essential. Worse, regardless of softness, the relative precision with which $\phi$ can be measured is around 6\% ~\cite{poon2012}. Again, Fig. ~\ref{figAngell} indicates that, as far as the glass transition is concerned, such uncertainty is quite sufficient to yield data quantitatively meaningless. We thus argue that any data from colloidal experiments that pertains to phenomena whose behaviour is highly dependent on state point should be supported by simulation. Any data not thus supported should be treated with extreme caution.

\subsection{Common model systems}
\label{sectionCommonModelSystems}

While acknowledging that glasses are formed in a great many systems, we consider three kinds of popular model system. Atomic and molecular systems are frequently modelled with the Lennard-Jones interaction. The one-component Lennard-Jones model readily crystallises (\emph{i.e.} is an extremely poor glass-former), therefore binary mixtures are usually employed. Two such mixtures which see frequent use are the Wahnstr\"{o}m ~\cite{wahnstrom1991} and Kob-Andersen ~\cite{kob1995a} models. In both, the two species of Lennard-Jones particles interact with a pair-wise potential,

\begin{equation}
u_{\mathrm{LJ}}(r_{ij}) = 4 \varepsilon_{\alpha \beta}\left[\left(\frac{\sigma_{\alpha\beta}}{r_{ij}}\right)^{12}-\left(\frac{\sigma_{\alpha\beta}}{r_{ij}}\right)^6\right]
\label{eqLJ}
\end{equation}

\noindent where $\alpha$ and $\beta$ denote the atom types $A$ and $B$, and $r_{ij}$ is the separation. The Wahnstr\"{o}m mixture is equimolar (50:50) and the energy, length and mass values are $\varepsilon_{AA}=\varepsilon_{AB}=\varepsilon_{BB}$, $\sigma_{BB}/\sigma_{AA}=5/6$, $\sigma_{AB}/\sigma_{AA}=\frac{11}{12}$ and $m_A=2m_B$ respectively. The Kob-Andersen binary mixture is composed of 80\% large (A) and 20\% small (B) particles possessing the same mass $m$~\cite{kob1995a}. The nonadditive Lennard-Jones interactions between each species, and the cross interaction, are given by $\sigma_\text{AA}=\sigma$, $\sigma_\text{AB}=0.8\sigma$, $\sigma_\text{BB}=0.88\sigma$, $\varepsilon_\text{AA}=\varepsilon$, $\varepsilon_\text{AB}=1.5\varepsilon$, and $\varepsilon_\text{BB}=0.5\varepsilon$. The Kob-Andersen system is loosely based around the metallic glassformer Ni$_{80}$P$_{20}$ but using Lennard-Jones interactions ~\cite{grest1987,ernst1991} with the non-additivity where the cross-attraction is stronger than the attraction between each species ~\cite{weber1985}. We note here that ``soft-sphere'' or inverse power law systems are also frequently employed. This refers to taking the first term in Eq. \ref{eqLJ} only, sometimes with an exponent other than 12.

Silica glass is frequently modelled with ``BKS'' silica, introduced by van Beest, Kramer and van Santen~\cite{vanbeest1991}. This is a parameterisation of \emph{ab initio} and experimental data.

\begin{equation}
u_{\mathrm{BKS}}(r_{\alpha\beta}) = \frac{q_\alpha q_\beta}{r_{\alpha\beta}} + A_{\alpha\beta} \exp \left( -b_{\alpha\beta} r_{\alpha\beta} \right) - \frac{c_{\alpha\beta}}{r_{\alpha\beta}^6}
\label{eqBKS}
\end{equation}

\noindent where $q$ is electrostatic charge and $A$, $b$ and $c$ are species-specific constants. Silicon and oxygen are denoted by $\alpha$ and $\beta$. Large-scale simulations show good agreement with experimental data, at least at the two-body level of the static structure factor (see section \ref{sectionTwoPointStructure} for a discussion of the merits of two-point correlations)   ~\cite{horbach1999}.

In recent years, colloidal dispersions have come to play an increasing role in studies of the glass transition. Colloids are epitomised (if only approximately \emph{realised} \cite{royall2013myth,poon2012,bryant2002}) by the hard sphere model. The requirement here is that no particles may overlap with one another, \emph{i.e.} be separated by a distance less than $\sigma$. Otherwise the interaction energy is zero. In this athermal system, in the absence of potential energy, entropy is the only contributor to the free energy, which determines the equilibrium phase.

The phase behaviour of hard spheres as a function of $\phi$ is similar to those of simple atoms, as a function of $1/T$ ~\cite{pusey}. Thus hard spheres feature a glass transition analogous to that which simple atoms exhibit as a function of $1/T$ (see Fig. ~\ref{figAngell}). However, a direct analogy may be obtained from the compressibility parameter $Z=\Pi / (\rho k_BT)$ where $\Pi$ is the (osmotic) pressure ~\cite{berthier2009witten}, but this is hard (though not impossible ~\cite{williams2013}) to measure in colloid experiments.

Hard spheres have a crystalline phase. To form a glass, crystallisation is avoided by the use of two (or more) species of particles of different sizes (similar to the Lennard-Jones mixtures outlined above), or polydisperse systems, where the particle size follows a distribution. In hard spheres, when the standard deviation of this distribution exceeds around 6\% of the mean, crystallisation is not observed ~\cite{henderson1996,kofke1999}. Here we shall often refer to temperature $T$, but as Fig. ~\ref{figAngell} shows, a suitable parameter for hard spheres is $1/\phi$ \footnote{In fact a more formally equivalent parameter is the compressibility factor $Z$ ~\cite{berthier2009witten}}.

A basic model of anisotropic interactions is that of Stillinger and Weber, originally introduced to describe silicon ~\cite{stillinger1985}. The model features 2- and 3-body terms. The former takes a much softer form than the Lennard-Jones model, which is appropriate to the semimetallic nature of silicon. The latter has a minimum at bond angles appropriate to tetrahedral bonds. This model can then be parameterised to silicon ~\cite{stillinger1985} and has more recently been applied as a monatomic model for water (mW) with considerable success ~\cite{molinero2009}.

\subsection{Dynamics approaching the glass transition : the Angell Plot}

\begin{figure}[!htb]
\centering 
\includegraphics[width=80mm]{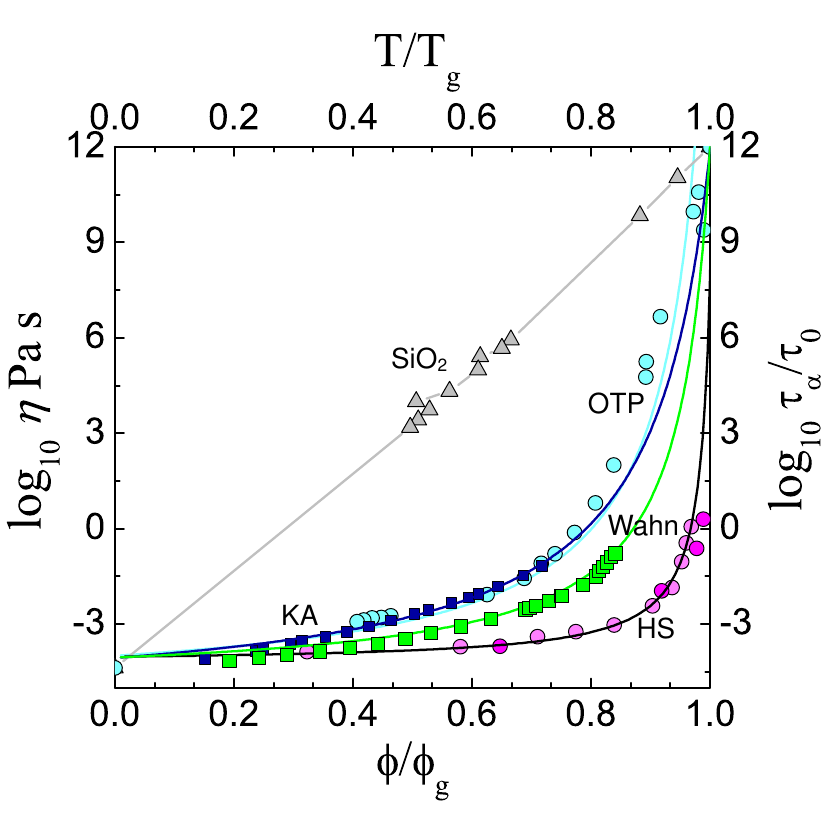} 
\caption{The Angell plot : Arrhenius representation of liquid viscosities, with inverse temperature scaled by $T_{{\rm g}}$. \textit{Strong liquids} exhibit Arrhenius behaviour characterised by an approximately straight line, indicative of a temperature-independent activation energy. \textit{Fragile liquids}, on the contrary, reveal super-Arrhenius behaviour, where activation energy grows as temperature decreases ~\cite{angell1988}. Data for SiO$_2$ and orthoterphenyl (OTP) are quoted from Angell ~\cite{angell1995} and Berthier and Witten ~\cite{berthier2009witten}. The other data concern model systems discussed in section ~\ref{sectionCommonModelSystems} : KA denotes Kob-Andersen and Wahn denotes Wahnstr\"{o}m binary Lennard-Jones systems while HS denotes hard spheres where the control parameter is the volume fraction $\phi$. In these cases, $\tau_0$ is scaled to enable data collapse at $T_g/T = \phi/\phi_g$ ~\cite{royall2014}.
\label{figAngell} }
\end{figure}

\begin{figure}[!htb]
\centering 
\includegraphics[width=88mm]{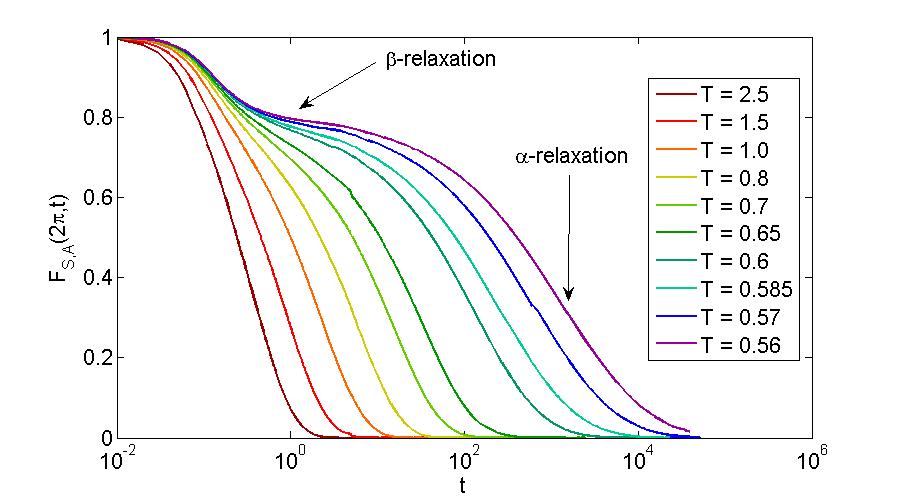} 
\caption{Intermediate scattering functions at different temperatures in the Wahnstr\"{o}m model. Note the characteristic two-step decay which distinguishes short-time $\beta-$ and long-time $\alpha-$relaxations ~\cite{malins2013jcp}.  }
\label{figGISF} 
\end{figure}

Approaching the glass transition, the single most remarkable feature is that the structural relaxation time $\tau_\alpha$ increases by many orders of magnitude over a modest change in temperature (or density in the case of systems such as hard spheres). One can quantify the dynamic slowdown using the (self) intermediate scattering function (ISF) $F(t,k)$ which is the Fourier transform of the self van Hove function $G(r,t)$ (Fig ~\ref{figGISF}).

\begin{equation}
F(\mathbf{k},t)=\frac{1}{\rho}\langle \rho_{{\bf k}}(t)\rho_{-{\bf k}}(0)\rangle=\int d{\bf r}\; G({\bf r},t)\exp\left(-i{\bf k}\cdot{\bf r} \right)
\label{ISF}
\end{equation}

This characterises time decay of density-density correlations on the lengthscale related to a certain wavevector $\mathbf{k}$ which is often taken close to the first peak in the static structure factor $S(k)$. The ISF typically exhibits a stretched-exponential decay which can be fitted as

\begin{equation}
F(t,\mathbf{k})=c \exp \left(-(t/\tau_{\alpha})^{b} \right)
\label{eqKWW}
\end{equation}

\noindent the Kohlrausch-Williams-Watts (KWW) law with $0<c\leq1$ and $0<b\leq1$. The KWW form is used to fit the ISFs in Fig. \ref{figGISF}. At temperatures lower than the glass transition point, long-time $\alpha$-relaxation does not occur on the experimental timescale. This is accompanied by complete arrest of structural relaxations and only local ``in-cage'' motions remain. Different relaxation regimes are illustrated in Fig.~\ref{figGISF} showing typical ISFs from MD simulations of the Wahnstr\"{o}m model.

Angell ~\cite{angell1988} made the assumption that the $\alpha$-relaxation \index{relaxation!$\alpha$-} timescale follows an Arrhenius-like behavior.

\begin{equation}
\tau_{\alpha}\propto \exp\left( \frac{E_{\mathrm{a}}}{k_{\mathrm{B}}T} \right),
\end{equation}

\noindent where $E_{\mathrm{a}}$ may be viewed as an activation energy. Then in the plot shown in Fig.~\ref{figAngell}, the logarithm of the viscosity which, (neglecting Stokes-Einstein breakdown ~\cite{berthier2011}), %\textcolor{blue}{should we discuss SE breakdown?}
is proportional to $\tau_{\alpha}$ should be represented by a straight line. For some materials, notably silica, it is indeed close to a straight line. However, for many a \emph{fragile} or super-Arrhenius behavior is found. The increase in relaxation times is often well described by the semi-empirical Vogel-Fulcher-Tamman (VFT) law,

\begin{equation}
\tau_{\alpha}=\tau_{0}\exp\left[ \frac{\tilde{E}_{{\rm a}}}{k_{{\rm B}}(T-T_{0})} \right]
\label{eqVFT}
\end{equation}

\noindent where $T_{0}~(\simeq T_{{\rm K}}$) is rather lower than the experimental glass transition temperature $T_{{\rm g}}$ and $\tilde{E}_{{\rm a}}$ is a measure of the \textit{fragility} -- the degree to which the relaxation time increases as $T_g$ is approached. Furthermore, Eq. ~\ref{eqVFT} can be rationalised by both Adam-Gibbs (section ~\ref{sectionAdamGibbs}) and random-first order transition (section ~\ref{sectionRFOT}) theories. Remarkably, $T_0 \approx T_K$ for a large number of glassformers, fuelling the idea of a thermodynamic phase transition at $T_K$. See Cavagna  ~\cite{cavagna2009} for an interpretation of $T_0 \approx T_K$.

As successful as Eq. ~\ref{eqVFT} has been, it is not the only option. Dyre and co-workers showed that it is not the best fit to experimental data \cite{hecksler2008}, and other forms, for example one also related to the Adam-Gibbs theory ~\cite{mauro2009} provide better agreement with experimental data. Other approaches include identifying an activation energy $E_\mathrm{act}(T)$ related to the temperature ~\cite{tarjus2000},

\begin{equation}
\tau_{\alpha}=\tau_{0}\exp\left[E_\mathrm{act}(T)\right].
\label{eqFragilityGilles}
\end{equation}

\noindent This approach is discussed in more detail in section ~\ref{sectionGeometricFrustration}, for now we note that at higher temperature, $E_\mathrm{act}(T) \rightarrow E_\infty$ where $E_\infty$ is a constant leading to Arrhenius (strong) behaviour in $\tau_\alpha$. For fragile behaviour at lower temperature, the theory of geometric frustration predicts the behaviour of $E_\mathrm{act}(T)$.

\subsection{Dynamic heterogeneity and dynamic length scales}
\label{dynamicHeterogeneity}

\begin{figure}[!htb]
\centering \includegraphics[width=80mm]{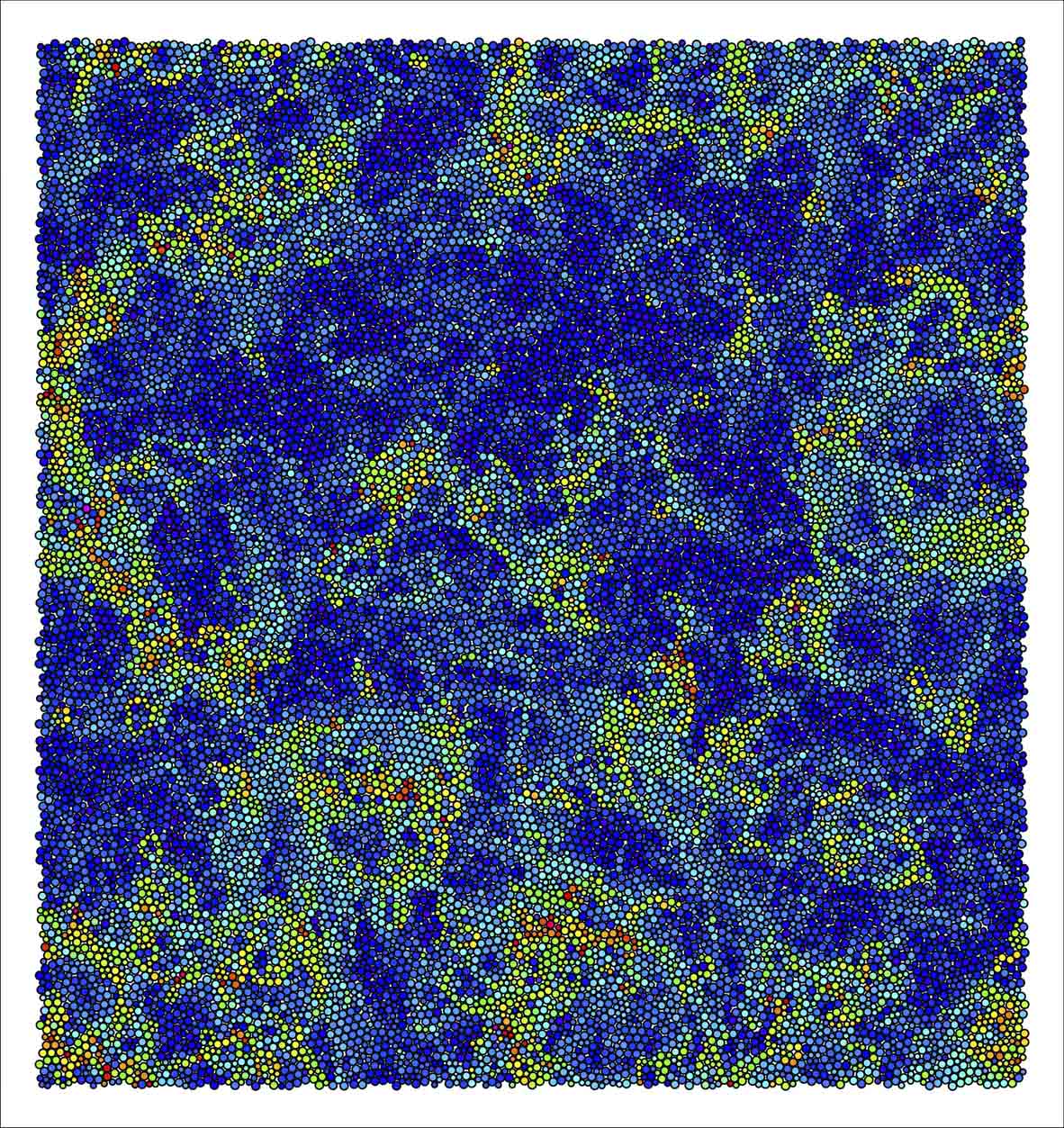} 
\caption{Dynamic heterogeniety in a simulation of binary hard discs with a size ratio of 1:1.4. Blue particles have moved the least (mean squared displacement $\langle r^2 \rangle <0.01\sigma_\mathrm{large}^2$) and red have moved the most ($\langle r^2 \rangle>\sigma_\mathrm{large}^2$). Here the timescale is taken over the structural relaxation time $\tau_\alpha$.}
\label{figDynamicHeterogeneity} 
\end{figure}

\begin{figure}[!htb]
\centering \includegraphics[width=70mm]{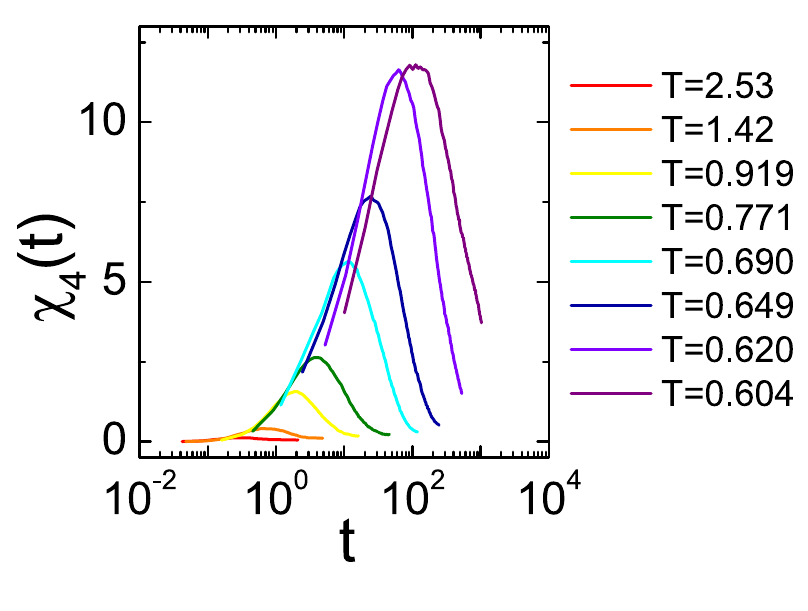} 
\caption{The dynamic susceptibility $\chi_{4}(t)$ in the Wahnstr\"{o}m binary Lennard-Jones glassformer. Colours correspond to different temperatures
~\cite{malins2013jcp}. }
\label{figChi4} 
\end{figure}

Approaching the glass transition, \textit{dynamic heterogeneity} becomes a significant feature: Particles move in an increasingly cooperative manner creating dynamically correlated mesoscopic domains ~\cite{perera1996,kob1997}. A \textit{dynamic length scale} can be associated with the increasing \index{dynamic heterogeneity} dynamic heterogeneity -- a measure which characterizes the size of growing cooperative motion. This lengthscale can be quantified by using so-called four-point correlation functions: A time-dependent order parameter which measures the overlap between two configurations separated by time $t$ is represented by the following binary correlation function:
\begin{equation}
W(t)=\int d\mathbf{r}\rho(\mathbf{r},t)\rho(\mathbf{r},0)=\sum_{i,j}^{N}\delta\left[\mathbf{r}_{i}(t)-\mathbf{r}_{j}(0)\right].
\end{equation}
The quantity $\langle W(t)\rangle / N$ is equal to the probability of finding a particle at time $t$ at a given position that was occupied by a particle at time zero. Initially we have $\langle W(0)\rangle=N$ which starts to decay rapidly once the particles have moved more than some lengthscale, often $0.3a$ where $a$ is the particle interaction radius. If we extend this length then $\langle W(t)\rangle$ decays rapidly towards zero. Thus $W(t)$ is sensitive to whether the particles are slow-moving or not. The fluctuations are then characterised by the \emph{dynamic susceptibility},
\begin{equation}
\chi_{4}(t)=\frac{1}{N \rho k_{{\rm B}}T}\left[\langle W^{2}(t)\rangle-\langle W(t)\rangle^{2}\right],
\label{eqChi4}
\end{equation}
\noindent which is obtained by integrating a four-point time-dependent density correlation function over volume \cite{lacevic2003}. For a given temperature $T$ (or packing fraction $\phi$), the fluctuations (i.e., the susceptibility) attain a maximum at certain $t=\tau_{h}$, and then die away, as illustrated in Fig.~\ref{figChi4}. The time where $\chi_{4}(t)$ is maximised $\tau_{h}$ has a value similar to $\tau_{\alpha}$.

\begin{figure}[!htb]
\centering \includegraphics[width=85mm]{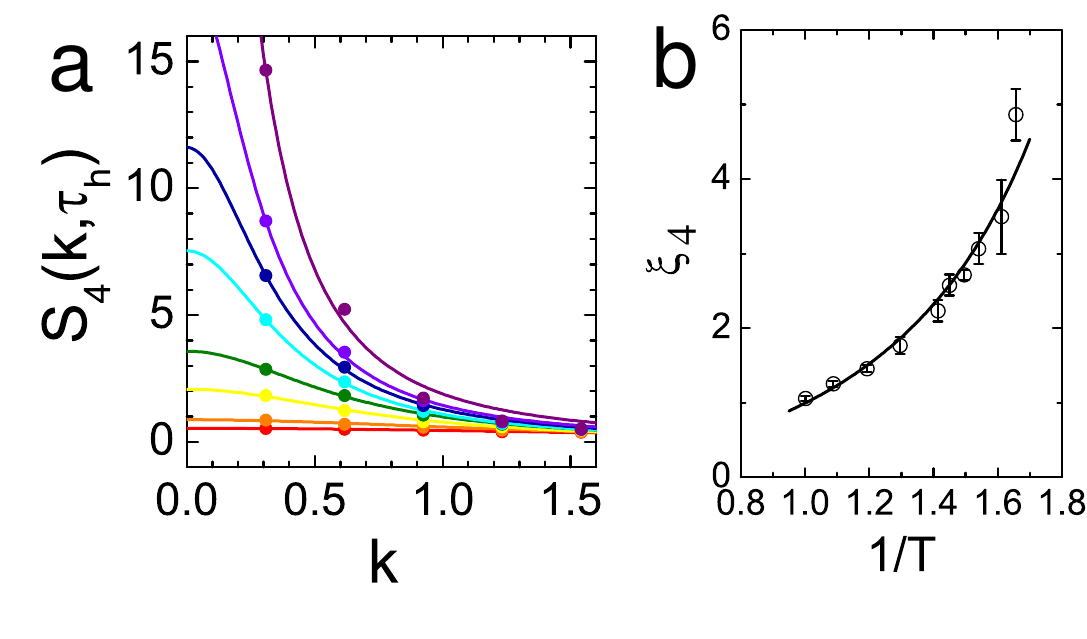}
\caption{Determining the four-point dynamic lengthscale. (a) The ``four-point'' structure factor $S_{4}$ in the Wahnstr\"{o}m binary Lennard-Jones glassformer. $\tau_{h}$ is the time corresponding to the peak in the dynamic susceptibility $\chi_{4}$ and is close to $\tau_{\alpha}$ . Colours are different temperatures as in Fig. \ref{figChi4}. (b) the dynamic correlation length $\xi_{4}$ as fitted to $S_{4}$ [lines in (a)] following Eq. \ref{eqOZcrit}. ~\cite{malins2013jcp}. }
\label{figGS4squareXi4} 
\end{figure}

The exact nature of the increase of the dynamic length scale remains unclear, and depends upon the system under consideration.  In simulations of the Wahnstr\"{o}m binary Lennard-Jones glassformer, this increase of a dynamic lengthscale was likened to divergence of the static density-density correlation length observed in liquid-gas critical phenomena ~\cite{lacevic2003}. In this spirit, one can define a four-point structure factor $S_{4}(k,t)$, so that the dynamic heterogeneities are maximised at $t=\tau_{{\rm h}}$. The four-point dynamic structure factor $S_4(\textbf{k},t)$ reads

\begin{eqnarray*}
S_4(\textbf{k},t) &  =  & \frac{1}{N\rho} \langle \sum_{jl} \exp[i \textbf{k} \cdot \textbf{r}_j(0)]w(|\textbf{r}_j(0)-\textbf{r}_l(t)|)  \\
&\times & \sum_{mn} \exp[i \textbf{k} \cdot \textbf{r}_m(0)]w(|\textbf{r}_m(0)-\textbf{r}_n(t)|) \rangle,
%\label{eqS4}
\end{eqnarray*}

\begin{equation}
\label{eqS4}
\end{equation}

\noindent where $j$, $l$, $m$, $n$ are particle indices. For time $\tau_h$, the orientationally averaged version is $S_4(k,\tau_h)$. This may then be fitted with the Ornstein-Zernicke relation to obtain a dynamic correlation length $\xi_{4}$,

\begin{equation}
S_{4}(k,t_{{\rm h}})=\frac{S_{4}^{0}}{1+(k\xi_{4})^{2}},
\label{eqOZcrit}
\end{equation}

\noindent where $S_{4}^{0}$ is a fitting parameter. Equation \ref{eqOZcrit} is thus fitted in Fig. \ref{figGS4squareXi4}(a) and the resulting $\xi_{4}(T)$ is plotted in Fig. \ref{figGS4squareXi4}(b), showing the increase with $1/T$. $\xi_{4}(T)$ also has been fitted to critical-like exponents where the Mode-Coupling transition (see section ~\ref{sectionMCT}) was taken as the transition temperature ~\cite{lacevic2003}. We emphasise that a number of other ``scalings'' are possible with divergences at $T=0$~\cite{whitelam2004}, $T=T_0$~\cite{tanaka2010} and scalings which couple to the relaxation time ~\cite{flenner2009,flenner2011,flenner2013,kim2013}, thus divergence of $\xi_4$ can be coupled to where one imagines $\tau_\alpha$ to diverge. Such a variation reflects the fact that the observable range over which $\xi_4$ varies is less than one decade, so discussions of divergence require a considerable degree of faith. In other words, measuring $\xi_4$ from simulation data does not enable discrimination between different theories. We return to this discussion in section \ref{sectionStaticAndDynamicLengths}.

We remark that Fig. \ref{figDynamicHeterogeneity} indicates that a single relaxation time - $\tau_{\alpha}$- may be too simple to describe such a complex system where different regions relax on different timescales. Indeed, it is this local variation in relaxation times which means that the intermediate scattering function (Fig. \ref{figGISF}) effectively becomes a superposition of many different local ISFs, each with its own $\tau_{\alpha}$. This superposition leads to stretching in the overall form, and the stretching exponent $b$ in Eq. \ref{eqKWW} can be used to determine the degree of dynamic heterogeity. One may also enquire as to whether it is appropriate to suppose that a single dynamic correlation length properly describes a deeply supercooled liquid.

\subsection{Crystal nucleation and the limit of metastability}
\label{sectionSpinodalNucleation}

\begin{figure}[!htb]
\centering \includegraphics[width=75mm]{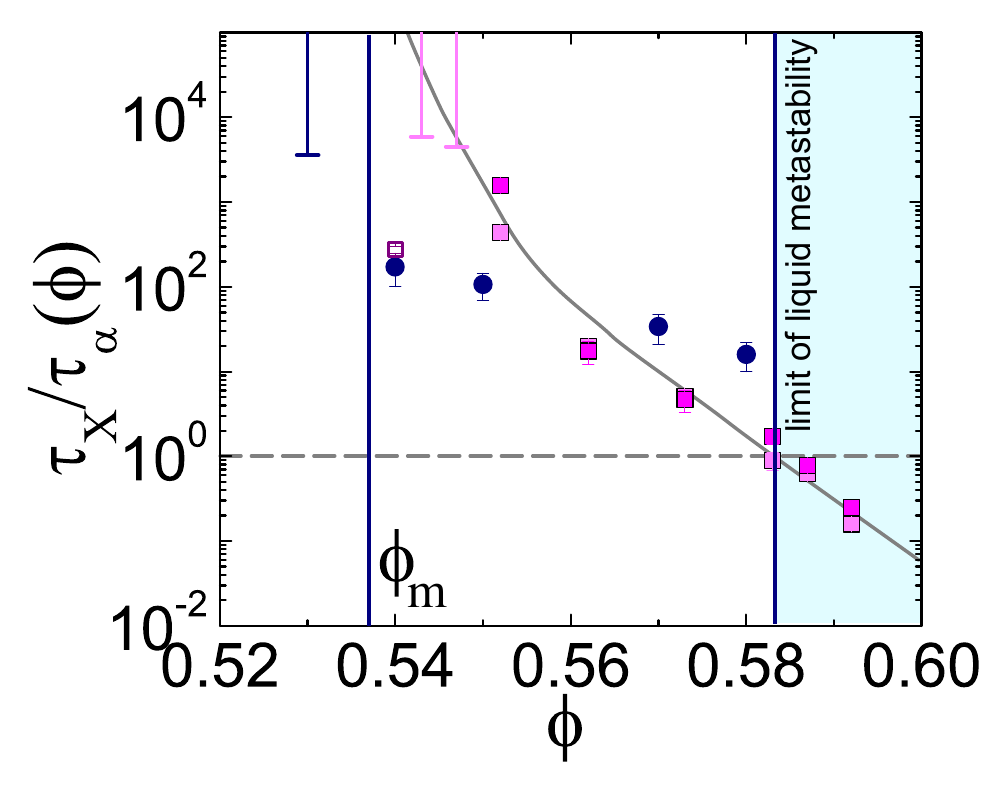} 
\caption{The limits of metastability in a (nearly) hard sphere fluid. Above volume fraction $\phi\approx0.58$, the time to crystalise $\tau_{X}$ falls below the structural relaxation time $\tau_{\alpha}$ (shaded region). In other words, it becomes impossible to prepare a metastable fluid. Here $\phi_m$ denotes the melting volume fraction, which for the hard sphere system does not coincide with freezing ($\tau_{X}$ diverges at the latter)
 ~\cite{taffs2013}. }
\label{figXtalTimes} 
\end{figure}

To form a glass, it is necessary that the system does not crystallise upon cooling (or compression). Upon sufficient cooling below the freezing point, in some marginally stable liquids crystallisation on timescales shorter than the relaxation time can be observed. This has been found for the one-component Lennard-Jones model ~\cite{trudu2006}, hard spheres as shown in Fig. ~\ref{figXtalTimes} ~\cite{zaccarelli2009xtal,sanz2011,valeriani2012,taffs2013}, BKS-silica ~\cite{saikavoivod2009} and suggested for the Stillinger-Weber model ~\cite{molinero2006}. When crystallisation occurs on such short timescales (relative to $\tau_{\alpha}$), it has been termed ``spinodal crystallisation''.

This balance of crystallisation and vitrification is crucial to the exploitation of metallic glassformers (where crystallisation places a fundamental limit on the size of metallic glass items), and in materials based on chalcogenide glassformers where it underlies their function. The competition between crystallisation and vitrification is explored further in section \ref{sectionCrystallisationVersusVitrification}.

\subsection{Dynamic arrest and soft matter}
\label{sectionGTSM}

A common misconception is that the \emph{experimental} glass transition $T_g$ is somehow absolute. This is connected to a drastic difference in timescales between soft matter and molecular fluids, and can be illustrated with Fig.~\ref{figAngell}. Molecules are said to vitrify when the relaxation time exceeds 100~s, which is an entirely anthropocentric quantity. This is around 14 orders of magnitude longer than relaxation timescales in high-temperature liquids. On the other hand, for instance colloids can have relaxation times of $\sim100$~s without exhibiting any traces of slow dynamics. This obviously presents a major problem: If we employ for colloids the same criteria as those applied to molecular fluids, we find that the ``molecular'' 100~s corresponds to $\sim10^{8}$~\emph{years}. In other words, in order to complete the equivalent \emph{dynamic range} of 14 orders of magnitude in particle-resolved techniques on colloids, the measurement would have to have commenced in the Jurassic period!

This dynamic disparity between soft matter and molecular systems has led to states being termed ``glasses'', which have similar relative relaxation times to moderately supercooled liquids (with the viscosity of honey ~\cite{gasparoux2008}, for example) ~\cite{cipeletti2005}. None the less the typical correlation functions for soft matter glass formers do exhibit the characteristic glassy two step decay (as a function of logarithmic time), albeit over a narrower time window. Based on results from dynamic light scattering experiments being fitted to power law behaviour consistent with mode coupling theory (MCT), a packing fraction of around $\phi_\mathrm{MCT}\simeq0.58$ was widely accepted ~\cite{vanmegen1994}. It turns out this narrower time window coincides with the mode-coupling transition, which led to an identification of MCT as some kind of absolute transition (as might occur in mean-field).

More recent experiments, which nominally ~\cite{royall2013myth} accessed higher packing fractions than previous work, concluded that colloidal hard spheres could relax at densities higher than the mode-coupling transition ~\cite{brambilla2009}. However these systems had a rather larger polydispersity than the earlier studies ~\cite{vanmegen1998}  and appear to slow down far less dramatically at the highest volume fractions. It was found that the power-law divergence of the mode coupling theory did not fit the data, while a generalised VFT form did. The importance of this work is that it showed convincing deviation from MCT in a colloidal system. This is in line with what has long been known in the case of molecular glasses where dynamic ranges of $10^{14}$ are routine ~\cite{berthier2011,angell1988,angell1995}. The detailed extent to which these findings are affected by the large polydispersity is worthy of further experimental investigation. It is important to know to what extent these conclusions stand up as the polydispersity is reduced down to something more like 6\% and whether longer time scales are needed to observe the break down of MCT as the polydispersity is reduced. As discussed in section ~\ref{sectionTowardsAStructuralMechanism} polydisperse systems are less fragile which is consistent with the findings of  ~van Megen and Williams ~\cite{vanmegen1998} and Brambilla \emph{et al.} ~\cite{brambilla2009}.

\subsection{Gelation} 
\label{sectionGelation}

Before concluding this discussion of the phenomenology of dynamical arrest, we mention a mode of arrest predominantly exhibited by multicomponent soft materials ~\cite{zaccarelli2007}. We shall be dealing with gelation of colloidal particles which pertains to a dynamically arrested network. Gelation of colloids comes in two ``flavours'', arrested phase separation (non-equilibrium gelation) and a network stabilised by anisotropic or competing interactions (for example short range attraction and long-range repulsion) which may be regarded as equilibrium gelation. This latter scenario, particularly in the case of anisotropic interactions ~\cite{bianchi2011,chaudhuri2010,ruzicka2010}, where particles form a network, can be thermodynamically similar to network glasses such as silica  ~\cite{sciortino2008,saikaVoivod2011}. Such systems may behave as strong liquids (Fig. ~\ref{figAngell}) as shown recently in experiments on limited valence DNA oligomers ~\cite{biffi2013}.

In the former case, non-equilibrium gels are formed when a system becomes thermodynamically unstable to demixing into colloid-rich (``colloidal liquid'') and colloid-poor (``colloidal gas'') phases, but the ensuing spinodal decomposition is arrested by vitrification of the dense phase ~\cite{zaccarelli2007,verhaegh1997,lu2008,testard2011}.

\subsection{Jamming and the glass transition} 
\label{sectionJammin}

\begin{figure}[!htb]
\centering \includegraphics[width=40mm]{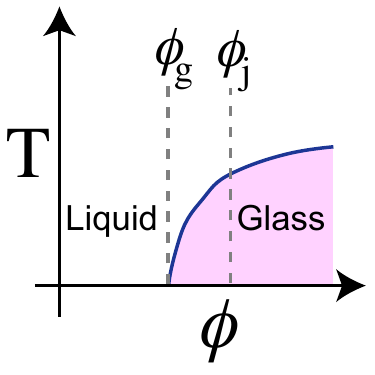} 
\caption{The relation of jamming to the glass transition for soft spheres. Jamming lies within the dynamically arrested region (shaded), i.e. the glass transition occurs without jamming.
 Based on ~\cite{berthier2009witten}. }
\label{figJammin} 
\end{figure}

The jamming transition is a 20-year-old phenomenon of athermal systems. It occurs when sufficient contacts between particles have been established that the system becomes rigid ~\cite{cates1998,liu2010,torquato2010}. For spheres in three dimensions, \emph{isostatic} jamming corresponds to six contacts per particle.  Berthier and Witten ~\cite{berthier2009witten} investigated the relationship between jamming and the glass transition in a system of soft spheres, which become hard spheres in the limit that temperature $T\rightarrow0$. Jamming occurs around random-close-packing, $\phi_j\approx\phi_\mathrm{RCP}\approx0.64$. The ``around'' we will expand upon below, but pertains to the fact that random close packing is ill-defined ~\cite{torquato2010}. In the case of hard spheres, a key question was whether the glass transition, by which we refer to the packing fraction predicted by for example VFT, fits to an ``ideal glass transition'' at $\phi_0=\phi_\mathrm{RCP}$ or  $\phi_0<\phi_\mathrm{RCP}$. The significance of the results, that the latter case held, is that jamming is decoupled from the glass transition. At finite temperature, the glass transition occurs at higher packing fraction, however, as shown in Fig. \ref{figJammin} the jamming transition $\phi_j$ lies within the glass.

This means that jamming is an intrinsically nonequilibrium phenomenon. Consequently $\phi_j$ is protocol dependent and even repeats of the same protocol give different values. However, they are all close to $\phi_j\approx0.64$ ~\cite{chaudhuri2010,kamien2007}. Soft spheres can be compressed above $\phi_j$. Approaching $\phi_j$ from the high density side, a number of scaling laws emerge, and parallels may be drawn with the normal mode analysis discussed in section ~\ref{sectionIsoconfigurational} ~\cite{wyart2005,xu2007}. Jamming is thus distinct from vitrifcation, moreover any role of local structure in the transition has seen relatively little attention. We refer the interested reader to the recent review on jamming by Liu and Nagel ~\cite{liu2010} and packing by Torquato and Stillinger ~\cite{torquato2010}.

\section{Theories of the liquid-to-glass transition}
\label{sectionTheories}

\begin{figure*}[!htb]
\centering \includegraphics[width=120mm]{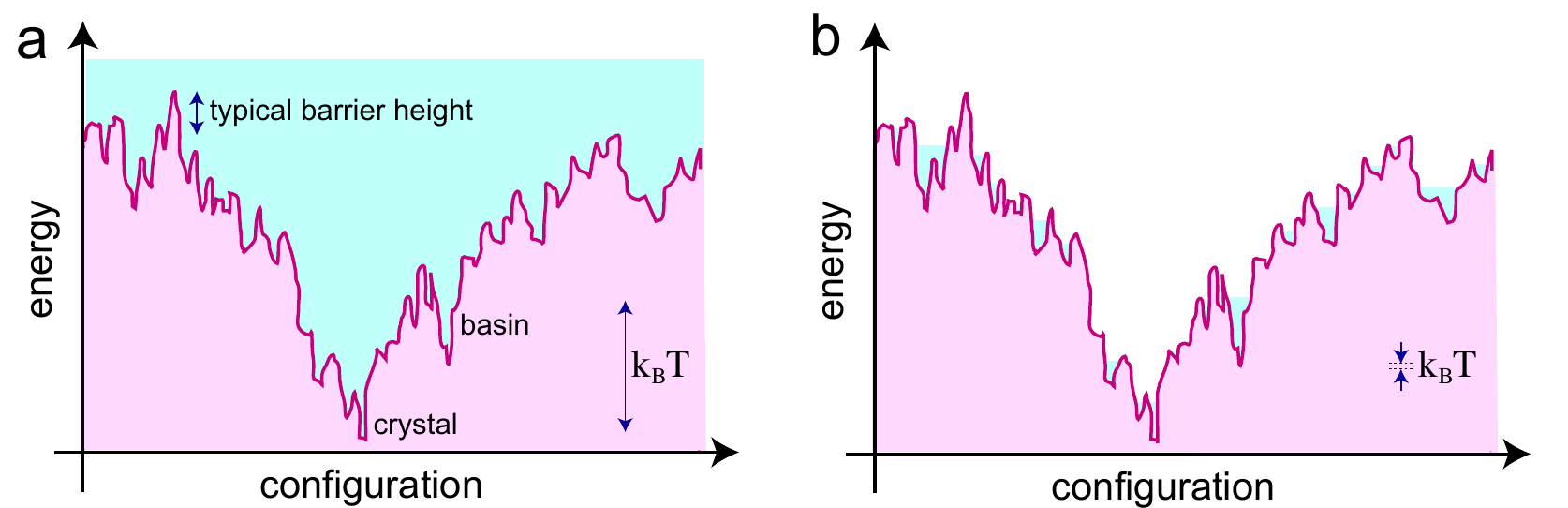} 
\caption{A schematic illustration of the potential energy landscape. The $x$-axis represents configurations of all $3N$ coordinates. (a) High temperature liquid, the typical barrier height is less than the thermal energy, and all configurations can be accessed as indicated in blue. (b) Low temperature glass. The barrier height between basins is now much higher than $k_BT$, and the system is no longer able to explore all configurations and has become a glass but can only access local states, as indicated by the blue shading.
\label{figEnergyLandscape} }
\end{figure*}

Here we mention a few key theories of the glass transition, and refer the reader to the relevant reviews for each. We seek only to give the briefest essence of each theory, and in particular its relation to structure, in order to facilitate the review. Some of these theories link explicitly to the thermodynamics of the system (and thus are coupled to the structure). These include the energy landscape, Adam-Gibbs and random first order transition theories. Some, such as geometric frustration, make explicit reference to local structure. Others are decoupled from the thermodynamics, such as dynamic facilitation.

\subsection{Mode-coupling theory}
\label{sectionMCT}

The mode-coupling theory (MCT) of G\"{o}tze and co-workers ~\cite{goetze} provides a description of dynamics which is based on liquid state theory. The idealised MCT deals with the dynamics of simple liquids.  A self-consistent treatment of the cage effect of particles prevented from moving by the proximity and immobility of their neighbours, leads to a closed set of integro-differential equations for the density correlator $\Phi_{k}(t)=F(k,t)/S(k)$, \emph{i.e.}, for the intermediate scattering function  (ISF) (Eq. \ref{ISF}) normalised by the structure factor $S(k,t)$. By taking the static structure factor as its input, MCT is thus coupled to the (two-body) structure of the liquid. We explore this connection further in section \ref{sectionStructure}. At a certain temperature (or packing fraction), the density correlator no longer decays to zero, which signifies the mode-coupling transition to a non-ergodic state (a glass).

Regarding systems such as those outlined in Section \ref{sectionCommonModelSystems}, MCT has been directly tested in the Kob-Andersen ~\cite{kob1995a} and Wahnstr\"{o}m ~\cite{lacevic2003} mixtures, colloidal hard sphere experiments ~\cite{brambilla2009,vanmegen1998} and for BKS silica  ~\cite{voightmann2008}. In hard spheres, MCT predicts a transition at a critical packing fraction of $\phi_{\mathrm{MCT}}\simeq0.52$, which can be rescaled to ``fit'' with experimental data at $\phi_\mathrm{MCT}\simeq0.58$ (this fit holds only for the first few decades of dynamic slowing, see the discussion in section \ref{sectionGTSM}).

In its basic form MCT is a mean-field theory. Thus the transition it predicts takes no account of local fluctuations but in mean field there is a true dynamical transition to a state where the density correlator does not decay as a function of time Fig. \ref{figDynamicHeterogeneity} ~\cite{charbonneau2005}. In $d=2,3$, it turns out that local fluctuations, such as dynamical heterogeneity, only really start to matter when the relaxation time increases by about four decades above the value of the normal liquid. Thus, MCT provides a reasonable description of the \emph{onset} of slow dynamics, but at deeper supercoolings, deviations are found in simulations ~\cite{flenner2011} and experiments ~\cite{brambilla2009,cheng2002}. MCT is, moreover, the only analytic route to predict a dynamical transition given an interaction potential. More recently extensions to incorporate fluctuations have been made~\cite{biroli2006,szamel2012}. For a review see ~\cite{charbonneau2005}.

\subsection{The energy landscape}
\label{sectionEnergyLandscape}

Goldstein's energy landscape picture of the equilibrium dynamics of a deeply supercooled liquid is so intuitive, that it is almost obvious in retrospect~\cite{berthier2011,goldstein1969}. The influence of this piece of work on the understanding of the glass transition cannot be overstated. Goldstein put the emphasis on the evolution of the system in phase space, \emph{i.e.} the space of all the configurational degrees of freedom, thus establishing an intimate link between structure and the glass transition. In the case of a simple liquid in three dimensions, this is the space of all 3N coordinates of the particles. The total potential energy is then $U(r_0,r_1..r_N)$ where the $r_i$ are the coordinates of the $N$ particles which comprise the system.  Each configuration is represented by a point in phase space, and the dynamics of the system can be thought of as the motion of this point over this \emph{potential energy landscape}. In other words, such a classical system is entirely described by its energy landscape ~\cite{cavagna2009}.

The local minima of the potential energy correspond to locally stable configurations of the particle system. One of these is the crystal, and this will usually be the absolute minimum (Fig. \ref{figEnergyLandscape}). Introducing defects, dislocations, not to mention other polymorphs represent other minima. Beyond the crystal-related minima, there will be many local minima corresponding to particle arrangements that lack crystalline order. These are amorphous, or glassy, minima, and have a potential energy that is \emph{extensively} larger than the crystal one.

Below some crossover temperature ($T_x$ in Fig. \ref{figCavagna}), a supercooled liquid explores the phase space mainly through activated hops between different amorphous minima, which are separated by potential energy barriers. Note that the system is here in (meta) equilibrium (metastable to crystallisation), so the potential energy is constant in time (we assume there is no aging) and the system is ergodic  ~\cite{cavagna2009}. In an idea that links to the random first order transition theory (RFOT) outlined below in section \ref{sectionRFOT}, one should note that these hopping events involve only a few particles (the typical number of dynamically correlated particles can be estimated by $\chi_4(t)$, Eq. \ref{eqChi4}, Fig. \ref{figChi4}).  Particles far from  rearrangement events are essentially unaffected by the hopping event. Significantly, the number of particles involved in the re-arrangement, $n$ is sub-extensive, so the (free) energy barrier to re-arrangement is also sub-extensive, and thus the system can relax ~\cite{cavagna2009}. It has recently been shown that the energy landscape (for hard spheres) splits in a fractal way ~\cite{charbonneau2014}. That is to say each basin has sub-basins and these have smaller basins within them and so on, until one reaches the force chains related to the contacts between the particles. This connection to force chains then allows a direct relation to be made to the local structure and to jamming. The fractal nature of the energy landscape means that, where it comes into play (for deep quenches) it is not clear that it is meaningful to talk of single relaxation mechanisms.

There is little doubt that the energy landscape approach provides a very comprehensive picture ~\cite{wales}. Moreover its connection with the structure of the liquid, exemplified in the ideas of Sir Charles Frank (section ~\ref{sectionFrank}), is so clear as to need no emphasis. The catch is that it is challenging, to put it mildly, to tackle $3N$ position coordinates in all but the smallest (low $N$) systems.

\subsection{Adam-Gibbs theory}
\label{sectionAdamGibbs}

The theory of Adam and Gibbs is among the first modern theories of the glass transition ~\cite{adam1965tdr}. The basic idea is the assumption that liquids in the supercooled state organise themselves into cooperatively re-arranging regions whose size $\xi(T)$ increases as the temperature goes down. Next, it is supposed that each cooperative region has a few preferred states $M$ independent of $\xi$. Rearrangement between these states then corresponds to $\alpha$ relaxation. Then the configurational entropy per unit volume is given by

\begin{equation}
S_{{\rm conf}}(T)\sim k_{{\rm B}}\xi^{-3}\ln M.
\end{equation}

\noindent The final assumption is that an energy barrier $E_{{\rm a}}$ for rearranging a cooperative region -- which enters the Arrhenius scaling of the $\alpha$-relaxation \index{relaxation!$\alpha$-} time -- is proportional to region's volume, $E_{{\rm a}}=C_{0}\xi^{3}$. This yields a super-Arrhenius (VFT-form) form for the relaxation time,

\begin{equation}
\tau_{\alpha}=\tau_{0}\exp\left[\frac{C_{0}\ln M}{TS_{{\rm conf}}(T)}\right],
\end{equation}

\noindent leading to reasonable agreement with experiments~\cite{bouchard2004}. Due to several strong assumptions made in the Adam-Gibbs theory it cannot be considered as a completely quantitative approach: Indeed, it is unlikely that $M$ does not depend on $\xi$, and that the energy barrier is proportional to $\xi^{3}$. Nevertheless, the Adam-Gibbs theory follows a simple and intuitive framework and introduces several key concepts, such as cooperative relaxation and a growing dynamic length scale. Like the energy landscape picture, the Adam-Gibbs theory is intimately related to the structure of the supercooled liquid.

\subsection{Random first-order transition theory}
\label{sectionRFOT}

\begin{figure*}[!htb]
\centering \includegraphics[width=120mm]{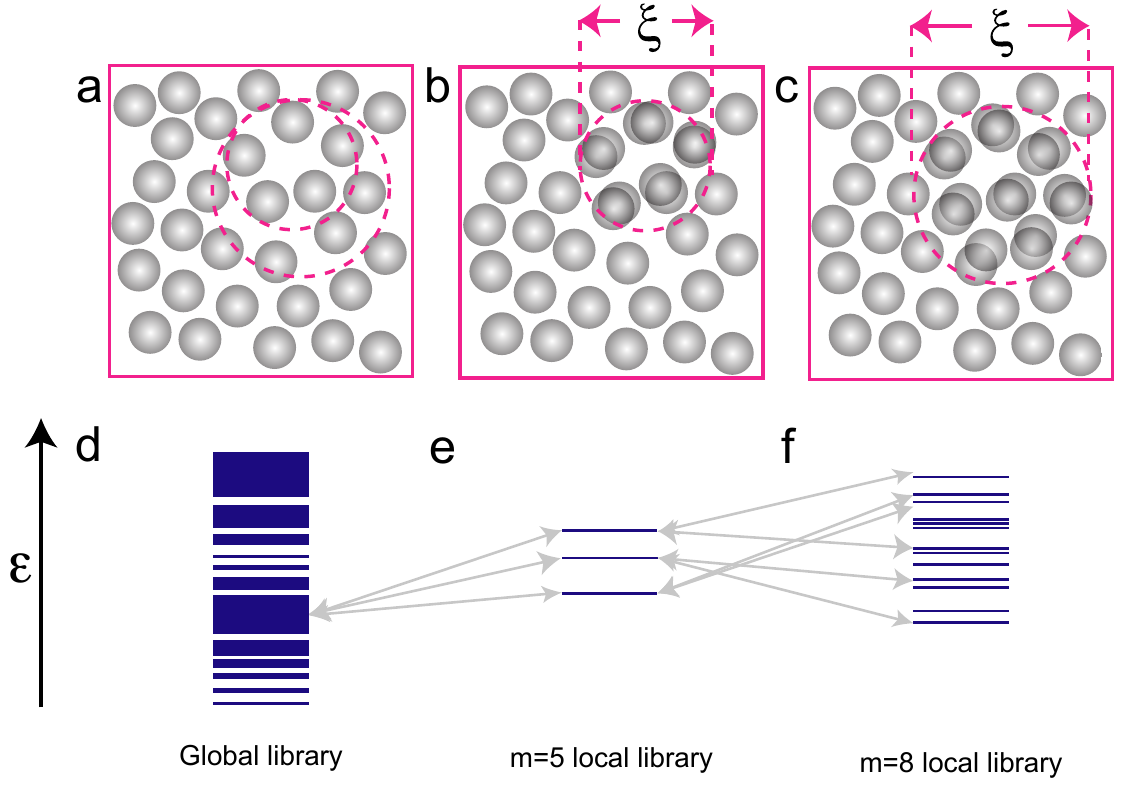} 
\caption{The ``local library'' as a mechanism for entropically-driven re-arrangements.
(a) A configuration from a global energy landscape. 
(b) A cooperatively re-arranging region of  $m=5$ particles. (c) A larger CRR of $m=8$ particles. (d) The huge (global) density of states (DOS) of the system prior to re-arrangement. (e) The DOS found in the local library originating from a given initial state with $m=5$ particles moving locally. The resulting energy typically corresponds to an increase over the original state owing to the mismatch between the two structures. (f) The larger DOS with the $m=8$ particles re-arranging. As the library grows in size, we still find the states as a whole at higher energies, but the width of the distribution grows. Eventually with growing $m$, we find a state within an energy range of order of the thermal energy, thus in practise we expect CRRs to be typically of this size corresponding to energy changes of order $k_BT$. Based on~\cite{lubchenko2007}. 
\label{figLocalLibrary} }
\end{figure*}

An alternative view, proposed by Kirkpatrick, Thirumalai and Wolynes~\cite{kirkpatrick1989} is random first-order transition theory. RFOT provides a mechanism for activated relaxation in a temperature regime between a low temperature ($\sim T_{K}$), where there is a thermodynamic phase transition to a true glass (which is a solid and whose dynamic correlation functions do not relax on any timescale) and a higher crossover temperature $T_A$, at which activated dynamics become the dominant mechanism of relaxation. This occurs around the mode-coupling temperature $T_{\mathrm{MCT}}$ ~\cite{lubchenko2007}. Thus, around $T_A$, the system crosses over to an energy-landscape dominated regime (see section \ref{sectionEnergyLandscape}). Rather than a \emph{first-order} transition to a crystal (as would occur in the freezing transition), RFOT envisages that the transition is to a vast number of \emph{random}, aperiodic states. However there is a good deal in common with the freezing transition, not least that particles become localised under the Lindemann criterion ~\cite{lindemann1910} which states that around melting, the mean square displacement saturates at around $a^*\sim0.1a_L$ where $a_L$ is the crystal lattice constant. For materials such as glasses the ``lattice constant'' $a_L \sim \sigma$ ~\cite{lubchenko2007}.

To explain how entropy can drive relaxation in this \emph{mosaic state} of cooperatively re-arranging regions (CRRs) where relaxation is activated, the concept of local and global libraries of states is introduced, Fig. \ref{figLocalLibrary}. Here the CRRs have the same interpretation as in the preceding section, that is, a rearrangement in a CRR corresponds to $\alpha$ relaxation. The global library corresponds to the spectrum of states accessible to a large system. Now consider a small cooperatively re-arranging region of that system, of 5 particles (Fig. \ref{figLocalLibrary}). Re-arranging a few particles leads to mostly highly overlapped configurations, which are energetically unfavored, thus only a few states are energetically accessible --- the local library. Larger regions, in this case 8 particles (Fig. \ref{figLocalLibrary}), lead to rather more states being energetically accessible, \emph{i.e.} larger local libraries. Sufficiently large local libraries have accessible states within $\sim k_BT$ of the state the system is in, so relaxation via re-arrangements becomes possible.

Thus the supercooled regime corresponds to a mosaic of cooperatively rearranging regions  (reminiscent of the Adam-Gibbs picture, section \ref{sectionAdamGibbs}), in which on timescales much less than the structural relaxation time $\tau_\alpha$, the particles are localised much like those in a crystal ~\cite{lubchenko2007}. In higher dimension, where mean field theories become more accurate, the transition from the liquid to the mosaic state is sharp, in low dimension, it is blurred out to become a crossover ~\footnote{Alas MCT fails in higher dimension, possibly due to the trivial nature of 2-point density correlations ($S(k)\rightarrow1$ in higher dimension).}.

The size of a CRR is then determined by the balance of entropy gain (the increase in accessible states from the local library) and effective surface tension. For a CRR of size $\xi$ the  gain in free energy due to the configurational entropy is $TS_{\mathrm{conf}}(T)\xi^{3}$. Meanwhile the free energy loss, which is due to mismatch between the nucleating and ambient states, is proportional to an effective surface tension which expresses the energetic penalty of rearrangement $\gamma_{{\rm eff}}$ and scales as $\gamma_{\mathrm{eff}}\xi^{\theta}$. Note that, because some states in the library are better matched to the cavity in the sense that the potential energy is smaller due to fewer overlaps $\theta\leq2$ (in $d=3$). See ~\cite{lubchenko2007} for more details. The characteristic length of this mosaic state is determined from the balance of the two processes,

\begin{equation}
\xi_{{\rm a}}\sim \left( \frac{\gamma}{TS_\mathrm{conf}(T)} \right)^{1/(3-\theta)}.
\end{equation}

\noindent The resulting $\alpha$-relaxation time is then given by a generalized VFT law which coincides with the Adam-Gibbs results for $\theta=3/2$. Both Adam-Gibbs and random first-order transition theories lead directly to dynamic heterogeneity ~\cite{berthier2011,lubchenko2007,bouchard2004}.

\textit{Pinning --- } On the basis of RFOT, Biroli and coworkers have argued that the existence of a length-scale $\xi$ over which motion is coupled enables a new approach to access the ``ideal'' glass transition (around $T_{0}$). A cavity can be identified, outside which all particles are frozen or \emph{pinned}. The pinning of the particles on the edge of the cavity reduces the configurational entropy of those inside, biasing the system in favour of arrest. If the radius of the cavity is larger than $\xi$, the particles at the centre remain mobile, if the cavity is smaller than $\xi$, the entire cavity is immobilised ~\cite{biroli2008}.

Rather than pinning all particles outside a cavity, one can instead pin a certain proportion throughout the system ~\cite{berthier2012}. The concentration of this pinned population of particles $c$ leads to a mean separation $d_{m}=c^{-1/d}$ (where $d$ is dimension). When $d_{m}<\xi$, it is argued that an ideal glass transition corresponding to that predicted by RFOT around $T_K$ is found ~\cite{cammarota2012,berthier2013overlap}. It is important for our purposes to note that pinned configurations are (with sufficient sampling) \emph{formally structurally indistinguishable} from their unpinned counterparts at the same temperature \footnote{Crudely speaking this can be seen from the fact that the pinned system interacts via the same Hamiltonian as does the unpinned system, and that pinned particles are chosen to be frozen in an equilibrium supercooled liquid ~\cite{jack2014}}. However, that the concentration of pinned particles required to drive the glass transition decreases upon cooling indicates that the CRRs become larger as predicted by RFOT. It is thus inescapeable that upon cooling something must change in the structure in atomistic glassformers to enable this reduction in configurational entropy.

%Another interesting piece of work is the suggestion of spinodal-like phase separation into regions whose dynamics exhibit and high- or low- overlap with the starting configuration. In other words that these were either ``slow'' or ``fast'' ~\cite{cammarota2010}. Upon sufficient supercooling, it was possible to make a Maxwell construction suggesting a phase separation of these regions around the mode-coupling temperature. Such an approach pointed towards a surface tension between the CRRs as postulated by RFOT.

\subsection{Montanari-Semmerjian and the rationale for a static lengthscale}
\label{sectionMontanari}

\begin{figure}[!htb]
\centering \includegraphics[width=40mm]{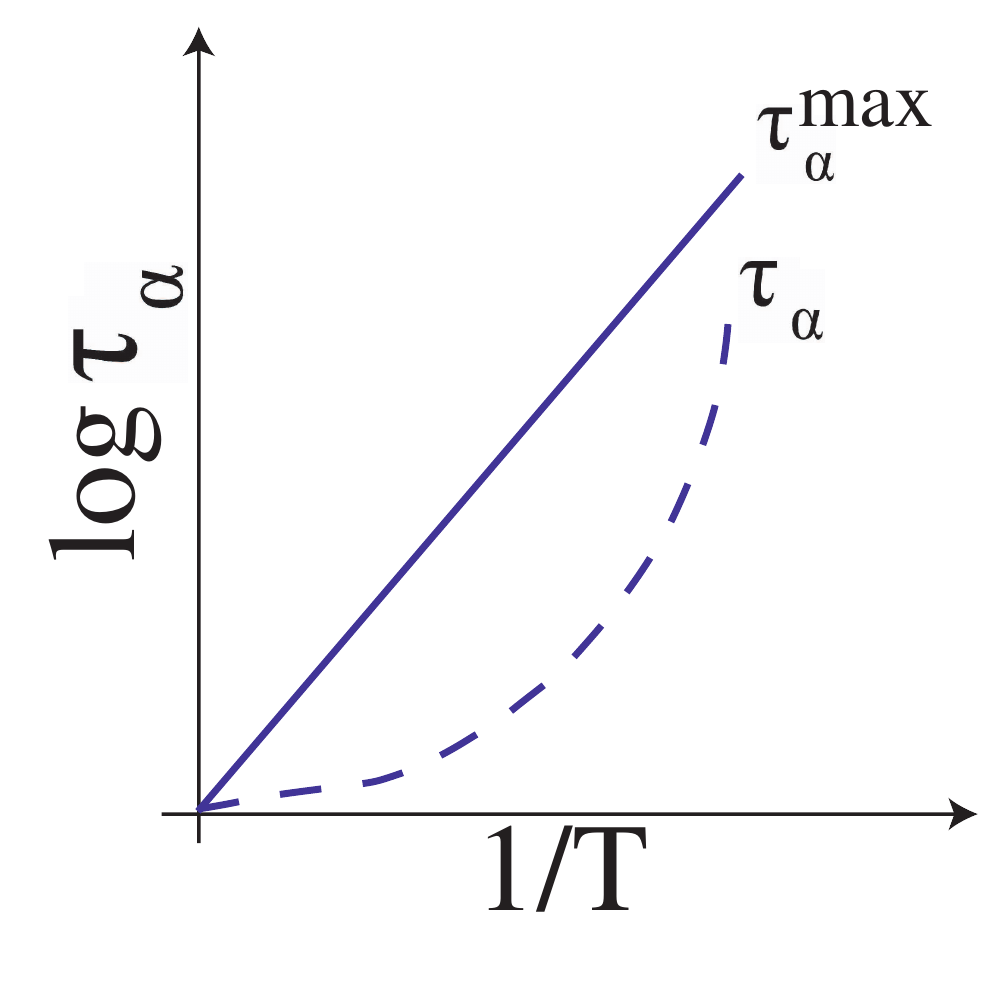} 
\caption{Montanari and Semerjian's argument for an increasing lengthscale. Were the cavity to have a fixed size $\xi$ (i.e. the dynamic correlation length is fixed), then the maximum possible relaxation time $\tau_\alpha^{\mathrm{max}}$ would be Arrhenius (solid line). Of course, the system may relax much faster than this, but, if the increase is super-Arrhenius (dashed), it \emph{must} cross the solid line at some very low but \emph{finite} temperature.}
\label{figMontanari} 
\end{figure}

Montanari and Semmerjian imagined a cavity in which re-arrangements are possible, but which is pinned by immobile particles outside in much the same way as the RFOT ideas outlined above ~\cite{montanari2006}. Now, at the lengthscale --- the size of the cavity --- where re-arrangements are just possible (see section ~\ref{sectionRFOT}), they occur and $\tau_\alpha$ is finite, \emph{i.e.} a supercooled liquid. The requirement for such a cavity is simply that there is more than one way to arrange the particles in the cavity.

Under these conditions, the system \emph{must} be able to relax. There is some maximum possible relaxation time $\tau_\alpha^{\mathrm{max}}$, which corresponds to the transition between these different configurations. Now let us cool the system down. This cavity has a fixed lengthscale, and, under these assumptions, $\tau_\alpha^{\mathrm{max}}$ should be Arrhenius, as we have that there are multiple configurations inside the cavity and there is some activation barrier --- whose height fixes $\tau_\alpha^{max}$, as shown in Fig. \ref{figMontanari}.

Of course, especially in the liquid $T>T_m$, the system relaxes much faster than $\tau_\alpha^{\mathrm{max}}$. However, if $\tau_\alpha$ follows a super-Arrhenius behaviour, it must, eventually, cross $\tau_\alpha^{\mathrm{max}}(T)$. This would mean that the assumption of a cavity in which the number of configurations available is fixed with respect to temperature, is flawed. \emph{In other words, under super-Arrhenius dynamics some structural change must occur}.

\subsection{Frank and frustration}
\label{sectionFrank}

We now turn to theoretical approaches which explicitly consider particular structures in connection with dynamical arrest. It is conceptually helpful to begin with the ideas of Sir Charles Frank \cite{frank1952}. Frank showed that the ground state of 13 atoms of the monatomic Lennard-Jones model was the icosahedron [see Fig. \ref{figBaka}(c) and (d)], rather than the FCC or HCP crystal. Thus, upon cooling, such atoms should first form icosahedra rather than crystals. Now icosahedra are five-fold symmetric, and thus do not tile Euclidean space. This introduces geometric frustration, and the system cannot reach thermodynamic equilibrium. Put another way, what Frank did was to consider the energy landscape, not of the whole $N$-particle system, but of a subset of $m=13$ particles. As mentioned in section \ref{sectionEnergyLandscape}, the limitation of the energy landscape is that it is intractable for large $N$ but one way around this is to make $N$ small (as we shall see below in section \ref{sectionTCC}).

With Kasper, Frank built upon these ideas with the objective of better understanding complex alloy crystal structures now known as Frank-Kasper phases. Much of the content of their original paper ~\cite{frank1958} contains ideas and results that are also relevant to amorphous phases and have been an important influence for subsequent work on geometric frustration. A system of atoms was considered in terms of a bond network with an atom located at each vertex and the various edges between  bonded pairs of atoms (i.e. nearest neighbours) forming faces, which in turn form polyhedra. Based on arguments about the efficient packing of equally or nearly equally sized spheres it was assumed that all faces were triangles. Further, upon the same basis, it was assumed that all atoms had from 12 to 16 bonds (\emph{i.e.} coordination number, or number of nearest neighbours). They then considered the network between each set of nearest neighbours in isolation and again, based on the same packing efficiency arguments, assumed that each of the vertices in this reduced subset could have only either of 5 or 6 edges, termed 5-fold and 6-fold vertices, $S_5$, $S_6$. Given this they were able to limit themselves to 4 possible topological structures in terms of the nearest neighbour network. The first of these is the icosahedron which has no $S_6$ vertices, thus for coordination number $Z = 12$ we have $n_6 = 0$, there is no $Z = 13$ satisfying the above assumptions, there is one possible $Z = 14$ with $n_6 = 2$,  a $Z = 15$ with $n_6 = 3$, and finally a $Z = 16$ with $n_6 = 4$. In all cases the $S_6$ vertices are arranged in a specific relation to each other, making each of these structures unique. These structures all obey the simple relation $n_6 = Z - 12$.

From here the implications on the entire condensed phase were considered. The atoms with a coordination number of 12 are termed minor sites and those with more are termed major sites. An atom at a minor site has no neighbour for which it has a further 6 neighbours in common, therefore the edge between any two atoms which have 6 neighbours in common is termed a major ligand. The entire systems bond network may then be specified in terms of the major ligands. The major ligands obey a valence type principle, as used in chemistry. However there is no case with $n_6 = 1$ and thus the major ligands cannot be terminated. This means that long strings of major ligands must run through the condensed phase which are termed the \emph{major skeleton}. In terms of crystal phases the possibility of long strings, which come around and close themselves, was ruled improbable and it was therefore concluded that layering of the major skeleton must occur. This leads directly to the Frank-Kasper phase.

\subsection{Geometric frustration}
\label{sectionGeometricFrustration}

\begin{figure}[!htb]
\centering \includegraphics[width=65mm]{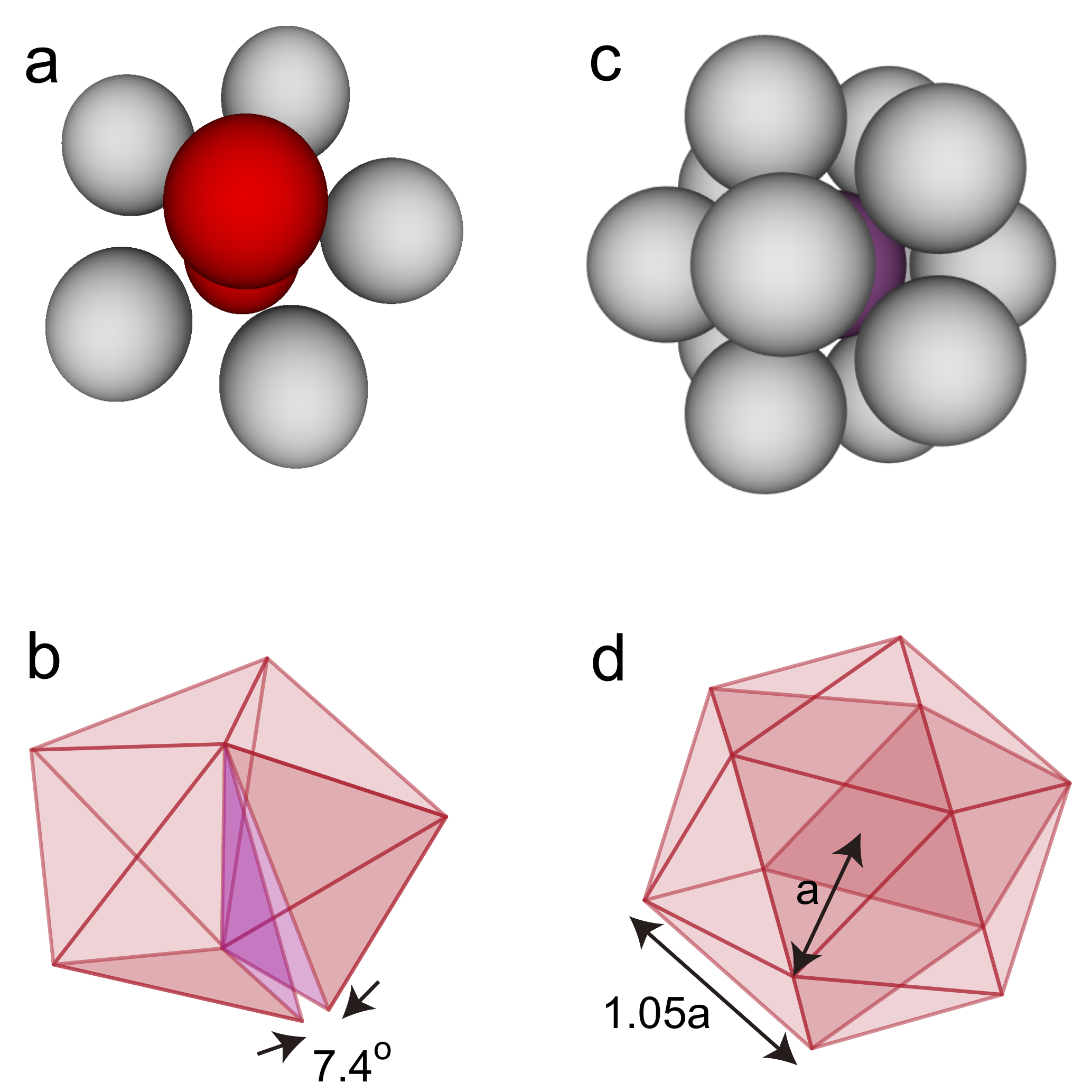} 
\caption{Frustration in polytetrahedra in Euclidean $d=3$. (a) The pentagonal bipyramid ``7A'' is constructed with five tetrahedra (b) which leads to a gap of $7.4^\circ$. (c) 13-membered icosahedron. (d) Icosahedra have bonds between particles in the shell stretched by 5\% with respect to the bonds between the central particle and the shell particles. 
\label{figBaka}}
\end{figure}

The theory of geometric frustration develops Frank's ideas in a more formal manner ~\cite{tarjus2005,kivelson1995}. In essence, it is assumed that the system ``wants'' to form locally favoured structures, such as icosahedra but considerations of Euclidean geometry prevent a transition to a fully icosahedral state. However, in curved space it is possible to have a fully icosahedral state ~\cite{nelson}. So one can consider a system in curved space, and extrapolate back to Euclidean space. Geometric frustration boils down to the following ~\cite{tarjus2005} :

\begin{enumerate}

\item A liquid is characterised by locally favoured structures (LFS) which minimise the (free) energy of a small number of particles (13 in the ``canonical'' case of icosahedra).

\item These LFS do not tile Euclidean space. The growth of domains of LFS, which induces strain, is thus the essence of frustration. This strain (free) energy suppresses
 the formation of an ``LFS-state''.

\item There exists a reference system (which may not be experimentally realisable, for example it may reside in curved space) in which the LFS do tile space. In that system there is a ``critical point'' $T_c$ at temperatures below which there is a transition (which is continuous or weakly first-order) to the LFS-state.

\end{enumerate}

Let us explore the meaning of these concepts by analogy to known systems ~\cite{tarjus2005}. Locally favoured structures adopted in the liquid become prevalent below $T_c$. Examples of systems where the LFS is unfrustrated in Euclidean space include discs in 2d, where the 2d simplex, a triangle of discs, tesselates to form the hexatic phase in a weakly first-order transition from the disordered fluid  ~\cite{bernard2011,engel2013}. Another example is cubes whose orientation is constrained to follow Cartesian axes. In that case the freezing transition is continuous ~\cite{jagla1998}.

Now the simplex in 3d is the tetrahedron. As Fig. \ref{figBaka} shows, placing 5 tetrahedra together to form a pentagonal bipyramid leads to a gap of $7.4^\circ$. In the case of icosahedra, the separation of particles in the first shell is around 1.05 times that of the central particle and the particles in the shell. As a consequence, strain is induced, and in fact for short-ranged interactions (such as those prevalent in colloidal systems \cite{asakura1954,likos2001}), the strain is sufficient that the icosahedron is not the minimum energy configuration ~\cite{doye1995,taffs2010}. For the longer-ranged Lennard-Jones interaction, icosahedra and variants such as Mackay icosahedra are the minimum energy stable structure for clusters of up to $\sim 500$ atoms. For larger sizes, strain leads to the adoption of FCC clusters ~\cite{wales,baletto2005}.

In curved space, the situation is rather better. As described by Coexeter ~\cite{coexeter,coexeterpolytope} and later by Sadoc with reference to glassformers ~\cite{sadoc1981}, 600 perfect (strain-free) tetrahedra comprising 120 particles can be embedded on the surface of a four-dimensional hypersphere. Each particle in this 4d Platonic solid or ``polytope'' is at the centre of a 12-particle icosahedrally coordinated shell. This polytope is denoted ${3,3,5}$ because three equililaterial triangles meet at every vertex with five tetrahedra wrapped around every bond ~\cite{nelson,nelson1983,straley1984}. Mapping from curved space back to Euclidean space, defects in the tiling of icosahedra are introduced.

It is convenient to explore frustration in dimension $d=2$. This has been implemented elegantly by Modes and Kamien ~\cite{modes2007} and Tarjus and coworkers ~\cite{sausset2008,sausset2010,tarjus2010,sausset2010pre} as described in Section \ref{sectionTowardsAStructuralMechanism}. As noted above, in $d=2$ Euclidean space, discs are not frustrated. Frustration is introduced with hyperbolic curvature (equal to $-\kappa^2$), which (unlike curving onto a spherical surface) enables the thermodynamic limit to be tackled. One can express the free energy of defects in the curved space in a Ginzburg-Landau form, where the curvature $-\kappa$ is on a lengthscale much larger than the particle size, and the degree of frustration is ``small'' ~\cite{nelson1983prb,tarjus2005}. Frustration in this curved $d=2$ space amounts to the free energy of growing a \emph{defect-free} domain of lengthscale $\xi_D$. Thus if we impose perfect order within a domain of LFS and take leading order contributions to the free energy it can be shown that the free energy scales as $\kappa^4 \xi_D^4$ ~\cite{tarjus2005} and this 
$\xi_D^{d+2}$ scaling holds in $d=3$ also.

\begin{figure}[!htb]
\begin{center}
\includegraphics[width=45mm]{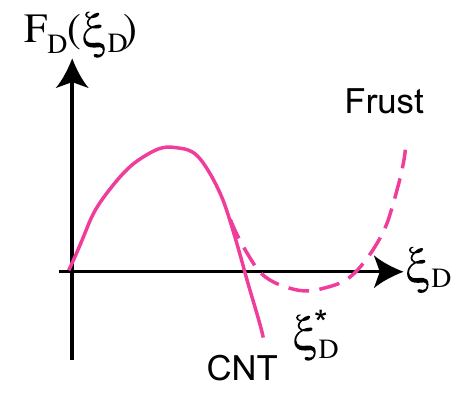} 
\caption{Schematic of geometric frustration limiting the growth of domains of locally favoured structures. Solid line is conventional CNT with the first two terms in Eq. \ref{eqCNTFrustration} which would occur in the non-frustrated case. Dashed line denotes the effect of the third term, leading to a preferred lengthscale for the LFD domain 
 $\xi_D^*$.\label{figFrustration}}
\vspace*{-12pt}\end{center}
\end{figure}

\begin{figure}[!htb]
\centering \includegraphics[width=50mm]{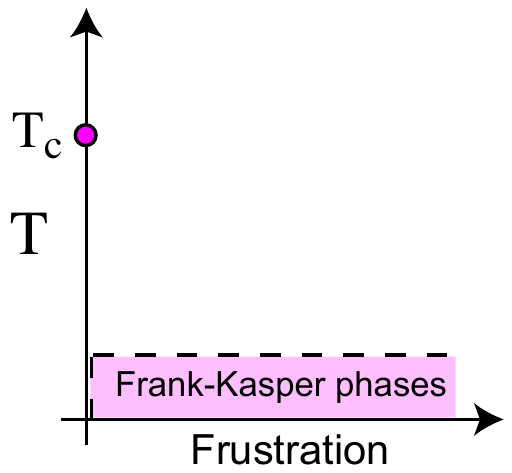} 
\caption{Avoided criticality. In the case of no frustration (here the $y-$axis), the system transforms to an ordered phase below $T_c$. When frustration is present, ``critical'' point is avoided, and there is no transition: the frustration-limited domain growth scenario. Frank-Kasper phases are formed at low temperature.
\label{figAvoided}}
\end{figure}

\textit{Frustration-limited domain theory. --- } These arguments lead to scaling relations for the growth of domains of locally-favoured structures. Geometric frustration imagines an \emph{avoided critical point}, at $T_c$ in the unfrustrated system (Fig. \ref{figAvoided}). At temperatures below this point, growth of domains of the LFS in the frustrated system would follow a classical nucleation theory (CNT) like behaviour, with an additional term to account for the frustration. Since the free energy cost induced by frustration scales as $d+2$, in $d=3$ we have

\begin{equation}
F_{D}(\xi_D,T)= \gamma(T)\xi_D^\theta-\delta \mu (T) \xi_D^3 + s_{\mathrm{frust}}(T)\xi_D^5
\label{eqCNTFrustration}
\end{equation}

\noindent where the first two terms express the tendency of growing locally favoured order and they represent, respectively, the energy cost of having an interface between two phases and a bulk free-energy gain inside the domain. In 3d, $\theta \leq 2$, and may be related to Adam-Gibbs theory or RFOT (sections \ref{sectionAdamGibbs} and \ref{sectionRFOT}). As shown in Fig. \ref{figFrustration}, the inclusion of the third term leads to a preferred domain size. Without the last term long-range order sets in at $T =T_c$. Geometric frustration is encoded in this third term which represents the strain free energy resulting from the frustration. This strain free energy is responsible for the fact that the transition is avoided and vanishes in the limit of zero frustration ~\cite{tarjus2005}.

\textit{Avoided Criticality. --- } The existence of a ``critical point'' is argued in terms that the frustration-induced lengthscale implicit in Eq. \ref{eqCNTFrustration} should diverge as frustration vanishes. Now defects resulting from the frustration of the LFS usually interact in a Coulombic or screened Coulombic fashion from which a critical scaling can be determined, for varying degrees of frustration ~\cite{tarjus2005}. In the case that the critical point is narrowly avoided (weak frustration), the LFS domains exhibit an increasing lengthscale as $T_c$ is approached from either above or below in a manner reminiscent of liquid-gas critical phenomena. The reader is referred to Tarjus \emph{et al.} for a more complete discussion of the expected scaling~\cite{tarjus2005}.

In the spirit of RFOT (section \ref{sectionRFOT}), one can imagine a ``mosaic'' of domains of characteristic size $\xi_D^*$. Assuming their interactions are weak, one can obtain a free energy density ~\cite{tarjus2000}

\begin{equation}
\frac{F(\xi_D,T)}{\xi_D^3}=\frac{\gamma(T)}{\xi_D}-\phi(T)+s_\mathrm{frust}(T)\xi_D^2
\end{equation}

\noindent where $\gamma$, $\phi$ and $s_\mathrm{frust}$ are renormalised with respect to the case of an isolated domain. Minimising leads to $\xi_D^*(T)\sim(\gamma/s_\mathrm{frust})^{1/3}$. In addition to $\xi_D^*(T)$, the correlation length associated with the critical point in the unfrustrated system, $\xi_D^0(T)$ may also influence the lengthscale of domains of LFS, in the case that criticality is narrowly avoided, or that frustration is weak. Note that decreasing temperature below $T_c$, $\xi_D^0(T)$ should decrease while $\xi_D^*(T)$ increases.

Now structural relaxation requires the re-arrangement of the domains comprising the mosaic, and the associated energy should scale as $\Delta E^*(T) \sim \gamma(T)\xi_D^*(T)^2$. If we consider Eq. \ref{eqCNTFrustration}, in the mean field regime, the surface tension should vanish as $\gamma(T) \sim (\xi_D^0)^{-2}\sim T_c-T$  is approached in analogy with liquid-gas critical phenomena ~\cite{onuki}. Thus

\begin{equation}
\Delta E^*(T) \sim (T_x^{c(0)}-T)^\psi
\end{equation}

\noindent where for the simple mean-field picture $\psi=3$. Including an Arrhenius contribution to the structural relaxation within the domains, around $T_c$ one expects a crossover to a super-Arrhenius dependence in the relaxation time as

\begin{equation}
\tau_\alpha(T)=\tau_\infty \exp \left( \frac{E^*(T)+E_\infty}{k_BT} \right)
\end{equation}

\noindent where $E=\infty$ is the Arrhenius contribution. The super-Arrhenius contribution then follows

\begin{equation}
\Delta E^*(T)=\begin{cases}
0 & \text{for}~T>T_c,\\
 k_BT_c \left(  1-\frac{T}{T_c}  \right)^\psi & \text{for}~T\leq T_c,\end{cases}.
\label{eqSuper}
\end{equation}

We close this summary of geometric frustration by emphasising that it makes explicit reference to experimentally (or computationally) measurable structural entities. In section ~\ref{sectionTowardsAStructuralMechanism}, we shall see that frustration appears to be strong, at least in the regime of supercooling accessible to computer simulation and particle-resolved experiments.

\subsection{Crystal-like and competing ordering}
\label{sectionCrystalLikeOrdering}

\begin{figure}[!htb]
\centering \includegraphics[width=70mm]{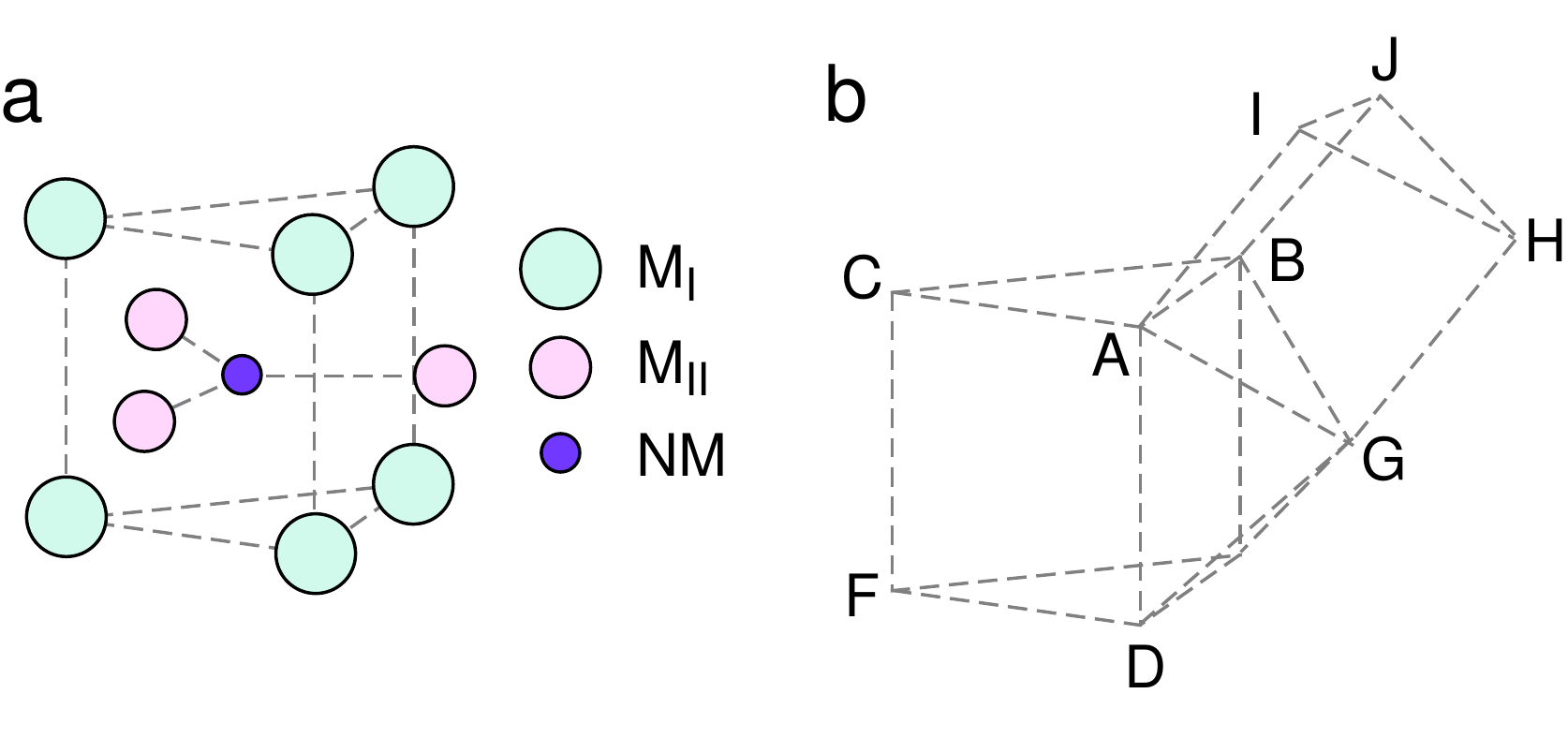} 
\caption{Gaskell's picture of local structural motifs in metallic glasses, with local structure of the crystal. Shown in (a) is the tricapped trigonal prism structure centred on a non-metal (NM) with metal atoms at the corners (M$_I$) and a further three next-nearest neighbours (M$_{II}$). Such motifs can connect together in a non-crystalline fashion to form the glass, as shown in (b), with M$_I$ atoms at the vertices indicated by the letters. Based on \cite{gaskell1978}.
\label{figGaskell}}
\end{figure}

A related, though distinct approach has been advanced by Gaskell ~\cite{gaskell1978,gaskell1979} and also more recently by Tanaka ~\cite{tanaka2005a,tanaka2005b,tanaka2005c,tanaka2011,tanaka2012}. Gaskell considered metallic glasses with one or more metallic species and a non-metal. In particular, in the case of Fe$_3$C the tricapped trigonal prism unit cell of the crystal shown in Fig. ~\ref{figGaskell}(a) can link together to form non-crystalline arrangements in the glass ~Fig. \ref{figGaskell}(b). Thus Gaskell placed a strong emphasis on the chemical nature of the system, as of course different materials have different unit cells. While Gaskell's approach certainly was appropriate for a number of materials, such as certain metal oxides ~\cite{gaskell1978,gaskell1979}, by virture of the emphasis placed on specific materials it was not intended as any kind of universal description. Indeed more recent developments show that some metallic glasses have amorphous order even at local lengthscales (see section ~\ref{sectionGlassformingAbilityMetallic}) ~\cite{sheng2006}.

\textit{Competing ordering : crystal versus amorphous  }
Using his two-order parameter model, Tanaka has placed more emphasis on model systems than the more specific materials considered by Gaskell ~\cite{tanaka2005a,tanaka2005b,tanaka2005c,tanaka2011,tanaka2012}. The two-order parameter model emphasises the relationship between competing forms of ordering, one of which being the underlying crystal order which competes with liquid order and other possible (amorphous) locally favoured structures. The crystal order may organise into regions of ``medium-range crystalline order'' (MRCO). These MRCO regions are distinct from the crystal in that there is no long-ranged translational order. The system is prevented from freezing by frustration, though this can be more general than the curved space scenario considered in section \ref{sectionGeometricFrustration}. At sufficient supercooling (perhaps below $T_{\mathrm{MCT}}$, see section ~\ref{sectionTowardsAStructuralMechanism}), the lengthscale of these regions of medium-range order should scale with the dynamic correlation length.

\begin{figure}[!htb]
\centering \includegraphics[width=45mm]{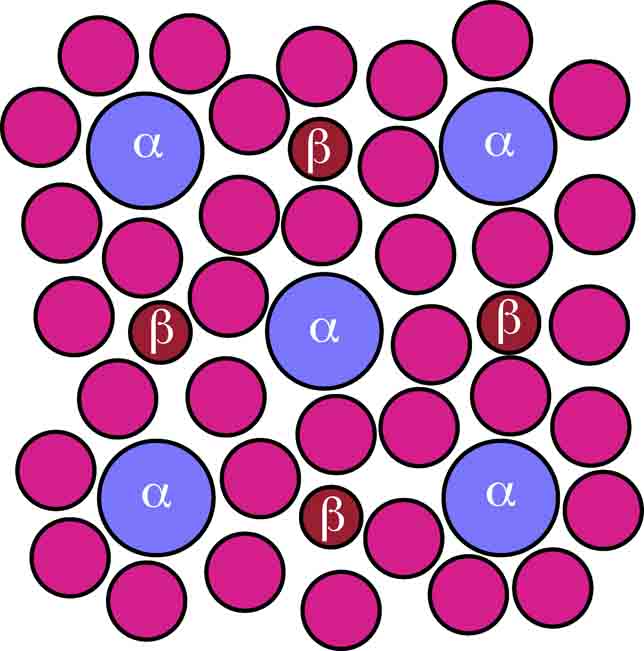} 
\caption{Miracle's local structural motifs in metallic glasses, with amorphous local structure and regular medium range order. Illustrations of portions of a single cluster unit cell for the dense cluster packing model. 2d representation of a dense cluster-packing structure in a (100) plane of clusters illustrating the features of interpenetrating clusters and efficient atomic packing around each solute ~\cite{miracle2004}.
\label{figMiracle}}
\end{figure}

More recently Miracle devised another model for structure particularly relevant to metallic glasses ~\cite{miracle2004,miracle2006}. In a sense, this turned Gaskell's and Tanaka's ideas on their head. Rather than local crystal-like order randomly arranged, Miracle proposed local amorphous order arranged at medium range (a few atomic spacings) in an ordered fashion as is illustrated schematically in Fig. \ref{figMiracle}. This was found to reproduce certain metallic glasses at the level of two-body structural correlations. See the discussion in section \ref{sectionTwoPointStructure} regarding two-point structural correlations. However, as visually appealing as Miracle's approach, for certain size ratios, it has been shown to be unstable to amorphous arrangements of the ordered particles, and thus its utility in describing the structure of glasses and supercooled liquids might be called into question \cite{alcaraz2008}.

\subsection{Quasispecies}
\label{quasispecies}

The quasispecies approach is a recently developed structural method for reconstructing the system in a tractable statistical mechanical model ~\cite{boue2009,lerner2009prl}. One decomposes the system into quasispecies (for example Voronoi polyhedra). These are then measured as a function of temperature. This enables construction of a predictive model of quasispecies populations, which is tractable under the significant assumption that the quasi species do not interact. The model can then predict the populations at all temperatures. Then one observes that certain (solid-like) quasispecies become poplar at low T and other liquid like ones become popular at high T. Remarkably, the model even predicts the potential energy of the original system with considerable accuracy ~\cite{boue2009}. A structural correlation length $\xi_\mathrm{qs}$ can then be constructed based on the distance between the reducing population of liquid-like quasispecies in the deeply supercooled state. This resulting correlation length can then correlated with the alpha relaxation time as $\tau_\alpha=\tau_0 \exp [\beta \mu \xi_\mathrm{qs}(T)] $ where $\mu$ is a constant whose value is around 0.3. The fits of this equation with numerical data are excellent.  The quasispecies picture can be related to dynamic facilitation in the sense that the population of liquid-like quasispecies (excitons in the language of facilitation) drops upon cooling. Like dynamic facilitation (see next section), the quasispecies picture of Procaccia and coworkers suggests no thermodynamic glass transition, just continuous dynamic slowing to T=0 ~\cite{boue2009,lerner2009prl}.

\vspace{0.8cm}

\subsection{Facilitation and dynamical phase transitions}
\label{sectionFacile}

Before concluding this section, we emphasise that the proposition that structure plays an important role in the glass transition is far from universally accepted. The dynamic facilitation approach posits that the glass transition is a predominantly dynamical phenomenon, and is not dependent upon a thermodynamic or structural component. Facilitation places much emphasis on the mobility of particles. It notes that motion in deeply supercooled liquids occurs through local events termed excitons  whose timescales are much shorter than the (overall) relaxation time. Upon cooling, these relaxation events become rarer, but remain essentially unchanged. Thus, an increasing dynamic lengthscale $\xi_\mathrm{fac}$ corresponds to larger separations between these events as their population falls. Within this picture there is no actual glass transition as such, as absolute zero is approached, the number of events drops to zero ~\cite{chandler2010}.

\begin{figure}[!htb]
\centering \includegraphics[width=70mm]{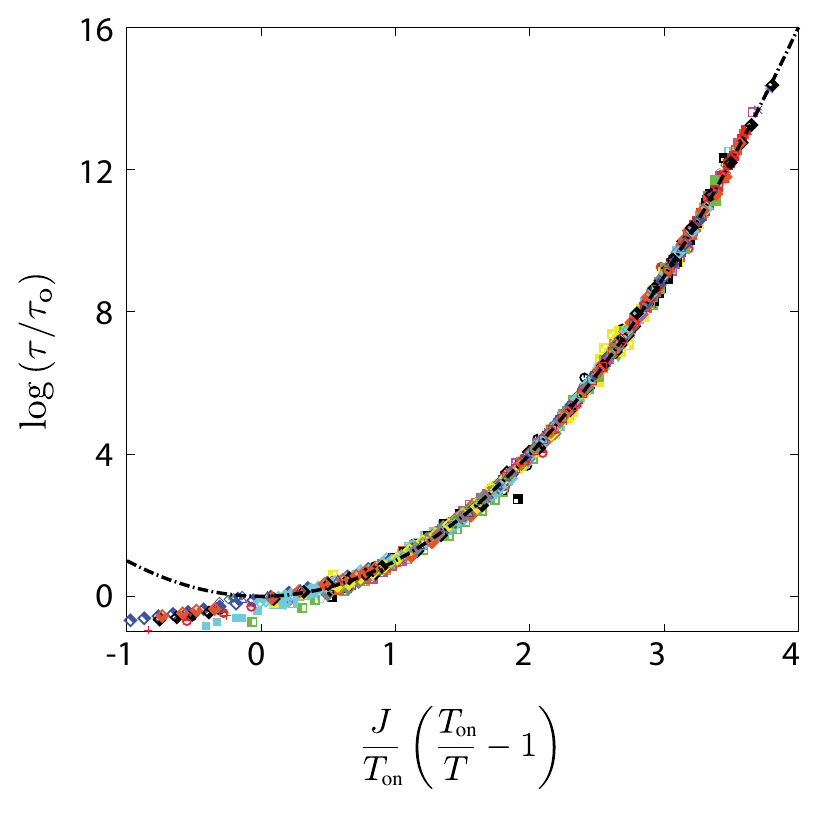} 
\caption{The Elmatad-Garrahan-Chandler form for the structural relaxation time. Equation ~\ref{eqElmatad} describes relaxation dynamics at least as well as VFT in the temperature regime of interest and collapses all glassformers onto a single curve (unlike the Angell plot, Fig. \ref{figAngell}) ~\cite{elmatad2009}.
Reprinted with permission from Y. Elmatad \emph{et . al. J. Phys. Chem. B} \textbf{113} 5563 (2009) Copyright {2009} American Chemical Society.
}
\label{figElmatadChandler} 
\end{figure}

Coupling between excitons is achieved through ``surging'' events, which are long-ranged, string-like motions of very small displacements (around $0.1 \sigma$). The motion is very often reversed, such that many surging events are required before another exciton is ``facilitated''. Through surging events, excitons are coupled to one another logarithmically such that the activation energy for relaxation follows $E_\mathrm{fac} \sim \log \xi_\mathrm{fac}$ ~\cite{chandler2010}. Now the Boltzmann factor suggests that the concentration of excitons goes like $c \sim \exp(- E_\mathrm{fac}/k_BT)$ so $\log c\propto 1/T$. The mean separation between excitons is $\xi_\mathrm{fac} \approx c^{-1/d}$. Thus to leading order the activation energy scales as $1/T$ and the timescale for relaxation $\log \tau_\alpha \sim E_\mathrm{fac}/k_BT\sim\exp(1/T^2)$. These arguments underlie the Elmatad-Garrahan-Chandler form for the relaxation time ~\cite{elmatad2009}

\begin{equation}
\log \tau_\alpha = \left( \frac{J}{T_\mathrm{on}}  \right)^2  \left(\frac{T_\mathrm{on}}{T}-1\right)^2
\label{eqElmatad}
\end{equation}

\noindent where $T_\mathrm{on}$ is the onset temperature for slow dynamics and $J$ is a parameter to scale the activation energy. A number of glassformers, with varying fragility, are shown in Fig. \ref{figElmatadChandler}. All collapse onto a single curve described by Eq. \ref{eqElmatad} which fits the data at least as well as the VFT fit (Eq. \ref{eqVFT}).

Facilitation is a mechanism by which kinetically constrained models relax. That these are often ideal gases indicates that facilitation can be decoupled from the thermodynamic aspects of the glass transition ~\cite{berthier2011} (there is no potential energy landscape, for example). Facilitation has been investigated in atomistic models ~\cite{keys2012}, where similar behaviour to the kinetically constrained models has been found. However, facilitation need not necessarily rule out a role for structure. One can enquire as to whether excitons have different local structure from the rest of the system, which may facilitate motion. Indeed it is hard to believe that if certain regions are predisposed to relax, such predisposition is not somehow encoded in the structure.

\begin{figure}[!htb]
\centering \includegraphics[width=80mm]{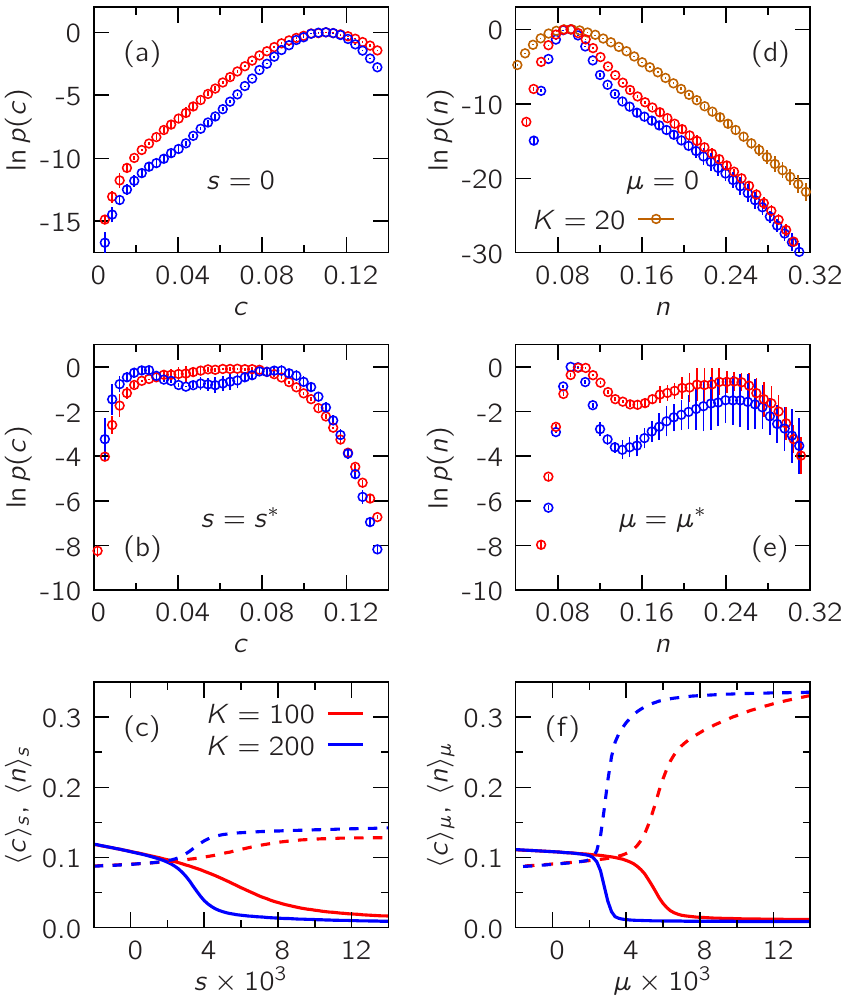} 
\caption{Phase transitions in trajectory space.
Left column: $s$-ensemble (a)~Probability distributions $p(c)$ for the density of mobile particles $c$ for two trajectory lengths. The non-concave shape indicates a phase transition in trajectory space as becomes obvious from the bimodal distribution~(b) at the field $s^\ast$ that maximises the fluctuations $\langle{c^2}_s\rangle-\langle{c}_s\rangle^2$. (c)~Average fractions of mobile particles (solid lines) and bicapped square antiprism cluster population (dashed lines) \textit{vs.} the biasing field $s$. Right column: (d-f)~as left column but for the $\mu$-ensemble (see section ~\ref{sectionMu}). Throughout, red and blue lines refer to $K=100$ and $K=200$, respectively. $K$ denotes the length of the trajectory. Here $K \approx 0.2 \tau_\alpha$ ~\cite{speck2012}.
\label{figSMu} }
\end{figure}

\textit{The $s$-ensemble --- }
Following developments with kinetically constrained models ~\cite{garrahan2007}, Chandler and Garrahan and coworkers were able to identify a phenomenon which may have significant implications for our understanding of the glass transition ~\cite{hedges2009}. They used small systems ($N\sim100$ particles) of the Kob-Andersen model to demonstrate a first-order \emph{dynamic} phase transition in trajectory space. To demystify this sentence, what is meant is that, upon weak supercooling, trajectories (of the entire system) of length a few $\tau_\alpha$ are analysed. The distribution of mobilities in trajectories which amounts to the sum of the mean-squared displacement, is not Gaussian, rather there are many more slow trajectories than expected as shown in Fig. ~\ref{figSMu}(a). Such ``fat tails'' in the distribution are akin to a thermodynamic transition in a small system (say the liquid-vapour transition) where the system samples configurations from both phases with some probability.

In this \emph{dynamical} ensemble, rather than density (liquid or vapour), it is the mobility $c$ of the trajectories which form the two ``phases''. The fraction of mobile particles, $c$, are those which have moved further than some criterion. The distribution in Fig. \ref{figSMu}(a) shows that low-mobility trajectories are much more likely than expected, and is taken as evidence for an ``inactive phase''. The implication is that dynamical heterogeneity is a manifestation of these two phases, fluctuating in and out of existence like density fluctuations in liquid-vapour critical phenomena. Histogram reweighting can be carried out, such that two peaks are found [Fig. \ref{figSMu}(b)], which represent the two phases. This reweighting is achieved through the application of the $s$ field (and the ensemble is thus termed the $s$-ensemble), where trajectories are biased towards those with slow dynamics. In other words, the dynamical transition occurs under non-equilibrium conditions, and whether it occurs in the $s=0$ case relevant to experiments remains unclear, although there is some evidence supporting an $s=0$ transition in certain spin-glasses ~\cite{jack2010}.

\section{Identifying structure in amorphous systems}
\label{sectionStructure}

\begin{figure*}
\centering
\includegraphics[width=120mm]{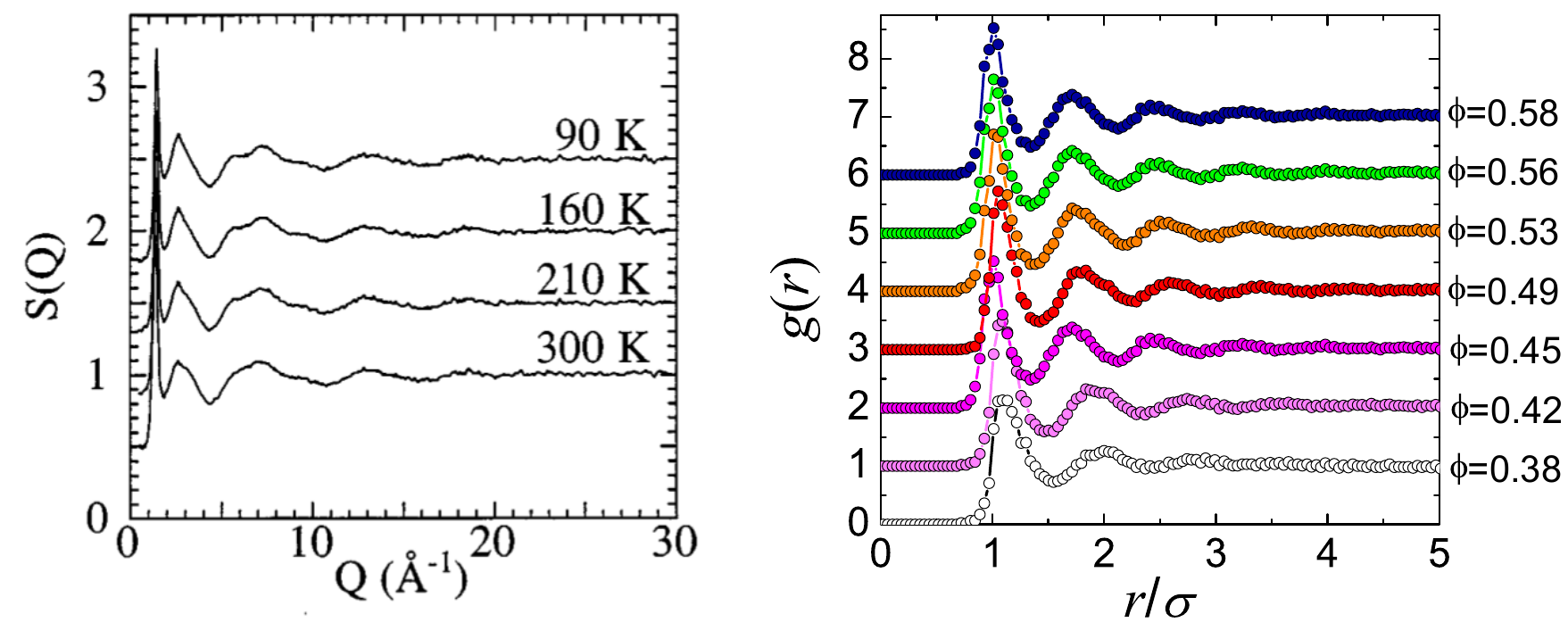} 
\caption{Two-point measures of structure. (a) The static structure factor $S(k)$ measured in propylene glycol, where $T_g \approx 160 $ K ~\cite{leheny1996}
Reproduced with permission from American Institute of Physics Copyright 1996. (b) The complete radial distribution function $g(r)$ measured in a suspension of colloidal ``hard'' spheres. Indicated is the \emph{effective} volume fraction ~\cite{royall2013myth}.
\label{figG}}
\end{figure*}

\subsection{Two-point correlation functions}
\label{sectionTwoPointStructure}

We now review methods to measure and interpret structure in amorphous systems. The principle means of studying structure in atomic and molecular systems is by scattering of x-rays or neutrons, which enables the two-point correlation function, the static structure factor $S(k)$ to be measured. Its real-space counterpart $g(r)$ may be obtained in real space from \textit{particle-resolved} systems, where the coordinates of the constituent particles are known. This is the case for the ``ball-bearing'' models of liquids in the pioneering work of Bernal ~\cite{bernal1959,bernal1960}, Scott ~\cite{scott1960}, Finney  ~\cite{finney1970} and Bennett ~\cite{bennett1972}. Of course for such large particles, $k_BT$ is meaninglessly small and thus the structures formed are in no sense equilibrated. On the other hand colloidal systems obey the laws of statistical mechanics ~\cite{ivlev}, and although it must be driven and is thus far from equilibrium, nonetheless granular matter can exhibit ``thermal'' behaviour \cite{candelier2010,berthier2013nphys}. One further approach which delivers fully particle-resolved data is of course molecular simulation. Here we focus on some more recently developed and popular measures and experimental techniques rather than attempting to provide an exhaustive list. Other measures may be found for example in ~\cite{torquato2010}.

\begin{figure}[!htb]
\centering \includegraphics[width=80mm]{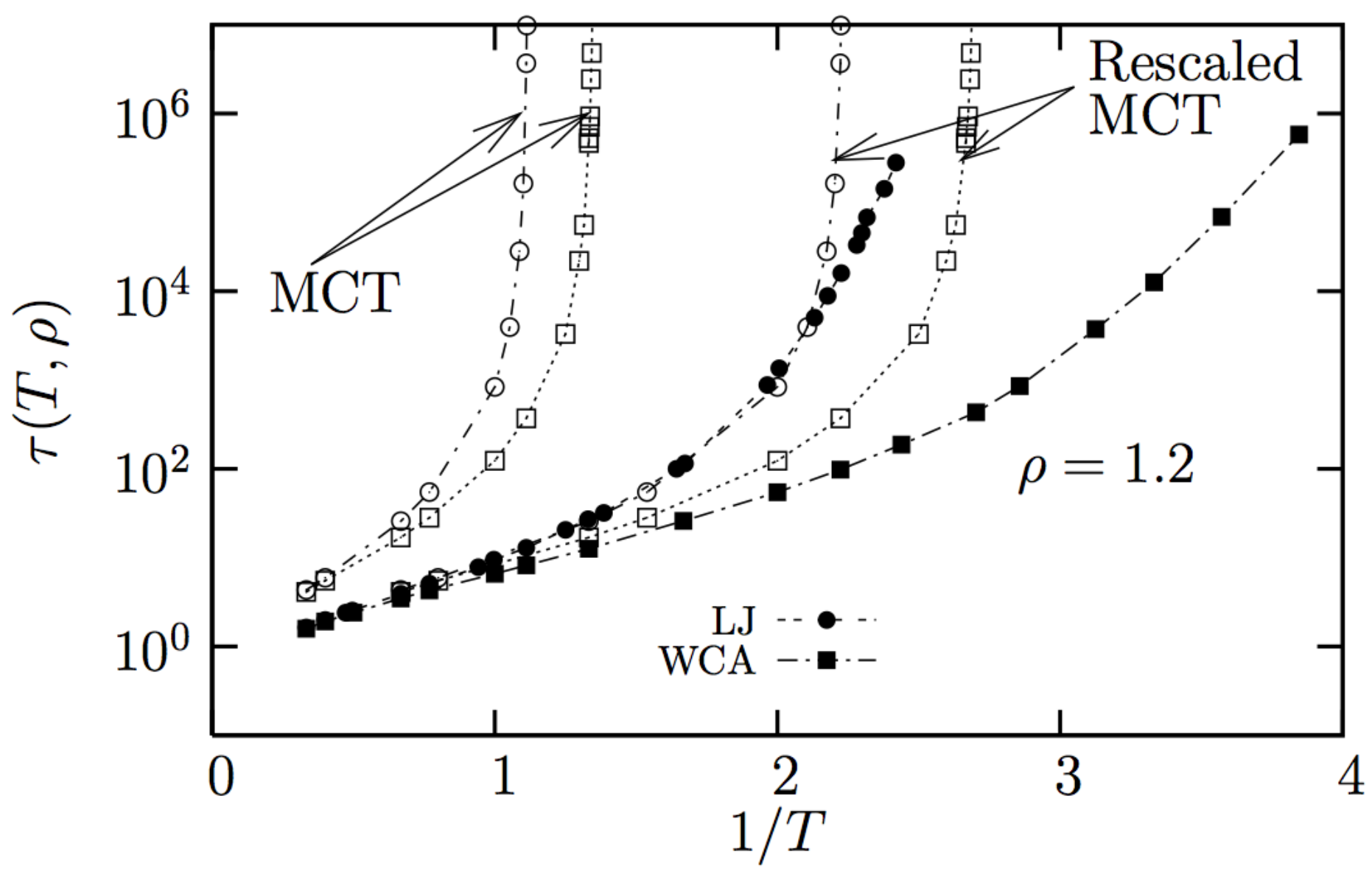} 
\caption{Comparison of the relaxation times obtained for the LJ and WCA models. Note that WCA is ``faster''. MCT predicts this speedup in a qualitative fashion (but fails quantitatively)~\cite{berthier2010pre}. Reproduced with permission from American Physical Society Copyright 2010.}
\label{figLudovicGilles} 
\end{figure}

An example of $S(k)$ is shown in Fig. \ref{figG}(a) for deuterated propylene glycol, a molecular glassformer. Here it is clear that the two-point structure measured by neutron scattering shows very little change going through its glass transition ($T_g=160$ K) ~\cite{leheny1996}. However, although there is little change in $S(k)$, that does not mean there is \emph{no} change. Indeed, since mode coupling theory takes $S(k)$ as its input, there \emph{must} be some change and that change, given the considerable success of MCT in predicting the initial stages of dynamical arrest, should somehow be linked to the dynamics. Indeed, MCT even predicts, qualitatively, the dynamic speedup upon removing the attractive part of the Lennard-Jones interaction (Fig. \ref{figLudovicGilles}) ~\cite{berthier2010pre}. The fact that removing the attraction from the Lennard-Jones model even leads to any measurable changes in dynamics at all is surprising. This is because attraction-less Lennard-Jones, better known as the Weeks-Chandler-Andersen (WCA) model ~\cite{weeks1971}, is notable for reproducing many properties, including the two-body structure, of the full Lennard-Jones liquid with remarkable accuracy. However, although qualitatively in the right direction (WCA is faster than Lennard-Jones), MCT fails quantitatively to predict the change in $T_\mathrm{MCT}$ with density. Other microscopic theories also exhibit significant shortcomings \cite{berthier2012epje}. Regardless of the limitations of MCT in reproducing some aspects of the dynamical behaviour (see section ~\ref{sectionMCT}), it is clear that looking at the two-point $S(k)$ or $g(r)$ is a challenging (though fruitful, see section \ref{section2PointDetail} ~\cite{salmon2013}) way to tackle the question of how structure changes in the liquid to glass transition because the two-point structural measures change so little. However, one clear feature in $g(r)$ that is common as a liquid is cooled to a glass is the emergence of a split second peak ~\cite{rahman1976,hiwatari1980}. There are also occasions where the pair correlation functions can reveal significant effects in structure. One such is the re-entrant melting of soft sphere glasses. Usually, upon increasing density, $g(r)$ becomes more sharply peaked. Certain classes of soft systems, harmonic spheres in this case, actually show a drop in the $g(r)$ peak, which is related to particles overlapping at sufficient density. This drop in the peak is accompanied by a strong increase in mobility, \emph{i.e.} melting ~\cite{berthier2010presoft}.

Another two-body property which may be obtained from two-point correlation functions is the so-called structural entropy ~\cite{truskett2000}:

\begin{equation}
s_2 = - \frac{k_BT \rho}{2} \int_0^\infty dr \left[ g(r) \ln g(r) - \{g(r)-1\} \right].
\label{eqS2}
\end{equation}

Using a slightly modified version, Truskett \emph{et al.} showed that confined hard spheres lost structural entropy upon approaching the glass transition ~\cite{mittal2006}, and that the change was markedly stronger than other two-point structural measures such as $g(r)$ ~\cite{mittal2006,krekelberg2007}. It is then possible to relate this to such entropic measures to theories such as Adam-Gibbs or RFOT (sections ~\ref{sectionAdamGibbs} and ~\ref{sectionRFOT}) and also to obtain structural correlation lengths ~\cite{tanaka2010,dunleavy2012,leocmach2013}, but their interpretation is not always straightforward ~\cite{dunleavy2012}.

\begin{figure}
\centering
\includegraphics[width=80mm]{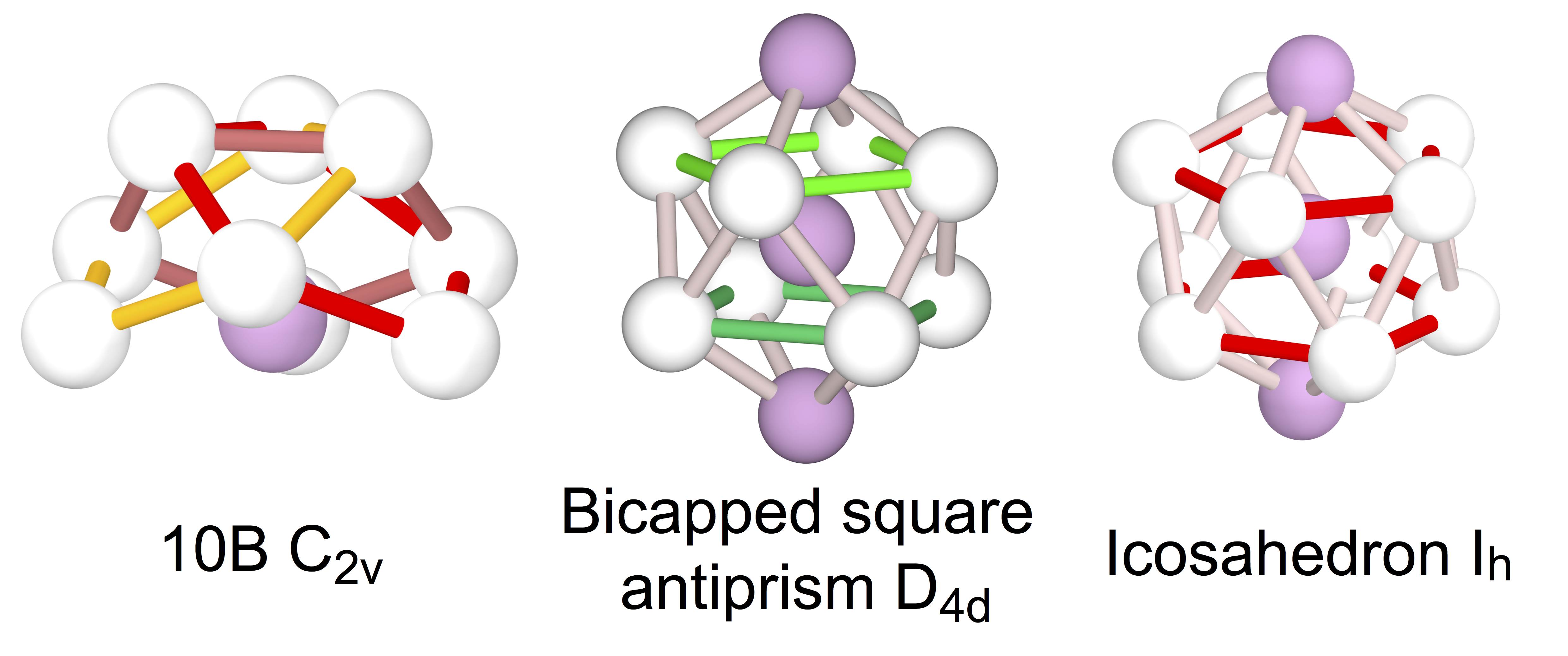} 
\caption{Popular structural motifs in glass forming systems and their point group symmetries. The ``10B'' cluster is found often in hard sphere systems ~\cite{royall2014,taffs2013} and colloidal gels ~\cite{royall2012}. The bicapped square antiprism is found in the Kob-Andersen Lennard-Jones model (section ~\ref{sectionCommonModelSystems}) ~\cite{coslovich2007,malins2013fara}. The  icosahedron found in the Wahnstr\"{o}m Lennard-Jones model ~\cite{malins2013jcp,coslovich2007}.
\label{figFavourite}}
\end{figure}

We now detail some popular higher-order measures for structure in amorphous systems, focussing on particle-resolved approaches. Then in sections ~\ref{section2PointDetail}, ~\ref{sectionHigherOrderReciprocal} and ~\ref{sectionLudovicLength} we discuss ways in which higher-order data may be extracted from reciprocal space data appropriate to molecular systems. The next highest order from pair correlations is three body correlations.  While these show stronger changes in structure approaching the glass transition ~\cite{coslovich2013}, see also section ~\ref{sectionHigherOrderReciprocal}, here we will focus on still higher-order correlations, specifically those intended to extract structural motifs such as icosahedra. Before proceeding we note three popular structural motifs which are illustrated in Fig. \ref{figFavourite}. The ``10B'' cluster, bicapped square antiprism are found in hard spheres  ~\cite{royall2014,taffs2013}, the Kob-Andersen model  ~\cite{coslovich2007,malins2013fara} and the Wahnstr\"{o}m model  ~\cite{malins2013jcp,coslovich2007} respectively. Despite their differences, these structures are related : the icosahedron is the ``canonical'' local structure predicted by Frank (section ~\ref{sectionFrank}) which minimises the interaction energy for 13 particles if a variety of systems ~\cite{doye1995}. The 10B is actually an icosahedron missing three particles and thus may be thought of as an imperfect icosahedron. The bicapped square antiprism may be thought of as an icosahedron with two  four- rather than five-membered rings. As such the shell is therefore tighter and this structure can form in the case of non-additive interactions, as is the case of the Kob-Andersen model and certain metallic systems discussed in section \ref{structureInMetallicGlassformers}.

\subsection{Voronoi polyhedra}
\label{sectionVoronoi}

Section \ref{sectionTwoPointStructure} makes it clear that something more than the two-point structural measures --- $S(k)$ and $g(r)$ --- is desirable to fully understand the change in structure as a liquid is cooled towards a glass. For example, we should like to be able to directly count the icosahedra predicted by Frank. Clearly, counting icosahedra requires a higher-order structural analysis than the two-point $S(k)$ or $g(r)$. This is possible using the Delauney triangulation to connect the centres of spheres and the subsequent Voronoi decomposition which divides space into polyhedra such that each polyhedron encloses the volume around each particle which is closer to that particle than any other.

Such Voronoi polyhedra were identified in the context of liquids by Bernal \cite{bernal1959,bernal1960}, and Finney in ``ball-bearing'' models \cite{finney1970} and in computer simulation \cite{finney1970mc}. Both Bernal and Finney observed a tendency for five-membered rings, i.e. pentagonal faces of the Voronoi polyhedra. Bernal also emphasised the nature of the holes in the centres of the Voronoi polyhedra. The differing shapes of polyhedra gave rise to holes of differing size, with tetrahedra featuring the smallest holes. Tanemura \emph{et al.} ~\cite{tanemura1977} categorised the faces of the Voronoi polyhedra as a sequence of numbers counting the vertices of polyhedra of 3, 4, 5...sides. Now, as shown in Fig. \ref{figBaka}(d) the icosahedron is expressed as [20,0,0] for its faces are comprised of 20 triangles. However, the Voronoi polyhedron containing the central particle has 12 faces (one for each neighbour) which are pentagonal, thus an icosahedral environment is expressed as [0,0,12].

One point to note is how perfect a polyhedron need be. The Voronoi construction provides robust topological criteria for the identification of certain structures. However in dense assemblies of hard spheres for example (almost) every particle will be identified as the the vertex of a tetrahedron. By considering \emph{regular} tetrahedra it is possible to discriminate even the simplex and we find an increase in regular tetrahedra with packing density \cite{anikeenko2007,anikeenko2008}. The ``perfectness'' of the tetrahedra was used as a measure of configurational entropy (see sections  \ref{sectionAdamGibbs},  \ref{sectionRFOT} and \ref{sectionEvaluatingTheEnergyLandscape}).

\subsection{Bond-orientational order parameters}
\label{sectionBOO}

An alternative to the Voronoi polyhedra was introduced by Steinhardt  \emph{et al.} in 1983 ~\cite{steinhardt1983} in the form of bond-orientational order parameters. These decompose a particle's bond symmetry onto spherical harmonics $Y_{lm}$ :

\begin{equation}
q_{lm}\equiv\frac{1}{Z_j}\sum_{j=1}^{N_{b}(i)}Y_{lm}(\mathbf{r_{ij}}).
\end{equation}

\noindent where $Z_j$ is the number of bonds (coordination number) of particle $j$. We discuss ways to define bonds in section \ref{sectionStrategies}.

In isotropic systems, one needs rotationally invariant measures

\begin{equation}
Q_{l}=\frac{4 \pi}{2(l+1)} \Sigma_{m=-l}^l | \langle q_{lm} \rangle |^2
\label{eqQ}
\end{equation}

\noindent and

\begin{equation}
W_{l}\equiv\sum_{m_{1},m_{2},m_{3},m_{1}+m_{2}+m_{3}=0}
\left(\begin{array}{ccc}
l & l & l\\
m_{1} & m_{2} & m_{3}
\end{array}\right)\bar{q}_{lm_{1}}\bar{q}_{lm_{2}}\bar{q}_{lm_{3}}
\label{eqBond}
\end{equation}
\noindent where the coefficients 

\[
\left(\begin{array}{ccc}
l & l & l\\
m_{1} & m_{2} & m_{3}\end{array}\right)
\]

\noindent are the Wigner 3J symbols.

Steinhardt \emph{et al.} \cite{steinhardt1983} used $W_6$ to produce a kind of ``order spectroscopy'' to distinguish icosahedra ($W_6=-0.18$) and FCC or HCP crystalline geometries ($W_6 \approx 0$). Bond orientational order (BOO) parameters thus provide a means to investigate how ``icosahedral'' a supercooled liquid is, as discussed in section \ref{sectionEarlyMeasurements}. Lechner and Dellago ~\cite{lechner2009} have recently emphasised the importance of considering the environment of neighbouring particles and including multiple values of $l$ in the use of BOO to identify particular local structures.

In 2D, BOO parameters may be defined in an analogous fashion. Since local order is typically hexagonal, an often-used BOO is $\psi_6$ defined as follows:

\begin{equation}
\psi_6 = \left| \frac{1}{Z_j} \sum_{m=1}^{z_j} \exp \left(i 6 \theta_m^j \right) \right|
\label{eqPsi6}
\end{equation}

\noindent where $Z_j$ is again the co-ordination number and $\theta_{m}^{j}$ is the angle made by the bond between particle $j$ and its $m \mathrm{th}$ neighbour and a reference axis. For perfect hexagonal ordering, $\psi_6=1$, while disordered but locally hexagonal systems the orientations to the reference axis of each particle add incoherently such that the sum of $\psi_6$ for the whole system tends to zero. This is analogous to the role of the rotational invariant $W_{l}$ as opposed to the $Q_l$ 3d BOO parameters.

\subsection{The common-neighbour analysis}
\label{sectionCNA}

Honeycutt and Andersen introduced another development in structural analysis of amorphous systems in the form of the common neighbour analysis (CNA) ~\cite{honeycutt1987}. Here one considers the number of neighbours common to two particles. A common motif is the pentagonal bipyramid [Fig. ~\ref{figBaka}(a) and (b)]. In the CNA nomenclature, this is 1551. The first 1 specifies that there is one bond between the two ``spindle'' particles, the first 5 refers to the 5 common neighbours between the two spindles and the second 5 is the number of bonds between the neighbours.

\subsection{The topological cluster classification}
\label{sectionTCC}

\begin{figure*}[!htb]
\centering \includegraphics[width=150mm]{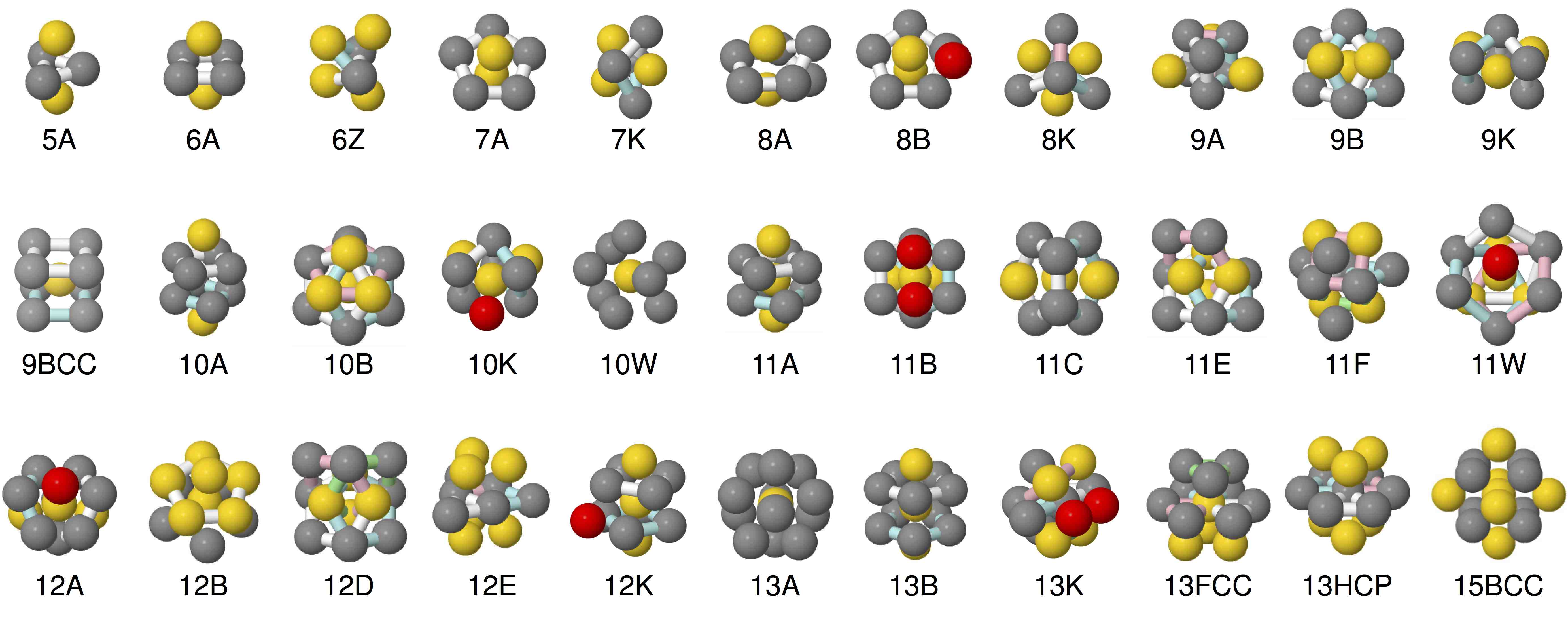} 
\caption{Some of the structures detected by the TCC. Letters correspond to different models, numbers to the number of atoms in the cluster. K is the Kob-Andersen model, W is the Wahnstr\"{o}m model \cite{malins2013jcp}. Other letters correspond to the variable-ranged Morse potential, letters at the start of the alphabet are long-ranged interactions, letters at the end are short-ranged interactions, following Doye ~\emph{et al.} ~\cite{doye1995}. Also shown are common crystal structures.}
\label{figTCC} 
\end{figure*}

Frank's conjecture that icosahedra should be formed in the Lennard-Jones model, because this structure minimises the potential energy for 13 atoms was extended in the topological cluster classification (TCC) ~\cite{williams2007,royall2008,malins2013tcc}.  In other words, rather than considering only the $m=13$ icosahedron for the monodisperse Lennard-Jones model, the TCC considers all clusters of size $m \leq 13$ particles in a variety of model systems. Unlike the methods outlined above, which are purely geometric in nature, the TCC considers the ground state clusters of the system. In the TCC bonds are identified, for example by using a Voronoi construction. The TCC detects all 3, 4 and 5 membered shortest-path rings within the network of bonded particles, and from these the clusters which minimise the potential energy for a variety of model systems  ~\cite{doye1995,doye1996science,doye2005} are identified. These clusters are shown in Fig. \ref{figTCC}.

The main assumption the TCC makes is that isolated clusters are relevant for bulk systems. Although this need not be the case, Mossa and Tarjus ~\cite{mossa2003} investigated the effect that immersion in a bulk mean-field Lennard-Jones liquid had on the energy landscape of 13 Lennard-Jones atoms. The icosahedron remained the locally favoured structure, but the difference in potential energy with respect to the crystal was less than that of an isolated cluster.

The TCC typically identifies a number of different clusters illustrated in Fig. ~\ref{figTCC} in a given system. In dense liquids, almost all particles are part of one cluster or another, thus it is desirable to identify those clusters which might be related to slow dynamics. By measuring the cluster lifetime, with the \emph{dynamic} TCC it is possible to identify long-lived clusters which can then be correlated with regions of low mobility (i.e. with dynamic heterogeneity) ~\cite{royall2014,malins2013jcp,malins2013fara}.

\subsection{Strategies for identifying a bond network}
\label{sectionStrategies}

The above methods, Voronoi polyhedra, bond-orientational order parameters, the common neighbour analysis and the topological cluster classification are based on the identification of a bond network between the particles. Here we briefly discuss strategies by which this bond network may be identified. In the case of the TCC the reader is directed to ~\cite{malins2013tcc} in which a detailed discussion is presented.

There are two common methods, one is a Voronoi construction and the other is to set a bond length and consider particles closer than the bond length to be bonded. In the latter, so-called \emph{simple bond} approach, clearly whether particles are bonded depends on the bond length. Typically one aims to consider all particles in the first neighbour shell as bonded, thus the bond length is typically set to be a little larger than the first maximum on the radial distribution function, and the first minimum is often chosen. Of these two methods, the Voronoi construction is better defined and leads to a unique bond network under certain criteria, on the other hand in the case of the simple bond the network is clearly dependent on the bond length. In practise this often makes little difference. One can combine the Voronoi method with a simple bond, in particular to define a maximum bond length. This can be very useful in the case of colloidal gels, for example, where particles far apart on opposite sides of a void for example would be considered to be bonded. In dense liquids as well the use of a maximum bond length can be useful to accelerate the computational time to analyse the data. We have found that there are no bonds longer than $1.8\sigma$ in a monodisperse Lennard-Jones liquid and that there are very few longer than $1.4\sigma$ ~\cite{malins2013tcc}.

\vspace{10 mm}

\subsection{Order-agnostic approaches}
\label{sectionOrderAgnostic}

All the methods discussed so far have some specific structure in mind. However, in general one does not know the locally favoured structure \emph{a priori}, and these are only known in a few systems in any case. The dynamic TCC outlined in the previous section provides one means to identify structures relevant to dynamic arrest from a pre-specified set of clusters (Fig. \ref{figTCC}). Other more general ``order-agnostic'' methods have been developed, which rely not on a specific structure but on generic properties. Often these methods amount to extracting a structural lengthscale. Kurchan and Levine ~\cite{kurchan2009} used the Renyi entropy to identify a lengthscale where the structure changes from non-extensive (regime of local order) to extensive. Dunleavy \emph{et al.} ~\cite{dunleavy2012} used mutual information to determine structural lengthscales in 2D glassformers, Cammarota and Biroli used related methods in a lattice (placquette) model ~\cite{cammarota2012}. Mutual information is an information theoretic quantity based on the Shannon entropy $H$~\cite{shannon1948}. The Shannon entropy of a random variable $X$ with a probability distribution $p(x)$ over a support (here the support is the configurational phase space) $\mathcal{X}$ is given by $H(X) = - \sum_{x \in \mathcal{X}} p(x) \log_2 p(x)$. This quantity is larger for a uniform probability distribution over a broad support and smaller when the distribution becomes more peaked. It is a measure of the uncertainty of the outcome of drawing a sample from the distribution.

Mutual information can be used to investigate how the structure in one part of a system affects the structure in another. In other words, it is a general means of accessing structural correlations. If the structure, $X$, in some part of the system influences that in another part, $Y$, then it will be the case that when $X$ is held constant the range of possible values of $Y$ is smaller than when $X$ can take any value. This reduction in uncertainty can be quantified by treating configurations as random variables and taking the mutual information \cite{cammarota2012,dunleavy2012,cover1991}. The mutual information between two random variables measures the entropy difference between the marginal probability distribution of a variable, and its conditional distribution.

\begin{align}
I(X;Y) &= H(X) - H(X|Y) = H(Y) - H(Y|X) \\
&= H(X) + H(Y) - H(X,Y) \label{eqWeuse}\\ 
&= \sum_{x \in \mathcal{X}, y \in \mathcal{Y}} p(x,y) \log_2 \frac{p(x,y)}{p(x)p(y)} \label{eqMutual}
\end{align}

The mutual information between two regions can then be computed as a function of separation. In amorphous systems the mutual information decays as a function of distance, and the rate of this decay enables a structural correlation which requires no \emph{a priori} knowledge of the structure to be computed ~\cite{dunleavy2012}.

Another approach is to analyse a network (for example the bond network) with graph theoretic methods. This method seeks to divide the network into an ideal gas of non-interacting isolated communities. Multiple lengthscales are considered, and lengthscales which minimise interactions between communities lead to the identification of clusters. These clusters are not specified beforehand, thus the method may be regarded as order-agnostic \cite{ronhovde2012}.  Alternatively, one may consider the local structure around each particle and seek a measure of the typical structure \cite{sausset2011,fang2011}. This local structural similarity can expressed as ``patches'' from which a ``patch correlation length'' can be defined. This lengthscale has been shown to exhibit similar behaviour to other lengthscales in the system \cite{sausset2011}.

Pinning, or point-to-set measures as they are also known, can similarly be used to extract structural lengthscales ~\cite{hocky2012,kob2011non} (see section \ref{sectionRFOT}). However, these pinning techniques are ``invasive'', in that particles must be pinned, \emph{i.e.} immobilised.

A further method is to introduce an external, static perturbation in the form of an affine deformation of coordinate data ~\cite{mosayebi2010,mosayebi2012}. 
A configuration may be affinely deformed, and both non-deformed and deformed configurations are subjected to a steepest descent energy quench to yield an inherent structure. Displacement vectors between the two inherent structures are defined. The displacement vector for the $j$th particle is then $\mathbf{d}_j=\mathbf{r}_j^{\mathrm{deform}}-\mathbf{E}\cdot\mathbf{r}_j^{\mathrm{0}}$ where the $0$ denotes the undeformed case and $E=\mathbf{1} + \gamma \mathbf{e}_1 \mathbf{e}_2$. Here $\mathbf{e}$ denotes a Cartesian unit vector and $\gamma$ is the amplitude of the deformation. Averaging over neighbours then yields regions with similar displacement vectors which are then structurally correlated in the sense that they are ``relatively solid'' because across the region the response to the deformation is coupled. Clearly the quantity extracted here is structural because it depends only upon configurational information. Its nature is not as clear as the structures identified by Voronoi polyhedra, for example. We see that ``something'' in the structure makes the material ``more solid'' and it is possible to extract an associated structural lengthscale with this  ~\cite{mosayebi2010,mosayebi2012}.

\begin{figure}[!htb]
\centering \includegraphics[width=80mm]{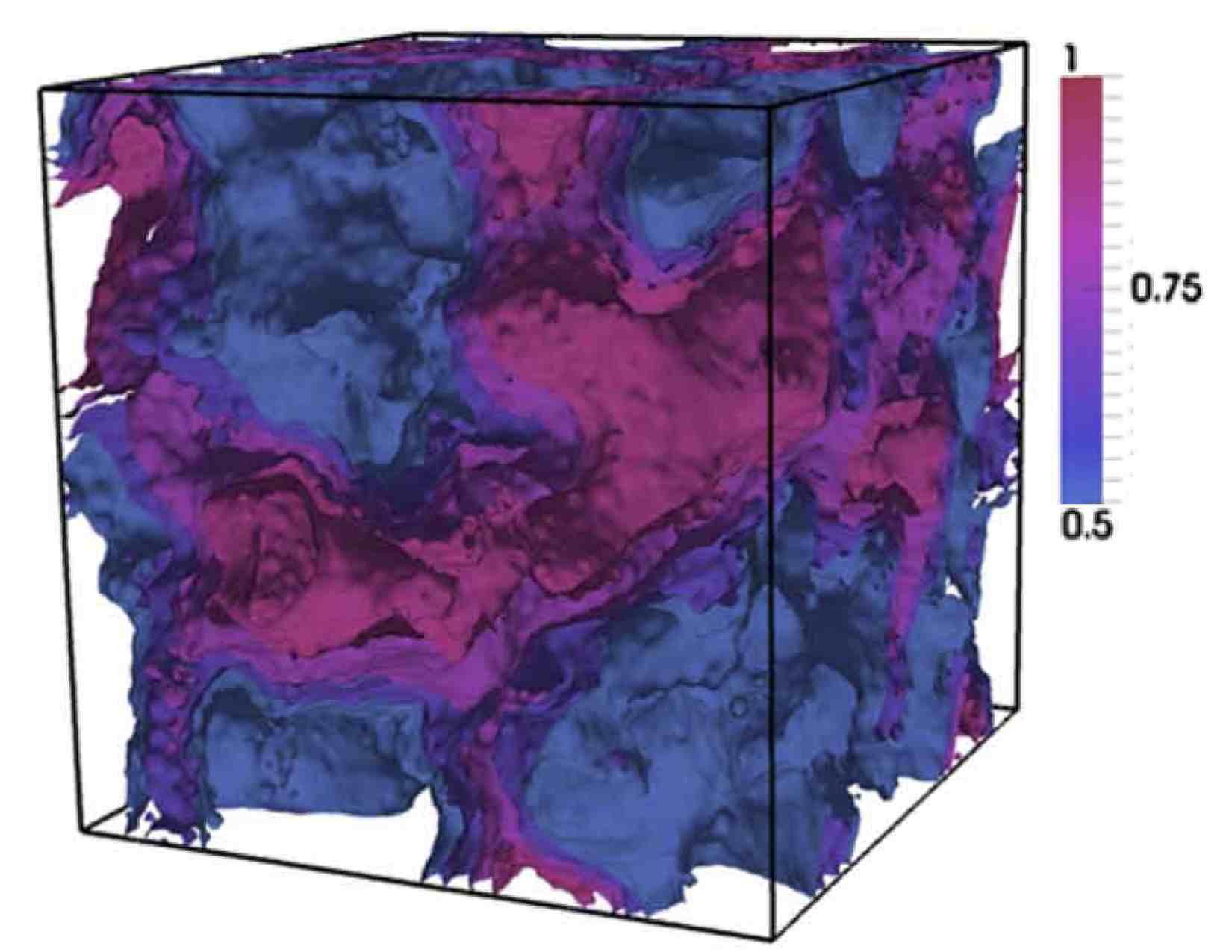} 
\caption{Regions of similar displacement vector (as indicated by the colour bar right)
in the deeply supercooled Kob-Andersen mixture. This provides a static lengthscale found to increase significantly with supercooling.
\cite{mosayebi2010}.  Reproduced with permission from American Physical Society Copyright 2010.}
\label{figMosayebi} 
\end{figure}

\subsection{Details in two-point correlation functions : analysis of reciprocal space data}
\label{section2PointDetail}

The methods discussed in sections ~\ref{sectionVoronoi}-\ref{sectionTCC}, based on particle-resolved studies, have the attraction that such data delivers a remarkable degree of information. However, these approaches are restricted to model systems such as colloids ~\cite{ivlev}, and computer simulations. For atomic and molecular systems in the supercooled liquid state, scattering of x-rays and neutrons can access the relevant lengthscales (Fig. ~\ref{figG}). Using isotope substitution, with neutron scattering it is possible to ``highlight'' the correlations in one particular species. Salmon and Zeidler have recently published a review of reciprocal space techniques, particularly isotope substitution ~\cite{salmon2013}. Of course, such an approach is also open to particle-resolved studies, in particular the correlations between the minority species $g_{BB}(r)$, have been shown to show a comparatively large change in the Kob-Andersen model using the $s$-ensemble to access very deeply quenched glassy states (see section \ref{sectionFacile}) ~\cite{speck2012jcp}.

Combined with intuition as to the nature of structures the system is likely to form upon cooling (tetrahedra for example ~\cite{salmon2013}), analysis of two-point correlation functions can reveal whether those tetrahedra share edges or corners. Of course, this kind of approach requires some prior knowledge of the likely structural motifs, which is typically itself gained in real space (simulation or models). Another approach is to calculate the static structure factor $S(k)$ for some expected structural motifs, and to assume that the measured $S(k)$ is a superposition of these structures~\cite{wette2009}. Related to this is the identification of the first sharp diffraction peak in $S(k)$. If this is significantly less than the wavevector associated with the atomic size ($\sim 2 \pi \sigma^{-1}$), then the first sharp diffraction peak provides evidence for structure on longer lengthscales, i.e. medium range ordering. Such approaches are often associated with more complex glassforming systems than the simple models outlined in section ~\ref{sectionCommonModelSystems} such as oxides and metallic glassformers ~\cite{salmon2013}.

A further powerful (if brave) approach is to combine two-point structural correlations with computer simulation through ``reverse Monte Carlo'' or RMC ~\cite{mcgreevy2001,sheng2006,evans1990}. Here configurations in computer simulations, often with a simple model interaction, are successively refined by comparison with two-point (or in principle higher-order) experimental data. Thus one can ``extract'' particle-level data from atomic and molecular systems. As rich as such data is, enabling the kind of analysis otherwise available only to simulation and colloid experiment, one should not overlook that fact that the ``true'' experimental data is typically of a two-body level. In principle, this only determines the system to a two-body level (there is no more information) leaving higher-order correlations underdetermined ~\cite{hansen}, moreover experimental uncertainties lead to limitations in the accuracy with which even this two-body interpretation can be made. In short, there are many higher-order correlations which give two-body correlations which are indistinguishable, as recently explicitly demonstrated ~\cite{coslovich2013,malins2013isomorph}.

\textit{A cautionary tale --- } 
That reciprocal space methods only measure the intensity of the scattered radiation and not the phase can lead to multiple interpretations of real space data. Additionally a two body distribution analysis can result in the same underdetermined outcome.  For example, the relationship between the wavevector of the first sharp diffraction peak $k_\mathrm{FSDP}$ and atomic volume $v_a \sim \sigma^3$ are related by $k_\mathrm{FSDP} v_a^\eta=C$ where $C$ is a constant. In crystalline solids, $\eta=1/3$. The finding that $\eta=0.433$ in a range of metallic glasses was interpreted as the fractal nature of their structure ~\cite{ma2009}. However it was later shown via a pole analysis which describes decay of two-point correlation functions at large distances ~\cite{evans1994,dijkstra2000pole}, that such scaling was consistent with a more conventional non-fractal structure ~\cite{chirawatkul2011}. A second example of the limited accuracy of two-point correlation functions to describe structure in amorphous materials is the work of Treacy and Borisenko ~\cite{treacy2012}, who used fluctuation electron microscopy (see section ~\ref{sectionFEM}) to resolve higher-order structure in silica. Although silica is often thought of as a random network, their work showed some small nanocrystalline regions which were not apparent in bulk, averaged, two-point data.

\subsection{Higher-order structure from reciprocal space data}
\label{sectionHigherOrderReciprocal}

Beyond the analysis of two-point correlation functions, higher order measures of structure in atomic and molecular glassformers have been developed, though may require special analysis or sample geometry. Total internal reflection of x-rays was used to create evanescent rays in liquid lead close to an interface. The in-plane scattering of the evanescent rays in the liquid gives higher order structural information, and was used to infer five-fold local symmetry in liquid lead ~\cite{reichert2000} and more recently in AuSi ~\cite{schulli2010}. However computer simulation of hard spheres by a wall ~\cite{heni2002}  found no evidence for five-fold symmetry, inferred through bond-orientational order parameters, unless a strong (unphysical) field was applied to artificially increase the layering by the wall. More recent simulations  ~\cite{godonoga2010} of a variety of models at a free liquid-vapour interface found a \emph{suppression} of five-fold symmetry (evidenced by the population of pentagonal bipyramids identified with the topological cluster classification, section ~\ref{sectionTCC}). However those five-fold symmetric structures that were identified close to the interface tended to oriented their five-membered rings parallel to the interface. This may provide some explanation for the experimental result ~\cite{reichert2000}, which explicitly probed structure in the layer by the wall and did not average over all orientations.

X-ray absorption spectroscopy (XAS) is an indirect technique by which higher-order structure may be inferred. In XAS, a photoelectron excited from a deep core level behaves as a local structural probe which interacts with the surrounding atoms leading to scattering effects which are observed as oscillations in the absorption cross section ~\cite{dicicco2003,filipponi2001}. The XAS signal is given as

\begin{eqnarray}
\langle \chi(k) \rangle = \int_0^\infty dr 4 \pi r^2 \rho g_2(r) \gamma_2(r,k) \\ \nonumber
+  \int_0^\infty dr_1 dr_2 d \theta 8 \pi^2 r_1^2 r_2^2  \times \\ \nonumber
 \sin \theta \rho^2 g_3(r_1,r_2,\theta) \gamma_3(r_1,r_2,\theta,k) + \dots
\label{eqXAS}
\end{eqnarray}

\noindent where $\theta$ is a bond angle and the $\gamma$ are XAS $n$-body signals and the $g_n$ are $n-$body density correlation functions ~\cite{dicicco2003}.

\begin{figure}[!htb]
\centering \includegraphics[width=65mm]{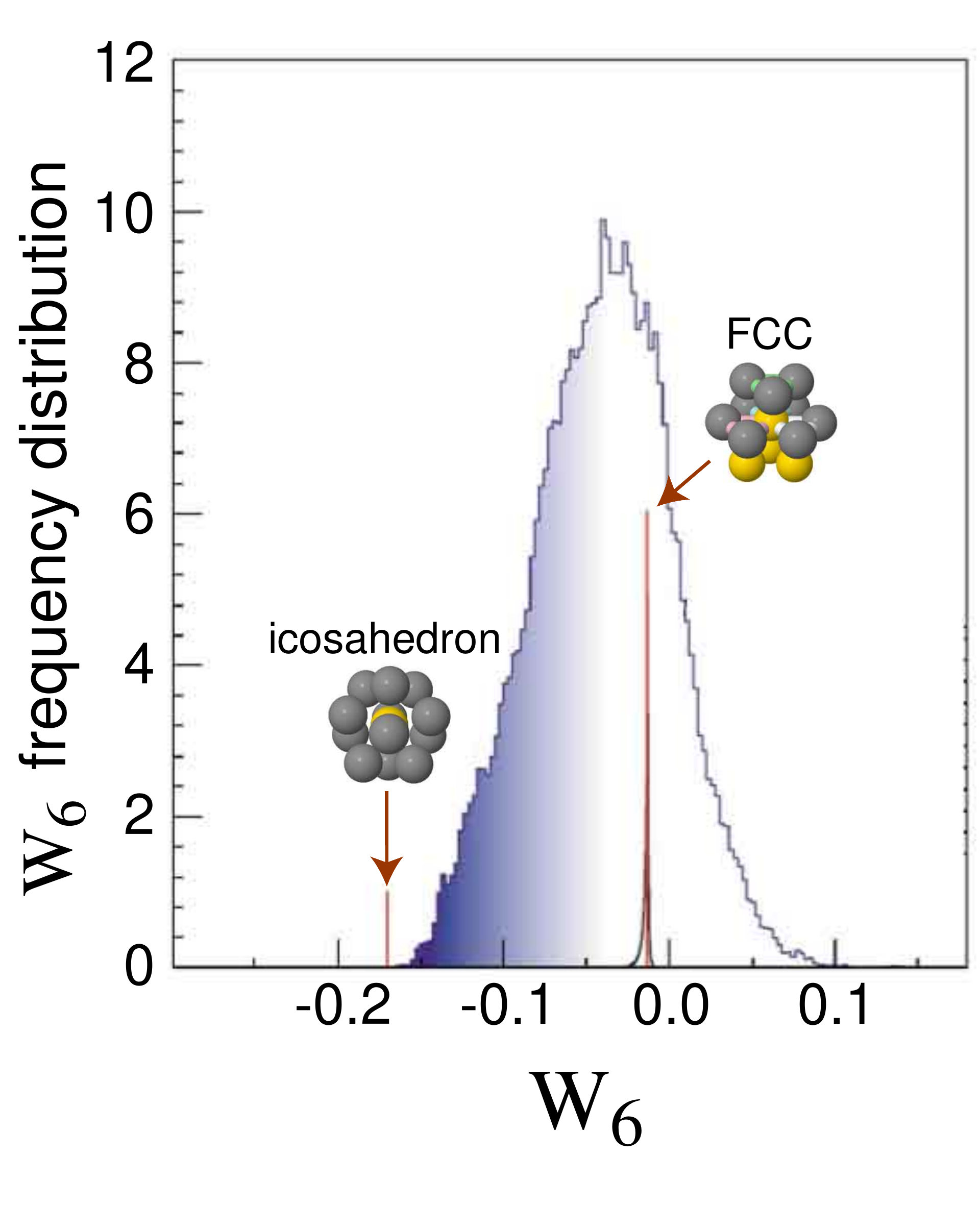} 
\caption{Frequency distribution of the $W_6$ invariant supercooled liquid Cu. The histogram related to the FCC solid Cu, limited to a very narrow region around $W_6=0.013$ (corresponding to a FCC 13-atom cluster), is shown. The $W_6$ distribution is found to be very broad and includes a region (shaded area) corresponding to fivefold symmetry (icosahedra have $W_6=-0.18$). ~\cite{dicicco2003}. Reproduced with permission from American Physical Society Copyright 2003.}
\label{figDiCicco} 
\end{figure}

Since there is an explicit contribution from higher-order terms, which is dominated by the three-body term, it is possible to carry out RMC on both two- and three- body contributions \emph{separately}  ~\cite{dicicco2003}. With the three-body information, in particular the bond angle distribution, it is possible to construct the $W_6$ bond-orientational order  invariant  (section ~\ref{sectionBOO}). As shown in Fig. ~\ref{figDiCicco}, the $W_6$ distribution was indeed found to show a negative trend associated with five-fold symmetry ~\cite{steinhardt1983}. This considerable achievement aside, the complete understanding of the XAS signal remains a challenging theoretical problem involving the many-body-electron response ~\cite{filipponi2001}.

\subsection{Fluctuation electron microscopy}
\label{sectionFEM}

Measuring the fluctuations in transmission electron microscopy (TEM) allows mixed real- and reciprocal space higher-order correlations to be measured ~\cite{treacy2007}. Essentially the fluctuation electron microscopy (FEM) method combines electron diffraction with microscopy, such that local fluctuations in diffraction can be measured. Thus the technique removes one of the limitations of many reciprocal space techniques in that they can often be sensitive to bulk, rather than locally varying, structure. The principle quantity is the intensity of the scattered electron signal as a function of wavevector and real space position in the sample $I(\mathbf{k},\mathbf{r})$. For example one may measure the normalised variance :

\begin{equation}
V(\mathbf{k},\mathbf{r}) =\frac{ \langle I^2(\mathbf{k},\mathbf{r})  \rangle }{\langle I(\mathbf{k},\mathbf{r})  \rangle^2}-1.
\end{equation}

Two lengthscales are important, the resolution of the incident beam, and the scattering vector. If the beam is much larger than the structures of interest, one recovers the bulk scattering scenario and the technique is reduced to two-point scattering information. If on the other hand, the incident beam is very narrow, the signal-to-noise ratio is poor. Thus by tuning the size of the incident beam (and sweeping through the appropriate wavevectors) one can identify the size of structural units by seeking the maximum in the normalised variance $V(\mathbf{k},\mathbf{r})$.

Fluctuation electron microscopy is suitable for accessing structure at the 1 nm lengthscale, and is particularly good for picking out small crystal-like structures ~\cite{kwon2007,lee2009}. However, modelling of $V(\mathbf{k},\mathbf{r})$ shows that fully amorphous structures (such as icosahedra and domains of icosahedra) exhibit a much weaker signal, suggesting that FEM may be more suited to those glassformers whose local structural motifs contain crystalline symmetries ~\cite{wen2009}.

\subsection{Nanobeam electron diffraction}
\label{sectionNanobeamElectronDiffraction}

\begin{figure}[!htb]
\centering \includegraphics[width=80mm]{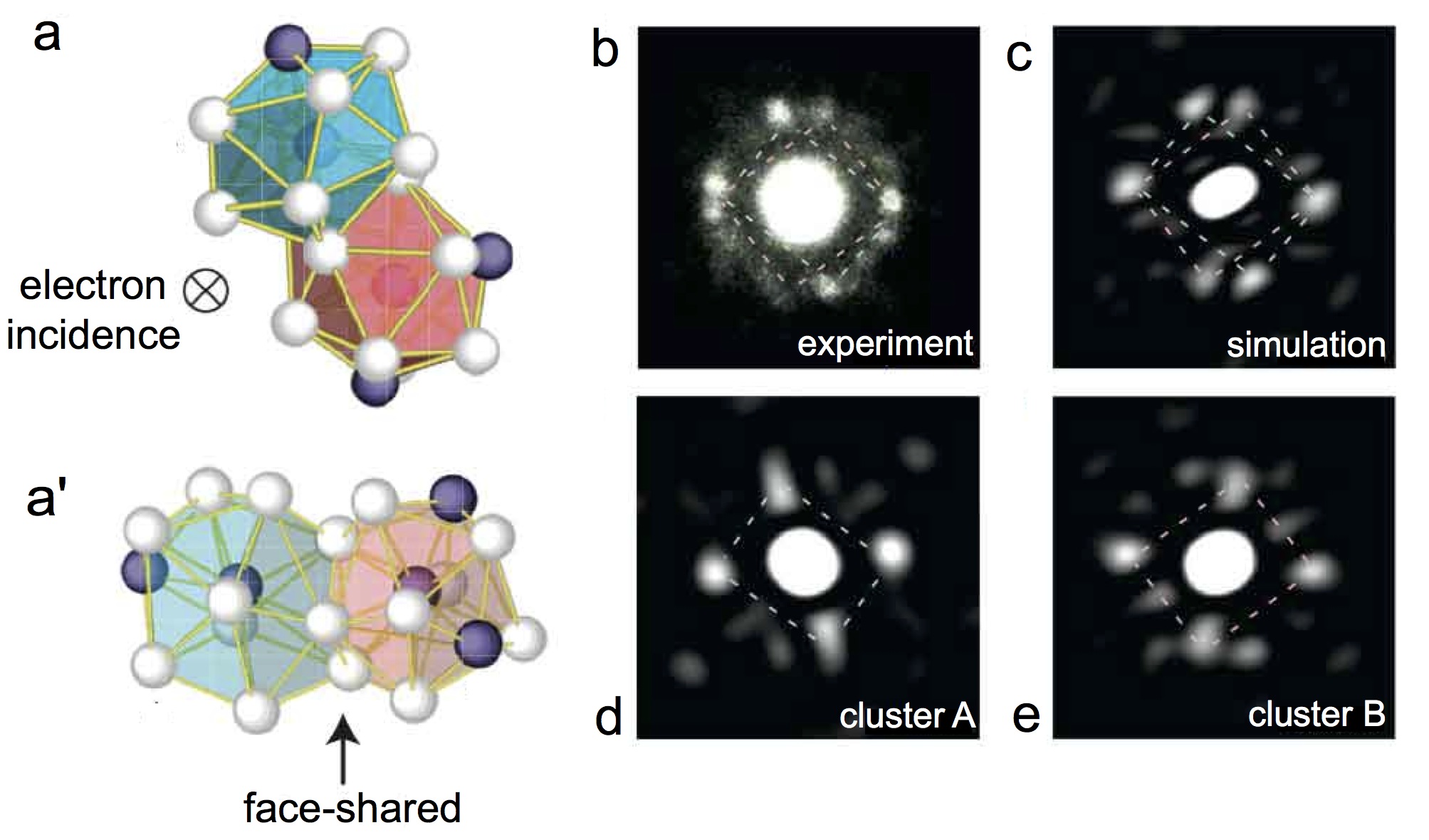} 
\caption{3d amorphous structure resolved with nanobeam electron diffraction. Here two face-sharing polyhedra are identified in Ni$_{66.7}$Zr$_{33.3}$, a model metallic glassformer (see section ~\ref{structureInMetallicGlassformers}). (a), Two face-sharing [0,2,8,1] Voronoi polyhedra with a common on-axis orientation for Bragg diffraction. (a') The cluster pair viewed from a direction showing the face-sharing configuration. (b) Experimental NBED pattern including two sets of possible rectangle diffraction patterns.(c) Simulated NBED pattern obtained from the super-cluster shown in a with the on-axis electron incidence. (d) Simulated NBED pattern obtained only from cluster A in the super-cluster. (e) Simulated NBED pattern only from cluster B.
~\cite{hirata2010}.  Reproduced with permission from Nature Publishing Group 2010.}
\label{figNBED} 
\end{figure}

Another solution to the problem that reciprocal space techniques provide only averaged information has been demonstrated very recently ~\cite{hirata2010}. Conceptually beautiful in its simplicity, nanobeam electron diffraction (NBED) circumvents the bulk averaging problem by using a remarkably finely focussed electron beam, whose full-width half-maximum is a mere 0.72 nm ~\cite{hirata2010}. Like fluctuation TEM, NBED must still be combined with simulation to extract the details of the structure, but it is a powerful technique as Fig. ~\ref{figNBED} makes clear. One observation regarding NBED is that it must take quite some time to scan a bulk sample.

\subsection{Measuring lengths close to $T_g$ in molecular glassformers}
\label{sectionLudovicLength}

\begin{figure*}[!htb]
\centering \includegraphics[width=150mm]{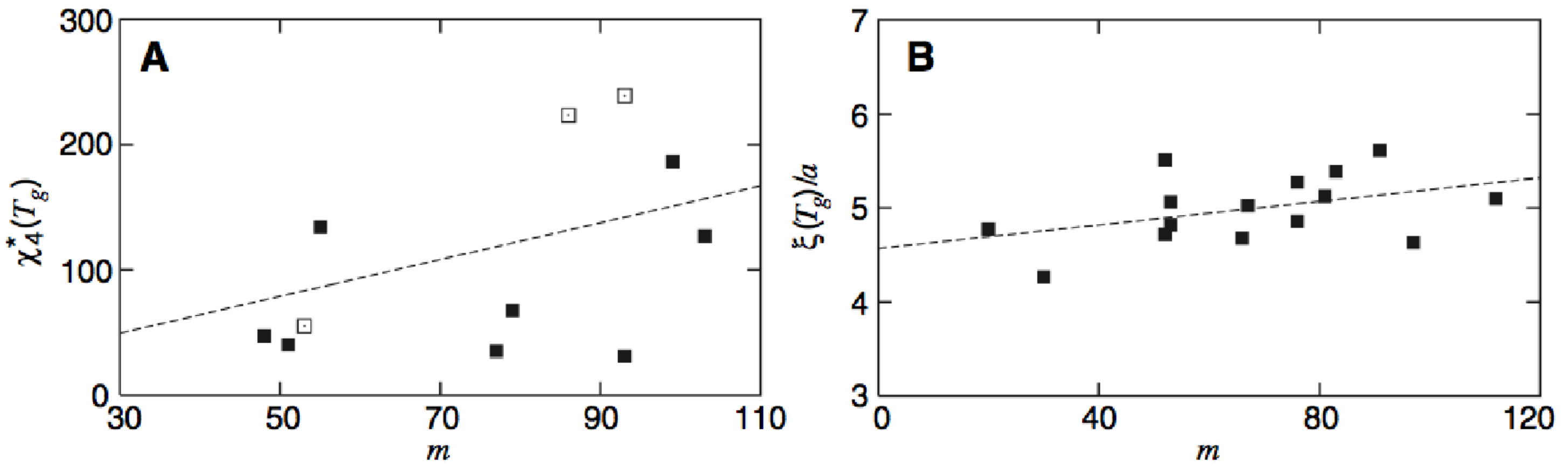} 
\caption{Lower bounds for $\chi_4$ and $\xi_{dyn}$ obtained from experiments on molecules at $T_g$. $m$ is fragility obtained from VFT fits to Angell plots.
\cite{berthier2005}. Reproduced with permission from The American Association for the Advancement of Science Copyright 2005.}
\label{figLudovicLength} 
\end{figure*}

In addition to structure, dynamical lengthscales have been probed in molecular glassformers. In 2005, Berthier \emph{et al.} ~\cite{berthier2005} used a sequence of arguments to obtain estimations of dynamic susceptibility $\chi_4$ (Eq. \ref{eqChi4}, Fig. \ref{figChi4}) from experimentally accessible quantities such as the dielectric susceptibility. The following inequality was derived

\begin{equation}
\chi_4(t) \geq \frac{k_B}{c_P} T^2 \chi_T^2(t).
\end{equation}

\noindent where $\chi_T$ is an experimentally accessible quantity. Thus a lower bound for $\chi_4$ was obtained. Recall that in Fig. \ref{figChi4}, we saw that $\chi_4(t)$ has a characteristic peak value, $\chi_4^\mathrm{max}$. With some more assumptions ~\cite{berthier2005} one expects that

\begin{equation}
\chi_4^\mathrm{max} \approx \left( \frac{\xi_\mathrm{dyn}}{\sigma/2}\right)^\zeta
\end{equation}

\noindent where $2 \lesssim \zeta \lesssim 4$. Thus one can find an estimate for the lower bound of the dynamic correlation length in \emph{molecular} experiments via dielectric spectroscopy. As shown in Fig. \ref{figLudovicLength}, the length extracted is not in any sense large, compared to expectations from Fig. \ref{figGS4squareXi4}, for example. Recall, in simulation or colloid experiment $\xi_4$ is typically measured around $T_{\mathrm{MCT}}$, so a large increase might be expected by $T_g$. Now of course, the data presented in Fig. \ref{figLudovicLength} only represents a lower bound, but the difference is so large that it might lead one to think there is a different regime between $T_{\mathrm{MCT}}$ and $T_g$ compared to that for $T>T_{\mathrm{MCT}}$.

The method we have outlined is by no means the only possibility for detecting dynamic lengthscales in molecular systems. It is possible to deduce such lengthscales indirectly from heat capacity measurements ~\cite{tatsumi2012} and more directly from measurements of fluorescent molecules embedded in the glassformer ~\cite{cicerone1995} and even by direct imaging ~\cite{ashtekar2012}, though the latter technique is limited to the surface, whose dynamics may not be representative of the bulk. As indirect as these techniques are, all give very comparable values that $\xi_\mathrm{dyn}\lesssim 10 \sigma$ at $T_g$. We return to this issue in section ~\ref{sectionStaticAndDynamicLengths}.

\section{Structure in model glassformers}
\label{sectionStructureModel}

Having discussed various techniques to measure structure (and dynamics) in glassforming systems, we now describe the broad picture of structural changes in model glassformers approaching dynamical arrest. Such a picture was conceptualised by Richert and Angell ~\cite{richert1998}, who suggested that more fragile glassformers should be expected to exhibit a greater change in structure, as evidenced by the jump in specific heat capacity upon falling out of equilibrium at $T_g$.

Another way of looking at this is to consider the Angell plot (Fig. \ref{figAngell}). In the case that there is no structural change in the system approaching arrest, one might expect Arrhenius-like behaviour (a strong liquid, a straight line on the Angell plot)  ~\cite{ito1999,ngai1999,martinez2001}. Indeed strong liquids such as silica ~\cite{coslovich2009} and GeO$_2$ ~\cite{salmon2013} exhibit little structural change upon cooling. Other more fragile systems exhibit more change ~\cite{coslovich2007,malins2013jcp}. However, that kinetically constrained models, which are thermodynamically ideal gases, can exhibit fragile behaviour ~\cite{pan2004} as do higher-dimensional systems where structure is less relevant ~\cite{charbonneau2013pre} suggests that structure need not be invoked to observe fragility, and indeed the interpretation of some molecular glassformers is not trivial ~\cite{ito1999,ngai1999,martinez2001}. Furthermore, different classes of molecular glassformers are found to exhibit differing correlations between the jump in specific heat at $T_g$ and fragility ~\cite{huang2001} and the number of particles involved in dynamic correlations inferred from dielectric spectroscopy has been found by some to be independent of fragility \cite{dalleferrier2007} although more recent work does suggest a link between the size of correlated regions and fragility \cite{bauer2013}. In any case the more simple classes of materials, metal-oxide and metallic alloy glassformers do exhibit the expected positive correlation between a jump in specific heat (a proxy for configurational entropy) and fragility ~\cite{richert1998,huang2001}. Most of the early work on measuring structure in glassformers was carried out on hard spheres or Lennard-Jones models. Assessing the fragility of these models can be limited by the dynamically accessible range, but the degree of structural change is correlated with fragility ~\cite{royall2014,coslovich2007,coslovich2007ii}.

\subsection{Early measurements in amorphous systems}
\label{sectionEarlyMeasurements}

We have already noted the pioneering work of Bernal ~\cite{bernal1959,bernal1960}, Scott ~\cite{scott1960}, Finney ~\cite{finney1970} and Bennett  ~\cite{bennett1972}  of building models of hard spheres. These established a tendency for the formation of five-membered rings, in the form of pentagonal faces of Voronoi polyhedra. Finney also carried out computer simulations ~\cite{finney1970mc} which showed similar behaviour upon supercooling.

Simulations showed evidence for a split second peak in $g(r)$ ~\cite{rahman1976,hiwatari1980,finney1970mc}, characteristic of metallic glasses, yet absent from liquids ~\cite{cargill1970}. Indeed the radial distribution function calculated from hard spheres gave reasonable quantitative agreement with experimental data on NiP, at the expense of having the model at an unphysically high density to fit the experimental data ~\cite{cargill1970}.  As significant as this early result undoubtedly was, that \emph{monodisperse} hard spheres gave such good agreement with NiP may also be interpreted as a reason to treat interpretations from two-body data with a pinch of salt. An improvement was obtained by Polk ~\cite{polk1970,polk1972} who considered the possibility of fitting atoms into the holes in the polyhedra considered by Bernal ~\cite{bernal1959,bernal1960} in his random packings of hard spheres. Polk observed that interactions between metals and metalloids such as Ni and P are weaker than those between each species, in other words that such interactions are non-additive \footnote{This is the basis for the non-additivity in the Kob-Andersen model, section ~\ref{sectionCommonModelSystems}.}. In this way it might be possible to fit more of the metalloid atoms into the system than would be the case for additive hard spheres, which would relieve the density discrepancy found by Cargill ~\cite{cargill1970}. The next step was to take hard sphere coordinates and to relax them using a more realistic (Morse) interatomic potential ~\cite{vonheimendahl1975,finney1977}.

Pioneering work on \emph{binary} systems, usually crucial to prevent crystallisation, was carried out by Sadoc \emph{et al.} ~\cite{sadoc1973}. In 1984, Hiwatari \emph{et al.} ~\cite{hiwatari1984} decomposed the soft-core (repulsive part of the Lennard-Jones interaction, Eq. ~\ref{eqLJ}) glasses into Voronoi polyhedra. Almost at the same time, the bond-orientational order parameters were introduced by Steinhardt et al. ~\cite{steinhardt1983} in considering the Lennard-Jones model. These two pieces of work appear to be the first which systematically consider structure upon passing from a liquid to a glass. Both find increasing five fold symmetry approaching the glass. However Steinhardt \emph{et al.} went even further, and determined an orientational correlation length $\xi_{m=6}$ by exponential fits to $G_6$ defined as

\begin{equation}
G_l(r)=\frac{4 \pi}{2l+1} \frac{\sum_{m=-l}^{l} \langle Q_{lm}(\mathbf{r})  Q_{lm}(\mathbf{0}) \rangle }{G_0}
\label{eqGl}
\end{equation}

\noindent where

\begin{equation}
G_0(r)=4 \pi \langle Q_{00}(\mathbf{r}) Q_{00}(\mathbf{0}) \rangle.
\end{equation}

\begin{figure}[!htb]
\centering \includegraphics[width=80mm]{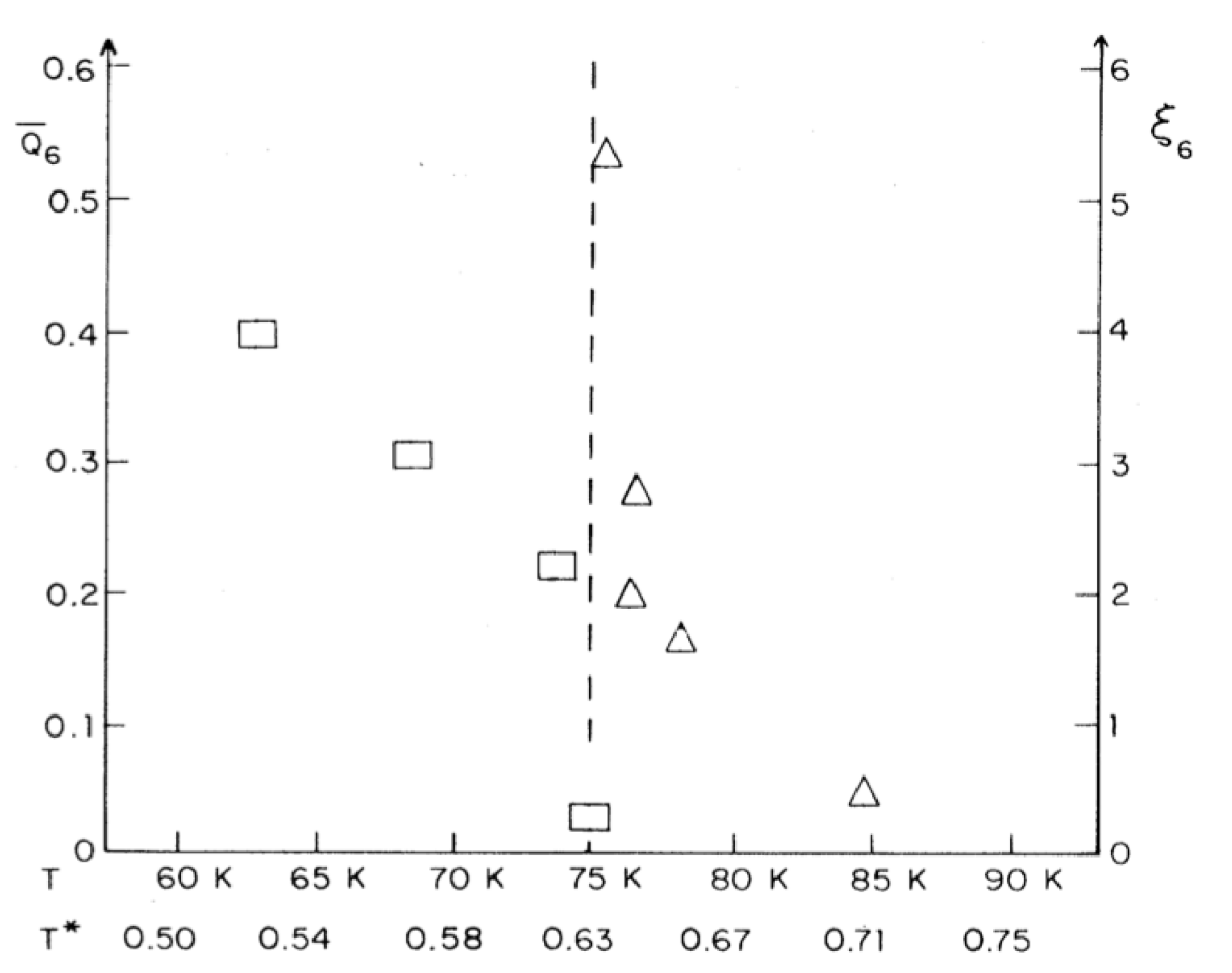} 
\caption{
An early measure of amorphous order. $Q_6$ increases upon cooling (squares), as does the orientational correlation length $\xi_6$ (triangles). The ``divergence'' of this length is likely related to finite-size effects - the system size was $N=864$. 
~\cite{steinhardt1983}. Reproduced with permission from American Physical Society Copyright 1983.}
\label{figSteinhardt} 
\end{figure}

As shown in Fig. ~\ref{figSteinhardt}, the correlation length $\xi_6$ increases upon cooling. This was interpreted by the authors as a possible transition to an icosahedral phase. However, given their use of a monodisperse Lennard-Jones, which is very prone to crystallisation (and some of their runs did indeed crystallise) some of the deeper quenches would not have been able to equilibrate prior to crystallisation (see section ~\ref{sectionSpinodalNucleation}). Moreover subsequent work using a similar approach in a binary system found no evidence of an increasing structural lengthscale ~\cite{ernst1991}.

\begin{figure}[!htb]
\centering \includegraphics[width=80mm]{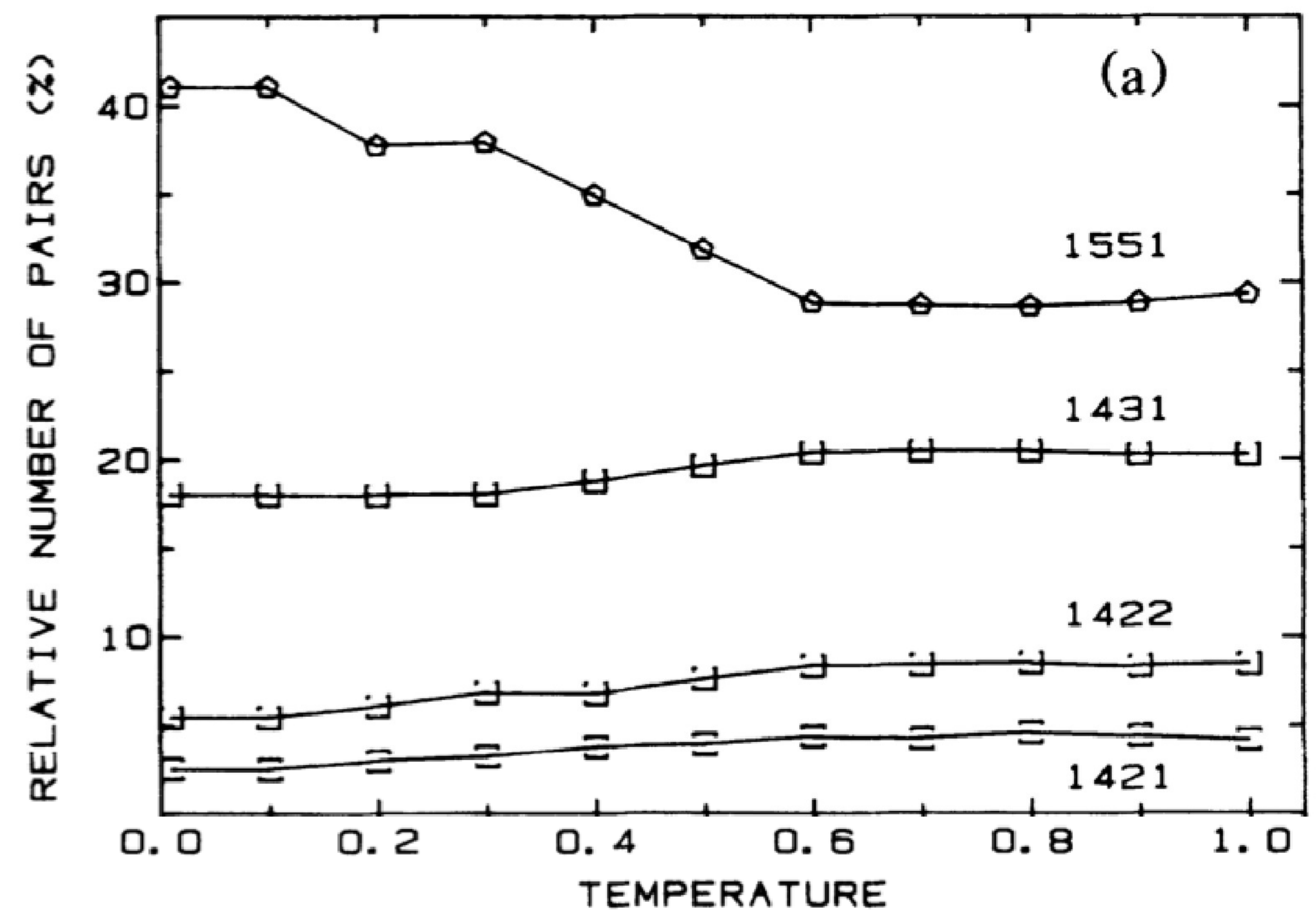} 
\caption{Populations of pentagonal bipyramids (1551) and related structures in a binary Lennard-Jones glassformer ~\cite{jonsson1988}. Reproduced with permission from American Physical Society Copyright 1988.}
\label{figJonssonA} 
\end{figure}

With the advent of the common neighbour analysis ~\cite{honeycutt1987}, another structural tool was available to seek structural changes in a model glassformer. This coincided with the use of binary systems to prevent the crystallisation that must have plagued early work on supercooled liquids. Jonsson and Andersen used an additive binary Lennard Jones model and investigated its structure approaching the glass transition. As shown in Fig. \ref{figJonssonA}, the population of 1551 pentagonal bipyramids increases upon cooling in a binary Lennard-Jones system ~\cite{jonsson1988}. That an icosahedron is two 1551's sharing a spindle means that the CNA provided strong evidence in support of Frank's conjecture. While it is not always \emph{a priori} clear when the icosahedron is the minimum energy structure for a binary Lennard-Jones system, it is the minimum for moderate size asymmetry ~\cite{doye2005}. Jonsson and Andersen further identified a network of icosahedra spanning their simulation box (recall icosahedra do not fill space). Tomida and Egami ~\cite{tomida1995} used a model for iron, and by detecting local fivefold symmetry using BOO parameters also found evidence for an icosahedral network.

The early 90s marked a number of developments. The first was the development of real-space analysis of colloidal glasses in 3d with confocal microscopy by van Blaaderen and co-workers ~\cite{vanblaaderen1992,vanblaaderen1995}. In these hard sphere suspensions, evidence was found, corroborating the earlier simulation work, of five-fold symmetry in the form of a negative $W_6$ distribution. More recently in 2d experiments on a binary colloidal system with dipolar interactions, bond-order parameters have been used to identify a variety of local structures associated with vitrification ~\cite{mazoyer2011,konig2005}.

Another development was Dzugutov's introduction of a monatomic model (though admittedly somewhat idealised) which was designed not to crystallise. This Dzugutov potential has a repulsive hump where the second neighbours of an FCC lattice lie, and thus crystallisation is suppressed ~\cite{dzugutov1992}. Indeed it was believed instead to form quasicrystals ~\cite{dzugutov1993}, although it is thought these may be quasicrystal approximants ~\cite{keysPersonal} and in any case the equilibrium state is a regular crystal ~\cite{roth2000dzugutov}. Whatever its equilibrium state, the Dzugutov model certainly forms icosahedra efficiently ~\cite{dzugutov2002}. Before moving to the search for a structural mechanism for vitrification, we note the work of Pan \emph{et al.} ~\cite{pan2011} who used a simulation of CuZr to correlate Voronoi polyhedra with the split second peak of the pair correlation function long associated with vitrification.

\subsection{Towards a structural mechanism?}
\label{sectionTowardsAStructuralMechanism}

\begin{figure}[!htb]
\centering \includegraphics[width=80mm]{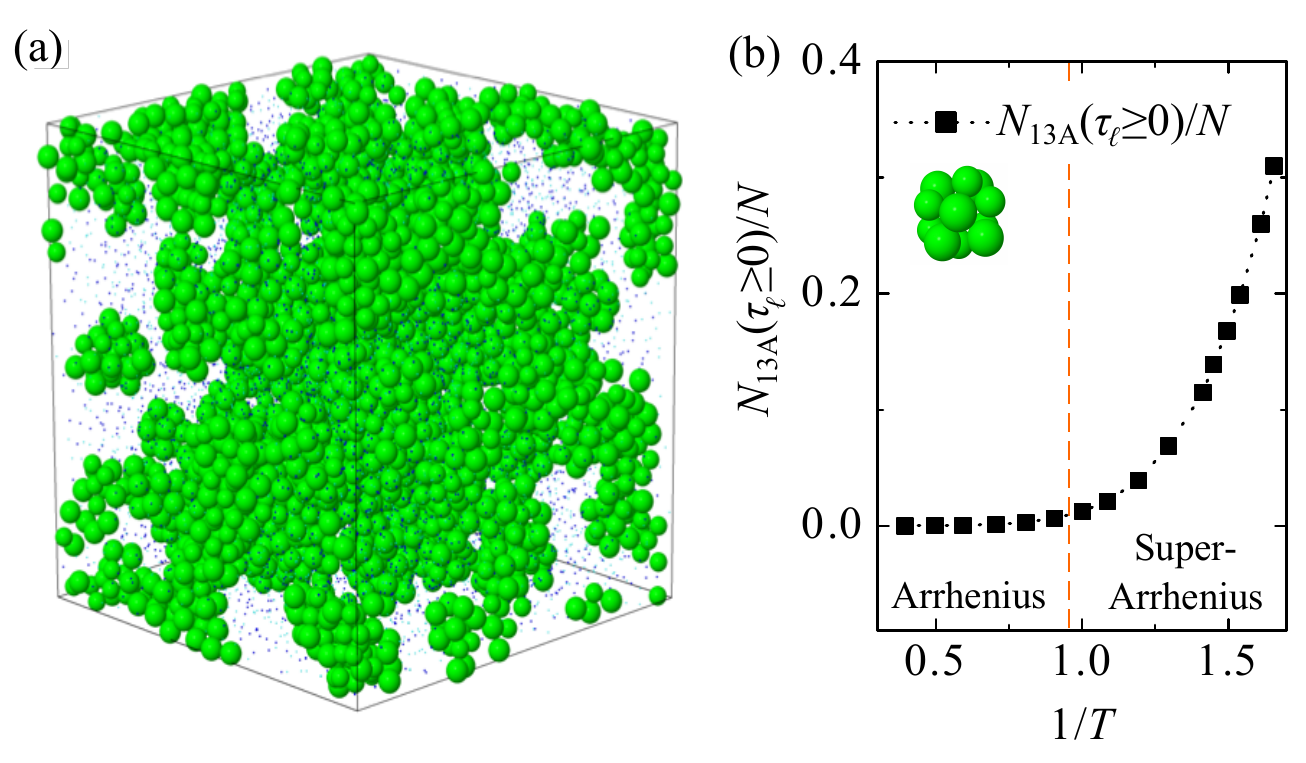} 
\caption{Frank's conjecture demonstrated. (a) A network of icosahedra in the Wahnstr\"{o}m Lennard-Jones model formed upon supercooling. (b) The increase in particles in icosahedra upon cooling  \cite{malins2013jcp}.}
\label{figAlexNetwork} 
\end{figure}

Thus far, it had only been possible to identify structural changes upon supercooling. No suggestion of a mechanism had been made, beyond a verification of Frank's original conjecture, which focussed on the avoidance of crystallisation, not on a mechanism for dynamical arrest. That changed in the 1990s, with the discovery of dynamical heterogeneity by Harrowell and co-workers in simulation ~\cite{perera1996,hurley1995}, and by Schmidt-Rohr and Speiss in experiment ~\cite{schmidtrohr1991}. It was now possible, if one could correlate the dynamically slow particles with those in icosahedra (for example) to suggest that the formation of such structures was somehow responsible for dynamical arrest. Among the first to do this were Dzugutov \emph{et al.} ~\cite{dzugutov2002} in 2002 who identified icosahedra with dynamically slow particles. They were followed by Coslovich and Pastore ~\cite{coslovich2007} whose careful Voronoi analysis of the small particles in a variety of binary Lennard-Jones systems found that not only was the icosahedron a locally favoured structure (for the Wahnstr\"{o}m model) but so was the bicapped square antiprism in the case of the Kob-Andersen model. This was also found using the topological cluster classification by Malins \emph{et al.} ~\cite{malins2013jcp,malins2013fara}. Pedersen \emph{et al.} ~\cite{pedersen2010} emphasized the role of Frank-Kasper bonds (here taking the form of hexagonal bipyramids) in the Wahnstr\"{o}m model and showed that particles within such structures had unusually long relaxation times. Interestingly, these Frank-Kasper bonds are implicated in the crystallisation of the Wahnstr\"{o}m model which forms a complex unit cell, equivalent to the MgZn$_2$ Laves phase. This crystal is based on icosahedra combined with Frank-Kasper bonds which have the effect of enabling the former to tile Euclidean space ~\cite{pedersen2010}. Moreover Pedersen \emph{et al.} showed that those particles involved in Frank-Kasper bonds had slower dynamics than the average. However, in the supercooled liquid the fraction of particles involved in Frank-Kasper bonds is around 5\% of the total and of these 83\% are also found in icosahedra. Indeed particles in icosahedra are also slower than the average and moreover account for around $1/3$ of the system. This suggests that icosahedra may be more significant in influencing the dynamics of the system ~\cite{malins2013jcp}. Conversely, working in 2d, Eckmann and Procaccia ~\cite{eckmann2008} used geometric motifs to argue that, although the population of these changed as a function of temperature, there was no evidence for any divergence at finite temperature (unlike the assumptions in theories such as Adam-Gibbs ~\ref{sectionAdamGibbs} and random first order transition~\ref{sectionRFOT}).

In colloid experiments also, correlations were found between slow dynamics and locally favoured structures using the topological cluster classification ~\cite{royall2008} and BOO parameters ~\cite{leocmach2012}. In 2d colloid experiments, a correlation has been found between locally favoured structures and a reduced mean squared displacement relative to the average ~\cite{tamborini2014,mazoyer2011}. Unusually for colloid experiments, where typically errors in particle coordinates prevent accurate determination of the potential energy for each particle ~\cite{royall2013myth,poon2012}, due to the large size of their colloids and the long-ranged interactions, Mazoyer \emph{et al.} were able to measure the potential energy of each particle and correlate this too with the mobility  ~\cite{mazoyer2011}.

\begin{figure}[!htb]
\centering \includegraphics[width=80mm]{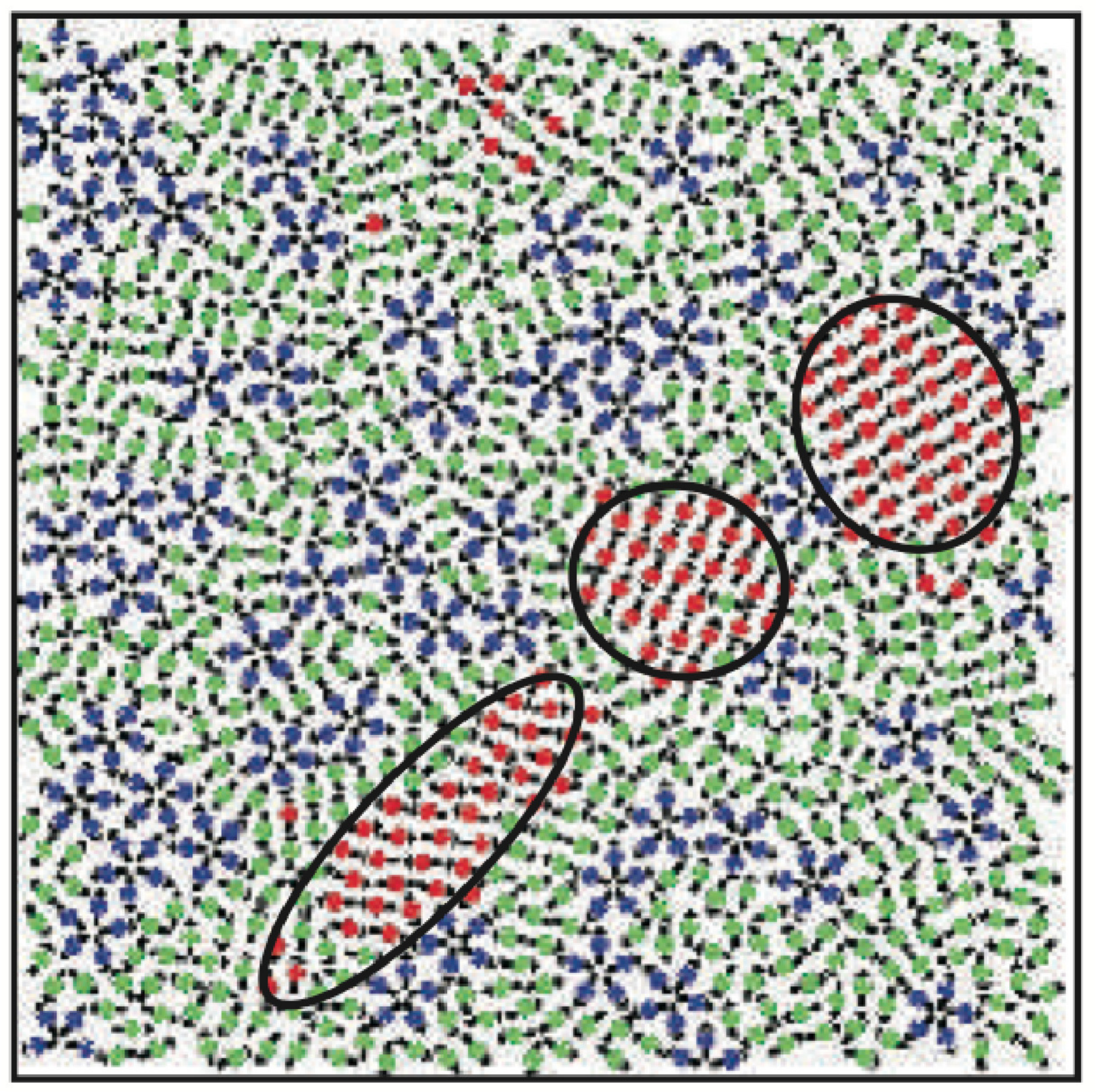} 
\caption{The frustrated spin model of Shintani and Tanaka ~\cite{shintani2006}.
In this model five-fold symmetric clusters (blue) compete with hexagonal symmetry (red).
Reproduced with permission from Nature Publishing Group Copyright 2006.}
\label{figShintani} 
\end{figure}

In 2006, Shintani and Tanaka ~\cite{shintani2006} devised a 2d model whose structure could be tuned from hexagonal to five-fold symmetric (Fig. \ref{figShintani}). In Tanaka's picture (Section \ref{sectionCrystalLikeOrdering}) the locally favoured structure can often have the same symmetry as the crystal. In 2d, this hexagonal structure (identified using BOO parameters) is also the locally favoured structure in the liquid [in the parlance of Geometric Frustration  (section \ref{sectionGeometricFrustration}), 2d systems are unfrustrated]. As a consequence, in 2d, the liquid locally favoured structure has to be suppressed to avoid crystallisation. In their tunable model, Shintani and Tanaka found that the more hexagonal system had a higher degree of structural change approaching the glass, \emph{and was more fragile}. This seems like a direct confirmation of Angell's suggestion of the relationship between structure and fragility. 2d hard discs, where the degree of hexagonal ordering was controlled through polydispersity showed a similar behaviour ~\cite{kawasaki2007}. However in 3d no obvious correlation between diffusivity and polydispersity was found ~\cite{zaccarelli2009xtal} although the response of the structural relaxation time \emph{to changes in volume fraction $\phi$} is sensitive to polydispersity  ~\cite{tanaka2011,kawasaki2010pnas,kawasaki2010jpcm}.

\begin{figure}[!htb]
\centering \includegraphics[width=80mm]{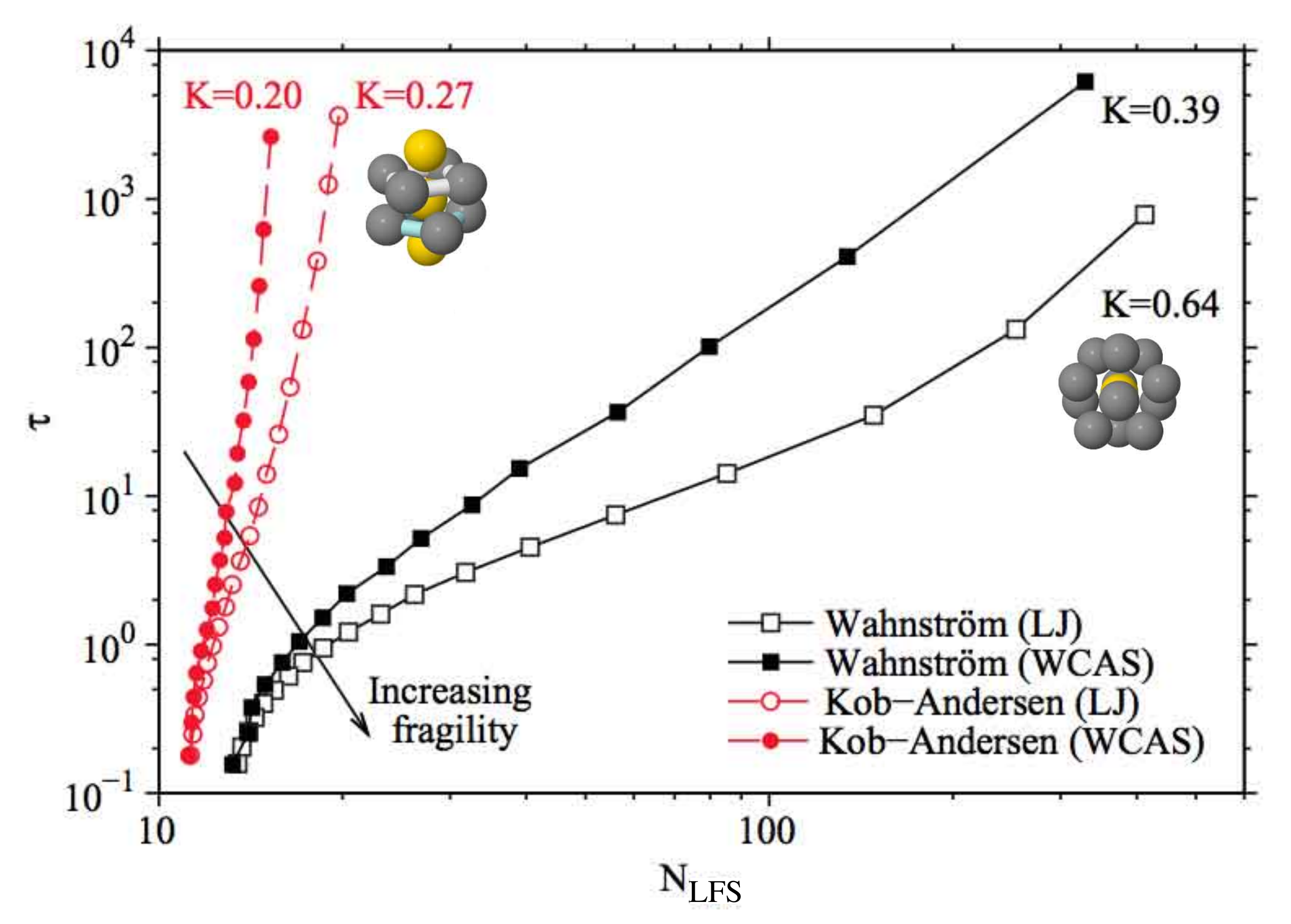} 
\caption{Structural relaxation time $\tau_\alpha$ as a function of the average number $N_\mathrm{LFS}$ of particles in locally favoured structures. $K$ denotes the fragility index and increases with fragility according to a fit to the VFT equation. LJ is the full LJ system while WCAS is the variant without attractions ~\cite{coslovich2011}. Reproduced with permission from American Physical Society Copyright 2011.
}
\label{figCoslovich} 
\end{figure}

This correlation between higher-order structure and slow dynamics enabled insight into previous observations. As mentioned in section ~\ref{sectionTwoPointStructure}, Berthier and Tarjus had found that very subtle changes in the two-point structure resulted in large changes in dynamics when attractions were removed from the Lennard-Jones interactions in the Kob-Andersen model via truncation at the bottom of the potential energy well. Having already identified the locally favoured structures for the full systems with attractions ~\cite{coslovich2007}, Coslovich proceeded to investigate the population of bicapped square antiprisms in the KA system with and without attractions, and carried out a similar analysis for icosahedra in the Wahnstr\"{o}m model ~\cite{coslovich2011}. In both models, the population of locally favoured structures was suppressed when attractions were removed, as shown in Fig. ~\ref{figCoslovich}, providing evidence for a link between structure and fragility. Coslovich has also shown that significant differences are found at the three-body level between the Lennard-Jones and the WCA variant ~\cite{coslovich2013}.

This direction of study was  extended by the Roskilde group ~\cite{pedersen2010} whose work explores the concept of ``highly correlated'' or ``Roskilde'' liquids ~\cite{bailey2008,bailey2008a}. This provides a protocol by which liquids at different macroscopic states may be described as ``isomorphic'', \emph{i.e.} having indistinguishable structure. The isomorphic concept can be traced back to early work on inverse power law (IPL) potentials ~\cite{hoover1970}, $\phi_{ij}(r_{ij})=\epsilon (\sigma/r_{ij})^n$, where it was shown that the excess thermodynamic properties are a function of the property $\rho^{n/3}\epsilon/k_BT$ only. If we follow a line in macroscopic state space given by $\rho^{n/3} = a k_BT$, where $a$ is some arbitrary constant, then the reduced equilibrium structure remains invariant, i.e. we have an isomorphic line. In the limit where $n\rightarrow\infty$ we recover the hard sphere potential and the isomorphic line becomes a line of constant density as the temperature is increased. That is we recover the result that hard spheres are athermal. Indeed at low pressure and temperature, the dynamical behaviour and equation of state of soft spheres can be mapped onto hard spheres with high accuracy \cite{xu2009}.

\begin{figure*}
\begin{centering}
\includegraphics[width=16 cm]{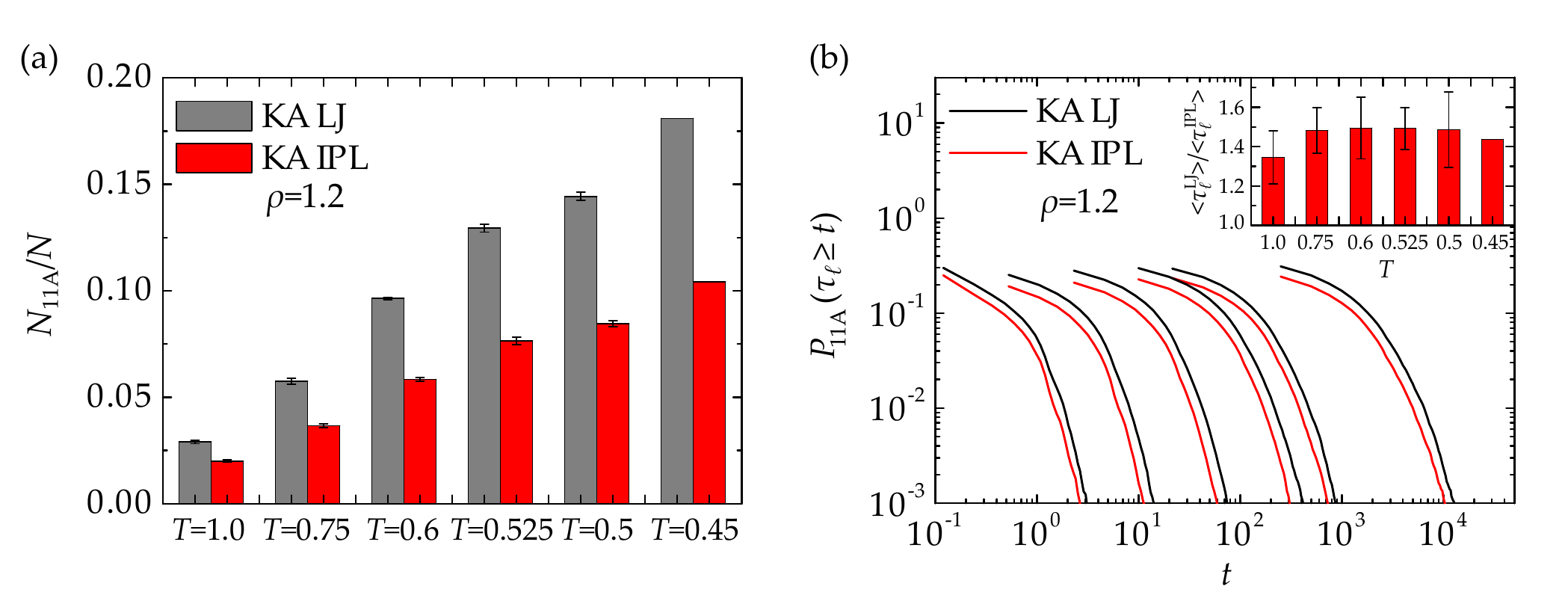}
\par\end{centering}
 \caption{Analysis of the statistics and dynamics of bicapped square antiprism (11A) clusters in the Kob Andersen Lennard-Jones and KA IPL mixtures for $T=\{1.0,0.75,0.6,0.525,0.5,0.45\}$. (a) 11A population $N_\mathrm{11A}/N$ and (b) $P_\mathrm{11A}(\tau_\ell \ge t)$. Inset shows the ratio of the lifetimes of 11A clusters $\langle\tau_\ell^\mathrm{LJ}\rangle/\langle\tau_\ell^\mathrm{IPL}\rangle$ obtained by scaling the curves in the main figure onto one another at the temperatures indicated on the abscissa ~\cite{malins2013isomorph}. %\{magenta}{NB eps file needs sorting this is a MALINS figure....}
\label{figGCompare}}
\end{figure*}

These ideas were extended by considering the correlation between the potential energy, $U$, and the virial component of the pressure, $W$,  in the NVT ensemble ~\cite{gnan2009}, the latter being given by

\begin{equation}
W(\mathbf{r}^N)=-\frac{1}{3}\sum_{i=1}^N \sum_{i<j}^N r_{ij}u'(r_{ij}).
\label{eqVirial}
\end{equation}

If the Pearson's correlation coefficient in the fluctuations of $U$ and $W$ exceeds 0.9, the liquid is termed a ``strongly correlated'' (``Roskilde'') liquid ~\cite{bailey2008a,dyre2013}. For the special case of an IPL the correlation will be unity and the exponent given by $\gamma = n/3$ while more generally for a strongly correlated liquid the exponent may be obtained as

\begin{equation}
\gamma = \sqrt{\frac{\langle \Delta W^2 \rangle}{\langle \Delta U^2 \rangle}}.
\label{eqExpon}
\end{equation}

A key idea here is that the structure of a liquid is dominated by the short range repulsive part of the potential, as established by Barker and Henderson ~\cite{barker1976}. This may be rationalised upon considering how many more particles there are in a spherical shell upon increasing the shell's radius. The equilibrium distribution function depends explicitly upon the energy, and due to the law of large numbers the contribution to the energy of the particle pairs in close proximity to each other dominates the fluctuations in the energy. This close range repulsive interaction can be accurately modelled in terms of the IPL potential ~\cite{hoover1970} leading us directly to strongly correlated liquids and the resulting isomorphic states. Two states which are on the same isomorphic line may have their static and dynamic properties mapped onto each other through the use of the appropriate reduced units ~\cite{gnan2009}, thus enabling the collapse of a two-dimensional phase diagram to one dimension. This approach is far from limited to simulation of model systems. For example, such density-scaling has been used to describe the dynamical behaviour of number of molecular glassformers with a single parameter ~\cite{gundermann2011}

Pedersen \emph{et al.} ~\cite{pedersen2010iso} employed this approach to investigate the effect of removing attractions from the Lennard-Jones Kob-Andersen model, not under the Weeks-Chandler-Andersen approach which truncates the interaction at the bottom of the energy well, but instead by mapping to an inverse power law reference system. The result was intriguing : where Berthier and Tarjus ~\cite{berthier2009,berthier2010pre,berthier2012epje} found that the two-point structure was essentially identical between the Lennard-Jones and WCA variant, but the dynamics were markedly different, Pedersen \emph{et al.} ~\cite{pedersen2010iso} found \emph{both} structure and dynamics were almost identical between the full Lennard-Jones and the WCA variant. As successful as this approach was, only two-point dynamics in the form of intermediate scattering functions and two-point structure were considered.

Malins \emph{et al.} \cite{malins2013isomorph} considered higher-order dynamics in the form of the four-point dynamic susceptibility $\chi_4$ (Eq. \ref{eqChi4}). The $\chi_4$ revealed increasing discrepancies between the IPL and full Lennard-Jones approaching vitrification. As shown in Fig. ~\ref{figGCompare}, higher-order structural correlations, such as the population of bicapped square antiprism ``11A'' clusters, the LFS for the Lennard-Jones system were found to exhibit population differences increasing to some 40\% between the LJ and IPL systems. Furthermore, these clusters were significantly longer-lived in the full Lennard-Jones system at an equivalent state point. This work underscores the importance of higher-order correlations than two-point, both in structure and dynamics to have a clear picture of how similar two systems are. This further casts doubt over the use of methods such as reverse Monte-Carlo to obtain structure of higher-order than two-body.

Another approach has been to directly test geometric frustration in 2d (see section \ref{sectionGeometricFrustration}) ~\cite{modes2007,sausset2008}. Effectively this implements the reverse of the 3d case, where \emph{curved} space is commensurate with the locally favoured structure. In 2d, Euclidean space is commensurate with the (hexagonal) locally favoured structure, thus one must curve space to introduce geometric frustration. This can be done by curving in hyperbolic space, where the degree of curvature can be continuously varied. Weakly curved systems had a strong tendency to hexagonal ordering, which was controllably frustrated by the curvature. However, all correlation lengths were dictated by the curvature in this system. Thus both static, ie structural, and dynamic, ie $\xi_4$, (Eq. \ref{eqOZcrit}) correlation lengths scaled together. So the frustration-driven term in Eq. \ref{eqCNTFrustration}, $s_{\mathrm{frust}}(T)L^5$, prevents the growth of any LFS domains and the dynamic correlation length beyond the lengthscale set by the curvature ~\cite{sausset2010,sausset2010pre}. Thus no divergent structural correlation lengths could be found of the type imagined in the Montanari-Semmerjian picture  (section ~\ref{sectionMontanari}).

\subsection{Static and dynamic correlation lengths : coupled or decoupled?}
\label{sectionStaticAndDynamicLengths}

To recapitulate the preceding section, we have seen in sections \ref{sectionEarlyMeasurements} and \ref{sectionTowardsAStructuralMechanism} that, at least for systems of the Lennard-Jones, hard and soft sphere type, there are considerable changes in higher-order structure upon approaching dynamical arrest and that these can be measured in particle-resolved studies. However this does not indicate that local structure is \emph{responsible} for dynamical arrest, merely that there is a correlation. Some more evidence in favour of a structural mechanism is given by the finding that particles in locally favoured structures, be they amorphous or crystalline, are dynamically slower than the average ~\cite{pedersen2010,malins2013jcp,tanaka2010,coslovich2007,malins2013fara,royall2008,leocmach2012,tamborini2014,shintani2006,kawasaki2007,kawasaki2010pnas,kawasaki2010jpcm}. These and other approaches to determine lengthscales, both static and dynamic have recently been extensively discussed an a review by Karmakar \emph{et al.} ~\cite{karmakar2014} to which we direct the interested reader.

One key piece of evidence in favour of a structural mechanism would be dynamic lengthscales (for example $\xi_4$) scaling with some structural lengthscale [for example $\xi_6$ (Fig. \ref{figSteinhardt}) ]. To identify structural lengthscales, a variety of approaches have been pursued. Over a limited range, $\xi_6$ 
%as introduced by Steinthardt \emph{et al.} ~\cite{steinhardt1983} 
used now to measure crystalline order
has been found to scale with $\xi_4$ in Tanaka and Shintani's frustrated 2d system ~\cite{shintani2006}, 2d polydisperse hard discs ~\cite{kawasaki2007}, 3d polydisperse hard spheres ~\cite{kawasaki2010jpcm} and 3d polydisperse Lennard-Jones ~\cite{tanaka2010}. On the basis of these results, a link was made with Ising critical scaling, using the same exponents, where the critical point was the VFT temperature $T_0$, or packing fraction $\phi_0$ in the case of hard spheres or discs. However, others have not found that $\xi_6$ scales with $\xi_4$ in simulations of 2d binary hard discs ~\cite{dunleavy2012} nor with a dynamic correlation length based on correlations in the mean square displacement in 2d polydisperse colloids ~\cite{tamborini2014}.

It is also possible to define a structural correlation length via a point-to-set, or pinning approach (section \ref{sectionRFOT}) ~\cite{biroli2008}. This is an order-agnostic approach, so it should be insensitive to any particular locally favoured structures. The exact protocol varies ~\cite{berthier2012}, but a consistent picture has emerged from a number of different groups ~\cite{hocky2012,kob2011non,karmakar2014}: the structural correlation length extracted from pinning increases much more slowly than the dynamical length $\xi_4$. The point-to-set length is also compatible with lengthscales extracted from the quasispecies approach ~\cite{karmakar2012,biroli2013prl}. Other order-agnostic methods based on mutual information, which also pick up whatever structural correlations are present, also produce structural lengths which do not scale with  $\xi_4$ ~\cite{dunleavy2012,cammarota2012epl}.

Another approach to obtain a static length is the lengthscale at which a crossover occur between two contributions to the eigenvalues of the Hessian (see section \ref{sectionIsoconfigurational} ~\cite{karmakar2012}. The contributions correspond to plastic events (which occur at short lengths) and elastic contributions at longer lengths. The resulting lengthscale follows others in that it increases by a small amount in the dynamically accessible regime  ~\cite{karmakar2012,biroli2013prl} and turns out also to be the lengthscale required to collapse the configurational entropy as a function of system size  ~\cite{karmakar2009}. Very recently, this length has been measured at exceptionally deep supercoolings generated by swapping particles ~\cite{gutierrez2014}. The increase in the static length seems comparable with measurements of dynamic lengths in molecular system at the experimental glass transition [see section ~\ref{sectionLudovicLength}]. One final order-agnostic length to mention is that developed by Mosayebi \emph{et al.} ~\cite{mosayebi2010,mosayebi2012} (section \ref{sectionOrderAgnostic}) which found critical-like scaling for a structural correlation. Like Tanaka and coworkers ~\cite{tanaka2010}, the lengthscales over which the correlation length was measured was a factor of $\sim 5$  ~\cite{mosayebi2010,mosayebi2012}.

Charbonneau and Tarjus have investigated hard spheres in a variety of dimensions and with a number of structural measures ~\cite{charbonneau2013pre,charbonneau2012,charbonneau2013jcp}. Appealing to geometric frustration (section ~\ref{sectionGeometricFrustration}), they investigated the spacing between defects. In the case that the curved space tiled by the LFS is close to Euclidean space, one expects that defect density might be small, and that the space between defects might become large leading to a long structural correlation length ~\cite{tarjus2005}. This and other measures of the structural correlation length however showed little increase in the simulation-accessible regime. This points to geometric frustration being ``strong'', and the third term in Eq. ~\ref{eqCNTFrustration} preventing any significant growth of domains of LFS. %In higher dimension, hard hyperspheres have less structure (with $S(k)\rightarrow1$ as $d\rightarrow \infty$) such that it is clear that structure should matter much more in lower dimension (including $d=3$). 

\begin{figure}[!htb]
\centering \includegraphics[width=60mm]{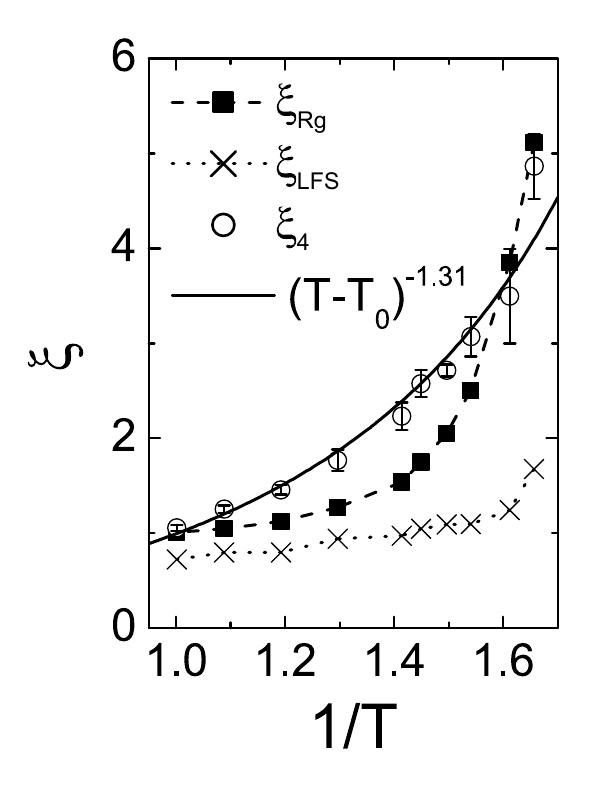} 
\caption{Scaling of various lengthscales in the Wahnstr\"{o}m binary Lennard-Jones glassformer. $\xi_4$ is the dynamic length from Eq. \ref{eqOZcrit}. $\xi_{\mathrm{LFS}}$ is the same measure applied to particles in icosahedra, the LFS for this system (Fig. \ref{figAlexNetwork}), $\xi_{Rg}$ is the radius of gyration of clusters of icosahedra \cite{malins2013jcp}.}
\label{figGlengths} 
\end{figure}

Other techniques to extract a structural lengthscale include an Ornstein-Zernike fit to particles in a structure factor comprised of particles in locally favoured structures $S_\mathrm{LFS}(k)$. This yields a correlation length $\xi_{\mathrm{LFS}}$ in the same spirit as $\xi_4$ in Eq. \ref{eqOZcrit}

\begin{equation}
S_\mathrm{LFS}(\textbf{k})=\frac{1}{N\rho} \langle \sum_{j=1}^{N_\mathrm{LFS}} \sum_{l=1}^{N_\mathrm{LFS}} \exp[-i \textbf{k} \cdot \textbf{r}_j(0)]\exp[i \textbf{k} \cdot \textbf{r}_l(0)] \rangle,
\label{eqSLFS}
\end{equation}

\noindent where $j$, $l$ index particles that are detected within a locally favoured structure and $N_{\mathrm{LFS}}=N_{\mathrm{LFS}}(\tau_\ell\ge0)$ is the total number of particles in LFS. The correlation length $\xi_{\mathrm{LFS}}$ is found by fitting the Ornstein-Zernike (OZ) equation ~\cite{royall2014,malins2013jcp,malins2013fara}. However plots such as Fig. \ref{figGlengths} make it clear that the linear $\xi_{\mathrm{LFS}}$ does not scale in the same fashion as $\xi_4$. However, although the actual lengthscale of the LFS domains themselves is less than that of $\xi_4$, it is worth noting that LFS retard the motion of surrounding particles ~\cite{malins2013jcp,malins2013fara}, so the discrepancy may not be as stark as Fig. ~\ref{figGlengths} indicates. Thus it is possible that particles stuck between two ``strands'' of LFS could be dynamically slowed, so in fact although the linear measure $\xi_{\mathrm{LFS}}$ does not increase much, in fact through the fractal nature of the LFS network, structure does lead to dynamically slow regions on the length scale of $\xi_4$ ~\cite{coslovich}. This could explain why the work by Mosayebi ~\emph{et al.} ~\cite{mosayebi2010,mosayebi2012} identified a larger structural lengthscale than most others.

\begin{figure}[!htb]
\centering \includegraphics[width=65mm]{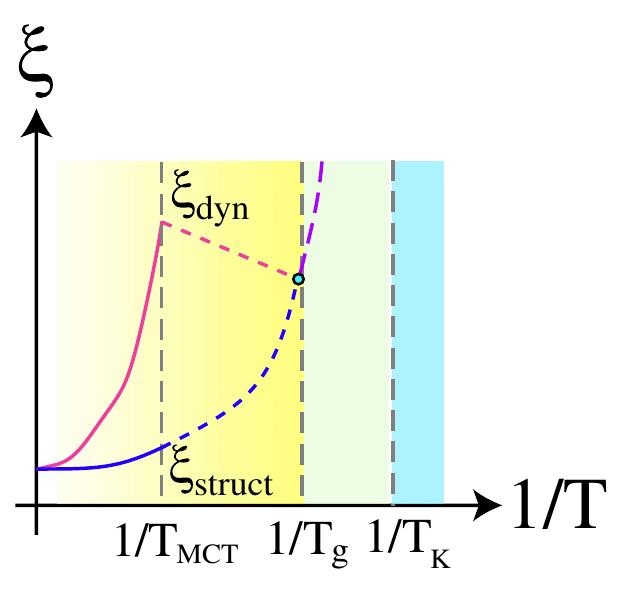} 
\caption{Schematic of possible behaviour of dynamic ($\xi_{\mathrm{dyn}}$) and static ($\xi_{\mathrm{struct}}$) lengthscales as the glass transition is approached. Particle-resolved studies data is available for $T\gtrsim T_{\mathrm{MCT}}$ (solid lines). Dashed lines represent a possible scenario at lower temperatures, extrapolated from recent simulations ~\cite{kob2011non}. Green dot is dynamic length $\xi_3$ deduced from molecular experiments close to $T_g$ (section \ref{sectionLudovicLength}). Purple dashed line represents coincidence of structural and dynamic lengths of cooperatively re-arranging regions envisioned by RFOT and Adam-Gibbs.}
\label{figXi} 
\end{figure}

If we focus on the lengths only, the picture that emerges is one of disparity between $\xi_4$ and the majority of structural lengths, as illustrated in Fig. \ref{figXi}. This leaves at least three possibilities :

\begin{enumerate}
\item Dynamic and structural lengths decouple as the glass transition is approached. And thus although structural changes are observed in many fragile glassformers, these are not the only mechanism of arrest and that other mechanisms play an important role.
\item $\xi_4$ is not representative of dynamical lengthscales. %Or its increase as a function of supercooling is not sustained. 
Or dynamics are perturbed at lengthscales beyond that of the LFS.
\item The vast majority of data so far considered is in the range $T>T_\mathrm{MCT}$ and thus is not supercooled enough for RFOT or Adam-Gibbs-type cooperatively re-arranging regions to really matter.
\end{enumerate}

We believe that a combination of all three, weighted differently depending on the system, is the most likely outcome. Some evidence for the first scenario is given by the fact the kinetically constrained models ~\cite{pan2004}, and hyperspheres in high dimension ~\cite{charbonneau2013pre} undergo arrest. In sufficiently high dimension, all structure is lost as the system becomes more mean-field like, although fragility increases ~\cite{sengupta2013}. Other mean-field like systems include very soft ``gaussian core'' potentials which allow core overlap for a moderate energy cost. Thus each particle interacts with many neighbours and little correlation between structure and dynamics is found ~\cite{hocky2014}. The system is moreover well described by MCT and thus may be expected to be very fragile ~\cite{ikeda2011}. If one accepts these (some admittedly abstract) models, structure cannot be a universal mechanism for dynamical arrest, although for the higher-dimensional work we are numerically restricted to $d \lesssim 12$ which may still be low enough.

%Further evidence in support of scenario one is provided by Cammarota and Biroli ~\cite{cammarota2012} that pinning can drive ideal glass transition of the type envisioned by RFOT and Adam-Gibbs theory, namely that configurational entropy vanishes. Under the pinning field, no change in structure occurs (subject to certain constraints) as a function of pinned particles, but a bona-fide glass transition as described by RFOT does  ~\cite{cammarota2012}. One possibility is to note that, as temperature drops a lower concentration of pinned particles is required for this pinning glass transition and that the transition is somehow driven by a combination of structure and pinning. Moreover, the separations between the pinned particles in the simulation accessible regime can approach one or two particle diameters ~\cite{kob2013}, suggesting rather small cooperatively re-arranging regions.

However, the comparatively rapid increase in $\xi_4$ is not without question ~\cite{harrowell2011}. Firstly, as indicated in Fig. \ref{figXi} $\xi_4$, seems to increase rapidly in the regime accessible to particle-resolved studies above $\sim T_{\mathrm{MCT}}$ ~\cite{harrowell2011}. Indeed, a free fit to measurements of $\xi_4$ for the Kob-Andersen model yielded \emph{divergence} close to the Mode-Coupling temperature for this model ~\cite{malins2013fara}. Moreover among early papers involving $\xi_4$, La\v{c}evi\'{c} \emph{et al.} showed divergence of this dynamic length around $T_\mathrm{MCT}$ ~\cite{lacevic2003}. This may indicate an echo in $d=3$ of the mode-coupling transition which would lead to a divergence of dynamic lengthscales in mean field. A recent paper by Kob \emph{et al.} ~\cite{kob2011non} indicates \emph{non-monotonic} behaviour of a dynamic correlation length based on pinning with a maximum around $T_{\mathrm{MCT}}$, as indicated in Fig. \ref{figXi}. These results are not without controversy ~\cite{flenner2012comment,kob2012reply}, but it has since been shown that, just below $T_\mathrm{MCT}$, at the limit of the regime accessible to simulations, $\xi_4$ at least exhibits a different scaling with temperature upon approach to the glass ~\cite{flenner2013}. Perhaps ironically, given the challenges of measuring $\xi_4$ in reciprocal space \cite{flenner2011,karmakar2009}, the early \emph{real} space measurements of La\v{c}evi\'{c} \emph{et al.} ~\cite{lacevic2002} show a very much weaker increase for the same system as their later (seminal) reciprocal space measurements ~\cite{lacevic2003}. Nor is this discrepancy between real and reciprocal space measurements unique : in conventional critical phenomena, the ``true'' correlation length is given in real space and only close to criticality do the real and reciprocal space calculations converge ~\cite{tarko1977} and in supercooled liquids, approach to any transition is challenging to say the least.

Further evidence that the dynamic correlation length might not diverge as fast as data from the $T> T_{\mathrm{MCT}}$ range would indicate is given by experiments on molecular glassformers close to $T_g$, an increase of some 8-10 decades in relaxation time compared to the particle-resolved studies. This approach measures a lower bound for the dynamic correlation length (section \ref{sectionLudovicLength}) ~\cite{berthier2005}. The lengths obtained by this approach correspond to a few molecular diameters ~\cite{berthier2005,dalleferrier2007,bauer2013,crauste2010,brun2011}. Other approaches include nuclear magnetic resonance ~\cite{tracht1998}, direct imaging of surface mobility in metallic glass ~\cite{ashtekar2012}, deductions from heat capacity and related quantities ~\cite{donth1982,tatsumi2012,yamamuro1998} and fluorescence measurements ~\cite{ediger2000,cicerone1995}. All suggest comparable lengthscales of a few diameters at $T_g$. Such a small dynamic correlation length (albeit a lower bound in some measures, section ~\ref{sectionLudovicLength}) certainly necessitates at the very least a drastic slowdown in the rate of increase of $\xi_{\mathrm{dyn}}$ followed by a levelling off.

Dalle-Ferrier \emph{et al.} \cite{dalleferrier2007} identified two regimes in the behaviour of the number of correlated particles. Experimental data on a number of molecular glassformers was considered. Above the mode-coupling temperature there was a relatively rapid increase in the number of particles undergoing correlated motion. $N_\mathrm{corr} \sim \tau_\alpha^{1/\gamma}$ with $\gamma=2-3$. This is consistent with mode-coupling theory. Below the $T_{\mathrm{MCT}}$ the increase in the number of correlated particles and thus the dynamic lengthscale seems much slower. However even in the high-temperature regime accessible to computer simulation, the lengthscales implied by these technique are very much less than $\xi_4$.

Theoretical work within the RFOT picture suggests that the picture of a unique length to characterise dynamic motion might obscure a change in the nature of co-operatively re-arranging regions ~\cite{stevenson2006}. Here it was shown that the CRRs should become more compact upon deeper quenching, with their fractal dimension approaching the dimension of the system. Around the mode-coupling temperature one expects more fractal CRRs, which is consistent with observations of strong-like motion ~\cite{kob1997,schroder2000,starr2013}. Around the mode-coupling crossover, it can be hard to distinguish individual CRRs, there can be connectivity and even percolation of these regions ~\cite{dunleavyThesis}. Thus defining a dynamic correlation length is not necessarily straightforward. We also note that domains of locally favoured structures have also been found have a fractal dimension $d_f \approx 2$ in a 3d system ~\cite{malins2013jcp}.

One possible way to reconcile these two regimes, $\sim T_g$ of molecular systems and $T \gtrsim T_{\mathrm{MCT}}$ of simulations and colloid experiments would be to extract correlation lengths from decay of pair correlations in molecular systems. Salmon and coworkers have shown that such decays may be described by a pole analysis (see section ~\ref{section2PointDetail}), and obtained a structural correlation length around 0.5 nm for GeO$_2$ (a strong liquid) below its experimental glass transition temperature ~\cite{salmon2006}. We further note that, small as these dynamic correlation lengths may be, recent work has linked the dynamic correlation length inferred from dielectric spectroscopy (section \ref{sectionLudovicLength}) with fragility, in that strong liquids are found to have essentially no change in their very small dynamic correlation length approaching $T_g$ whereas fragile liquids show a (modest) increase ~\cite{bauer2013}.

It is tempting to imagine that in the $T_{\mathrm{MCT}}>T>T_g$ range (or even in the regime below $T_g$), structural and dynamic lengths might scale together, corresponding to well-defined cooperatively re-arranging regions (Fig. ~\ref{figXi}). For now, however, the jury is well and truly out as to whether it is possible to demonstrate a structural mechanism for dynamical arrest via the coincidence of structural and dynamic lengthscales. The discrepancy observed by most between structural and dynamic lengthscales in the $T \gtrsim T_{\mathrm{MCT}}$ range is indicative that more is at play than structure, at least in the first few decades of dynamic slowing which are described by mode-coupling theory. However, it has been shown again and again ~\cite{pedersen2010,charbonneau2013pre,steinhardt1983,jonsson1988,dzugutov2002,shintani2006,kawasaki2007,kawasaki2010jpcm,tanaka2010,sausset2010,mosayebi2010,mosayebi2012,xu2012,charbonneau2012,charbonneau2013jcp,malins2013jcp,malins2013fara}
that the structure of deeply supercooled liquids is distinct from that of high temperature liquids and that particular locally favoured structures correspond to dynamically slow particles ~\cite{pedersen2010,malins2013jcp,tanaka2010,malins2013fara,leocmach2012,shintani2006}. Along with others~\cite{charbonneau2013pre}, we believe any coincidence in structural and dynamic lengthscales in the dynamical regime accessible to simulation may be obscured by the mode-coupling transition and that to resolve this question deeper supercooling is required. The recent work of Guti\'{e}rrez \emph{et al.} ~\cite{gutierrez2014} which studied static correlation lengths at exceptionally deep supercooling also produced lengthscales comparable with those dynamic lengthscales inferred in the experiments. We now turn to other approaches to investigate the role of structure in model glassformers.

\subsection{A structural-dynamical phase transition}
\label{sectionMu}

Inspired by the dynamical phase transition revealed in the $s$-ensemble, section ~\ref{sectionFacile}, one can enquire as to whether there is some structural difference between the inactive and active states observed. Evidence in this direction came from Jack \emph{et al.} ~\cite{jack2011} who used the $s$-ensemble to produce very stable states. These were apparent in that running normal (unbiased) dynamics from these configurations produced trajectories where essentially no relaxation was observed until ``melting'' to the normal (mobile) supercooled liquid. In other words, something in the configurations produced in the $s$-ensemble encodes slow dynamics.

Indeed the first-order transition may be seen in terms of structure as well as dynamics (Fig. \ref{figSMu}) ~\cite{speck2012}. Here the structural change takes the form of an increase in the population of bicapped square antiprisms which are known to be the locally favoured structure for the Kob-Andersen model ~\cite{coslovich2007,malins2013fara}. However, we may go further. Rather than sampling a biased distribution of trajectories with the mobility (the $s$-ensemble), one may instead apply similar methodology using structure rather than mobility to bias the ensemble of trajectories. This approach is termed the $\mu$-ensemble because it uses a ``dynamic chemical potential'' to drive the transition. There is a similar fat tail in the distribution of trajectory cluster population (Fig. ~\ref{figSMu} d) in the unbiased data. Biasing trajectories with the number of clusters  (Figs. ~\ref{figSMu} e,f) leads to a transition which appears more dramatic than the dynamic $s$-ensemble transition. Closer investigation revealed that both are in fact the same transition, but driven in different ways ~\cite{speck2012}. Thus \emph{time-averaged} structure may be used to drive a first-order phase transition to an inactive state. The $\mu$-ensemble may provide a way to prepare very stable glasses, which may be similar to those recently produced by chemical vapour deposition ~\cite{singh2013,lyubimov2013}. The $\mu$-ensemble is essentially a liquid-liquid transition in trajectory space (see section ~\ref{sectionPolyamorphism} for a discussion of conventional liquid-liquid transitions). Like pinning (section \ref{sectionRFOT}), at lower temperature a smaller value of the $\mu$ field is required for the transition. One may enquire as to the zero-field limit of the transition. Very recently evidence has been found in support of a liquid-liquid transition in the zero field case, i.e. a conventional LLT ~\cite{speck2014}.

\subsection{The isoconfigurational ensemble and normal modes}
\label{sectionIsoconfigurational}

The isoconfigurational ensemble was introduced as a means to probe predictability and the role of structure in an order-agnostic manner. As its name implies, in this ensemble many simulation trajectories are run from the same initial configuration, with randomised momenta sampled according to the Boltzmann distribution ~\cite{widmercooper2004,widmercooper2005,widmercooper2006}. This ensemble is not preserved by the equations of motion and as time progresses the differences between the configurations of the various ensemble members grow with the system relaxing towards the equilibrium specified by the equations of motion. Widmer-Cooper and Harrowell used this ensemble to define the \emph{propensity}, a measure that ascribes a value to each particle in the system, typically the mean square displacement for each particle at $1.5$ $\tau_\alpha$ averaged across the ensemble of trajectories. They found that the propensity in their new ensemble was highly inhomogenous in a binary soft disc model upon supercooling ~\cite{widmercooper2006}. Because the simulations are run from the same initial configuration, that some particles have a high propensity (are inclined to move) and some have a low propensity (are inclined to stay put) provides strong evidence that something in the structure is responsible. The question is what? Not the free volume around each particle ~\cite{widmercooper2005}, nor the potential energy ~\cite{widmercooper2006jnonxsol,matharoo2006}. We are unaware of any locally favoured structures for the 1:1.4 size ratio binary soft disc system studied, though a similar 1:1.4 size ratio binary hard disc system showed a characteristic discrepancy between structural and dynamic correlation lengths ~\cite{dunleavy2012}. Moreover in the Kob-Andersen system, Berthier and Jack found that structure at longer lengthscales than the particle-level was much more effective in predicting the ensuing dynamics ~\cite{berthier2007}. Jack \emph{et al.} ~\cite{jack2014} and Hocky \emph{et al.} ~\cite{hocky2014} have recently revisited the degree to which local structure predicts dynamics and found a limited degree of correlation.

Propensity has been identified with low-frequency ``soft'' modes ~\cite{widmercooper2008}. Soft modes had been studied previously in the context of the jamming transition and have recently been connected to glasses ~\cite{xu2007}. Soft modes refer to low-frequency modes, obtained from a normal mode analysis on the Hessian matrix ~\cite{feynman} $\mathbf{H}$ which is given in terms of the second derivative of the potential energy $H_{ij}=\partial^{2}\Phi/\partial q_{i}\partial q_{j}$ where there are $dN$ coordinates $q_{i}$ representing the positions $x_{i}$, $y_{i}$ etc. with $d$ the dimension of the system. The Hessian is a symmetric matrix and so it has orthogonal eigenvectors which are given by the orthonormal coordinates $\mathbf{e}_{n}$ and the corresponding eigenvalues $\lambda_{n}$. A displacement along any direction specified by one of these eigenvectors results in a Hookean force in the opposite direction. Because the eigenvectors are all mutually orthogonal we obtain an independent vibrational mode in each of the directions given by the eigenvectors $\mathbf{e}_{i}$. The eigenvalues $\lambda_{n}$ form the spring constant for each of these modes with the frequency $\omega_{n}=\sqrt{\lambda_{n}/m}$ where $m$ is the particle mass . The low frequency modes have a weak spring constant, $\lambda_{n}$, and are termed soft modes.

\begin{figure}[!htb]
\centering \includegraphics[width=80mm]{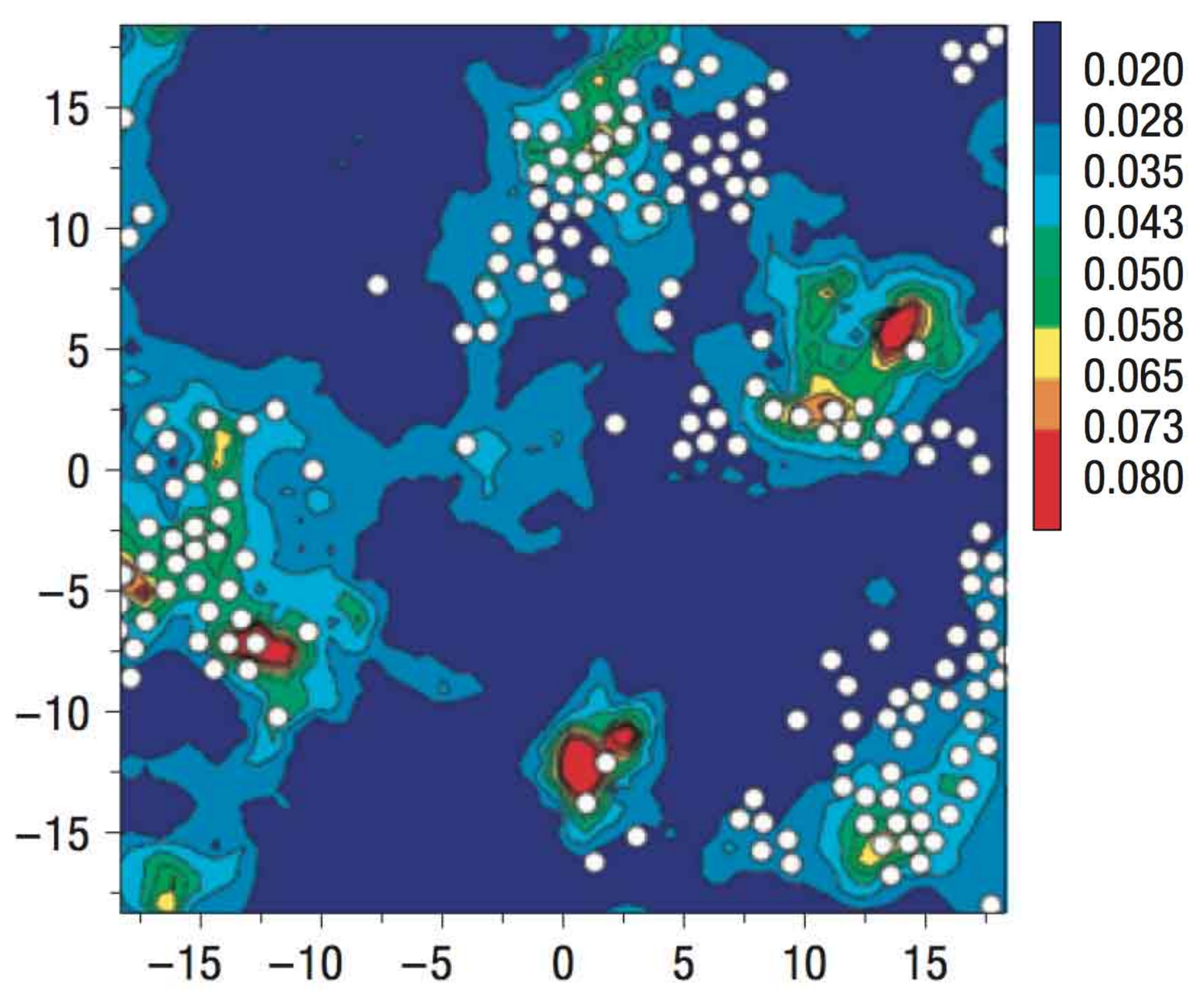} 
\caption{Correlation between normal-mode structure and propensity in the isoconfigurational ensemble. Contour plots of the low-frequency mode participation, overlaid with the location of particles (white circles) where the iso-configurational probability of losing four initial nearest neighbours within 0.3$\tau_\alpha$  is greater than or equal to 0.01. 
 \cite{widmercooper2008}. Reproduced with permission from Nature Publishing Group Copyright 2008.
\label{figNormalModes} }
\end{figure}

In reference ~\cite{widmercooper2008} a connection between the soft modes and a new variable closely related to the propensity was made using contour plots for the $d=2$ simulation results. The key point being that by taking the initial configuration, a connection or causal relationship was found for what would happen when the equations of motion were used to propagate the system forward in time. To do this the participation fraction of each particle in a given mode $p_{ij}$ was considered. This was given as the square of the component due to $j^{th}$ particle in the $i^{th}$ mode. So given $e_{ij}=\mathbf{e}_{i}\cdot\hat{\mathbf{q}}_{j}$ where $\hat{\mathbf{q}}_{j}$ is the unit vector in the direction given by summing the $d$ unit vectors in the Cartesian directions given by $j^{th}$ particle, then the participation fraction is $p_{ji}=e_{ij}^{2}.$ Contour maps were produced for the participation fraction of the various particles in the 30 softest modes. One of these maps is shown in Fig. \ref{figNormalModes} where a comparison is made with a variation in the \emph{propensity}. Here the particles that lose 4 nearest neighbours at $1.5$ $\tau_\alpha$ with a probability of more than 0.01 using the isoconfigurational ensemble are also plotted. This shows a causal relationship between the particles which contribute most substantially to the softest modes at the initial time and the likelihood to loose nearest neighbours in the future, which is closely related to a larger dynamic propensity. Thus soft small oscillations are causally related to how likely the particles are to undergo large excursions in the future. This was explored further by Candelier \emph{et al.} ~\cite{candelier2010} who connected normal modes with ``avalanches'' which are cooperative relaxation events spanning many particle diameters. Thus structure can influence dynamics over long lengthscales.

The connection between modes and propensity is given further significance by Coslovich and Pastore ~\cite{coslovich2006,coslovich2007ii}, who showed that the modes undergo a transition to localised saddles in the energy landscape around the mode-coupling temperature. Furthermore local structural measures (in the form of two-point correlation functions) were correlated with propensity. Higher-order structure in the form of LFS have also been quantitatively correlated to propensity by Jack \emph{et al.} and Hocky \emph{et al.} along with the normal modes ~\cite{jack2014,hocky2014}. The latter are strongly predictive of particle propensity at short times $< \tau_\alpha$ ~\cite{jack2014}, while LFS are correlated with the propensity at longer times $\sim \tau_\alpha$  ~\cite{jack2014,hocky2014}. Normal modes have proved very useful in connection with sheared systems. In particular they have been shown to correlated with so-called shear transformation zones where the first plastic events occur when a glass yields. We discuss these in section \ref{sectionShear}.

\subsection{A structural origin of the Boson peak ?}
\label{sectionBosonPeak}

\begin{figure}[!htb]
\centering \includegraphics[width=60mm]{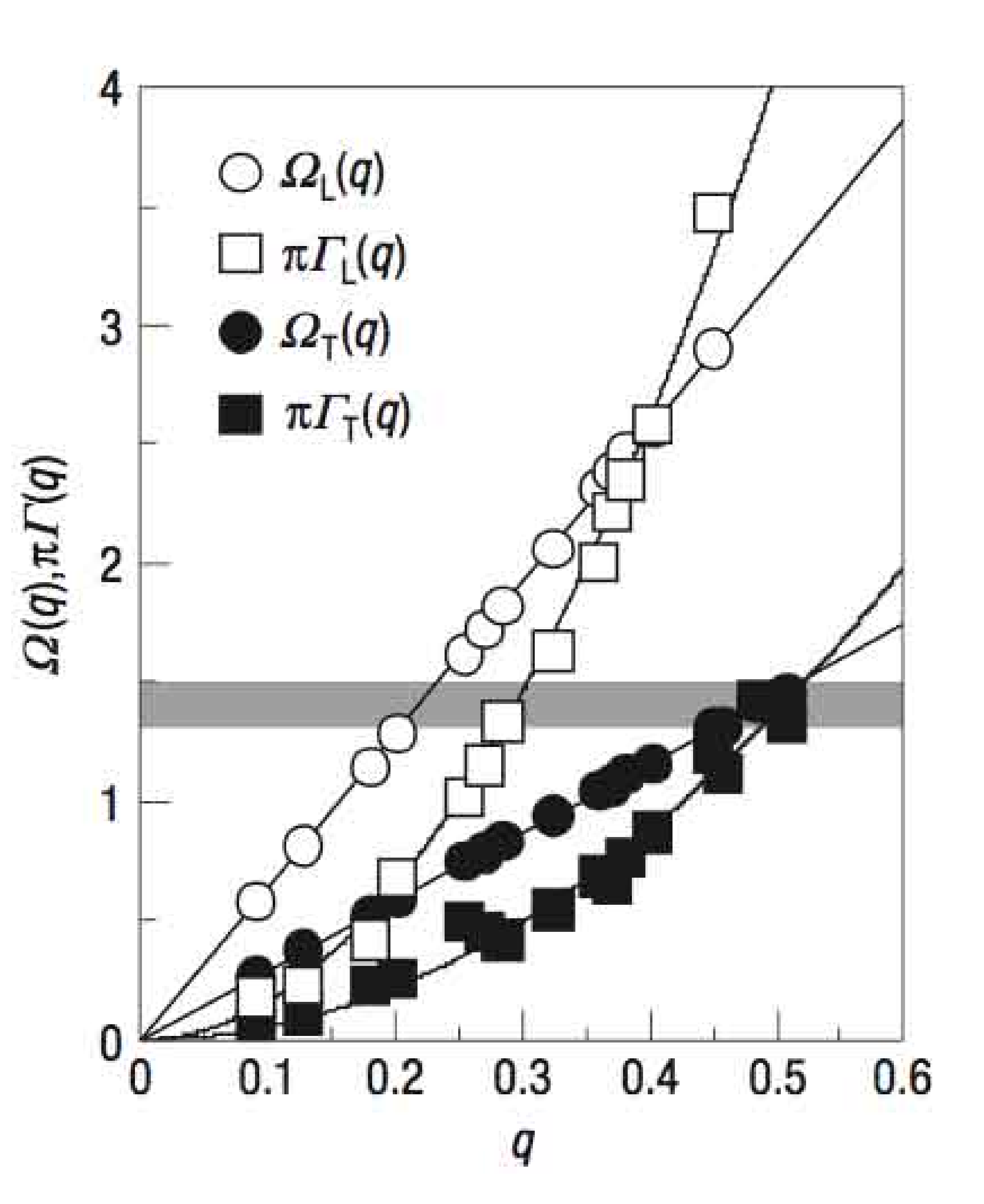} 
\caption{Phonon driving frequencies $\Omega$ (circles) and full-width-half maxima $\pi\Gamma$ (squares) with the Boson peak frequency shown as the grey bar. Unfilled symbols are longitudinal modes, filled are transverse modes. Note the conicidence of the Boson peak frequency with the transverse Ioffe-Regel limit above which phonons no longer propagate.
 \cite{shintani2008}. Reproduced with permission from Nature Publishing Group Copyright 2008.
\label{figShintaniIoffe} }
\end{figure}

The Boson peak is one of the mysteries of glasses. Note that here we consider solid glasses, rather than supercooled liquids, thus full structural relaxation does not occur on the experimental timescale. Now the Boson Peak refers to an excess vibrational density of states, at low frequency, over and above the usual Debye $\omega^2$ scaling, where $\omega$ is the vibrational frequency, which is typical of solids ~\cite{berthier2011}. Xu ~\emph{et al.} investigated normal modes in the Kob-Andersen model and a WCA model ~\cite{xu2007}. They found an excess of soft modes (the same kind of soft modes as Widmer-Cooper ~\emph{et al.} considered in the preceding section) which they  correlated with the Boson peak. This low-temperature extension of glasses ultimately crosses over into the zero-temperature domain of jamming, which we discuss further in section~\ref{sectionJammin}.

With their classical model of NiZr (see section ~\ref{structureInMetallicGlassformers}), Guerdane and Teichler ~\cite{guerdane2008} investigated the relationship between local structure and the boson peak. In their model, atoms with a high tendency to be involved in ``local environment transitions'' (bond network re-arrangement) contributed much more to the Boson peak that other atoms. These atoms tended to be located in [0,3,6,0] Voronoi polyhedra of one atom and nine neighbours.

Shintani and Tanaka used their 2d model (discussed in section \ref{sectionTowardsAStructuralMechanism} and Fig. \ref{figShintani}) to investigate the structural origins of the Boson peak. They found that the peak frequency for their model corresponded to the Ioffe-Regel limit. This latter quantity is the upper limiting frequency at which phonons propagate and is given by $\Omega(k)=\pi\Gamma(k)$ where $\Omega$ is the phonon excitation frequency and $\Gamma$ is the full-width half maximum of the phonon spectrum. Thus the Boson peak corresponds to the upper limit for the propagation of transverse phonons ~\cite{shintani2008}. 

Knowing the local structure from their previous work ~\cite{shintani2006}, Shintani and Tanaka correlated the amplitude of the particle vibrations at the Boson peak frequency. There are three kinds of local structure in this system, crystal-like structure (which contributed little to the vibrations at the Boson peak frequency), local fivefold symmetry and liquid-like disorder. The latter two quantities were found to contribute to the excitations at the Boson peak frequency ~\cite{shintani2008}, similar to the work of Guerdane and Teichler ~\cite{guerdane2008}.

Related behaviour has been observed in molecular glassforming systems. Carini \emph{et al.} ~\cite{carini2013} showed that upon decreasing the charactistic structure in B$_2$O$_3$ glass, groups of three corner sharing BO$_3$ triangles, the boson peak was reduced. However, whether such structures play a similar role as the local structures in Shintani and Tanaka's model in the supercooled liquid regime ~\cite{shintani2006} is unknown for these  molecular glassformers.

\subsection{Evaluating the energy landscape}
\label{sectionEvaluatingTheEnergyLandscape}

In recent years, improvements in computational processing power have enabled development of techniques to investigate some of the predictions of the energy landscape concept (section ~\ref{sectionEnergyLandscape})  ~\cite{goldstein1969,sciortino2005}. Among the first developments ~\cite{stillinger1983,stillinger1984} was to establish the ``inherent structure''. The inherent structure is the result of a steepest-descent quench to $T=0$, which thus takes the system to its nearest local minimum in the potential energy landscape (Fig. \ref{figEnergyLandscape}). Stillinger and Weber ~\cite{stillinger1983,stillinger1984} further showed that for low temperatures (where the system is confined to a particular basin for long periods), the entropy could be expressed as a sum of a configurational and vibrational part, $S=S_{\mathrm{conf}}+S_{\mathrm{vib}}$. This holds for spherically symmetric simple liquids such as the model systems discussed in section \ref{sectionCommonModelSystems}. The vibrational part can be treated harmonically, and combined with the inherent state contribution $S_\mathrm{conf}$, it is possible to ``numerically evaluate the energy landscape'' ~\cite{heuer2008,karmakar2009,schroder2000,sastry1998,sciortino1999,sastry2001}.

Such investigations have largely confirmed Goldstein's energy landscape picture, that the system can be viewed as residing in basins for times of order of the $\alpha$ relaxation time before undergoing rapid transitions between basins ~\cite{goldstein1969}. In particular, Doliwa and Heuer ~\cite{doliwa2003} have related local structure in the basins (via the potential energy). Basins with high activation energy barriers for escape corresponded to low potential energy. They found that escape from a particular basin corresponds to a complex multistep process involving a succession of energy barriers. Interestingly from the point of view of local structure, some work has suggested that such transitions involve small compact clusters (which might correspond to co-operatively re-arranging regions) ~\cite{appignanesi2006}, unlike the string-like motion identified in earlier studies ~\cite{schroder2000,donati1998}.

Very recently the three-boy version of the structural entropy $s_2$ (Eq. \ref{eqS2}), $s_3$, was invoked to capture higher-order contributions to the entropy ~\cite{banerjee2014}. Banerjee \emph{et al.} found that, via the Adam-Gibbs theory, that the two-body structural entropy indicated a transition around the mode-coupling temperature, consistent with the idea that MCT describes two-body correlations (section \ref{sectionTwoPointStructure}). Meanwhile including $s_3$ indicated a vanishing of structural entropy at lower temperature, thus higher-order contributions to the structure, this time in the form of the configurational entropy, were invoked to explain the dynamical behaviour below the mode-coupling transition.

\begin{figure}[!htb]
\centering \includegraphics[width=60mm]{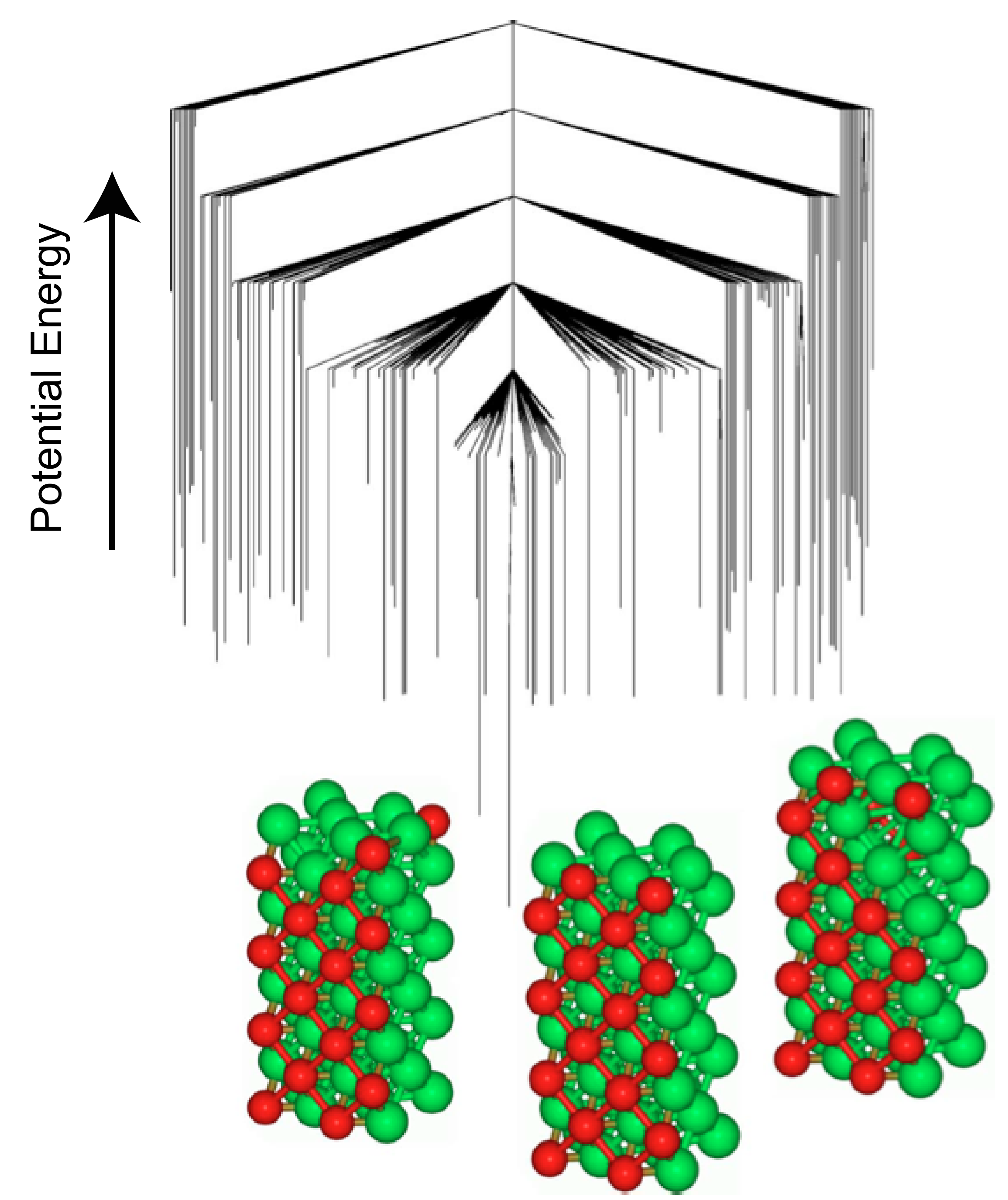} 
\caption{The energy landscape of a $N=320$ Kob-Andersen system represented as a disconnectivity graph. Three low-lying minima which have a crystal-like appearence are rendered ~\cite{middleton2001,calvo2007}. Note the similarity of these crystals to the bicapped square antiprism which is known to be locally favoured structure for this system ~\cite{coslovich2007,malins2013fara}. Reproduced with permission from American Institute of Physics Copyright 2007.
\label{figVanessa} }
\end{figure}

Wales and coworkers have developed powerful numerical techniques to enable the energy landscape of systems with a few hundred particles to be explicitly calculated ~\cite{wales}. This is sufficient to explore glassy behaviour, such that the energy of a representative set of basins in the energy landscape can be expressed in a ``disconnectivity graph'' (Fig. ~\ref{figVanessa}) ~\cite{calvo2007,desouza2008}. Such analysis yields the minimum energy structure, which, in the case of the Kob-Andersen system is crystalline  ~\cite{middleton2001} with a striking structural similarity to the bicapped square antiprisms which the locally favoured structure for this model ~\cite{coslovich2007,malins2013fara}.  Another means to evaluate the energy landscape, emphasising the dynamics, has been advanced by Ma and Stratt ~\cite{ma2014} through their geodesic approach to hard spheres. This approach emphasises dynamical aspects of the energy landscape. In particular, the distance between two configurations is considered, such that a non-overlapping pathways is found. The shortest path, or \emph{geodesic} distance becomes large upon the onset of slow dynamics. These geodesic paths may be related to the diffusivity with high precision. Given that the geodesic properties are encoded in the structure of the system, this approach would appear to allow a means to connect structure and dynamical arrest. One may think of it as quantifying the ``distance'' between two minima in the energy landscape.

\subsection{Gelation : a structural mechanism}
\label{sectionGel}

\begin{figure}[!htb]
\centering \includegraphics[width=80mm]{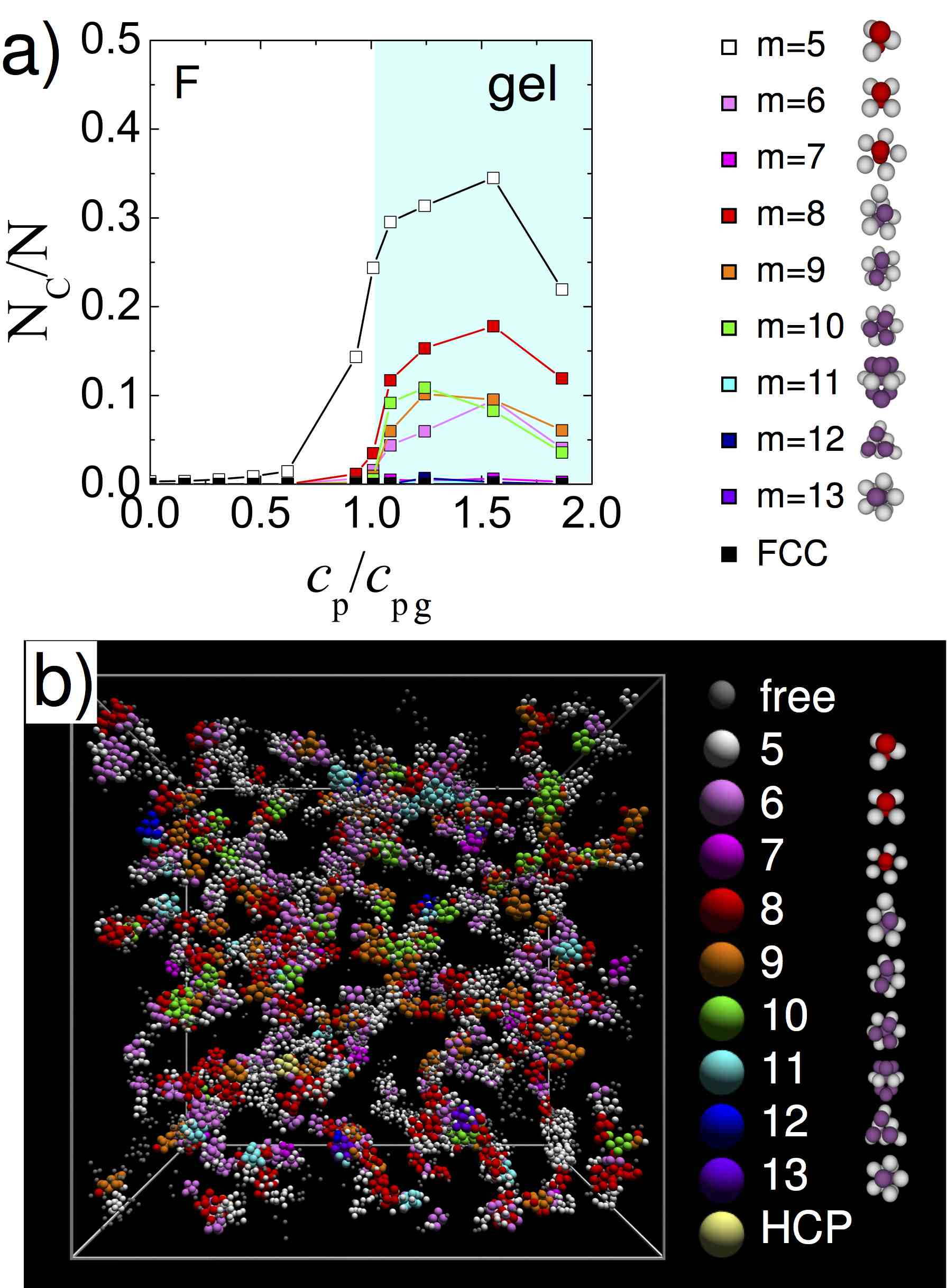} 
\caption{A colloidal gel decomposed into clusters (LFS) using the topological cluster classification (section \ref{sectionTCC}) ~\cite{royall2008}. (a) cluster population as a function of polymer concentration $c_p$ which plays the role of inverse temperature. The population increases suddenly at gelation ($c_{pg}$) leading to dynamically arrested networks of LFS.}
\label{figDilGel} 
\end{figure}

It is fair to say that, regarding the glass transition, the role of structure is not yet resolved. However some progress has been made regarding spinodal-type gelation (section \ref{sectionGelation}) ~\cite{royall2012,royall2008,royall2011c60}. Here the role of LFS is much more clear-cut. Upon condensing out of solution, as the system is quenched through the liquid-gas spinodal, the structures formed in the dilute colloidal suspension are precisely those small ($m<14$) structures identified by the topological cluster classification. These are geometrically resistant to motion and the resulting gel network (which is solid on experimental timescales) can be entirely decomposed into these minimum energy clusters (Fig. \ref{figDilGel}). Thus spinodal gelation corresponds to a local free energy basin via decomposition of the system into small groups of particles ~\cite{royall2012,royall2008}. Moreover while gels of this nature have typically been found in colloids, molecular gels (for example C$_{60}$) could also form by the same mechanism ~\cite{royall2011c60}.

\section{Structure in applied glassformers}
\label{sectionApplications}

So far we have mainly discussed structure in model glassformers. We now turn our attention to the vast number of materials which find application as glasses. Here we consider some pertinent structural aspects of such materials, noting first their many different classes: oxides ~\cite{salmon2013}, small molecules, metallic glasses ~\cite{greer2009,cheng2011,faupel2003},  chalcogenides ~\cite{wuttig2007,lencer2008} and polymer glasses. Here the focus is often on identifying structural motifs in the glass, rather than trying to relate structure to the emergence of solidity in the supercooled liquid (as in section \ref{sectionTowardsAStructuralMechanism}).

We will focus on metallic glasses, and also briefly discuss phase change materials based on chalcogenide glassformers, referring the reader to the reviews above for other classes of glassformers. Firstly, however, we recall the discussion at the beginning of section \ref{sectionStructureModel}, regarding fragility and jump in specific heat upon the material falling out of equilibrium at the glass transition~\cite{richert1998,ito1999,ngai1999,martinez2001}. Metallic glasses are among the materials where fragility is positively correlated with the jump in specific heat. The implication here is that more fragile materials require a higher degree of co-operativity to relax and are thus ``more structural'' in their relaxation mechanisms.

\subsection{Structure in metallic glassformers}
\label{structureInMetallicGlassformers}

Structure in metallic glassformers has been extensively reviewed by Cheng and Ma ~\cite{cheng2011}. Here we give a brief overview of studies which have focussed on local structure and correlated it with dynamical behaviour. The interactions of metal atoms are approximately spherically symmetric and have a reasonable amount in common with hard spheres and the Lennard-Jones systems. Both are often invoked to describe metallic glasses and indeed  the Kob-Andersen model is loosely based on Ni$_{80}$P$_{20}$ ~\cite{kob1995a}.  Metallic glassformers present a zoo of possibilities. Most elements are metallic, and may be combined with non-metallic elements to produce glasses which are themselves metallic. Given that there is no limit to the number of components a glass may have, the range of possible materials is immense. Here we shall consider a few well-studied systems. We refer the interested reader to the more in-depth analysis of Cheng and Ma ~\cite{cheng2011}.

Often more elaborate schemes than the simple models we described in section \ref{sectionCommonModelSystems} are used to model specific metallic glassformers. These include the embedded atom model (EAM) in which the interaction between the atom and electron charge density is included with pairwise terms~\cite{baskes1992} and full \emph{ab initio} simulations of the Car-Parrinello variety ~\cite{kuhne2012}. However, the primary difference between interactions in metallic systems and the model systems discussed in section \ref{sectionCommonModelSystems} is that metals are much softer ~\cite{chacon2001}.

Early insights into the structure of metallic glasses include those of Egami and Waseda ~\cite{egami1984} who considered that the minimum concentration of the minority species in binary alloys required to form a glass (to suppress crystallisation) was inversely correlated with the mismatch in atomic volumes $[v_A-v_B]/v_A$ where $v_\alpha=\pi \sigma_\alpha^3/6$ and $B$ is the minority species or solute. They attributed this result to solid solutions (crystals where lattice sites are substitued by another species) being destablised geometrically by the inclusions of the $B$ atoms.

Subsequently, Senkov and Miracle ~\cite{senkov2001} noted that for larger size ratios (larger than around $0.8$), smaller atoms could locate in the interstices between the larger atoms without disrupting the crystal and thus a higher concentration of the minority species would be required to suppress crystallisation as the size disparity increased further. The same group then went on to consider clusters of various geometries, noting the size disparity in a method reminiscent of the coordination $Z$ introduced by Frank and Kasper ~\cite{frank1958} discussed in section \ref{sectionFrank}. The idea was that when the minority species were larger than the hosts, more atoms were found in the shell, and that there were certain particularly favoured size ratios for each $Z$. In fact stable metallic glasses seem well-correlated with such stable size ratios ~\cite{miracle2003}, which can depend on the value of $Z$ ~\cite{miracle2006mt}. Miracle and coworkers subsequently developed the cluster model (see section \ref{sectionCrystalLikeOrdering})  ~\cite{miracle2004} to consider interconnected clusters of Bergman and Mackay icosahedra which are comprised of more layers on an $m=13$ icosahedron ~\cite{miracle2008}. At the level of pair distribution functions these larger icosahedra provide some agreement with experimental data.

\textit{Pure metals ---} 
Identification of icosahedral order began with neutron diffraction in (pure) Ni, Fe and Zr ~\cite{schenk2002} and Co ~\cite{hollandmoritz2002}. Here the static structure factor obtained from neutron scattering is compared to static structure factors calculated from specific structures. In this case, those calculated from icosahedra were found to provide good agreement with the experiments, thus evidence was presented in favour of icosahedral short-ranged order in (monatomic) metallic melts. Later, the same group went on to demonstrate similar behaviour (in terms of the shape of the static structure factor) in suspensions of charged colloids ~\cite{wette2009}, which underscores the similarity between colloids with soft electrostatic repulsions and metals. Other approaches include the X-ray absorption spectroscopy (XAS) of di Cicco \emph{et al.}  ~\cite{dicicco2003} (see section \ref{sectionHigherOrderReciprocal}). Subsequent \emph{ab initio} simulations of liquid Cu confirmed these results ~\cite{ganesh2006}. Li \emph{et al.} found a higher degree of icosahedral order [as determined by the $W_6$ BOO invariant (section ~\ref{sectionBOO}), which also picks up, for example pentagonal bipyramids] in embedded atom model simulations of Al ~\cite{li2008prb}. Other more complex scenarios are also observed. For example, in \emph{ab initio} molecular dynamics simulations Jakse and Pasturel found that Boron forms parts of empty icosahedra ``inverted umbrellas?? i.e. pentagonal bipyramids  (Fig. ~\ref{figBaka}) missing one particle ~\cite{jakse2014}.

\begin{figure}[!htb]
\centering \includegraphics[width=80mm]{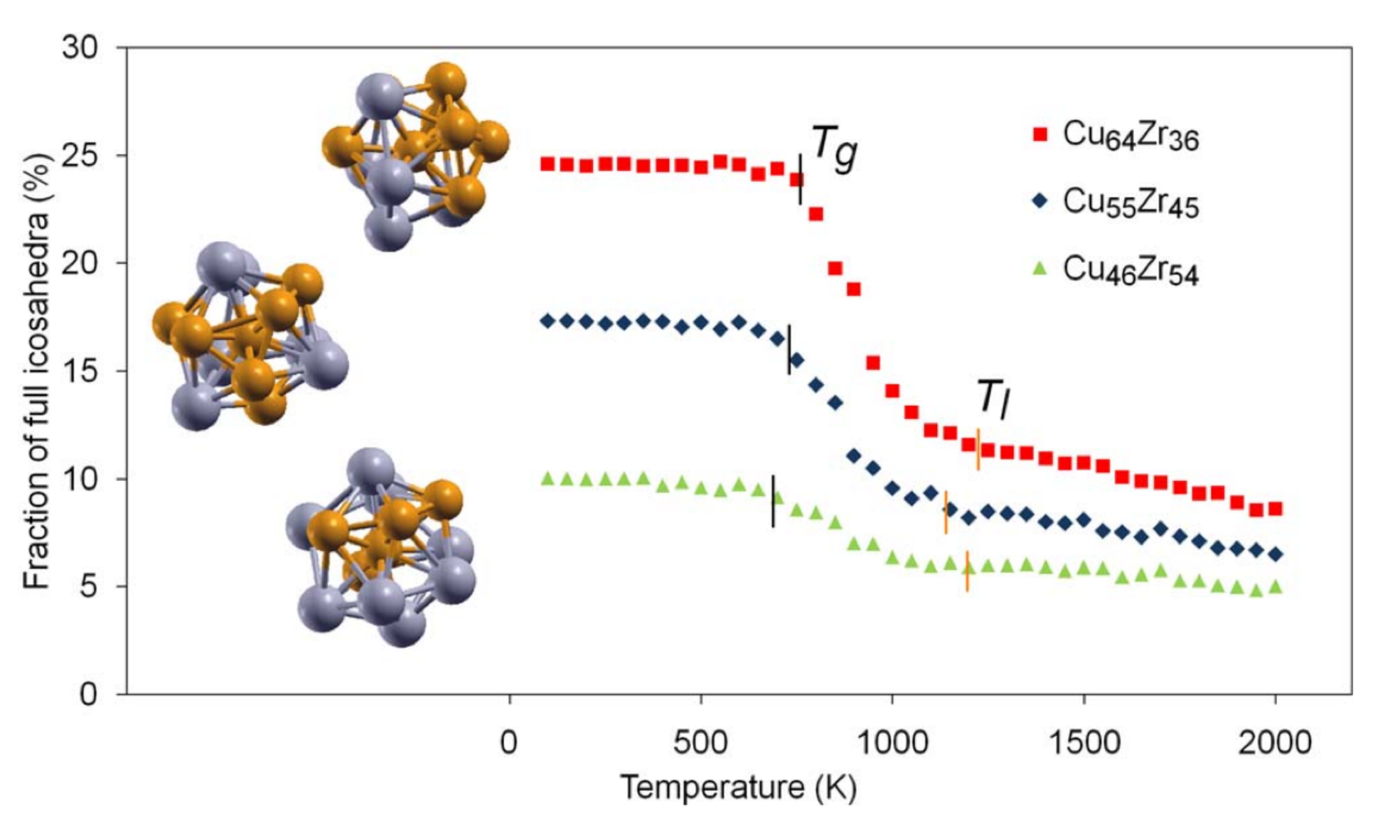} 
\caption{Increase in population of icosahedra upon cooling in CuZr ~\cite{cheng2008}. Note the similarity with Fig. \ref{figAlexNetwork}.
Reproduced with permission from American Physical Society Copyright 2008.}
\label{figMaIcos} 
\end{figure}

\textit{Copper-Zirconium : a model binary alloy ---} 
Amorphous alloys of early-late transition metals are often around 50-50 in composition. Here the prototype is CuZr (Cu is an early transition metal, Zr is late). The similarity in generic behaviour between this metallic glassformer and the model systems discussed in section \ref{sectionStructureModel} is exemplified in Fig. \ref{figMaIcos}. Here simulations of an embedded atom model of CuZr exhibit an increasing population of icosahedra upon cooling ~\cite{cheng2008} in very much the same way as does the Wahnstr\"{o}m model as shown in Fig. ~\ref{figAlexNetwork}  ~\cite{malins2013jcp,coslovich2007}. This generic behaviour for CuZr is further supported by \emph{ab-initio} simulations which also find a large number of icosahedra centred on Cu ~\cite{jakse2008}. The population of icosahedra appeared to peak around  Cu$_{0.64}$Zr$_{0.36}$ ~\cite{jakse2008prb}. Experimental data  (X-ray scattering) was fitted at the two-point level by reverse Monte Carlo. The RMC data were then compared against \emph{ab initio} simulations which were equilibrated at 1500 K and cooled to 300 K. In Cu-rich compositions strings of Cu-centred icosahedral clusters were formed ~\cite{li2009prb}.

A combined study of embedded atom model simulation and X-ray diffraction in the same system was made by Mendelev \emph{et al.}  ~\cite{mendelev2009}. Here it was found that there was a maximum in the diffusivity at a composition around Cu$_{0.7}$Zr$_{0.3}$, and the simulation revealed larger volumes around the Cu at this composition. This was thought to underly the change in $T_g$ with composition for CuZr. Peng \emph{et al.} ~\cite{peng2010} considered Zr-centred clusters. These took the form of [0,1,10,4] and [0,1,10,5] Voronoi polyhedra, and the atoms in these Zr-centered clusters were found to exhibit slower dynamics than the average and the presence of these clusters was correlated with the glass forming ability (GFA). The same group ~\cite{hao2010} then found a change in Zr-Zr order as Cu composition increased, and also found a drop in diffusivity under the same conditions.

The work of  Hao \emph{et al.} ~\cite{hao2010} is an example of a difference in perspective between those working in the metallic glass community and those working with model systems. Hao \emph{et al.} write ``our analysis for the dynamic properties of high temperature liquids of ZrCu metallic alloys shows that the icosahedral clusters together with some other pentagon-rich Voronoi clusters are responsible for slowing dynamics''. They based this statement on their observation that icosahedra last longer than other clusters, and that particles at the centre of icosahedra are slow. \emph{Qualitatively identical} findings were obtained by Coslovich and Pastore ~\cite{coslovich2007} and later by Malins \emph{et al.} ~\cite{malins2013jcp} for the Wahnstr\"{o}m Lennard-Jones model. Yet such is the emphasis laid upon the coincidence of structural and dynamic lengthscales (see section \ref{sectionStaticAndDynamicLengths}) that because the lengthscale associated with domains of icosahedra did not match the dynamic lenghscale of $\xi_4$ (Eq. \ref{eqOZcrit}), Malins \emph{et al.} concluded that the formation of icosahedra was not in itself the (only) origin of the slow dynamics.

\begin{figure}[!htb]
\centering \includegraphics[width=55mm]{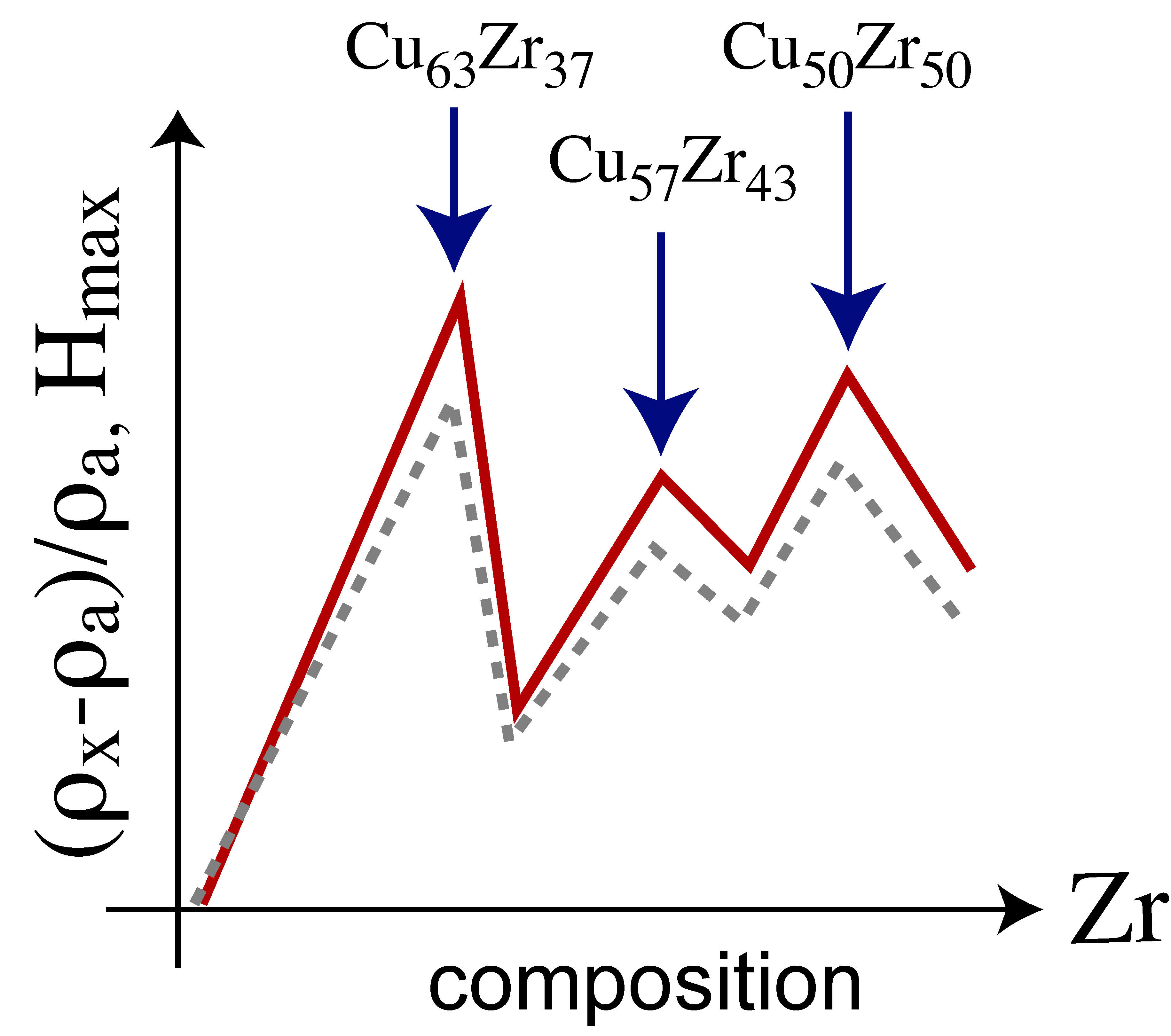} 
\caption{Correlation between density change between amorphous state and crystal $(\rho_x-\rho_a)/\rho_a$ and glass-forming ability for which the proxy is the largest thickness of material which does not crystallise. These quantities are almost perfectly correlated, and exhibit three clear peaks. Schematic based on Li \emph{et al.}  ~\cite{li2008science}. 
}
\label{figCuZr} 
\end{figure}

In CuZr, a near-perfect correlation was found between the density change between the crystal and amorphous form and the glassforming ability, as shown schematically in Fig. ~\ref{figCuZr} ~\cite{li2008science}. The GFA was taken as being determined by the maximum thickness which did not crystallise. Thus small changes in density between the glass and crystal (around 2\%) would indicate good glassforming ability (GFA), implying that the structure of the glass might be similar to that of the crystal. While a number of the papers cited above focussed on icosahedra in CuZr, the order-agnostic approach of Fang \emph{et al.} used their order-agnostic methods to identify so-called Bergman triacontahedra, which are combinations of icosahedra, dodecahedra. This work indicates structural lengthscales of order four particle diameters. While this is not a spectacular length in itself, it is rather larger than many of the lengthscales discussed in section \ref{sectionStaticAndDynamicLengths}.

\textit{Introducing impurities ---}
Using EAM potentials, Cheng \emph{et al.} ~\cite{cheng2009} found that inclusion of a small amount of Al enhanced the GFA. They attributed this to the shorter interatomic distances in the case of Al-centred icosahedra, which have the potential to reduce the strain as an icosahedron of like-sized atoms has a strained shell (see Fig. \ref{figBaka}). Fujita \emph{et al.} used \emph{ab initio} simulation and k-edge extended X-ray absorption fine structure (EXAFS) to investigate the addition of a small amount of Ag to CuZr. Their results indicated this led to local structural inhomogeneities which promoted GFA. In particular, they found Zr-rich clusters surrounding a pair or string of Ag atoms, along with Cu dominated icosahedra. The polyhedral nature of these clusters were held to suppress crystallisation. Similar conclusions were reached by Kang \emph{et al.} ~\cite{kang2009} using an EAM model, suggesting this tendency to microphase separation is at least reasonably robust.

\textit{Nickel-Zirconium --- }
In NiZr a pairwise, spherically symmetric pair potential treatment for the interactions between the atoms has been developed by Teichler and coworkers ~\cite{teichler2001jnon}which provides a reasonable description of the glassforming behaviour. For example using \emph{ab initio} simulations, Huang \emph{et al.} ~\cite{huang2011} found that for Ni$_{36}$Zr$_{64}$ the Ni follow an Arrhenius like behaviour and do not stop moving, but the larger Zr do, which is not predicted by standard mode-coupling theory, which instead predicts the dynamics of both scale together. The same behaviour was found in Teichler's classical model ~\cite{teichler2001jnon}. Interestingly, in another field, soft matter, such behaviour in a binary system is interpreted rather differently. NiZr is not a world away from a mixture of star polymers (which are treated as soft colloids) ~\cite{mayer2008}. There also one species can undergo arrest before the other. Indeed, Mayer \emph{et al.} ~\cite{mayer2008} went so far as to classify the case where one species arrested as a single glass, and where both arrested as a double glass. Moreover, precisely this kind of behaviour can be interpreted in terms of the multicomponent self-consistent generalized Langevin equation theory of dynamic arrest, which is related to MCT ~\cite{juarezmaldonado2008}.
Teichler and coworkers went on to perform a structural analysis of the classical model of NiZr at a composition of Ni$_{30}$Zr$_{70}$ in which $Z=9$ coordination polyhedra were the most plentiful ~\cite{guerdane2008}, however the analysis notation differed from that of Huang \emph{et al.} ~\cite{huang2011}. A direct comparison of these two models of the same system would be most attractive in order to assess how accurate the classical two-body model of Teichler  and coworkers is. Some moves in this direction were taken by Yang \emph{et al.} ~\cite{yang2007} who combined k-edge x-ray diffraction with RMC to investigate the structure of Ni$_{30}$Zr$_{70}$. Their RMC analysis suggested a combination of coordination number $Z$ around $11$ and $Z=12$ icosahedra. However, at a similar composition Ni$_{36}$Zr$_{64}$ with neutron scattering Holland-Moritz \emph{et al.} ~\cite{hollandmoritz2009} found a non-icosahedral mean coordination number of $<Z>=13.9$. 
Ye \emph{et al.} ~\cite{ye2011} showed with many-body classical simulation that the Ag-Ni-Zr system has an increasing population of Ni-centred icosahedra as a function of [NiZr] concentration.

\textit{Other alloys --- }
In NiAl, Ahn \emph{et al.} ~\cite{ahn2004} found local Ni induced ordering in experiments on Al$_\mathrm{87}$Ni$_\mathrm{7}$Nd$_\mathrm{8}$, in the form of a stronger first sharp diffraction peak. AlFe systems have poor glassforming ability. What little they do have seems related to the Al-Fe interaction which has a shorter range than the mean of the Al-Al and Fe-Fe, ie the system is non-additive and as a consequence the coordination number is 36\% less than assuming hard spheres with the same atomic radii ~\cite{saksi2005}.

Bicapped square antiprisms were found in parameterised models of FeM (where M is C,B,P) ~\cite{evteev2003}, but not in Ni$_{80}$P$_{20}$ RMC fitted data, which found a coordination of 9 for the P. Here random clusters with 9 neighbours were the most significant structure found ~\cite{lamparter1995}. This is perhaps ironic, given that Ni$_{80}$P$_{20}$ is the very material on which the Kob-Andersen model is based, and in the extensively studied model system, bicapped square antiprisms with coordination Z=10 are the locally favoured structure ~\cite{coslovich2007,malins2013fara}.

Using \emph{ab initio} simulations Jakse ~\emph{et al.} considered  AlMn and AlNi \cite{jakse2004}. Their treatment of Mn's magnetic moment indicated it was important in determing its (icosahedral) local order, meanwhile Ni had a close-packed local order. However at the two-body level, both systems appeared \emph{very} similar. This emphasises once more the importance of higher-order correlations, and also the need (for Mn) to carry out an \emph{ab initio} treatment ~\cite{jakse2004}. Sheng \emph{et al.} ~\cite{sheng2008} compared RMC and \emph{ab initio} simulations of Al-rich metallic glasses. They found the geometries of the clusters found were correlated with the Al-X bond length (X being the minority species). Improved glassforming ability was correlated with the use of more than one minority species. Minority species tended to be found at the centre of clusters, of geometry specific to the particular minority species. The use of multiple minority species thus led to clusters of multiple geometry, which suppressed crystallisation  better than clusters of one geometry, as would be the case with a single minority species.

This brings us to the point of noting that many metallic glassformers exhibit local icosahedral order \cite{cheng2008}. Others, particularly those in which the interactions are non-additive, with the cross interaction (in a binary system) being shorter-ranged than the mean of like-species interactions have bicapped square antiprisms as the local structure ~\cite{evteev2003}. Now this behaviour is captured by the additive Wahnstr\"{o}m model (icosahedra) and non-additive Kob-Andersen model (bicapped square antiprism) models ~\cite{coslovich2007,malins2013jcp,malins2013fara}. At least in these Lennard-Jones systems, the different LFS can be related to the interactions, as the 12 shell particles of the icosahedron enclose a larger central particle than do the 10 shell particles of the bicapped square antiprism. Both these clusters may be broken down into two layers of 5- and 4-membered rings with two ``spindle'' particles in the nomenclature of the topological cluster classification (see section \ref{sectionTCC}). The central particle is a shared spindle between both rings ~\cite{malins2013tcc}. In the case of hard spheres, such bipyramids as are formed by $m$-membered rings have been considered by Miracle and Harrowell ~\cite{miracle2009} who found that compact (all neighbours at contact) bipyramids with ratios $R>2/3$ exist only for pentagonal bipyramids. Given that compact quadrilateral bipyramids only occur for a size ratio 0.5 ~\cite{miracle2009} with small spindle particles, bicapped square antiprisms require some degree of non-additivity, strong size asymmetry or are not compact.

\textit{Gelation-like behaviour in metallic glass --- }
The range of metallic elements might be expected to produce rather varied behaviour. Although Inoue has noted that good metallic glassformers exhibit a negative enthalpy of mixing ~\cite{inoue2011}, (as captured in the Kob-Andersen model through its non-additivity, section ~\ref{sectionCommonModelSystems}), there are examples of more exotic behaviour. One such example is glass formation by a gelation-like process. Here, using molecular dynamics simulations with an embedded atom model, the phase-separating system of Cu$_\mathrm{50}$-Nb$_\mathrm{50}$ was found to undergo arrested phase separation, which appears to result from the formation of icosahedra on the \emph{interface} of the Cu and Nb phases ~\cite{baumer2013}. This is unlike normal colloidal gelation, where the arrest is associated with vitrification of the colloid-rich phase (section ~\ref{sectionGel}). In gels that would correspond to vitrification of either the Cu or the Nb, rather than the interface between them. However the soft material of bigels ~\cite{clegg2008} certainly shares some characteristics with these intriguing metallic glassformers.
One question that naturally arises is how universal this mechanism is, does it only occur in Cu-Nb? The presence of fivefold symmetry at liquid interfaces is a controversial topic, with indirect experimental evidence in support ~\cite{reichert2000} and simulations with a variety of models, including models of metals ~\cite{godonoga2010}, indicating no increase in five-fold symmetry at the interface except under the addition of an (unphysical) field  ~\cite{heni2002}, see section ~\ref{sectionHigherOrderReciprocal}. This stabilisation of metallic glasses during phase separation represents a complementary technique to already existing methods where (like the colloids ~\cite{royall2012,zhang2013,taylor2012}), phase separation can be controlled by a judicious selection of experimental parameters such as composition and quench rate ~\cite{dargaud2012}. Alternatively, in FeCo based alloys, spinodal demixing can act as a trigger for crystallisation ~\cite{may2004}.

Metallic glassformers present a wide range of possible behaviour. Our small survey suggests a strong emphasis on the identification of local structural motifs, perhaps guided by the practical desire to avoid crystallisation (see section ~\ref{sectionCrystallisationVersusVitrification}). Influenced by the work on model systems, where attempts have been made to link such local structures to dynamical arrest, we speculate that it may be fruitful to employ such methods in models of metallic glass forming systems.

\subsection{Structure in chalcogenide glassformers}
\label{structureInChalcogenideGlassformers}

\begin{figure}[!htb]
\centering \includegraphics[width=80mm]{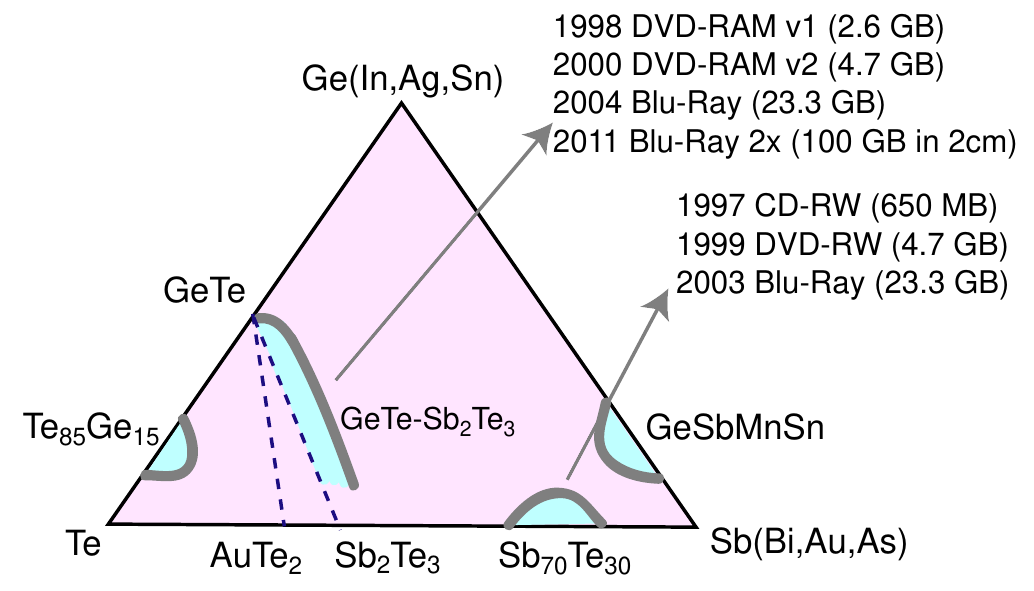} 
\caption{Compositions of phase change materials. Shaded blue regions are those from which phase change materials have been produced. Indicated are particular compositions which have been commercialised and the relevant technology.
Based on ~\cite{wuttig2007}}
\label{figPhaseChangeMap} 
\end{figure}

Chalcogenide glasses are based on elements such as Se and Te. These form the basis of \emph{phase change materials}, which exhibit a rather unusual combination of properties. They crystallise very rapidly, yet the change in their optical and electrical properties upon freezing is large. The former property enables rewriteable optical data storage, the latter makes phase change materials prime candidates for use as solid state memory in next-generation computers  ~\cite{lencer2008}. Regardless of the role of structure in vitrification, which is complex enough in better understood materials, Akola and Jones noted ``it is remarkable that phase-change materials could become the basis of commercially successful products with so much uncertainty about the structures of the phases involved'' ~\cite{akola2007}. A ``map'' of the compositions of phase change materials is shown in Fig. \ref{figPhaseChangeMap}.

The structural complexity of chalcogenides cannot be reproduced even qualitatively by spherically symmetric pair potentials of the type discussed in section \ref{sectionCommonModelSystems}. Often one uses full \emph{ab initio} simulations. While their timescales and system sizes are limited, crystalisation kinetics are so fast in phase change materials that \emph{ab initio} techniques can capture the crystallisation mechanisms ~\cite{loke2012}. Nevertheless, classical simulations can access far larger system sizes and timescales and for this purpose Bernasconi and coworkers developed their neural-net based $N$-body interaction approach. Here an $N$-body potential is mapped to DFT energy calculations for systems small enough to be tackled by full DFT. The resulting classical $N$-body potential then accurately reproduces the DFT result, both in the amorphous and crystal states~\cite{treacy2012}.

Before considering the Te-based phase change materials, we discuss Se-based glasses, which find use in for example infra-red optical equipment such as night sights. The principle consequence of the directional bonding is to reduce the coordination number. Rather than taking a Voronoi cell (see section ~\ref{sectionVoronoi}) to define the number of neigbours, here the coordination number is taken as the integral under the first peak of $g(r)$, which reflects covalent bonding rather than the packing considerations typical in model systems and metallic glasses.

\begin{figure}[!htb]
\centering \includegraphics[width=80mm]{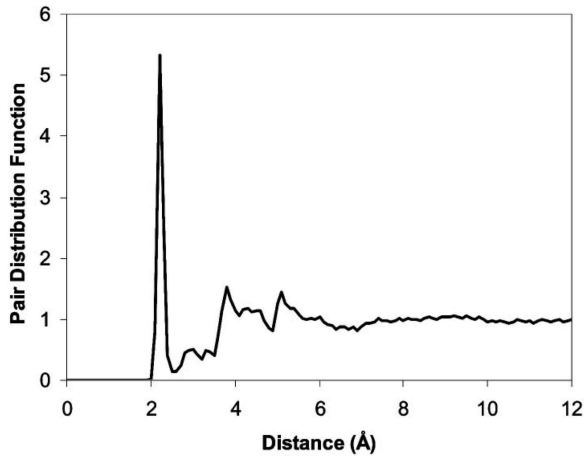} 
\caption{The pair-correlation function in Se. The coordination number is taken as the integral under the first peak of $g(r)$. Here it is equal to two, indiicating that Se tends to form chains
 ~\cite{mauro2005}. Reproduced with permission from American Physical Society Copyright 2005.}
\label{figMauro} 
\end{figure}

One example of the exotic behaviour of chalcogenides is pure Se. Here Molina \emph{et al.} ~\cite{molina1999} showed using simulations based on an \emph{ab intitio} approach that Se transitions from 2-fold to 3-fold coordination upon vitirification. Another example is that liquid As$_2$Se$_3$ has a structural transition. At low temperatures, As and Se have three- and two-fold coordination, while upon heating there is a structural transition in the liquid state to twofold connected chainlike structures. This structural transition is accompanied by a transition from a semiconductor to a metal upon heating ~\cite{shimojo1999}. Interestingly, the same group showed that such a structural transition is \emph{absent} from As$_2$Te$_3$ ~\cite{shimojo2002}, which is significant in underlining the different bejaviour of Se and Te, despite their chemical similarity. This is of particular interest given the latter's relevance to phase change materials. Before leaving the Se-based glassformers, we note the fragility (in the Angell sense) of GeSe glassformers is extremely low, and seems likely related to the rather open structures of these materials ~\cite{gunasekera2013}.

\textit{Phase change materials --- } Henceforth we consider chalcogenide glassformers based on Te, the phase-change materials. The origin of the difference in optical properties between glasses and crystals of phase change materials has been attributed to ``resonant bonding'' ~\cite{lencer2008,shportko2008}. Resonant bonding refers to covalent bonds with fewer than two electons. These can result from crystal lattices of Ge (with fourfold valency) and Te (with threefold valency) : the crystal lattice of GeTe, has insufficient valence electrons to go around, compared to a pure Ge lattice. Such resonant bonding leads to profound changes in reflectivity which underlies the use of phase change materials in rewriteable optical media.

Welnic \emph{et al.} ~\cite{welnic2007} used DFT to identify the orgin of the change in optical properties upon crystallisation in GeTe. They found that the changes in the number of bonds and in the local order lead to changes of the strength of electronic oscillations which in turn result in the unusually pronounced optical contrast between the crystal and glass. Thus Te based materials differ significantly from simple tetrahedral semiconductors such as Si or Ge, which retain the local order in the amorphous state and therefore only exhibit minor changes in their optical properties.

Crystalline Ge$_2$Se$_2$Te$_5$  (c-Ge$_2$Se$_2$Te$_5$) has the rocksalt (NaCl) structure. One of the FCC sublattices is formed of Te, so the stochiometry means the Ge-Se sublattice has 20\% vacancies. The Raman Spectrum broadens upon crystallisation due to the large number of vacancies. 73\% of Ge atoms were found to be in defective octahedra~\cite{sosso2011}. The same group went on to show that liquid GeTe features high mobility almost to the glass, ie that it is fragile ~\cite{sosso2012pss}. In fact, for many phase change materials $T_g$ lies within 10 K of the crystallisation temperature ~\cite{kalb2007}.

Kolobov \emph{et al.} ~\cite{kolobov2004} made a breakthrough in understanding Ge$_2$Sb$_2$Te$_5$, which is used in DVD-RAM. Using a combination of X-ray absorption fine-structure spectroscopy (XAFS) and \emph{ab initio} simulation they found that  crystalline Ge$_2$Sb$_2$Te$_5$ does not possess the (crystalline) rocksalt structure but rather randomly oriented domains consistent with cubic symmetry with a high number of voids due to the stochiometry. More specifically, crystalline Ge$_2$Sb$_2$Te$_5$ is rocksalt, amorphous is ``spinel'', a combination of octahedral (rocksalt) and tetrahedral bonding ~\footnote{Spinels are cubic (isometric) crystal systems, with the majority (here Te) arranged in a cubic close-packed lattice and the minority spieces (Sb and Ge) occupying some or all of the octahedral and tetrahedral sites in the lattice.}. These two structures have very similar ground state energies and compete with one another ~\cite{welnic2006}. Wuttig and coworkers used laser-induced amorphization which resulted in drastic shortening of covalent bonds and a decrease in the mean-square displacement, leading to an increase in the short-range ordering. This order-disorder transition was attributed to an ``umbrella-flip'' of Ge atoms from an octahedral position into a tetrahedral position without rupture of strong covalent bonds. Thus fast crystallisation is possible (without bond breaking), but significant change in structure and hence optical and electronic properties accompany the transition ~\cite{kolobov2004}. On the basis of RMC analysis, Kohara \emph{et al.} ~\cite{kohara2006} showed that amorphous Ge$_2$S$_2$Te$_5$ could be interpreted as being comprised of a combination of 4- and 6-membered rings. These even numbered rings then facilitated crystallisation (compared to 5-membered rings, which can suppress crystallisation). Other interpretations, such as amorphous Ge$_2$S$_2$Te$_5$ being built of four-membered rings are also possible \cite{yamada2012}.

Akola and Jones ~\cite{akola2007} carried out \emph{ab inito} simulations of amorphous GeTe and Ge$_2$Sb$_2$Te$_5$. Long-ranged order was found in the form of a four-membered ring with alternating atoms of types Ge or Sb and Te. The freezing transition took the form of organisation of these squares into a crystal lattice. The local environment of Ge and Sb was typically octahedral. Vacancies assist phase change, and comprise much more of the space in Ge$_2$S$_2$Te$_5$ (11\%) than in GeTe (6\%), which may explain why the former has superior properties. Kwon \emph{et al.} investigated Ge$_2$Sb$_2$Te$_5$ with fluctuation electron microscopy. They found that thermal annealing below the crystallization temperature increased the nanoscale order, with a decrease in crystallisation time, which they attributed to the increase in order ~\cite{kwon2007}. Such an increase in order prior to crystallisation seems to flout classical nucleation theory, where the actual nucleation and subsequent growth of nuclei is assumed to be very rapid. Mazzarello \emph{et al.} ~\cite{mazzarello2010} used a combination of \emph{ab initio} and bond polarisability model simulations to investigate glassy GeTe. Their work revealed both tetrahedral and defective octahedral structures, similar to previous work by the same group on Ge$_2$Sb$_2$Te$_5$ ~\cite{caravati2007}. Further profound differences were found in Ag$_{3.5}$In$_{3.8}$Sb$_{75.0}$Te$_{17.7}$ ~\cite{matsunaga2011}. Glassy Ag$_{3.5}$In$_{3.8}$Sb$_{75.0}$Te$_{17.7}$ shows a range of atomic ring sizes, unlike  Ge$_2$Sb$_2$Te$_5$ in which the rings are small and the number of cavities reaches 11\% ~\cite{akola2007}. The local environment of Sb in both crystalline and glassy Ag$_{3.5}$In$_{3.8}$Sb$_{75.0}$Te$_{17.7}$ is a distorted octahedron. Matsunaga \emph{et al.} ~\cite{matsunaga2011} suggested a bond-interchange model, where a sequence of small displacements of Sb atoms accompanied by interchanges of short and long bonds is the origin of the rapid crystallization of Ag$_{3.5}$In$_{3.8}$Sb$_{75.0}$Te$_{17.7}$, which is unlike that in Ge$_2$S$_2$Te$_5$.

\textit{Crystalisation kinetics in phase change materials --- } 
Among the challenges for phase change materials are their crystallisation times. These set an upper bound on data writing rates in memory applications, which is currently a key limitation in the exploitation of phase change materials for application in non-volatile hard disks. Kalb \emph{et al.}  carried out a study of the crystallisation kinetics in AgInSbTe,  Ge$_2$S$_2$Te$_5$ and Ge$_4$S$_1$Te$_5$ using \emph{a posteriori} atomic force microscopy ~\cite{kalb2004}. In the thin films in which phase change materials find application, nucleation is heterogeneous. Crystal growth rates in all three materials were found to be Arrhenius (though later work using direct measurements with high speed differential scanning calorimetry has imdicated that they are not ~\cite{orava2012}). Remarkably, growing crystallites in AgInSbTe were similar in size, while those in Ge$_2$S$_2$Te$_5$ and Ge$_4$S$_1$Te$_5$ showed a large size distribution. This is interpreted as a continuous formation of new crystals in the latter and the absence of any nucleating sites at longer times for the former. The same group went on to determine the critical work for nucleus formation, \emph{i.e.} the barrier height, as 76 $k_BT$ and 65 $k_BT$ for Ge$_4$S$_1$Te$_5$ and Ge$_2$S$_2$Te$_5$ respectively ~\cite{kalb2005}. By considering the role of film thickness in TeSb films, Martens \emph{et al.} ~\cite{martens2004} identified an optimum film thickness which minimised the height of the energy barrier to crystallisation. Elliot and coworkers ~\cite{loke2012,hedgedus2008} showed that four-membered rings could facilitate and accelerate crystallisation in the phase change material Ge$_2$Sb$_2$Te$_5$. Such four-membered rings, which form the basis of the rocksalt (NaCl) crystal structure are also present in the amorphous form, thus crystallisation can proceed rapidly without significant rearrangement of atomic positions. This kind of crystallisation behaviour is likely to be related to spinodal nucleation discussed in sections ~\ref{sectionSpinodalNucleation} and ~\ref{sectionCrystallisationVersusVitrification} which is exhibited by hard spheres and BKS-silica.

Caravati \emph{et al.} ~\cite{caravati2010} used \emph{ab initio} techniques to design phase change materials. They considered Sb$_2$Te$_3$ in which local geometries of Sb and Te atoms in glassy Sb$_2$Te$_3$, in terms of chain lengths of homopolar bonds were found to be similar to that found in Ge$_2$Sb$_2$Te$_5$ and GeTe.

In summary, the understanding of the role of structure in vitrification and crystallisation mechanisms in chalcogenide glassformers is a fast-developing field. The recent discovery of these materials and their extraordinary complexity make for a rich field of research. We expect that within a few years, it may be possible to carry out the kind of analysis to identity the dynamical role of local structural motifs as has been done for model systems.

\section{Crystallisation versus vitrification : what makes a good glassformer?}
\label{sectionCrystallisationVersusVitrification}

Practical glassformers by definition must not crystallise. High ``glassforming ability'' equates to a strong disinclination to crystallise --- a good glassformer. Poorer glsssformers require stronger quench rates to ensure vitrification. As Fig. ~\ref{figXtalTimes} in section ~\ref{sectionSpinodalNucleation} shows, crystallisation times can drop below $\tau_\alpha$ : if the quenching protocol is not fast enough, crystallisation may be unavoidable. Furthermore, crystallisation can be driven by nucleation, such that the rate-limiting step is the formation of a stable nucleus. This is of course more likely in a larger system. For this reason, the largest piece of ``bulk'' metallic glass yet produced is 7 cm in the longest dimension ~\cite{inoue2011}. Improved understanding of how to suppress crystallisation is thus essential for the exploitation of metallic glassformers.
Here we review crystallisation in glassforming systems and their variants. Beginning with model systems and we shall see that even in the most studied of these, hard spheres, that crystallisation is a bewildingeringly complex process. Even in this case crystallisation mechanisms tare much debated. Thus in producing for example metallic glasses with high GFA, we are still some way from general design rules or even guidelines.

\subsection{Crystallisation in model glassformers}
\label{sectionCrystallisationModel}

We begin this discussion by considering model glassformers, such as those discussed in section ~\ref{sectionCommonModelSystems}. Perhaps the most important observation about typical model systems is that they tend not to be good glassformers and a number have been crystallised on the \textit{simulation} timescale. 
Although the Kob-Andersen model itself is hard to crystalise, variants have been crystallised, and then extrapolated to the normal KA model. In this way it is suggested that the Kob-Andersen model crystallises by forming pure face-centered cubic crystals of the majority (large) component ~\cite{toxvaerd2009}. Meanwhile minimum energy calculations using the energy landscape techniques discussed in section ~\ref{sectionEnergyLandscape} indicated the importance of four-fold symmetric structures ~\ref{sectionEvaluatingTheEnergyLandscape} ~\cite{calvo2007}. It is possible these results reflect the difference in the system size, the latter calculations were carried out for a relatively small 320 particle system. Now at different compositions KA forms other crystal structures ~\cite{fernandez2003pre,fernandez2003jcp,fernandez2004,banerjee2013}. Therefore it is possible that the FCC which formed in  ~\cite{toxvaerd2009} is one of a number of competing crystal structures. The Wahnstr\"{o}m mixture forms a crystal of the MgZn$_2$ structure ~\cite{toxvaerd2009,pedersen2010}. When crystallising relatively quickly hard spheres form a random hexagonal close packed structure for polydispersities less than 8\% ~\cite{zaccarelli2009xtal,sanz2011} as the free energy favours FCC over HCP by a meagre $10^{-4}$ $k_BT$ per particle \cite{auer2004}. However at weak supercooling (and in microgravity) hard spheres form the slightly more stable FCC \cite{cheng2002prl}. The examples listed so far concern spontaneous crystallisation encountered during conventional simulations, without recourse to biasing techniques specifically adapted to calculate equilibrium phase diagrams. No such equilibrium phase diagrams exist for the two popular Lennard-Jones systems, but the challenging case of polydisperse hard spheres has been determined by Sollich and Wilding, who predict segregation into a hierarchy of coexisting FCC lattices with successively smaller polydispersity than the original system ~\cite{sollich2010,sollich2011}. As insightful as such calculations are, for practical purposes crystallisation can be neglected in hard spheres when the polydispersity is more than around 8\% ~\cite{poon2012,zaccarelli2009xtal,sanz2011}.

At high density (4.38 gcm$^{-3}$), even BKS silica cannot be equilibrated in the liquid state below 2800 K ~\cite{saikavoivod2009,saikavoivod2006}. This is significant because silica, as a strong liquid, exhibits relatively little structural change upon supercooling, and structural change has been negatively correlated with glassforming ability in some systems as we describe below. Before proceeding, we note that in the only system in which prediction and measurement of nucleation rates have been directly compared, hard spheres, wild discrepancies in excess of ten orders of magnitude have been found ~\cite{auer2004,auer2001,auer2001a}. A decade after this was discovered, the puzzle remains unsolved.

\subsection{Glassforming ability and structure}
\label{sectionGlassformingAbilityStructure}

While model systems are observed to crystallise (and necessarily on timescales much shorter than atomic and molecular glassformers), we now address any relationship between structure and glassforming ability. As Fig. ~\ref{figAngell} shows, three key model systems (the Kob-Andersen and Wahnstr\"{o}m binary Lennard-Jones glassformers and hard spheres) exhibit differing fragility ~\cite{berthier2009witten,royall2014,coslovich2007}, with the least fragile of the three, KA, showing similar behaviour to ortho-terphenyl, which lies at the most fragile end of molecular glassformers. Of the three model systems considered in Fig. ~\ref{figAngell}, hard spheres with more than 8\% polydispersity have never been crystallised, and a systematic study of crystallisation behaviour in the Wahnstr\"{o}m and KA models, while desirable, has yet to be carried out, perhaps due to the fact that crystallisation of glassformers while encountered, is not straightforward in simulation. Furthermore, except in the case of polydisperse hard spheres ~\cite{sollich2010,sollich2011}, the lack of an equilibrium phase diagram makes quantitative statements about nucleation rates very hard.

\begin{figure}[!htb]
\centering 
\includegraphics[width=65mm]{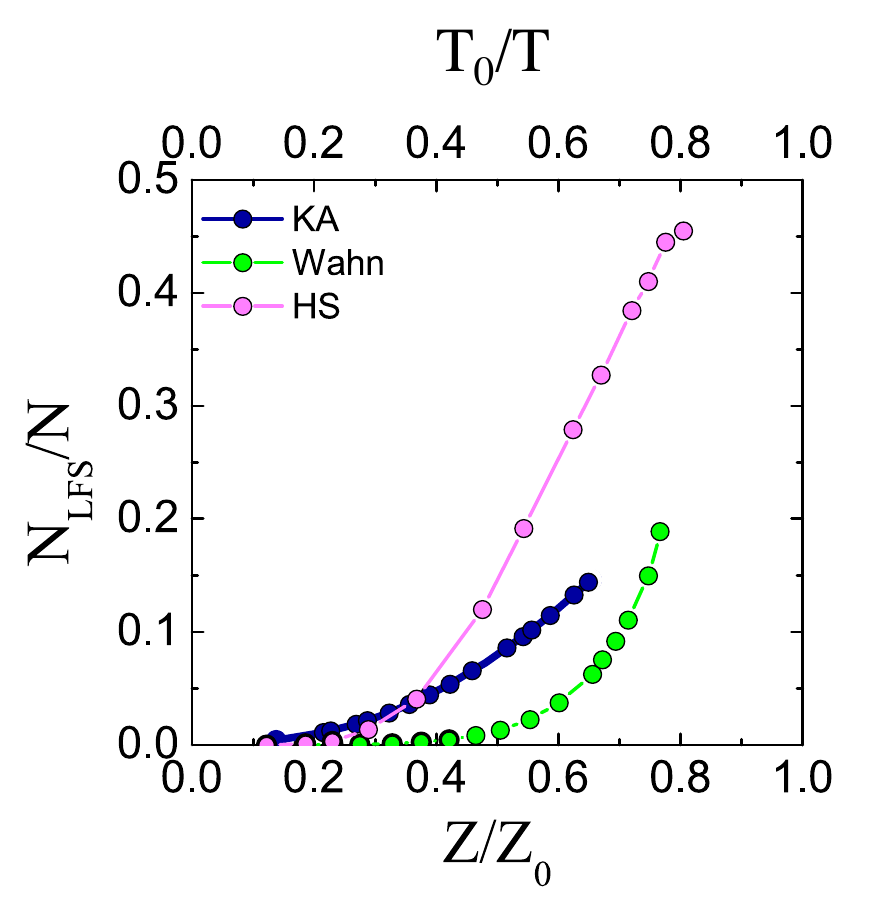} 
\caption{Fraction of particles in locally favoured structures $N_{\mathrm{LFS}}/N$ for three model systems. The locally favoured structure in each system was found with the dynamic topological cluster classification (see section \ref{sectionTCC}). HS denotes hard spheres, Wahn the  Wahnstr\"{o}m model and KA the Kob-Andersen model. The different growth rates of the population of particles in locally favoured structures reflects the differing fragilities of these systems (see Fig. ~\ref{figAngell}). Here, in the case of hard spheres, $Z$ is the compressibility parameter (see section ~\ref{sectionCommonModelSystems} ~\cite{royall2014}). 
\label{figNLFSN} }
\end{figure}

What we do know is that structural changes in these systems upon supercooling are correlated with the fragility : stronger liquids exhibit weaker structural change upon supercooling as shown in Fig. ~\ref{figNLFSN}. However, the mechanism of crystallisation is highly dependent on the model. The crystal the Wahnstr\"{o}m model forms incorporates the locally favoured structure of the system (the icosahedron) into a large unit cell of around 50 particles ~\cite{pedersen2010}, although the actual composition (the ratio of large and small particles) of icosahedra formed in the supercooled liquid only rarely corresponds to that of icosahedra found in the crystal ~\cite{malins2013jcp}. Whether this Frank-Kasper phase which forms in the Wahnstr\"{o}m model corresponds to equilibrium is of course open to question. Evidence has been presented for a closely related binary Lennard-Jones model (with size ratio 0.85 compared to 0.83333) freezing to an FCC at around T=0.5 ~\cite{hitchcock1999}. Ronceray and Harrowell ~\cite{ronceray2011} investigated the role of unit cell size and crystallisation with a lattice model whose local structure could be varied to produce eight distinct local structures of varying symmetry. Significantly, they found that low-symmetry structures were found at high concentration in the stable liquid, and could be associated with a large unit cell size,  mimicking the icosahedra formed in large unit cells in the crystallisation of the Wahnstr\"{o}m model ~\cite{pedersen2010}. Freezing tended to be more strongly first-order in the case of liquids with few LFS, while those liquids with many LFS tended to show only a weakly first-order transition upon crystallisation ~\cite{ronceray2011}.

We now consider the case of hard spheres, whose crystallisation mechanism has attracted considerable attention recently,  ~\cite{taffs2013,kawasaki2010pnas,kawasaki2010jpcm,omalley2003,schilling2010,williams2008,schilling2011,sandomirski2011,russo2012,russo2012sm}. A variety of mechanisms have been proposed, such as the formation of locally dense regions in which the crystal nuclei form ~\cite{schilling2010}, ordering to crystal-like structures prior to local densification ~\cite{russo2012} and a competition between local structures with a number of five-membered rings (``10B'' in Figs. ~\ref{figFavourite} and ~\ref{figTCC}) and crystal nuclei ~\cite{taffs2013}. These 10B are the LFS of the hard sphere system ~\cite{royall2014}. While the latter two mechanisms may be related to differing structural interpretations, the former seems to be at odds with them. One comment that can be made is that any density fluctuations \textit{without structural change} in the metastable hard sphere fluid would be expected to be small, due to its relatively high density and the total absence of a critical point in the hard sphere system. Having said that, nucleation rates are predicted to be extremely strong functions of density ~\cite{auer2004,auer2001,auer2001a}, so it is entirely possible that only tiny fluctuations are required for a strong impact on nucleation rate. Finally, the mechanism is likely to depend on supercooling. Schilling \emph{et al.} ~\cite{schilling2010} worked at comparatively weak supercooling (packing fraction $\phi=0.54$) where the nucleation rate is a stronger function of density than at higher degrees of supercooling as was the case for some of the other work ~\cite{taffs2013}.

Indeed at high supercoolings (above $\phi\approx0.57$), it was shown that crystallisation in monodisperse hard spheres occurred on timescales less than the structural relaxation time, ``spinodal nucleation'' ~\cite{zaccarelli2009xtal,taffs2013}. In a series of papers using event-driven molecular dynamics, Sanz and Valeriani and coworkers identified a mechanism for this crystallisation they termed ``avalanches'' ~\cite{sanz2011,valeriani2012,sanz2014}. Here some particles are predisposed to motion (high propensity, see section \ref{sectionIsoconfigurational}) and initiate avalanches which lead to crystallisation with very little motion of most of the particles. Interestingly, those more mobile particles appear not to be directly related to forming crystallites ~\cite{sanz2014}. These are somewhat similar to the avalanches found in supercooled liquids (section ~\ref{sectionIsoconfigurational}) ~\cite{candelier2010}. Before concluding our discussion of hard spheres, we emphasise that we have focussed on \emph{homogenous} nucleation. \emph{Heterogeneous} nucleation, where the crystal is formed at a wall or inclusion has also been considered in hard spheres and good agreement between experiment and Brownian dynamics simulation was found ~\cite{sandomirski2011}.

The Kob-Andersen model, like hard spheres, is expected to crystallise into a structure distinct from that of the liquid, an FCC crystal ~\cite{toxvaerd2009}. Now of these models we consider that KA is the strongest while hard spheres and the Wahnstr\"{o}m model appear more fragile ~\cite{royall2014,coslovich2007}. Given the correlation of fragility and structural change upon supercooling noted above, one may enquire of the structural change of stronger liquids upon supercooling. Silica is of course the canonical strong liquid, in which beyond two-point correlations, studies of structural changes are rather rare ~\cite{horbach1999}. A related (simpler) model of silica developed by Coslovich and Pastore showed little structural change upon cooling ~\cite{coslovich2009}. Such observations might suggest that the more fragile the liquid, the stronger the change in structure upon cooling. As noted in section ~\ref{sectionStaticAndDynamicLengths}, this is not always so. It has been shown that, in systems with effectively identical fragility, the change in structure upon cooling need not be same ~\cite{malins2013isomorph}.  In higher dimension, structure becomes less important, but fragile behaviour persists ~\cite{charbonneau2013pre}. Finally, some kinetically constrained models, which are thermodynamically equivalent to ideal gases by construction (see section ~\ref{sectionFacile}), exhibit fragile behaviour ~\cite{pan2004}.

This discussion illustrates that crystallisation mechanisms, which underlie the stability of the supercooled liquid and hence its glassforming abililty, are highly system-specific, and may even depend upon state point. Thus any kind of ``universal prescription'' seems very challenging. However, as we have seen above, and consider further below, there is some evidence that stronger liquids may be better glassformers, although this is not a universal phenomenon. We now move on to consider the interplay between vitrification and crystallsiation in metallic glassformers. This behaviour is intrinsic to phase change materials, and thus for chalcogenides, the reader is directed to section ~\ref{structureInChalcogenideGlassformers}.

\subsection{Glassforming ability in metallic systems}
\label{sectionGlassformingAbilityMetallic}

The commerical pressures to develop true ``bulk'' metallic glasses have led to a considerable amount of work in trying to understand how glassforming ability might be enhanced. Much of this work is empirical ~\cite{ding2014}. This is perhaps an inevitable consequence of the lack of generic mechanisms that might otherwise have come from the work on model systems noted above. In any case given the  system-dependent nature of crystallisation mechanisms, such an empirical approach seems reasonable. Much thinking harks back to Frank's original suggestion that icosahedra suppress crystallisation ~\cite{frank1952}. Some work has indicated a link between a high nucleation barrier to crystallisation [and thus a high glassforming ability (GFA)] with metallic glasses which crystallise into icosahedra-based quasicrystals, compared to those which crystallise into (hexagonal) Laves phases ~\cite{kelton2003,lee2005}. Other work equates strong liquid behaviour (i.e. a weak change in structure) with high glassforming ability ~\cite{mauro2012,mauro2013}.

\begin{figure}[!htb]
\centering 
\includegraphics[width=40mm]{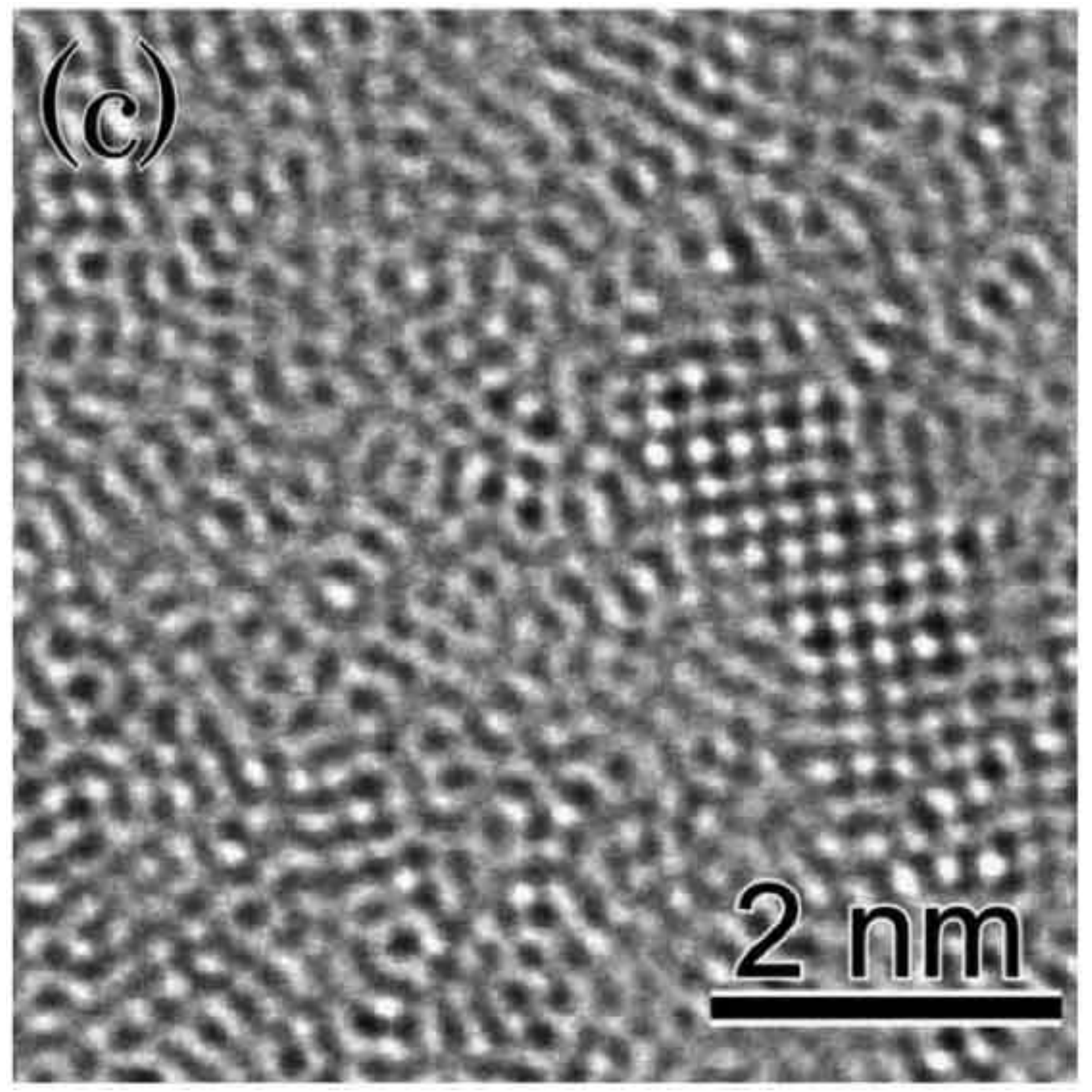} 
\caption{High-resolution TEM image showing crystalline region in Ni$_\mathrm{40}$-P$_\mathrm{20}$-Pd$_\mathrm{40}$ around 1 nm in extent. The reciprocal space data of this system shows an amorphous signal ~\cite{hirata2006}. Reproduced with permission from Elsevier Copyright 2006.
\label{figHirata} }
\end{figure}

Dense random packing (of hard spheres) with some short-range order provides good agreement with  x-ray scattering data for Pd-Ni-P ~\cite{egami1998}. The high GFA appears to stem from competing crystal structures. In particular a symmetric Pd-Ni composition suppresses crystallisation. Here therefore, the very high GFA is not related to an absence of a change in structure but rather how difficult it is to form the crystal. Similar conclusions regarding the same system were made by Alamgir \emph{et al.} ~\cite{alamgir2003} who additionally suggested that the structure of the high GFA Pd$_\mathrm{40}$P$_\mathrm{20}$Ni$_\mathrm{40}$ system could be treated as a linear interpolation of the Pd-rich or Ni-rich systems. \emph{In other words, nothing special was going on with the structure.} Rather, as Egami \emph{et al.} ~\cite{egami1998} had suggested, the weak Pd-Ni interactions made the mixed crystal unstable. Hirata \emph{et al.} ~\cite{hirata2006} then used high-resolution TEM to find (on the surface) nanoscale regions of FCC crystalline order, whose size was sufficiently small that the reciprocal space data nevertheless gave an amorphous signal (Fig. ~\ref{figHirata}). This system at least is able to tolerate some incipient crystallisation, far more than that contemplated in the small systems available to computer simulation. The same group ~\cite{guan2012} went on to show that the symmetric composition Pd$_\mathrm{40}$P$_\mathrm{20}$Ni$_\mathrm{40}$ appeared to contain \emph{both} icosahedra and P-centered trigonal biprisms suggested originally by Gaskell ~\cite{gaskell1978,gaskell1979} (section \ref{sectionCrystalLikeOrdering}) and previously surmised from experimental work on the same system ~\cite{park1999}.

In another metallic glassformer, Li \emph{et al.} showed that tiny amounts of Cu (0.3\%) could massively enhance the glass forming ability (suppress crystal nucleation) in Fe-rich (Fe$>$70 \%) Fe-C-Si-B alloys  ~\cite{li2013}. This is believed to be due to the need to redistribute the Cu in order to enable the nucleation of Fe into clusters of $\alpha-$Fe lattice that is the usual mechanism for crystallisation in these systems. Higher concentrations of Cu lead to Cu clusters which themselves act as nucleation sites for the $\alpha-$Fe. Cu has a positive enthalpy of mixing, ie Cu and Fe tend to segregate - precisely the opposite behaviour for normal attempts to suppress crystallisation, where mixing is sought.

Guerdane \emph{et al.} recently ~\cite{guerdane2013} used their classical two-body model of NiZr to investigate the relationship between crystallisation and local structure in the liquid. They found that the 100 crystal surface grew faster than the 111 surface due to bicapped square antiprism structures formed in the liquid whose geometry is commensurate with the 100 plane. Sch\"{u}lli \emph{et al.} have found in experiments on AuSi that crystallisation can be suppressed by Au which they suggest promotes five-fold symmetry in the plane of the interface ~\cite{schulli2010}. However given the uncertainties raised about the validity of experimental evidence for such five-fold symmetry ~\cite{heni2002,godonoga2010} (section \ref{sectionHigherOrderReciprocal}), simulations of this system would certainly be desirable.

\begin{figure}[!htb]
\centering 
\includegraphics[width=80mm]{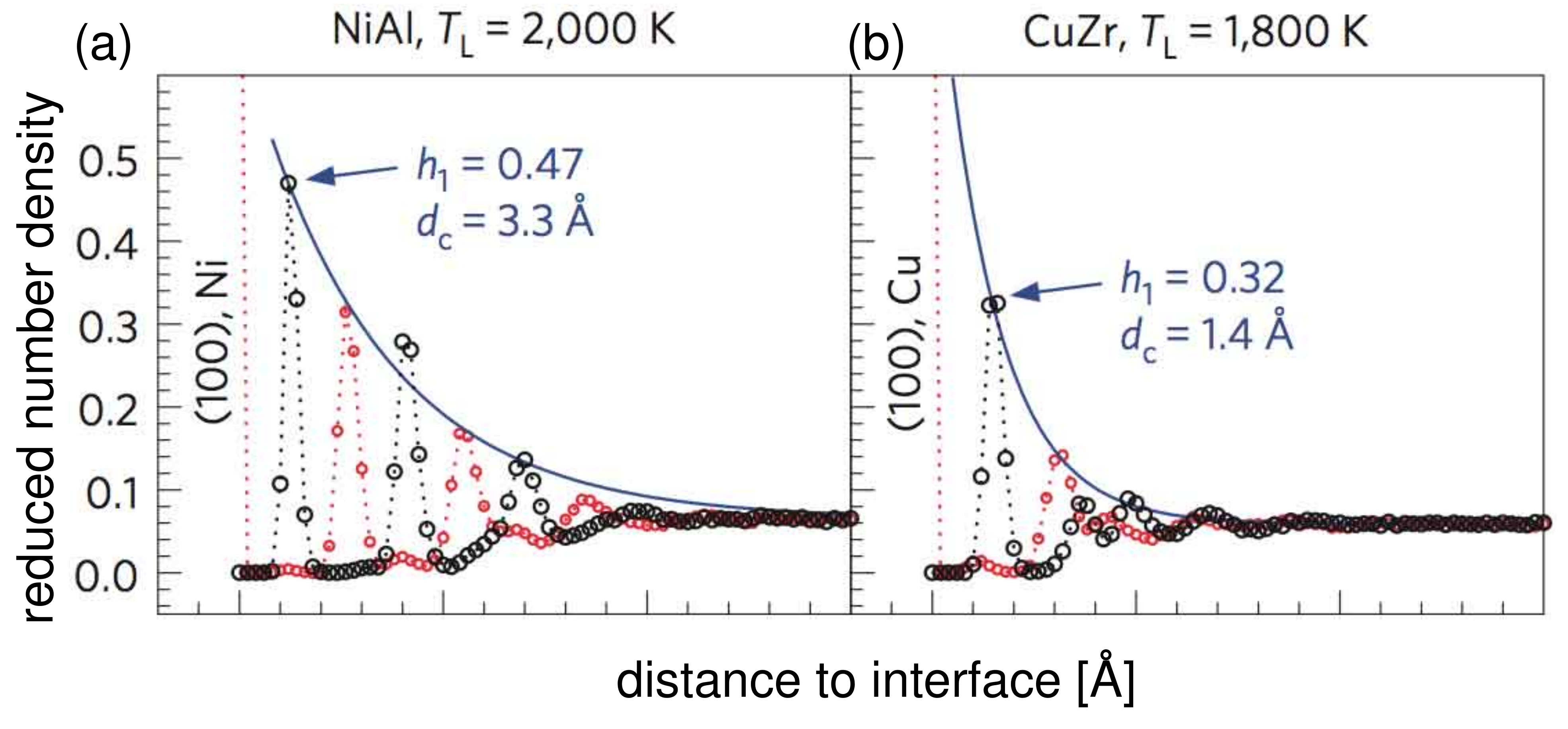} 
\caption{Atomic population in liquids with a frozen crystal interface. ``(100) Ni'' means the interface (0) is the Ni 100 plane (a) and similarly for the Cu crystal (b). $T_L$ is the liquid temperature, and scaled to similar undercooling for each system. Small (red) circles are for Ni or Cu, large (black) circles are for Al or Zr, and dashed lines are guides for the eye. $d_c$ is the decay length of the exponential fit whose prefactor is $h_1$
 ~\cite{tang2013}. Reproduced with permission from Nature Publishing Group Copyright 2013.
\label{figPeterComparison} }
\end{figure}

A very recent development in understanding of mechanisms of GFA has come from Tang and Harrowell ~\cite{tang2013}. Here two binary metallic glassformers, NiAl and CuZr with similar size ratios and identical crystal structures, yet very different GFA (CuZr is a much better glassformer) were investigated. By preparing a crystalline surface, it was possible to directly observe the growth mechanism in EAM simulation. As might be expected, the CuZr exhibited a much lower growth rate. More remarkably, the liquid structure around the NiAl liquid-crystal interface seemed to be much more perturbed than the ZnCu, and density oscillations persisted much further from the interface. The implication is that these somehow lead to a crystal-like structure penetrating into the liquid and facilitating growth.

\subsection{So what makes a good glassformer?}
\label{sectionSoWhatMakesAGoodGlassformer}

We now summarise how the question with which we opened this section may be addressed. In metallic glassformers, the plethora of systems and behaviours underlines the point that as yet there is no general prescription by which a high-GFA metallic system can be ``designed''. Perhaps the nearest to such a prescription is the correlation between limited change in structure upon cooling (few LFS) and high glass forming ability  ~\cite{mauro2012,mauro2013}. We have noted some systems which behave in a different manner, thus this correlation should be thought of as a rule of thumb.

One might reasonably imagine that if we sought some universal understanding of the relationship between LFS and crystallisation then we should turn to model systems (section ~\ref{sectionGlassformingAbilityStructure}). Let us summarise the situation. The Kob-Andersen model is thought to crystallise into structures distinct from the liquid ~\cite{toxvaerd2009} and there is some evidence of polymorphism ~\cite{middleton2001,fernandez2003pre,fernandez2003jcp,fernandez2004,banerjee2013}. Some possible polymorphs might even incorporate the bicapped square antiprism LFS ~\cite{middleton2001,fernandez2003pre}. The Wahnstr\"{o}m model on the other hand incorporates its icosahedral LFS into the crystal ~\cite{pedersen2010}. One imagines it is ``harder'' for the system to do this than to simply form an FCC crystal as in the case of the one-component Lennard-Jones model. So indeed it is possible to suggest that LFS influence the crystallisation mechanism much as Frank suggested ~\cite{frank1952}. In the case of hard spheres, Taffs \emph{et al.} ~\cite{taffs2013} discuss the competition between an LFS with five-membered rings and the nucleation of FCC crystals. However other work on the same system emphasises the role of crystal-like structures in the supercooled liquid ~\cite{kawasaki2010pnas,kawasaki2010jpcm,russo2012}.

A more generic approach has been pursued by O'Hern and coworkers who tackled the question of GFA metallic glasses using computer simulation of model systems. The authors emphasise that computer simulation can only access very high (several orders of magnitude higher than experiment) rates of cooling and thus any crystallisation corresponds to very poor glassformers. However they found that for additive binary Lennard-Jones models, the Eutectic lay at a 50\% mass fraction of each species. Interestingly, and counter to some of the prevailing ideas in the metallic glass community ~\cite{inoue2011}, they found that non-additivity had little effect on their cooling rates. That is to say the enthalpy of mixing has no effect. However they could only access weakly size-assymetric mixtures (size ratio >0.92) so it is possible that the enthalpy of mixing could play a more significant role for larger size ratios such as 0.8 ~\cite{zhang2013}. The same group also looked at binary hard spheres where they found two competing effects : larger size asymmetries (<0.8) suppress crystallisation but enhance the tendency to phase separate. Furthermore the densest crystal structure played no role in the GFA ~\cite{bertrand2014}.

In short, while some progress has been made in considering the role of structure in crystallisation, it can be highly system-specific. There seem to be at least two possibilities: either the LFS can be incorporated in the crystal (which then has a complex geometry and would require much organisation to nucleate) or the crystal can be very different from the liquid structure and thus frustrated by the LFS. In designing metallic glasses both mechanisms must be suppressed. A key question is which mechanism should one focus on for a given mixture? We believe that more work is required here to elucidate which of these approaches might be preferable if we are to become proficient in designing glassformers. A further key question to be tackled is to consider the stability of LFS in the supercooled liquid. High GFA would be correlated with stable (long-lived) LFS which are not incorporated into the crystal.

\subsection{Crystallisation in gels : self-assembly}
\label{sectionCrystallisationInGels}

Spinodal type gelation (sections ~\ref{sectionGelation} and ~\ref{sectionAgingInGels}) ~\cite{zaccarelli2007,lu2008} relates to a different perspective on the role of crystallisation : self-assembly. Contrary to the case of many glassformers, gelation is often undesireable. Here we consider the relationship between gelation and self-assembly.

The recent explosion in nanoparticles with tuneable properties ~\cite{glotzer2007} has led to considerable interest in ``bottom-up'' self-assembled nanodevices. However, for every case of successful self-assembly, there exist many more examples where kinetic arrest prevented formation of the desired product ~\cite{whitesides2002}. ``Sticky spheres'', with short-ranged attractions readily undergo gelation (section ~\ref{sectionAgingInGels}) which suppresses their self-assembly to face-centred cubic crystals. So strong is this suppression that locally crystalline gels have only recently been observed in experiment ~\cite{zhang2012cluster}. Thus gelation of these systems forms a suitable testbed for self-assembly. It had been known for some time that critical fluctuations [ie close to the (metastable) spinodal line] massively enhance crystallisation rates, by providing an alternative route to self-assembly ~\cite{tenwolde1997,fortini2008,savage2009}.

Deeper quenches were shown to suppress crystallisation, despite the increased thermodynamic driving force both in simulation ~\cite{fortini2008,klotsa2011} and experiment ~\cite{ilett1995,dehoog2001}. It was possible to interpret this in terms of the system being pushed far from equilibrium, locally. Such behaviour is thought to prevent successful self-assembly and thus to promote gelation or indeed vitrification ~\cite{whitesides2002}. Jack \emph{et al.} introduced the use of fluctuation-dissipation ratios as a means to quantify the extent to which a system deviates from local equilibrium ~\cite{jack2007}. Quantifiying this for sticky spheres ~\cite{klotsa2011} enabled predictions about the ability to optimally assemble this model system to be made. Another approach is to vary the attraction between the particles as a function of time ~\cite{royall2012,royall2011c60,taylor2012}. 

Now self-assembly is of course by its nature a non-equilibrium process, and the system evolves, which led Klotsa and Jack to introduce feedback to their methods to identify time-dependent optimal protocols for assembly ~\cite{klotsa2013}. Such methods could be implemented in colloidal systems where interactions between the particles can be tuned \emph{in-situ} ~\cite{taylor2012}.

\subsection{Quasicrystals}
\label{sectionQuasicrystals}

\emph{Somewhere between the amorphous glasses and rigidly regimented periodic crystals lie the quasicrystals} --- Steinhardt, ~\cite{steinhardt2008}. Such an introduction would imply that in the fine balance between crystallisation and vitrification, quasicrystals might be encountered ~\cite{shechtman1984}. Thus a discussion of the role of local structure in these materials which ``live on the edge'' is natural to our enquiry. Indeed, a number of metallic glassformers form quasicrystals ~\cite{kelton2003,li2000}, in the sense of a first-order transition to a solid rich in icosahedra occurring via homogenous nucleation ~\cite{saida1999}. Kelton and coworkers investigated nucleation in such materials experimentally using x-ray and TEM diffraction ~\cite{shen2009}. They found agreement with classical nucleation theory, but with a very low surface tension. This they interpreted as evidence that the solid form was structurally similar (dominated by local fivefold symmetry) to the liquid.

In simulations of model systems, the Frank-Kasper phase found in the Wahnstr\"{o}m model ~\cite{pedersen2010} discussed above in section ~\ref{sectionCrystallisationModel} shares many properties with quasicrystals (for example, the large number of icosahedra in the structure). Indeed in finite sized simulation boxes, it can almost be thought of as a quasicrystal forming system. Keys \emph{et al.} ~\cite{keys2007} considered the Dzugutov model, which we have noted forms a network of icosahedra (see section \ref{sectionEarlyMeasurements}) and was found to have a quasicrystalline phase ~\cite{dzugutov1993}, which is metastable to FCC and BCC phases ~\cite{roth2000dzugutov}. Keys \emph{et al.} ~\cite{keys2007} investigated the growth of the quasicrystal and found that the mechanism was the addition of icosahedral clusters which transformed into the dodecagonal symmetry of the quasicrystal with minimal rearrangement.

\section{Polyamorphism}
\label{sectionPolyamorphism}

Polyamorphism is the disordered equivalent of polymorphism, the formation of multiple crystal structures which is exhibited by most materials ~\cite{poole1997}. Polyamorphism is amongst the most controversial topics in condensed matter physics. Even demonstrating the existence of polyamorphism inspires heated debate in a number of systems. It is clear that more work is required before insights from polyamorphic systems can shed a significant amount of light on the role of structure in the glass transition. Nevertheless, the nature of polyamorphism inherently implies a defining role for structure in (supercooled) liquids and we therefore include some brief comments on the (supposed) polyamorphic glassformers pertinent for our purposes and refer the reader to McMillan \emph{et al.}'s review of polyamorphism ~\cite{mcmillan2007}, Tanaka's discussion emphasising the role of local structure ~\cite{tanaka1999jpcm,tanaka2000pre} and Angell's review of the relationship between polymorphic systems and dynamical arrest ~\cite{angell2008}. We consider polyamorphism in one-and multi-component systems, focusing on coexisting amorphous states. We then briefly discuss transitions between different liquids as a function of temperature and possible connections to transitions between fragile and strong dynamical behaviour.

\subsection{One-component systems}

\begin{figure}[!htb]
\centering \includegraphics[width=50mm]{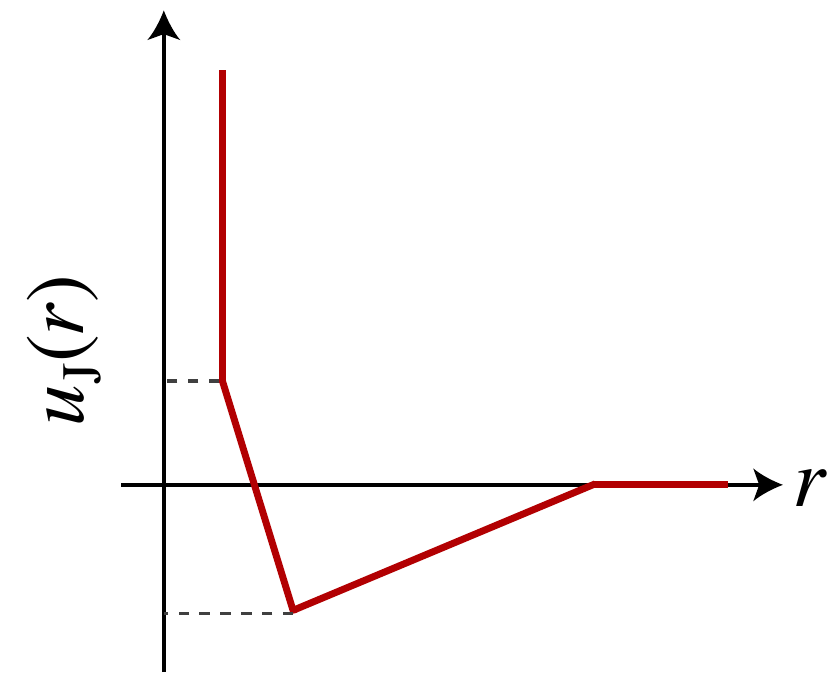} 
\caption{The Jagla potential. A two-scale potential with attractive and repulsive ramps which features coexisting liquids with differing densities.  ~\cite{xu2006}.}
\label{figJagla} 
\end{figure}

A handful of elements have been shown to be polyamorphic.  A moment's thought makes it clear that polyamorphism is not generally expected. In equilibrium, polyamorphism implies a liquid-liquid transition (LLT), which necessitates a second critical point, and this will be our focus, rather than non-equilibrium solid glasses of the same substance with more than one structure. Consider the Lennard-Jones potential. Formation of a liquid implies that atoms should attract one another sufficiently to undergo condensation, but not so much as to crystallise. In the case of the Lennard-Jones model, the typical pair separation is then around $2^{1/6}\sigma$, which is the minimum of the potential. It is thus difficult to imagine a situation which can give rise to any energetic gain that might favour two coexisting liquids presumed to be at different densities, which would necessitate differing mean interatomic separations.

There are two possibilities, either the potential must somehow introduce two lengthscales, facilitating the coexistance of two liquids at differing density, or some kind of directionality in the interaction is required. From a perspective of simulation, the former is more straightforward to implement and has been realised in the Jagla potential which is illustated in Fig. ~\ref{figJagla} and which exhibits two coexisting amorphous phases ~\cite{jagla2001,xu2006}. In atomic systems, directional bonding can lead to polyamorphism. Among the simplest models to capture directional bonding, Stillinger and Weber's treatment of silicon ~\cite{stillinger1985}, has been shown to possess an LLT in the supercooled state  ~\cite{sastry2003}. Experimental studies of the resistivity, optical and Raman spectra, along with x-ray diffraction indicate an LLT which features a semiconductor-metal transition, which is also supported by simulations with the Stillinger-Weber model ~\cite{deb2001,mcmillan2005}.

Other elements in which an LLT has been identified in experiments include phosphorous ~\cite{katayama2000}. The low-density liquid features tetrahedral structures while the high-density liquid is dominated by a polymeric structure, with a coordination of two. In carbon, $sp^3$ hybrization enables the liquid-liquid transition, the low-density liquid is $sp$ dominated, the high-density liquid is $sp^3$ dominated ~\cite{gloshi1999}. In molecular systems, evidence for an LLT in triphenyl phosphite (a glass former) has been obtained ~\cite{kurita2004}, and also in $n$-butanol ~\cite{kurita2005}. In triphenyl phosphite, the dense liquid structure is believed to be related to the molecules being able to stack on top of one another. In other words, such an LLT is thought to be driven by anisotropic interactions between the molecules. In triphenyl phosphite, the LLT competes with dynamical arrest, and whether there is any connection between structure and arrest in the two liquids would be an interesting line of enquiry.

\textit{Water --- } The existence or otherwise of an LLT in water is highly controversial. Any LLT that exists does so deep in the supercooled regime. The putative LLT was first predicted in computer simulations of the (classical) rigid three site ST2 model ~\cite{poole1992}. However, at ambient pressure liquid water is only metastable to -40 $^\circ$C, at lower temperatures it undergoes ``spinodal nucleation'' to form crystalline ice (see section \ref{sectionSpinodalNucleation}), and the supposed LLT lies deep within this ``no man's land'' region of the phase diagram. Thus any LLT is inaccessible to direct experiments, although crystallisation can be suppressed by working with amorphous ice ~\cite{mishima2002,cupane2014} which has the obvious detraction that even local equilibrium is not reached. Alternative approaches include seeking signs of the LLT in water at negative pressure ~\cite{pallares2013}. The experimental evidence for and against an LLT in water has recently been reviewed by Holten \emph{et al.} ~\cite{holten2012}. Other reviews include those of Debenedetti ~\cite{debenedetti2003} and Mishima and Stanley  ~\cite{mishima1998}. Given that direct observation in experiment is currently impossible it is hard to be sure  of an LLT to put it mildly. However the bulk of the experimental evidence seems to favour a transition (were crystallisation not to intervene).

One might imagine, especially given that the LLT in water was first reported following computer simulation, that this might form a suitable means to investigate the phenomenon. Here the situation is also very unclear, due in no small respect to the extremely glassy nature of water in the vicinity of the (supposed) second critical point. A study of criticality in the ST2 model with histogram reweighting identified the location of the second critical point ~\cite{liu2009}. Recently evidence has been presented by Limmer and Chandler that simulations are unstable against crystallisation in this regime ~\cite{limmer2011,limmer2013}, although work by other groups indicates sufficient stability to identify a metastable liquid on computational timescales ~\cite{sciortino2011,palmer2013,palmer2014}. The dispute continues ~\cite{chandler2014}.

High pressure liquid water has a structure similar to a simple liquid as shown in both simulations ~\cite{molinero2009} and experiment ~\cite{strassle2006}. Meanwhile the low density liquid is well-known to have a tetrahedral arrangement driven by hydrogen bonding. Both glasses exhibit corresponding structural differences ~\cite{chui2013}, and reproduce $S(k)$ determined in experiments ~\cite{finney2002}. The lengthscale of domains of tetrahedra in the low density liquid has been shown to grow, approaching the LLT transition temperature, which is accompanied by a strong dynamical slowdown. This might indicate a second critical point, but the exponent of the power law growth of the structural correlation length, 0.3, is much less than that associated with criticality. This might be related to the modest twofold increase in the correlation length of the tetrahedral domains ~\cite{moore2009}. Even such a twofold increase is markedly more than that usually observed approaching the glass transition (see section ~\ref{sectionTowardsAStructuralMechanism}). Recently, Russo and Tanaka ~\cite{russo2013arxiv} have introduced a new order parameter which quantifies the degree of order in the second shell around each molecule. A two-state model based on this parameter can explain the well-known anomalous behaviour of water, and is compatible with (but does not insist upon) the LLT. However, apart from the work of Moore and Molinero ~\cite{moore2009}, which in any case focusses more on the LLT, we are unaware of any studies investigating the role of structure in vitrification in water, in the spirit of those discussed in section \ref{sectionStructureModel}.

\subsection{Multi-component systems} 
\label{sectionMultiComponentSystems}

\emph{Liquid-liquid coexistence at fixed temperature.}

Since LLTs intrinsically imply a key role for local structure, here we review briefly LLTs exhibited by metallic glass forming systems. Soon after the first simulations indicating an LLT in water, Aasland and McMillan ~\cite{aasland1994} presented experimental results which puported coexistance of two liquid states of identical composition at temperatures around $T_g$ in Yttria-Alumina [(YO)$_x$-(AlO)$_{1-x}$]. Later, high-temperature experiments in the  Yttria-Alumina liquid ~\cite{greaves2008,greaves2009} found evidence of nanoscale fluctuations in the form of a rise in $S(k)$ at low $k$ which they probed with x-ray diffraction. Barnes \emph{et al.} ~\cite{barnes2009} investigated these findings with high energy x-ray diffraction, small angle neutron scattering, and pyrometric cooling measurements for [(YO)$_{20}$-(AlO)$_{80}$]. Their $S(k)$ are notably devoid of any rise at low $k$ which lead them to conclude that ``no evidence'' of any LLT could be found. Very recently, Wilding \emph{et al.} ~\cite{wilding2013} have performed x-ray diffraction at small temperature intervals also for [(YO)$_{20}$-(AlO)$_{80}$]. They find a discontinuous change in both position and intensity of the first peak in $S(k)$. Volume fluctuations in molecular dynamics simulations further support their claim that there is an LLT in Yttria-Alumina. To the best of our knowledge, detailed structural analysis of the simulation data for this system have yet to be carried out. In another system, Sheng ~\emph{et al.} provided experimental and \emph{ab initio} numerical evidence for polyamorphism in the Ce$_\mathrm{55}$Al$_\mathrm{45}$ metallic glassformer. In this case, the density change was attributed to f-electron delocalisation  ~\cite{sheng2007}.

\emph{Liquid-liquid transitions as a function of temperature.} Rather than coexistence similar to liquid-gas, another scenario for polyamorphism is that instead of freezing into a crystal as the temperature is lowered, some systems could ``freeze'' into another amorphous state. Recent years have seen growing evidence for such a phase transition between two amorphous states. This may be related to a fragile-to-strong transition where the supercooled liquid exhibits fragile behaviour at higher temperature and strong behaviour at lower temperature. Such a fragile-to-strong transition has been observed in some molecular glassformers ~\cite{ito1999,mallamace2010}. Amongst these is water ~\cite{ito1999} in which the fragile-to-strong transition is entirely distinct from the putative liquid-liquid transition at low temperature. Metallic glassformers based on Al and Gd or Pr, Ce and La have also been shown to exhibit such a transition ~\cite{zhang2010}.

Given the complex structural behaviour in metallic glassformers (section \ref{structureInMetallicGlassformers}), it is natural to expect that structure may be connected to the fragile-to-strong transition. Lad \emph{et al.} ~\cite{lad2012} considered the fragile-strong transition in NiZr. Some evidence was found of different regimes in diffusivity but in simulations it is hard to equilibrate the system below the mode-coupling transition. As the temperature is dropped towards $T_\mathrm{MCT}$ clustering of icosahedra is found in that these are more strongly grouped and have more neighbours.

Dzugutov and coworkers ~\cite{elenius2010} showed evidence for a transition to low-entropy liquid with \emph{fewer} icosahedra at lower temperature. By considering the entropic behaviour, Dzugutov and coworkers concluded that the transition was first-order. The structure of the low temperature liquid is based around tetrahedra condensed into spirals which form linear chains and is  somewhat reminiscent of colloidal gels ~\cite{campbell2005}. The data showed a clear structural liquid-liquid transition. One question arises as to whether this can be related to LLTs in metallic glasses? Dzugutov's one-component model is idealised, and is intended to capture Friedel oscillations (though these also lead to rather lower densities than Lennard-Jones models ~\cite{shi2006} and a ``gel-like'' behaviour). However the low temperature liquid is Arrhenius, while the high temperature one is super-Arrhenius, i.e. a fragile-to-strong transition is found. Of the other popular model systems, evidence for a LLT at very deep supercooling has been found in the Kob-Andersen model ~\cite{speck2014}. Details are discussed in section ~\ref{sectionMu} in the context of a ``dynamic'' LLT. Other systems with liquid-liquid transitions include Se ~\cite{molina1999} and As$_2$Se$_3$ ~\cite{shimojo1999}  (see section ~\ref{structureInChalcogenideGlassformers}).

\begin{figure}[!htb]
\centering \includegraphics[width=80mm]{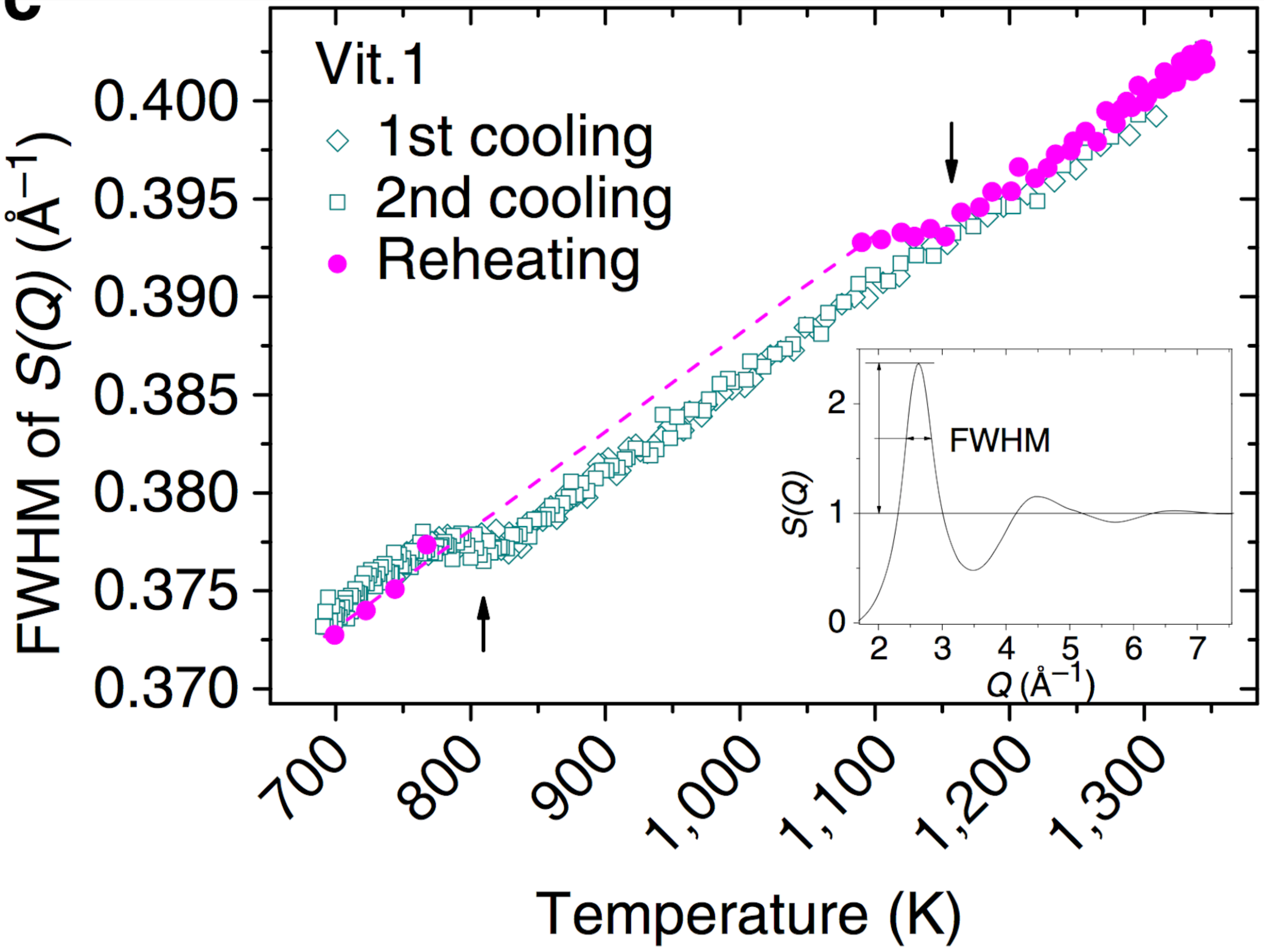} 
\caption{The full-width-half maximum (FWHM) of the 1st peak of $S(k)$ (see inset) versus temperature during thermal cycles of a metallic glass. The arrows indicate slope changes in the temperature range 760-830 K during cooling and 1,100-1,200 K upon reheating. The dashed line is the assumed heating data on reheating ~\cite{wei2013}.
Reproduced with permission from Nature Publishing Group Copyright 2013.}
\label{figRalfBusch} 
\end{figure}

Among the most convincing pieces of experimental evidence for a first-order LLT in metallic glasses is that of Busch and coworkers ~\cite{wei2013}. These workers considered  Z$_{41.2}$Ti$_{13.8}$Cu$_{12.5}$Ni$_{10}$Be$_{22.5}$ which is a strong metallic glass former. A fragile-to-strong transition and a peak in the specific heat capacity were identified, for which an LLT is overridingly best explanation. This LLT is accompanied by structural changes in the first peak of $S(k)$ (see Fig. ~\ref{figRalfBusch}), indicating a transition to a different kind of structural ordering.

On the other hand, such behaviour is far from universal, even in the subset of metallic glassformers. Kelton and coworkers ~\cite{mauro2013} have shown that CuZr shows anomalous behaviour in that the $S(k)$ peak continues to increase up to T$_g$, which is interpreted as indicating that it is fragile all the way, there is no fragile-to-strong transition, indeed the increase in structural change accelerates heading towards the glass. CuZr is moreover a poor glassformer. An interesting consequence of the fragile-to-strong transition is that structure need not play an important role in the strong regime ~\cite{richert1998,ito1999}. The considerations of Montanari and Semmerjian (section ~\ref{sectionMontanari}) make it clear that strong behaviour at low temperature does not require any structural change. It is possible that the fragile to strong transition corresponds to the system ``running out'' of possibilities to change its structure. For example the population of LFS is of course limited by the total number of particles and can saturate.

Finally we note that very recently, a LLT has been found in a simple lattice model system, by Ronceray and Harrowell  ~\cite{ronceray2014}. These authors used a lattice model with energetically favoured local enantiomers, with a chiral Hamiltonian. The system is stable in the (disordered) liquid to low temperature, and undergoes a LLT to a state where the populations of the two enantiomers is not equal, but there is not long-range order so it remains a liquid.

Liquid-liquid transitions seem to invite controversy. However it seems inescapable that LLT occur, the question is rather as to which materials exhibit them. We stress that by their very nature, LLTs must feature a strong structural element. Yet except in cases such as phosphorous ~\cite{katayama2000} with its polymerisation transition, the role of local structure in LLTs should be further explored, likely from simulation data.

\section{Aging}
\label{sectionAging}

\subsection{Aging phenomenonology}

An important yet challenging behaviour in the phenomenology of dynamical arrest occurs when the structural relaxation time exceeds the measurement time, and the system is classified as a glass (or gel). As is clear from Fig. ~\ref{figCavagna}, this experimental glass transition temperature depends on the cooling rate. In other words, the system falls out of of equilibrium at the experimental glass transition and its properties are non-stationary functions of (waiting) time. In the case of dynamical arrest, this variation is termed \emph{aging} ~\cite{berthier2011}. Here we consider the role of local structure in aging. In particular we focus on evidence for the system ``sinking lower'' in its energy landscape.

\begin{figure}[h!]
\begin{center}
\includegraphics[width=60 mm]{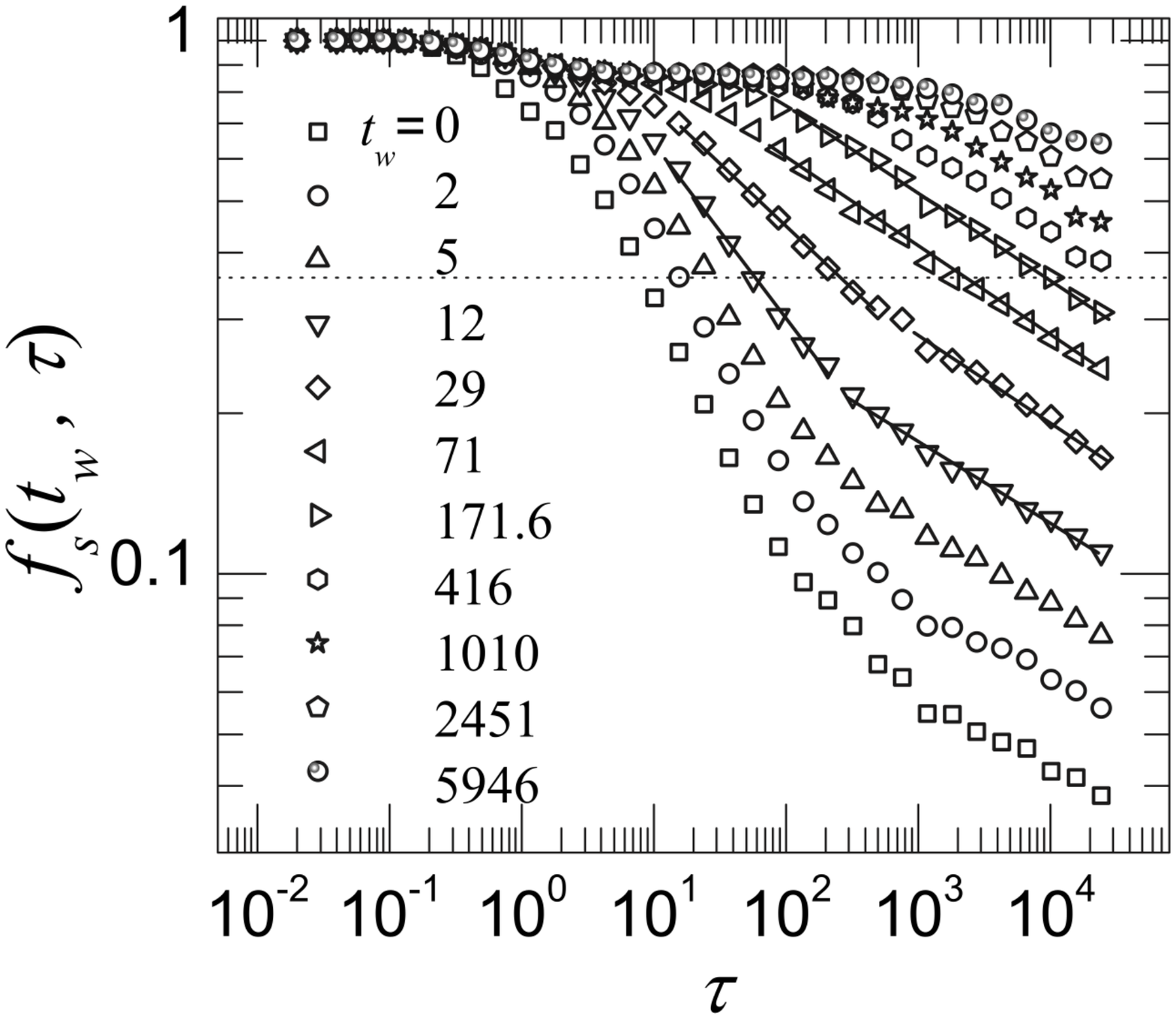}
\caption{Aging behaviour in structural relaxation. Upon increasing the waiting time (different symbols correspond to different waiting times), the intermediate scattering function decays more and more slowly as a function of time (here $\tau$). This corresponds to the system ``sinking lower and lower in its energy landscape''.
From Ref. ~\cite{elmasri2010}.}
\label{figLuca}
\end{center}
\end{figure}

Dynamics in aging systems have received a considerable amount of attention~\cite{elmasri2005,martinez2008,elmasri2010}. In particular, the relaxation time increases as a function of time as shown in Fig. \ref{figLuca}. At short times, in some systems such as ``hard'' spheres $\tau_\alpha \sim \exp(ct_w)$ where $c$ is a constant and $t_w$ is the waiting time. In the long-time asymptotic aging regime, $\tau_\alpha$ can increase linearly with waiting time. Relaxations are characterised by local events, and single-particle motion can be non-Fickian ~\cite{elmasri2010}.

Within the context of the energy landscape (section ~\ref{sectionEnergyLandscape}), aging corresponds to the system exploring deeper and deeper minima. Note that as Fig. ~\ref{figCavagna} suggests, at the experimental glass transition, in falling out of equilibrium the glass ceases to have a well-defined temperature. One can construct a ``fictive temperature'', the temperature the system would have were it in equilibrium in the context of some quantity. For example, Speck \emph{et al.} ~\cite{speck2012} suggested that the population of bicapped square antiprism clusters in their biased simulation corresponded to a fictive temperature of $T\approx0.35$ in the Kob-Andersen model, which is close to the VFT temperature, and certainly well below temperatures at which the system can be equilibrated in conventional simulations (around $T\approx0.43$).

\subsection{The role of structure in aging}

\begin{figure}[h!]
\begin{center}
\includegraphics[width=80 mm]{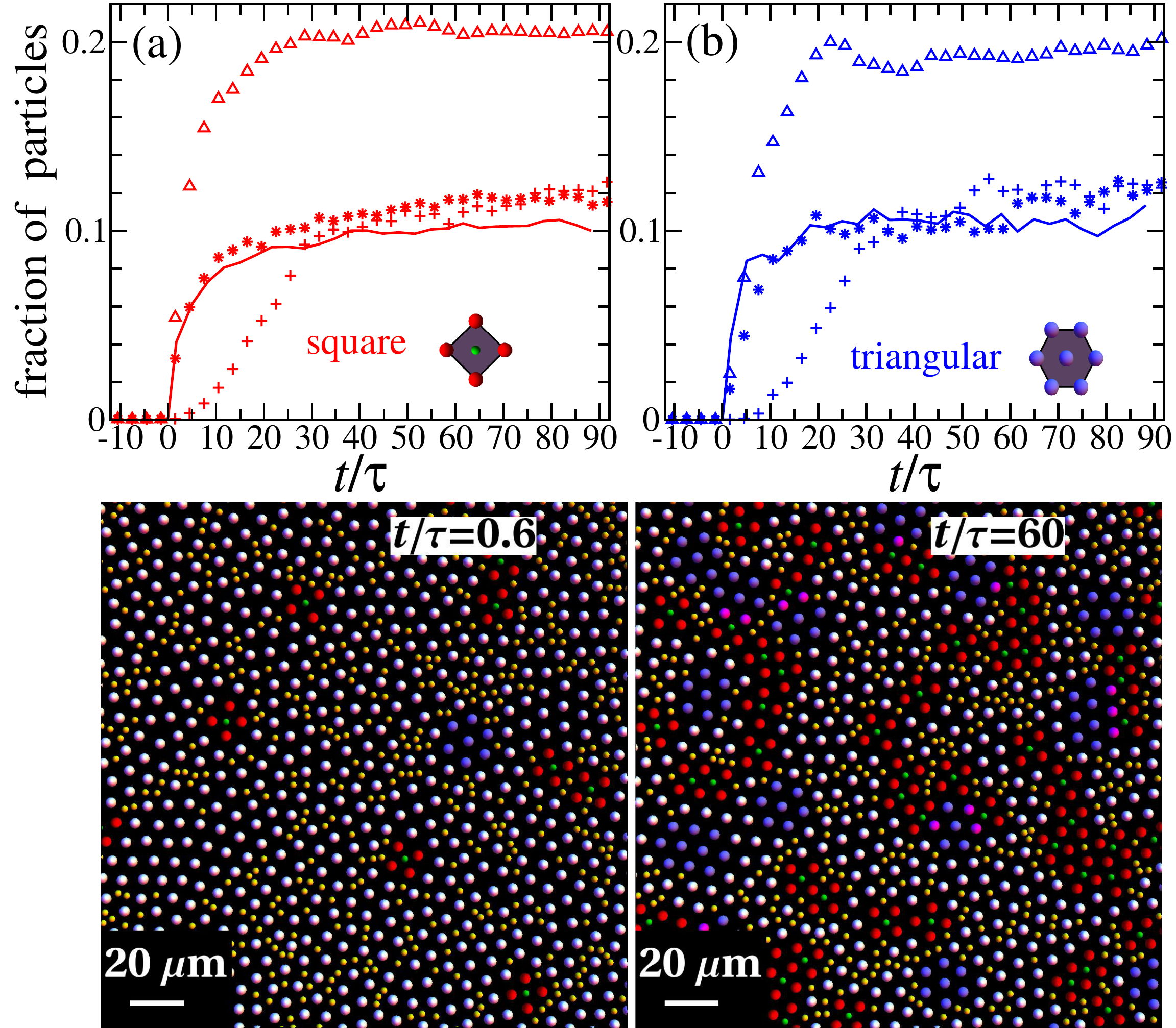}
\caption{Structural changes in an aging binary 2d colloidal model system with dipolar interactions. (a) Fraction of $B$ particles belonging to a crystalline square surrounding (see inset) and (b) fraction of $A$ particles belonging to a crystalline triangular surrounding (see inset) versus reduced time for an ultrafast quench. The lines are experimental data while the symbols ($\ast$) are data from BD simulations. Two experimental snapshots for a time $t/\tau=0.6$ just after the quench (left configuration) and a later time  $t/\tau=60$ (right configuration) are shown. Large particles are shown in blue if they belong to a triangular surrounding and in red if they belong to a square surrounding. All other large particles are shown in white color. The few big particles belonging to both triangular and square surroundings are shown in pink. The small particles are shown in green if they belong to a square center of big particles, otherwise they appear in yellow. From Ref. ~\cite{assoud2009}.}
\label{figAssoud}
\end{center}
\end{figure}

Relatively little work has been carried out concerning the role of structure in aging. Following the discussion above, one might expect that structures corresponding to low local free energy would become more prevalent as the system ages. Indeed, the mechanism of hard sphere crystallisation at very high volume fraction mentioned in section ~\ref{sectionCrystallisationModel} is entirely compatible with such a picture. Aging is an example where colloid experiments have an advantage over computer simulations.
Although the accessible timescale for quantities such as the structural relaxation time, which require continuous data acquisition, is around one or a few days, which leads to the same effective timescale (an increase over the normal liquid of 4 or 5 orders of magnitude) as that accessible to computer simulations run for months. However, leaving the experimental systems to equilibrate for months (or even years ~\cite{ruzicka2010}) enables much longer waiting times to be accessed than is currently possible in simulation.

Following the pioneering work of Courtland and Weeks, which demonstrated the applicability of particle-resolved studies to access structural changes during aging ~\cite{courtland2003}, we note four papers, all of which concern particle-resolved studies on colloidal systems. The first is also by the Weeks group ~\cite{cianci2006}, where 3d particle resolved studies were employed to consider both structural and dynamic behaviour of aging systems. The dynamics slowed. As for the structure, by decomposing it into tetrahedra, little change was observed, although a weak correlation between tetrahedra and mobility was found. The same group later considered a highly size-asymmetric mixture (1:2.1) in which the small particles were found to be much more mobile and appeared to facilitate the motion of the larger particles, although both species exhibited comparable aging dynamics ~\cite{lynch2008}.

A third study, this time in 2d by the Konstanz group ~\cite{assoud2009} looked at local structures relevant to their 2d system (Fig. ~\ref{figAssoud}), and noted systematic structural changes, both in their specific local structures and even in quantities such as the peak of the pair correlation function $g(r)$. Note that the latter can be accessed in reciprocal space measurements on atomic and molecular systems. Because in this system, these local quantities have been directly identified with low potential energy (see section ~\ref{sectionTowardsAStructuralMechanism}  ~\cite{mazoyer2011}), Fig. ~\ref{figAssoud} thus provides direct evidence in favour of the system settling deeper in its energy landscape during aging.

\begin{figure}[h!]
\begin{center}
\includegraphics[width=80 mm]{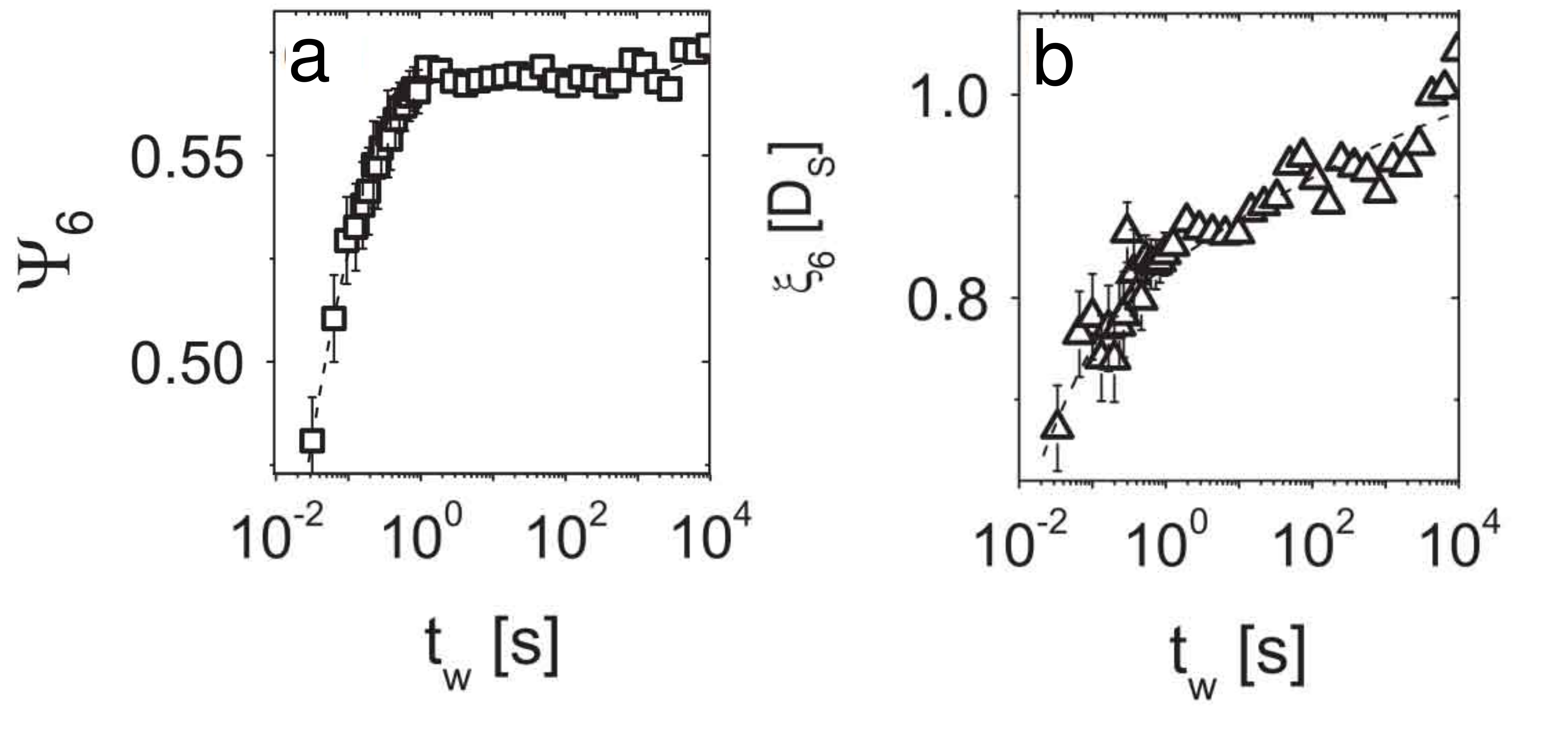}
\caption{Increase in local bond order parameter $\psi_6$ and structural correlation length of $\psi_6$-ordered regions $\xi_6$ in an aging 2d binary hard disc colloidal glass. From \cite{yunker2009}.}
\label{figPeterYunker}
\end{center}
\end{figure}

Also in 2d, with long-time experiments on a bidisperse hard-sphere-like microgel system, Yodh and coworkers ~\cite{yunker2009} identified irreversible structural relaxations, Fig. ~\ref{figPeterYunker}. These could be identified with the growing structural lengthscale $\xi_6^{2D}$, the 2d analogue of the structrial length defined in Eq. ~\ref{eqGl}, section ~\ref{sectionEarlyMeasurements}. Thus evidence was found of increasing structural co-operativity during aging. However, the discussion in section ~\ref{sectionStaticAndDynamicLengths} suggests that, if we believe that structural and dynamic lengths should be coupled, and that the latter are of order a few diameters even at the experimental glass transition, then the increase seen in Fig. ~\ref{figPeterYunker}(b) would be expected to plateau eventually. One noteworthy feature of Fig. ~\ref{figPeterYunker} is the temporal decoupling of the local structural order parameter $\psi_6$, which reaches an approximately constant value quickly, and the structural lengthscale.

\subsection{Aging in gels}
\label{sectionAgingInGels}

The effects of aging upon the structure of equilibrium gels (see section ~\ref{sectionGelation}) have been little investigated, while certain classes of non-equilibrium or ``spinodal'' gels have been studied in some depth. Among the systems to have received attention (and plenty of it) is laponite, which is an inorganic colloidal clay.

\textit{A model system for all eventualities --- }  Laponite consists of colloidal platelets, around 20 nm in diameter and 1 nm thick. These are highly polydisperse, in both size and shape. Moreover, the electrostatic interactions between the particles are poorly understood, although it is known that under certain conditions the rim and surfaces of the platelets can carry opposite charges. Furthermore, different batches of laponite can behave in different ways, leading to a bewildering range of phenomena. It has even been claimed (backed up with experimental evidence) that this system can take two distinct states --- glass or gel --- from indistinguishable starting conditions, an unusual behaviour for bulk materials ~\cite{jabbarifarouji2007}. What is clear is that laponite exhibits aging behaviour consistent with a glass (where the intermediate scattering function drops at short times, reflecting ``in-cage'' $\beta$-relaxation as in Figs. ~\ref{figGISF} and ~\ref{figLuca}) and with a gel where the attractive interactions lead to bonds which are much less than a particle diameter, thus any $\beta$-relaxation occurs on tiny lengthscales which are not apparent in the ISF. However in laponite, such gels are stablised by long-ranged electrostatic repulsions. Despite sometimes carrying charges of both signs, the particles nevertheless carry a net charge and thus experience an electrostatic repulsion ~\cite{jabbarifarouji2008}.

Less controversial are the results from computer simulations, where laponite has been treated as platelets with electrostatic interactions ~\cite{mossa2007}, in which a ``Wigner'' glass is formed. Wigner glasses are those formed through compression of systems which interact strongly through repulsive electrostatic interactions. Later work found quantitative agreement with experiment, at the two-body level ~\cite{ruzicka2010prl}.

Before leaving this intriguing system, we note that laponite was the first experimental system found to have an ``empty liquid'' state ~\cite{ruzicka2010}. Empty liquids are thermodynamically indistinguishable from liquids, but upon reducing the number of bonds per particle asymptotically towards two (in practise by mixing in a small number of particles with three bonds), the density of the liquid approaches zero ~\cite{bianchi2006,bianchi2011}. Laponite gels aged for years were found to have phase separated into a colloidal gas and ``empty liquid'', whose density, although very low, was nonetheless larger than that of the gas  ~\cite{ruzicka2010}.

In addition to laponite, the effects of aging in other stable gels have received some attention. In a system of competing interactions, Klix \emph{et al.} ~\cite{klix2010} identified a novel aging mechanism by which clusters underwent fission over time due to electrostatic charging.

\begin{figure}
\includegraphics[width=60mm]{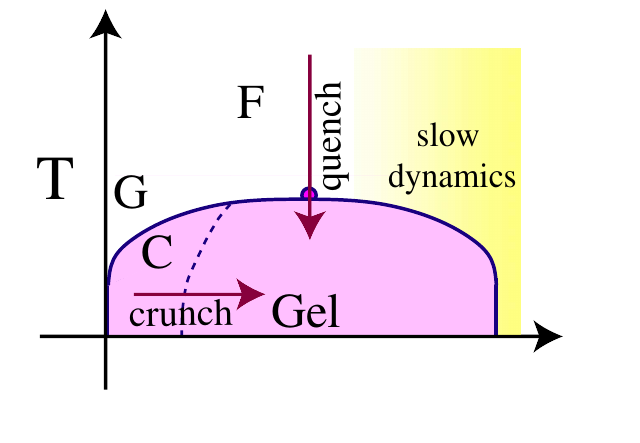}
\caption{Gelation in short-ranged attractive systems. Gelation results upon quenching, provided the quench rate is sufficient to avoid phase separation (pink region). Slow dynamics at high density may be tuned (shaded region) by quench depth and interaction range ~\cite{testard2011,zhang2013}. ``Crunching'' of isolated clusters provides an additional route to form a gel where accessible quench rates are too slow to prevent demixing, as is the case in molecular systems ~\cite{royall2011c60}. Here ``C'' denotes isolated clusters, ``G'' gas, ``F'' (supercritical) fluid.}
\label{figSchematicGel} 
\end{figure}

\textit{Aging in spinodal gels --- } Gels formed by quenching through the liquid-gas spinodal have an intrinsically different nature (Fig. \ref{figSchematicGel}) ~\cite{zaccarelli2007,verhaegh1997,lu2008,testard2011,zaccarelli2008}. Here, given sufficient time (in the absence of gravity, which can lead to sudden collapse of the gel ~\cite{starrs2002,manley2005,teece2011}), the gel network will gradually coarsen. In fact the process of gelation, as arrested spinodal decomposition, follows conventional spinodal decomposition for short times prior to arrest ~\cite{cahn1959}. If neither phase exhibits slow dynamics, complete fluid-fluid phase separation ensues. The mechanism of arrest thus relies on at least one of the emergent phases exhibiting slow dynamics. Indeed, were both to exhibit similarly slow dynamics, normal phase separation would proceed, albeit at a snail's pace ~\cite{tanaka2000}. In practise one phase, the colloid-rich phase in the systems we shall consider, is the phase which slows the demixing and leads to gelation. Thus the demixing is dynamically asymmetric, and is termed ``viscoeleastic phase separation''. This exhibits a zoo of different morphologies, notably the bicontinuous network characteristic of gels  ~\cite{tanaka2000,tanaka2005}. Note that spinodal gelation need not be restricted to multicomponent systems : phase separation of molecules, where the dense phase is dynamically arrested can similarly produce gels ~\cite{testard2011,royall2011c60}.

Once the initial network has formed, it ages following spinodal decomposition, with dynamics limited by the slower phase ~\cite{tanaka2000,carpinetti1992}. Colloid-polymer mixtures are a model system in which to explore this behaviour ~\cite{poon2002}. In these systems, the range of the effective attraction between the colloids can be tuned by varying the size of the polymers which induce the interaction ~\cite{asakura1954,asakura1958}. Now it is known that, upon decreasing the range of the attraction, that the density of the liquid increases (Fig. ~\ref{figSchematicGel}) ~\cite{elliot1999}. For long-ranged interactions (greater than around 0.3 colloid diameters), complete phase separation to a colloidal liquid and gas is found, but shorter ranged attractions lead to gelation ~\cite{ilett1995}. Since colloidal hard spheres undergo a glass transition at high volume fraction, this vitrification due to increased density of the colloid-rich phase has since been identified as a mechanism of gelation ~\cite{zhang2013}. Despite the slow dynamics, gels nonetheless age. The predominant structural change is an increasing thickness in the width of the ``arms'' of the gel. This has been observed to follow the same scaling as normal spinodal phase separation ~\cite{carpinetti1992}, and also a weaker, logarithmic increase ~\cite{lu2008,testard2011}. The dynamics of hard spheres is a strong function of their density (Fig. ~\ref{figAngell}). Thus quench depth can be used to control the density of the ``liquid'' and thus the rate of phase separation in colloidal ~\cite{zhang2013} and in principle molecular ~\cite{testard2011,royall2011c60} systems.

\section{Shear}
\label{sectionShear}

\begin{figure}[!htb]
\centering \includegraphics[width=80mm]{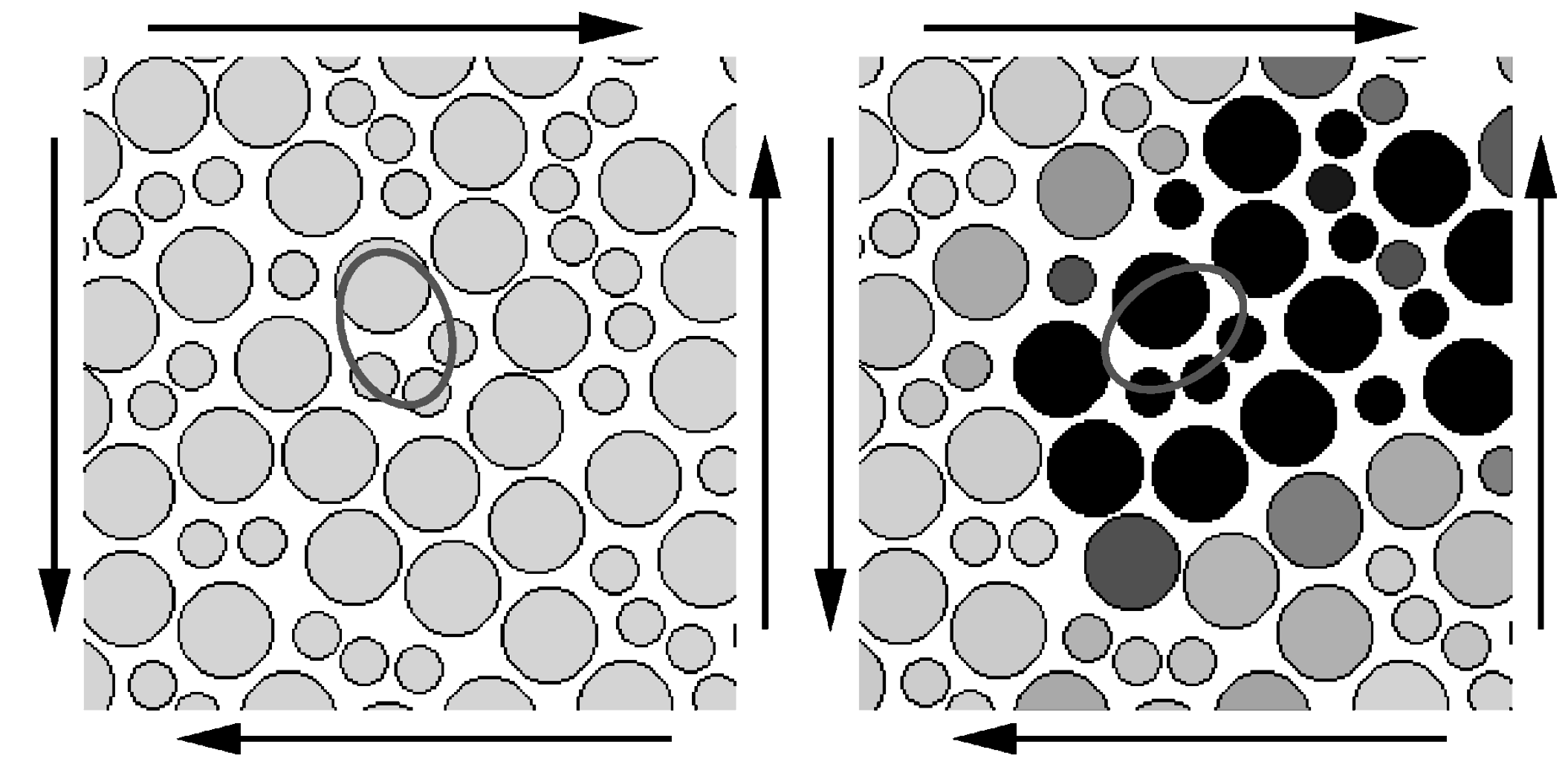} 
\caption{A schematic of a shear transformation zone. Particles are shaded depending on their deviation from affine deformation.
The direction of the externally applied shear stress is shown by the arrows. The ovals are guides for the eye. ~\cite{falk1998}.}
\label{figLanger} 
\end{figure}

Here we consider the role of local structure in the responses of glass forming systems to shear. The effect of shear is a vast field, and here we briefly focus on the interplay between shear and local structure. Much of the relevant literature concerns metallic glasses ~\cite{cheng2011}.  We shall first discuss the response of \emph{glasses}, where the material is already assumed to have undergone dynamical arrest and to be in the solid state. Arguably in this state the importance of structure will be most prevalent. We will also consider some work involving shear on deeply supercooled liquids. An area that we will focus on is the onset of plastic (irreversible) deformation in response to an applied strain. Here structure plays an important role. At some point after the initial plastic deformation the material yields or melts. The material property associated with this is the yield stress. This is an engineering type material property \emph{i.e.} it is defined in terms of a nonequilibrium process that depends upon the details of the protocol. It is hoped that given the protocol (e.g. the applied stress as a function of time) is slow enough then the response (the resulting strain, the amount of plastic deformation and the yield stress) is merely a function of the applied stress and not dependent on the rate of the protocol, i.e. how rapidly the stress has been increased, as would be the case in the quasistatic (equilibrium) limit. Obviously the very slow shear leading to a yield event is not thermodynamically quasistatic, but it is commonly assumed (with empirical justification) to be independent of the rate of the slow protocol.

At low temperatures the plastic deformation occurs non-uniformly in local regions known as shear transformation zones (STZ) (Fig ~\ref{figLanger}) ~\cite{cheng2011}. These are local regions which, upon the application of sufficient stress, have undergone a plastic deformation. Their definition remains vague and we note the following description \emph{``... plastic events are carried by some sort of microstructural defects referred to as `shear transformation zones' STZ. While the precise nature of these STZ or how to measure them experimentally or even in simulation has never been fully clarified, their existence as the source of `quanta' of plastic relaxation carried by a small number of atoms was taken as a basis for developing mean-field models of elastoplasticity'' } ~\cite{lerner2009}.

\textit{Direct imaging in colloids --- }
In colloidal dispersions, particle-resolved studies make it possible to access local phenomena.  3d imaging of sheared systems, which is perhaps the most technically challenging of all particle-resolved studies, has required the development of special techniques~\cite{besseling2007,besseling2009}. It is then possible to directly visualise strain at the single particle level ~\cite{schall2007}.

\begin{figure}[!htb]
\centering \includegraphics[width=80mm]{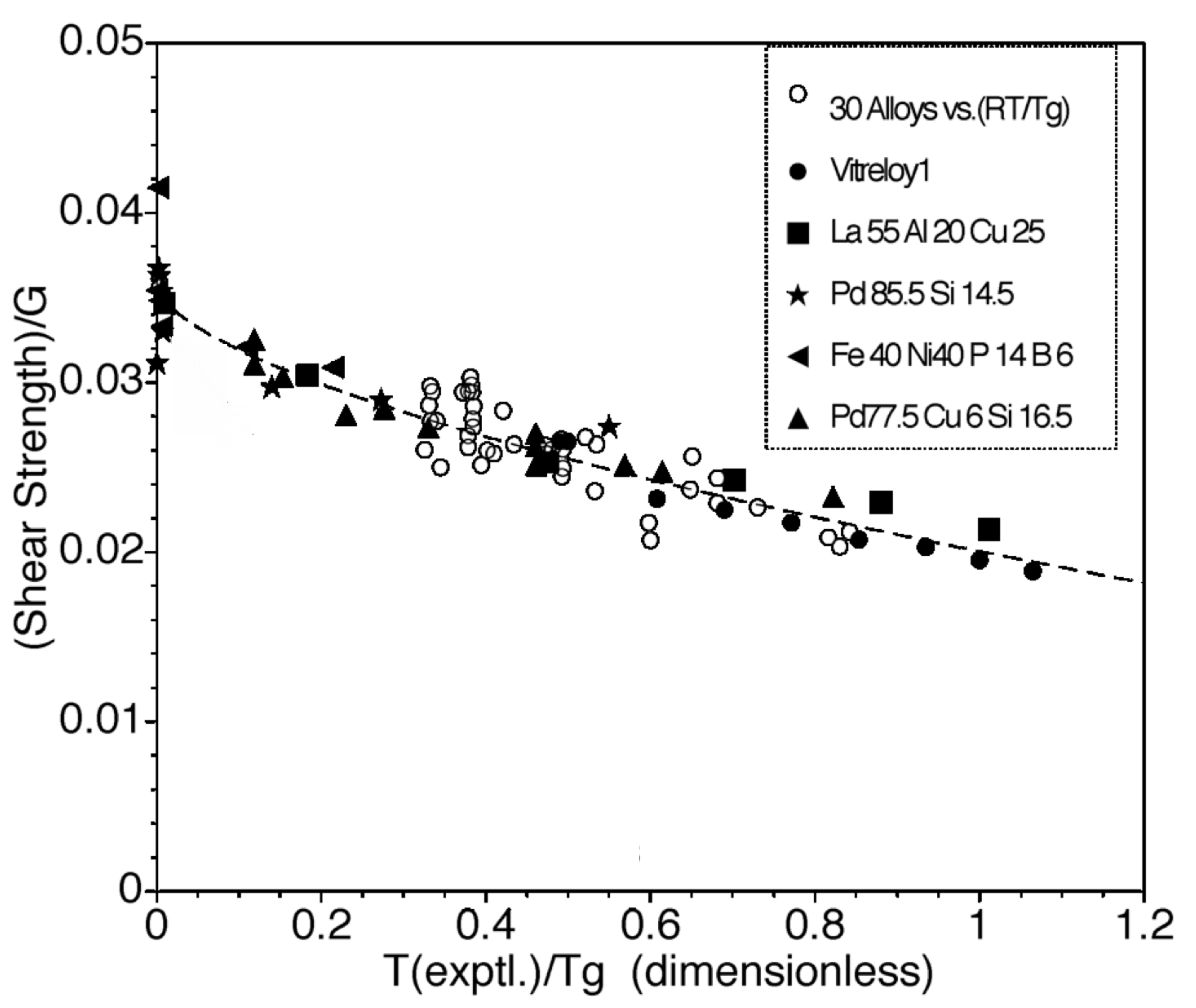} 
\caption{Experimental shear strain at yielding ($\tau_Y/G$) as a function of $T/T_g$. Small open circles show results at room temperature on 30 metallic glassformers. Solid symbols show the $T/T_g$ dependence of $\tau_Y/G$ for various individual alloys as indicated. Dashed line is Eq. \ref{eqSamwar}. Reproduced from ~\cite{johnson2005}.}
\label{figSamwar} 
\end{figure}

\subsection{Shear Transformation Zones}

Among the early treatments of the response of solid materials to strained inclusions (i.e. STZ) was that of Eshelby ~\cite{eshelby1957}. Eshelby showed that the stress field emanated from a (ellipsoidal) strained inclusion that had a quadrupolar form. Such strain fields have been identified in simulation ~\cite{chattoraj2013} and directly imaged in colloidal suspensions ~\cite{chikkadi2011,varnik2014}. Spaepen  ~\cite{spaepen1977} took a localised approach to homogenous (liquid) and inhomogeneous (glassy) flow using results from free volume theory ~\footnote{Free volume theory remains popular in the materials community. However, perhaps due to the fact that the free volume can only be rigorously defined in the case of hard interactions, it can be difficult to work with.} which was introduced long ago to describe vitrification in quiescent liquids ~\cite{cohen1959}. He identified competition between shear stress which leads to disorder, ripping regions apart and diffusion where they equilibrate and come back together again. Spaepen's approach was developed by Johnson, Samwar and coworkers ~\cite{johnson2005,demetriou2007} who considered plastic yielding of metallic glasses. Here the yield stress follows a simple linear form with respect to the shear modulus which led to the development of a simple model for yielding using the elastic energy per atom and assumed that plastic events correspond to cooperatively rearranging regions (see sections \ref{sectionAdamGibbs} and \ref{sectionRFOT}) to obtain an energy barrier density corresponding to the STZ. A rate of STZ yielding then followed given an attempt frequency in the barrier height which gave for the yield stress

\begin{equation}
\tau_{CT} = \tau_{C0} - \tau_{C0} \left[ \frac{k_B}{A} \ln \left( \frac{W_0}{c \dot{\gamma}}  \right)  \frac{G_{0}(T)} {G_0(T_g)} \right] ^{2/3} \left( \frac{T}{T_g} \right) ^{2/3}
\label{eqSamwar}
\end{equation}

\noindent where $A$ and $c$ are constants, $\tau_{C0}$ is the yield stress at $T=0$, $W_0$ is the elastic energy at $T=0$ and $G_0$ is the shear modulus of the unstressed material which has a weak temperature dependence. Treating the contents of the square brackets as a fitting parameter, this gave reasonable agreement with experimental data over a range of temperatures as indicated in Fig. \ref{figSamwar}. The model was later extended to obtain nonlinear viscosity (during shearing) for metallic glasses \cite{demetriou2007}.

Falk and Langer ~\cite{falk1998} carried out simulations at very low (almost zero) temperature of a 2d binary Lennard-Jones glassformer. Three regimes of response to an applied stress were identified: reversible plastic deformation, irreversible plastic deformation and unbounded flow. The local origin of deformation was via STZ, as shown in Fig. ~\ref{figLanger}. These hark back to ideas introduced in free volume theory ~\cite{cohen1959}. Falk and Langer introduced an equation of motion for the STZ ~\cite{falk1998}. This equation qualitatively reproduces the behaviour of the time-dependent onset of plastic deformation over a range of applied stresses below the yield stress and stress-strain hysteresis. Manning \emph{et al.} ~\cite{manning2007} have parameterised the STZ model to give good agreement with simulation data with physically reasonable parameters including a ``disorder temperature?? which as been tested in simulation by Shi \emph{et al.} ~\cite{shi2007} with good collapse of data assuming linear relationship between energy of atoms and disorder temperature in shear band. By considering a range of STZ energies, Procaccia and Langer argued that is possible to have plastic events even before the nominal yield stress \cite{bouchbinder2007}.  Procaccia and coworkers then extended this viewpoint to show that only infinitesimal strains have an elastic response. Otherwise, higher-order contributions to the elastic modulus dominate. Now the first of these does not converge to a well-defined value in the thermodynamic limit and others diverge. The upshot is that even for very small strains plastic events occur and thus solid-like (reversible under strain) behaviour is not found ~\cite{hentschel2011}.

\textit{The athermal quasistatic limit --- } The athermal quasistatic (AQS) limit holds the system on the local energy minimum upon applying a shear rate. As the system is sheared the periodic boundary conditions change, causing the local energy minimum to move. In a crystal that does not yield this follows affine transformation, but in a disordered system it does not. The AQS trajectory is advanced solely as a function of the changing strain, remaining slaved to the local energy minimum due to the extremely low temperature. This is different to the common usage of the term quasistatic (in thermodynamics) where the protocol is assumed to be so slow that the system is ergodic. Over a limited range the trajectory is reversible, but it eventually reaches a saddle point in the energy landscape. This is an unstable point and the trajectory drops to a new energy minimum causing a sudden change in the energy and a singularity in the stress ~\cite{lerner2009,karmakar2010pre}. The crossing of a saddle point is irreversible, with the trajectory falling onto a new energy minimum having different saddle points. We expect that this corresponds with (or at least is related to) the irreversibility of the STZ's studied by Falk and Langer discussed above. Obviously the crossing of the saddle points corresponds to some sort of yield stress and is responsible for plastic deformation in the AQS limit. The AQS simulations have shown that plastic deformation in solids is subextensive and not localised ~\cite{lerner2009}. This work supported the earlier work on the nonlocal nature of plastic events in amorphous solids by Maloney and Lema\^{i}tre ~\cite{maloney2006}. It has also been concluded, in the AQS limit, that the first nonlinear coefficient has anomalous fluctuations upon taking the thermodynamic limit and all higher coefficients diverge ~\cite{hentschel2011}. This calls into question the existence of solidity in amorphous solids at zero temperature. The AQS method does not necessarily sample the same regions of microscopic phase space that an equilibrated or well aged glass does. Whether this seriously affects the above conclusion about solidity is not known.

Now Eshelby's work suggested quadrupolar energy relaxations ~\cite{eshelby1957}, but these were not initially identified in simulation. In particular in Malandro and Lacks' \cite{malandro1998,malandro1999} energy landscape picture shear leads to the annihilation of local minima which are elementary events. With these considerations, Maloney and Lema\^{i}tre ~\cite{maloney2004} pondered whether the STZ should be local, working in the AQS limit \cite{maloney2004}. The picture Maloney and Lema\^{i}tre paint is one of localised events of plastic (irreversible) re-arrangements which are linked by reversible (elastic) branches. The plastic events can be interpreted as superpositions of quadrupolar energy fluctuations, or a cascade of elementary events leading to a superposition of quadrupolar fields. The distribution of numbers of particles participating in these events scales with system size. The energy dissipated apparently follows $E_0L$ where $E_0$ is an elementary excitation energy and $L$ is the box size. Now the earlier work of Malandro and Lacks ~\cite{malandro1998,malandro1999} had participation number independent of system size. Subsequently, Caroli and Lema\^{i}tre ~\cite{lemaitre2007} identified the plastic events as Eshelby-type quadrupoles which align with the shear field and soften. Here ``soften'' means that the non-affine field becomes concentrated in the lowest eigenvalue of the Hessian, i.e. in the softest normal mode (see section ~\ref{sectionIsoconfigurational}). Thus we find a connection between soft modes and dynamical heterogeneity. This is encapsulated by Goldstein's observation that ``\emph{rearrangements are of course occurring all the time in the absence of an external stress; the external stress, by biasing them, reveals their existence}'' ~\cite{goldstein1969,lemaitre2007}. We note that Lema\^{i}tre and coworkers worked in 2d ~\cite{maloney2004,lemaitre2007}, in which there is ``less dimensionality'' for the deformation to ``leak away'' while Malandro and Lacks worked in 3d. Maloney and Lema\^{i}tre's non-extensive scaling suggests cracklike patterns and cascading events. However,  Rodney and Schuh found little correlation between those events which percolated through the simulation box of event and the corresponding potential energy ~\cite{rodney2009}.

Rodney and Schuh ~\cite{rodney2009} considered STZs from the point of view of thermal activation. In a similar manner to the above they considered trajectories slaved to a local energy minimum at low temperature. However they also considered the rate at which trajectories could leave the minimum over the many possible saddle points using transition state theory. The traversal of the various saddle points is due to thermal fluctuations that become more important as the temperature is increased. This is still very much a low temperature approach and requires the computation of the barrier heights in the local energy landscape (section ~\ref{sectionEnergyLandscape}) between the energy minimum and the saddle point. The rates at which the various saddle points are traversed to escape the current inherent state are thus obtained and these rates change with the strain due to the strain dependence of the energy landscape. Plastic deformation in particular led to higher energy and lower stability. In order to evaluate the energy landscape for a sheared system, Rodney and Schuh used the activation-relaxation technique (ART Nouveau) of Malek and Mousseau ~\cite{malek2000}. In ART Nouveau one starts from an inherent structure, where the system is in a local energy minimum (see section ~\ref{sectionEnergyLandscape}) ~\cite{stillinger1983,stillinger1984}. Particles in a local region are displaced in such a way that a saddle in the energy landscape is reached, and the system is then allowed to relax. Relaxations back to the original minimum are discarded. With Valiquette, Mousseau applied ART Nouveau to amorphous silicon ~\cite{valiquette2003}. The activation events, at least in amorphous silicon, were found to be essentially local with a barrier height limited by the cost of breaking one single bond. Valiquette and Mousseau also found that there was very little variation in the entropic barrier despite a wide spread in energy. These observations might be reconciled with those of Maloney and Lema\^{i}tre because silicon is a strong liquid with relatively little structural change.

The athermal limit corresponds to granular matter (section ~\ref{sectionJammin}). While a serious discussion lies beyond our scope, we note that in the context of granular matter, Manning and Liu ~\cite{manning2011} argued that soft spots are the amorphous equivalent of dislocations in crystals. Their study of the normal modes showed that it is not clear how the modes in the quiescent system without shear correspond to the critical re-arrangement, in particular it is not always the lowest frequency mode. However a strategy for identifying which modes are important and how many particles participate in each mode was developed. Crucially, rearrangements were found to occur at soft spots whose lifetime is long compared to the time between re-arrangements. The soft spots are structurally distinct because the analysis by which they are identified is purely structural in nature with no dynamical component.  However (like the high-propensity particles in the isoconfigurational ensemble, section ~\ref{sectionIsoconfigurational}) the nature of the structural distinction is subtle.

These connectivity arguments chime with findings of Yamamoto and Onuki ~\cite{yamamoto1998}. These authors identified broken bonds to define a static structure factor determined from particles with such broken bonds $S_b(k)$. This enabled a dynamical correlation to be obtained in a similar spirit to the four-point structure factor $S_4(k)$ [Eq (\ref{eqS4})].  Like $S_4(k)$ their $S_b(k)$ is described well by a Lorentzian (appropriate for critical behaviour). Thus the regions of broken bonds increase in size with shear rate, so STZs are delocalised in this means of measurement.

Barrat, Bocquet and coworkers ~\cite{nicolas2014} developed a generalised coarse-grained model for flow in glasses. Their elastoplastic model emphasised spatial correlations and time fluctuations of local stresses. This gave local alternation between an elastic regime and plastic events during which the local stress is redistributed. Notably, at low shear the dynamic correlation length scaled with system size, in a similar way to the results of Maloney and Lema\^{i}tre  ~\cite{maloney2004}. Thus a connection is made with the energy landscape. At higher shear, the dynamic correlation length dropped, scaling as $\dot{\gamma}^{0.6}$.

Another key parameter in determining the strength of glasses is the \emph{quench rate}. Shi and Falk ~\cite{shi2005,shi2006} considered the role of quench rate on some model glassformers: the Wahnstr\"{o}m, Kob-Andersen and Dzugutov models. In all cases, for low quench rates the yield stress was higher, however plastic deformation appeared to occur at a rather constant stress. Quickly quenched systems exhibited homogenous deformation, with the stress-strain curve reaching a plateau without any significant peak. Slowly quenched systems showed a peak in the stress-strain curve at the yield stress, at higher strains the stress approached that of the quickly quenched systems. Shi and Falk also considered the local structure in the Wahnstr\"{o}m model, which was shown to exhibit a percolating network of Frank-Kasper polyhedra (section \ref{sectionFrank}) that was destroyed upon yielding ~\cite{shi2006}. In the KA model, no percolating network was found. However more recent studies \cite{coslovich2007,coslovich2011,malins2013fara} have found a network of locally favoured structures (distinct from those investigated by Shi and Falk) suggesting that the approach of Shi and Falk might be revisited in the case of the KA model, to investigate any generality of a network of LFS which is resistant to shear.

\subsection{Shear in supercooled liquids}

While most work focussed on the glass regime, \emph{i.e.} at temperatures far below those at which the system can be equilibrated, Chattoraj and Lema\^{i}tre ~\cite{chattoraj2013} considered the behaviour around the temperature at which equilibration can be achieved, $T_g^\mathrm{sim}$. Recall that this concerns simulation, so the relevant temperature is rather higher than would be the case for molecular experiments. In short, Chattoraj and Lema\^{i}tre showed that supercooled liquids behave as solids which flow. They obtained two major results : (i) Eshelby strains ~\cite{eshelby1957} are observed around and above $T_g^\mathrm{sim}$, in the Newtonian regime. This was a surprise, as such quadrupolar strains are usually thought to be peculiar to low-temperature, plastic, deformation; (ii) correlations between shear relaxation centres emerge as soon as the temperature is decreased enough that the system enters the shear-thinning regime. The Newtonian-to-shear-thinning crossover thus appears to be controlled by the competition between the flow events needed to relax shear and randomly oriented shear events triggered by thermal activation. Very recently Schoenholz \emph{et al.}  ~\cite{schoenholz2014} have extended the soft spot approach of Manning and Liu ~\cite{manning2011} to thermal glasses. They found that soft spots, identified via a normal mode analysis provided a reasonable correlation with the shear transformation zones identified following Falk and Langer ~\cite{falk1998}. The correlation extended well into the supercooled regime where the system could be equilibrated and the lifetime of the soft spots was comparable to $\tau_\alpha$. In colloid experiments, Schall and coworkers have investigated spatial correlations of non-affine displacements. These correlations are found to be anisotropic in the glassy regime (weak thermal fluctuation) and isotropic in the supercooled liquid regime (thermal fluctuations dominate shear) ~\cite{chikkadi2012}. Recently, Mosayebi and coworkers combined normal mode analysis with their non-affine displacements approach (see section~\ref{sectionOrderAgnostic}) ~\cite{mosayebi2014}. The non-affine displacements detect plastic events in a supercooled liquid. What is far from intuitive about their findings is that the normal modes related to non-affine displacements are distinct from those related to propensity, i.e. regions of the system inclined to undergo relaxation in the isoconfigurational ensemble (section~\ref{sectionIsoconfigurational}). In other words there are two populations of normal modes in the inherent structure related to solidity (the non-affine displacements) and mobility.

\begin{figure}[!htb]
\centering \includegraphics[width=50mm]{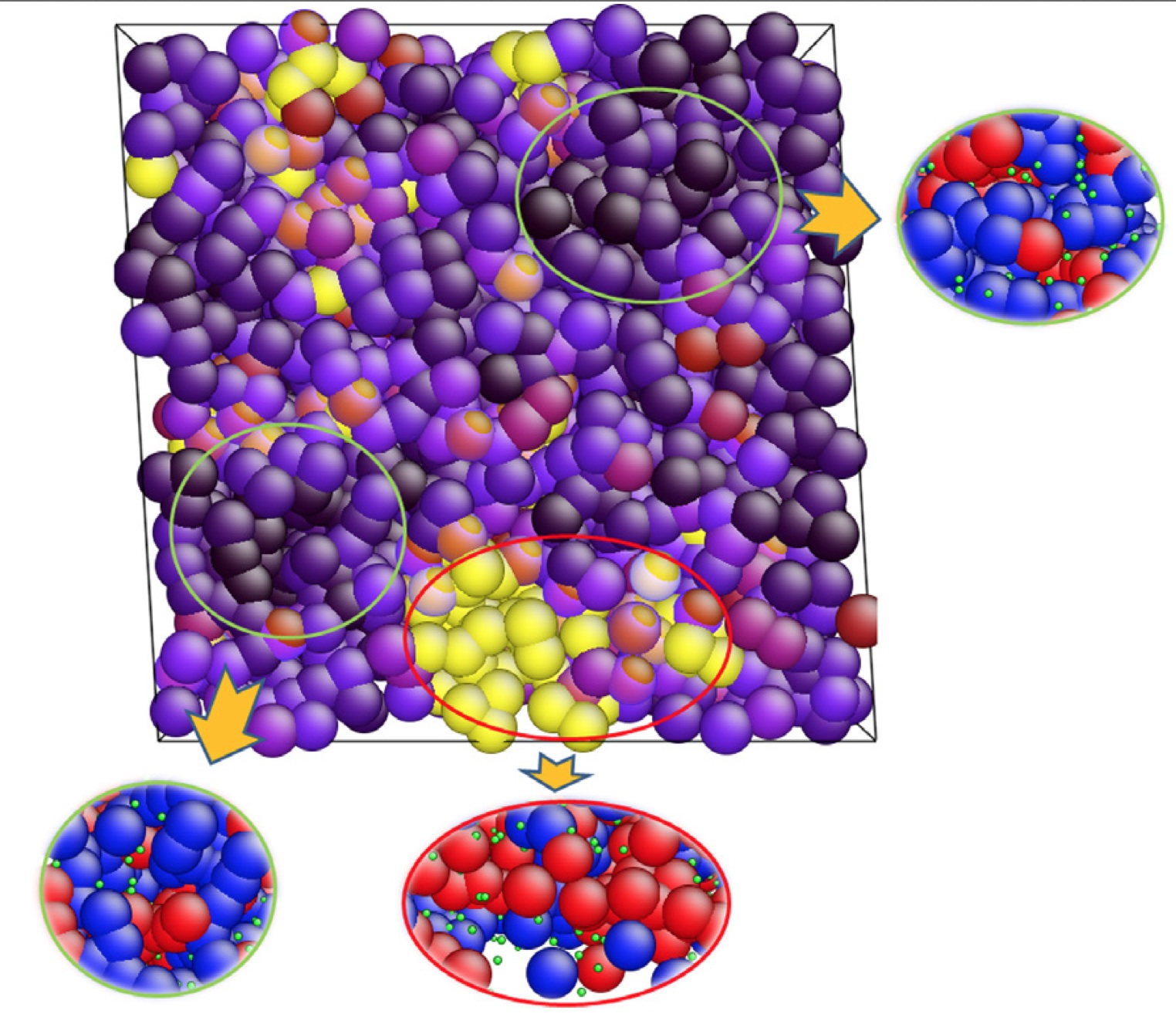} 
\caption{Shear transformation zones in Cu$_{64}$Zr$_{36}$ at a shear strain of 7\%. Cu atoms only are rendered. The colors are mapped to the various levels of non-affine (plastic) atomic strains. In particular, purple regions correspond to elastic deformation while yellow are STZs. Two elastic regions and one STZ are indicated and their local structure shown in the insets. For the elastic regions, full icosahedra (blue) are dominant, while in the STZ, more Cu atoms are found in non-icosahedral polyhedra (red). Smaller green dots are in small polyhedra  ~\cite{cheng2008acta}.}
\label{figShearIcos} 
\end{figure}

\subsection{Shear and local structure in metallic glass}

Cheng \emph{et al.} ~\cite{cheng2008acta} have shown that the local order in Cu-Zr and in Cu-Zr-Al can be controlled through the composition. The population of icosahedra increases with Cu content and on substitution of Zr with Al. Upon increasing the population of icosahedra, the resistance to the initial yielding is increased, as is the softening after the initial yielding and the inclination towards strain localisation. The critical stress (that before the onset of bulk plastic flow) was found to be proportional to the elastic modulus at different compositions. Cheng \emph{et al.}  went on to suggest that this might mean a universal critical strain, at least for the Cu-Zr-Al systems they considered ~\cite{cheng2008acta}. A snapshot (Fig. ~\ref{figShearIcos}) from Cheng \emph{et al.}'s work shows the relationship between icosahedra, which tend to resist plastic flow, and STZ. Regions undergoing elastic deformation are rich in icosahedra compared to STZ. These results may be contrasted with those of Shi and Falk ~\cite{shi2007} where nanoindentation of the  Wahnstr\"{o}m model in 2d was considered at low temperature, deep in the glass. Short range order, in the form of polygons, showed some correlation with regions which underwent high shear under indentation, but the correlation is weak. Considering local free volume ~\cite{cohen1959}, little correlation was found.

In experiment,  Das \emph{et al.} ~\cite{das2005} also investigated CuZr which, like a number of metallic glasses, has poor ductility. They found that CuZr has nano crystals, crystalline regions whose size was less than a nanometre but which accounted for up to 10\% of the volume. Introducing a small amount of Al is known to change the local structure (see section \ref{structureInMetallicGlassformers}). In this case, Cu$_{47.5}$Zr$_{47.5}$Al$_{5}$ was found to increase the yield strength and ductility. Upon failure, shear bands were observed. However these tended to intersect, providing a possible mechanism for suppression of catastrophic failure and increased strength. Also in CuZr, with embedded atom model simulations, the work of Peng \emph{et al.} \cite{peng2013} is among the few papers to directly focus on local structure and its role in deformation, in this case compression. They found a degree of correlation between pentagonal bipyramids and locally ``solid'' regions, i.e. those more resistant to deformation.

\section{Outlook}

We have considered a number of ways of measuring structure and its implications in various aspects of dynamical arrest. In closing, we consider what we believe to be the current state of the art regarding the role of local structure in arrest. In our introduction we noted that there were two main questions. The first being the role structure might play in the dynamical slowdown, and the second being its role in preventing crystallisation.

Perhaps the first task is to enquire as to which classes of materials there is enough evidence to meaningfully discuss the role of local structure. Of the model systems encountered in section ~\ref{sectionCommonModelSystems}, BKS silica and related models exhibit relatively little change in structure upon supercooling \cite{coslovich2009}. The others we have emphasised (Lennard-Jones models and hard spheres) do exhibit a considerable change in higher-order structure. This seems to be mirrored in variants of these models (such as soft spheres) in 2- and 3-dimensions. We have seen in section ~\ref{sectionStructureModel} that model systems often form closed-shell Voronoi polyhedra (such as icosahedra and bicapped square antiprisms and sometimes crystalline order) ~\cite{royall2014,malins2013jcp,coslovich2007,malins2013fara,leocmach2012}. Similar or identical structures are encountered in models of metallic glasses although we note that the latter field sometimes places emphasis on the Frank-Kasper polyhedra~\cite{cheng2011} (see section ~\ref{structureInMetallicGlassformers}). It appears that bicapped square antiprisms are associated with non-additive interactions between the species. This seems clear in the case of the Lennard-Jones models, where the non-additive Kob-Andersen mixture forms bicapped square antiprisms, while the additive Wahnstr\"{o}m mixture forms icosahedra  (section ~\ref{sectionCommonModelSystems}, Fig. ~\ref{figFavourite}). Bicapped square antiprisms are also encountered in metallic glassformers where interactions can often have a significant degree of nonadditivity  ~\cite{evteev2003}. This suggests that it may be possible to interpret metallic glassformers along the same lines as Lennard-Jones models and perhaps hard spheres.

We believe that, at the present time, too little is known of the structure in more complex systems, such as chalcogenides (see section \ref{structureInChalcogenideGlassformers}) to make similar statements. These materials are undoubtedly fascinating and as their interplay between crystallisation and vitrification is unravelled, we expect that parallels with the simpler systems based on spherical interactions may be made. But for now, it is unclear how exactly to think of structure during vitrification in chalcogenide glassformers, save to note that the process is more complex than the Voronoi polyhedra relevant to the model systems and metallic glassformers. Similar observations may be made concerning polyamorphic systems. At present, while their interactions may not exhibit the complexity of the chalcogenides, the current focus rests on demonstrating a liquid-liquid transition, rather than upon any relationship between local structure and slow dynamics. However, by definition, structure must play a role in polyamorphic systems, so these seem a rich field in which to investigate the role of structure in slow dynamics. Likewise structural changes in aging seem an attractive way of elucidating a system's route as it settles down its energy landscape, should one choose such an interpretation. We now summarise our opinion of the state of the art regarding the role of local structure in dynamical arrest.

\textit{Locally favoured structures --- } Supercooling leads to a high population of particles in locally favoured structures but this by itself does not demonstrate a complete structural mechanism for dynamical arrest. However evidence is provided by correlating those particles in locally favoured structures with dynamical heterogeneity in the sense that dynamically slow regions have a higher population of locally favoured structures than the dynamically fast regions.

Simultaneous increases of dynamical and structural lengthscales would provide strong evidence for a structural mechanism for dynamical arrest. However, as discussed in section ~\ref{sectionStaticAndDynamicLengths}, this simultaneous increase has certainly not universally been found, especially in $d=3$. Indeed, while some groups have identified a concurrent increase of structural and dynamic lengthscales ~\cite{tanaka2010,sausset2010,mosayebi2010,mosayebi2012,leocmach2012,kawasaki2007}, the majority seem to find that the dynamic correlation length \emph{as determined by the four-point correlation length $\xi_4$} increases faster than the static length, sometimes even in the same system ~\cite{charbonneau2013pre,royall2014,malins2013jcp,tamborini2014,dunleavy2012,hocky2012,kob2011non,karmakar2014,karmakar2009}.

Where does this leave the case for a structural mechanism for arrest? As discussed in section ~\ref{sectionStaticAndDynamicLengths}, there are at least three possibilities for the discrepancy between static and dynamic lengths, the first of which is that structure is not the dominant mechanism. Another possibility is that either the four-point dynamic correlation length $\xi_4$ or the LFS lengthscale are not the ``right'' length scale to consider. The latter point is rationalised by noting that the structure influences dynamics on lengthscales longer than the LFS themselves \cite{malins2013jcp,malins2013fara}, which could lead to arrest in non-LFS gaps between in the LFS network (Fig. ~\ref{figAlexNetwork}) ~\cite{coslovich}. Another possibility is to question the use of $\xi_4$ and to enquire as to whether simulation and colloidal experiments (the main techniques from which these quantities can be computed) actually access enough of the supercooled regime. As the Angell plot makes clear, (Fig. ~\ref{figAngell}), this runs to much more than the five or so decades probed by these particle resolved techniques.

Considering the use of $\xi_4$, we note that the mode-coupling transition marks, loosely, the end of the particle-resolved studies regime accessible to simulation and colloid experiment. We cite two pieces of evidence before speculating. The first is the observation that the divergence of $\xi_4$ correlated rather well with the mode-coupling temperature ~\cite{lacevic2003}, and indeed a free fit suggests divergence of $\xi_4$ at the mode-coupling temperature ~\cite{malins2013fara}. Furthermore, the behaviour of $\xi_4$ as a function of supercooling changes around the mode-coupling transition ~\cite{flenner2013,malins2013fara} as do certain other dynamic lengthscales ~\cite{kob2011non}. The second piece of evidence comes from (indirect) measurements of dynamic lengthscales around the molecular glass transition $T_g$, which are no bigger than $\xi_4$ in the accessible regime ~\cite{ediger2000,donth1982,cicerone1995,berthier2005,tatsumi2012,tracht1998,ashtekar2012,dalleferrier2007,yamamuro1998} and indeed appear rather \emph{smaller} than $\xi_4$ in this accessible regime \cite{dalleferrier2007}. This corresponds to nine decades change in relaxation time between the limit of particle resolved studies and the molecular glass transition ~\cite{harrowell2011}. This in our opinion casts doubt over the use of $\xi_4$ as \emph{the} dynamic correlation length unless the mode-coupling transition or some other phenomenon is invoked to explain a \emph{drop} in $\xi_4$ at deeper supercooling. Thus even if the structural correlation length does not exhibit the same scaling as $\xi_4$ in the regime accessible to particle-resolved studies $T\gtrsim T_\mathrm{MCT}$, this doesn't by itself rule out local structure as a mechanism for slow dynamics, especially in the $T<T_\mathrm{MCT}$ regime.

\begin{figure}[!htb]
\centering \includegraphics[width=76mm]{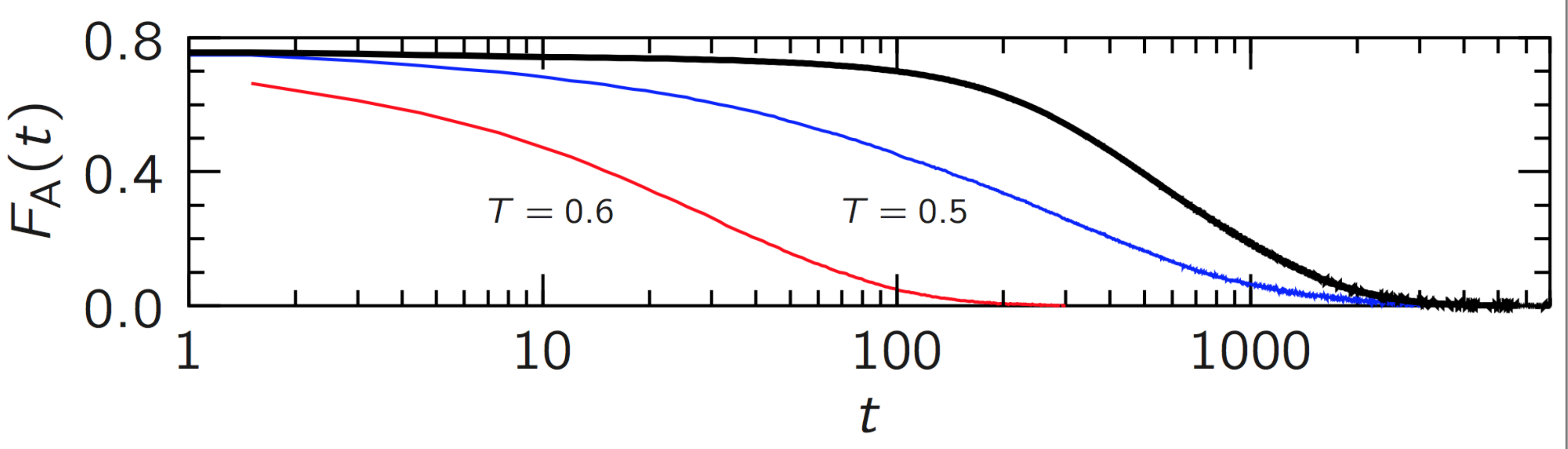} 
\caption{Structure encodes slow dynamics. Intermediate scattering functions for the Kob-Andersen mixture run at $T=0.5$ and $T=0.6$ is indicated. The thick black line corresponds to configurations prepared at $T=0.6$ in the $\mu$-ensemble (section ~\ref{sectionMu}) run with normal dynamics at $T=0.6$. Clearly, something in the configurations produced by the $\mu$-ensemble encodes slow dynamical behaviour.
 ~\cite{speck2012}. }
\label{figMelting} 
\end{figure}

Other pieces of evidence in favour of structure (somehow) controlling dynamics include the following.

\begin{enumerate}

\item The isoconfigurational ensemble (section ~\ref{sectionIsoconfigurational}). That the dynamics in the isoconfigurational ensemble are non-trivial ~\cite{widmercooper2006} demonstrates that \emph{something} in the structure can at least influence dynamic heterogeneity. If one believes that dynamic heterogeneity is related to vitrification, then it seems that something structural can, at the very least, influence vitrification. The question the isoconfigurational ensemble raises is exactly what aspect of the structure connects to the dynamics? The correlations of locally faroured structures so far identified with the dynamics seem somewhat weak ~\cite{jack2014,hocky2014,widmercooper2006jnonxsol}, normal modes are correlated with the dynamics more strongly ~\cite{widmercooper2008} but only at short times ~\cite{jack2014}.

\item The $s$- and $\mu$-ensembles  (sections ~\ref{sectionFacile} and  ~\ref{sectionMu}). Figure ~\ref{figMelting} illustrates the case that something in the structure of configurations produced by the ensembles which drive phase transitions in trajectory space influences the dynamics. Here configurations are run from the $\mu$-ensemble at $T=0.6$. The resulting intermediate scattering function (averaged over many independent runs) decays much slower even than conventional simulations even at the \emph{lower} temperature of 0.5 ~\cite{speck2012}. Given that the $\mu$-ensemble configurations are rich in bicapped square antiprisms (the locally favoured structure) it might be tempting to suggest these are responsible for the change in dynamical behaviour. However in reality the situation is not so straightforward. Firstly, similar behaviour has been observed in the $s$-ensemble which biases dynamics and also changes the structure ~\cite{speck2012,jack2011}. Now $s$-ensemble inactive configurations are also rich in bicapped square antiprisms ~\cite{speck2012}, so one could claim the same cause. However the fact that in the $\mu$-ensemble the trajectory length must be nontrivial to see any transition indicates that this effect is at least partly dynamical in origin.

Put another way, simply biasing a Monte-Carlo simulation to produce more bicapped square antiprisms without recourse to trajectory sampling leads to configurations rich in the LFS. Crucially, unlike the $\mu$-ensemble, these LFS-rich configurations exhibit similar dynamical behaviour to \emph{unbiased} simulations ~\cite{speck2012}. Thus, again, something in the structure controls the dynamics, but it is not as straightforward as being perfectly correlated with LFS. Perhaps there is something ``special'' in the LFS generated in the $\mu-$ensemble that is not generated upon structural biasing. Certainly, not all LFS are the same : they have a wide distribution of lifetimes ~\cite{royall2014,malins2013jcp,malins2013fara}. One possibility here is that the \emph{stochiometry} of the LFS may be important. As discussed in section ~\ref{sectionGlassformingAbilityStructure}, Malins \emph{et al.} ~\cite{malins2013jcp} emphasised the role of LFS stochiometry in the Wahnstr\"{o}m model.

\item Our final piece of evidence comes from tests of RFOT by pinning ~\cite{cammarota2012} and measuring overlaps ~\cite{berthier2013overlap} discussed in section ~\ref{sectionRFOT}. Both indicate a transition of a thermodynamic nature, and thus imply a role for structure. However, these approaches are order-agnostic (section ~\ref{sectionOrderAgnostic}) so do not rely on a particular type of structure. In this way, there is some similarity with observations from the isoconfigurational ensemble and indeed from the dynamical phase transitions discussed in the preceding paragraph : there is something in the structure but we do not yet know what.

\end{enumerate}

It is the authors' opinion that something in the structure plays a very significant role in determining the dynamics in systems undergoing arrest. Identifying the nature of the ``something'' remains a major challenge. Perhaps the closest we have to such a realisation at the moment, in the sense of tangible structures, are locally favoured structures, identified via Voronoi analysis, bond-orientational order parameters ~\cite{steinhardt1983}, the common neighbour analysis ~\cite{honeycutt1987} and the topological cluster classification ~\cite{malins2013tcc}. However, as our discussion makes clear, these do not yet form a complete description for dynamical arrest in glassformers. Finally the situation in spinodal gelation seems clearer where local structure can claim victory as a mechanism for dynamical arrest  ~\cite{royall2008}.

\textit{The role of structure in suppressing crystallisation --- }
The current state of the field appears to be a rather disparate collection of studies on individual systems. Even in the much-loved hard spheres, the absence of a consensus on a mechanism for crystallisation (not to mention the wild discrepancy on nucleation rates between experiment and simulation, section ~\ref{sectionCrystallisationModel}) suggests that identifying some kind of universal mechanism will be challenging. It is the authors' view that the crystallisation mechanism in each class of glass forming material needs to be treated separately, until such time as different classes of material may be identified. One possible way forward would be to investigate any link between common locally favoured structures in the liquid and crystal structures. However the work of Tang and Harrowell ~\cite{tang2013} (section ~\ref{sectionGlassformingAbilityMetallic}) already shows that care must be exercised with such an approach. One thing that would greatly further our understanding of the role of crystallisation would be the determination of full phase diagrams for model glassformers.

\vspace{2cm}
\section*{Acknowledgements} 

In connection with the preparation of this article, the authors would like to thank John Abelson, Ashraf Alam, Austen Angell, Ludovic Berthier, Giulio Biroli, Dwaipayan Chakrabarty, Patrick Charbonneau, Jeppe Dyre, Bob Evans, Jens Eggers, Peter Harrowell, Walter Kob, Matthieu Leocmach, Tannie Liverpool, Mina Roussenova, Phil Salmon, Grzegorz Szamel, Hajime Tanaka, Gilles Tarjus, Francesco Turci, Karoline Wiesner, Asaph Widmer-Cooper, Mark Wilson and Ryoichi Yamamoto for insightful discussions.
We also acknowledge, Andrew Dunleavy for the dynamical heterogeneity image and the $g(r)$ data, Rhiannon Pinney for her Angell plot data.
Alex Malins is thanked for various data and Daniele Coslovich, Peter Harrowell, Rob Jack, Rhiannon Pinney, John Russo, Thomas Speck, Jade Taffs and Ian Williams for a critical reading of the manuscript. Peter Crowther is acknowledged for his fine rendering of certain clusters.
CPR would like to acknowledge the Royal Society for financial support and European Research Council (ERC Consolidator Grant NANOPRS, project number 617266).
Finally we extend warm gratitude to Randy Kamien for giving us the opportunity to write this review.

%\bibliographystyle{plain}
%\bibliography{masterGlass}

\end{document}